\documentclass[a4paper, 12pt]{article}

\usepackage{varwidth}
\usepackage{lipsum}
\usepackage{threeparttable}
\usepackage{makecell}
\usepackage{amsfonts}
\usepackage{amsmath}
\usepackage{mathrsfs}
\usepackage{fancyhdr}
\usepackage{epsfig,morefloats}
\usepackage[authoryear]{natbib}
\usepackage{booktabs,multirow,multicol,longtable}
\usepackage{amsthm}
\usepackage{amssymb}
\usepackage{color}
\usepackage{lscape}
\usepackage{rotating}
\usepackage{caption}
\usepackage{subfigure}
\usepackage{graphicx}
\usepackage[compact]{titlesec}
\titlespacing*{\subsection}{0pt}{10pt}{0pt}
\titlespacing*{\subsubsection}{0pt}{10pt}{0pt}
\usepackage{lineno}
\usepackage{cancel}
\usepackage{multirow}
\usepackage{setspace}

\definecolor{mygray}{gray}{0.6}

\newcommand{\Date}[1]{\def\@Date{#1}}
\def\today{\number\day~\ifcase\month\or
	January\or February\or March\or April\or May\or June\or
	July\or August\or September\or October\or November\or December\fi~\number\year}

\def\be{\begin{equation}}
	\def\ee{\end{equation}}
\def\bea{\begin{eqnarray}}
	\def\eea{\end{eqnarray}}
\def\bd{\begin{displaymath}}
	\def\ed{\end{displaymath}}
\def\bda{\begin{eqnarray*}}
	\def\eda{\end{eqnarray*}}
\def\bsm{\begin{small}}
	\def\esm{\end{small}}

\def\ha1{\hat \beta_1}

\def\bnt{\begin{enumerate}}
	\def\ent{\end{enumerate}}

\def\bsc{\begin{scriptsize}}
	\def\esc{\end{scriptsize}}

\newcommand{\sgn}{{\mbox{\rm sgn}}}
\newtheorem{theorem}{Theorem}
\newtheorem{corollary}{Corollary}

\newtheorem{lemma}{Lemma}
\newtheorem{condition}{Condition}

\newtheorem{proposition}{Proposition}
\theoremstyle{definition}
\newtheorem{definition}{Definition}

\makeatletter
\newcommand{\figcaption}{\def\@captype{figure}\caption}
\newcommand{\tabcaption}{\def\@captype{table}\caption}
\makeatother

\def \E{{\mathcal{E}}}
\def \P{{\mathbb{P}}}
\def \R{{\mathbb{R}}}
\def \mE{{\mathbb{E}}}

\def \tr{{\rm tr}}
\def \vech{{\rm vech}}
\def \diag{{\rm diag}}
\def \c {{\rm c}}

\def\T{{\mathrm{\scriptscriptstyle \top} }}
\def \Rn{{ \mathcal{R}_n}}
\def \Rtheta{{\mathcal{R}(\btheta)}}

\def \0{0}
\def \TV{\mathrm{\scriptscriptstyle TV}}
\def \mF{\mathcal{F}}

\newcommand{\bGamma} {\boldsymbol{\Gamma}}
\newcommand{\bTheta} {\boldsymbol{\Theta}}
\newcommand{\btheta} {\boldsymbol{\theta}}
\newcommand{\bvartheta} {\boldsymbol{\vartheta}}
\newcommand{\bthetazero} {\boldsymbol{\theta}_0}
\newcommand{\hbthetan} {\hat{\boldsymbol{\theta}}_n}
\newcommand{\tbtheta} {\tilde{\boldsymbol{\theta}}}

\newcommand{\wLambadn} {\hat{\Lambda}_n}
\newcommand{\blambda} {\boldsymbol{\lambda}}

\newcommand{\bpsi} {\boldsymbol{\psi}}
\newcommand{\bdelta} {\boldsymbol{\delta}}
\newcommand{\bzeta} {\boldsymbol{\zeta}}
\newcommand{\bseta} {\boldsymbol{\eta}}

\newcommand{\bmu} {\boldsymbol{\mu}}
\newcommand{\bSigma} {\boldsymbol{\Sigma}}

\newcommand{\bfA}{{\mathbf A}}
\newcommand{\bfB}{{\mathbf B}}

\newcommand{\bfH}{{\mathbf H}}
\newcommand{\bfI}{{\mathbf I}}

\newcommand{\bfQ}{{\mathbf Q}}
\newcommand{\bfR}{{\mathbf R}}

\newcommand{\bfU}{{\mathbf U}}
\newcommand{\bfV}{{\mathbf V}}

\newcommand{\bfZ}{{\mathbf Z}}

\newcommand{\bfa}{{\mathbf a}}

\newcommand{\bfe}{{\mathbf e}}
\newcommand{\bfg}{{\mathbf g}}
\newcommand{\bfh}{{\mathbf h}}
\newcommand{\bfm}{{\mathbf m}}
\newcommand{\bfs}{{\mathbf s}}
\newcommand{\bft}{{\mathbf t}}
\newcommand{\bfu}{{\mathbf u}}
\newcommand{\bfw}{{\mathbf w}}
\newcommand{\bfx}{{\mathbf x}}
\newcommand{\bfy}{{\mathbf y}}
\newcommand{\bfz}{{\mathbf z}}
\newcommand{\bzero}{{\mathbf 0}}

\newcommand{\blind}{1}

\allowdisplaybreaks

\usepackage{amsmath, amssymb, amsthm, latexsym,color, natbib}
\usepackage{fancyhdr}
\usepackage{algpseudocode}
\usepackage{algorithm}
\usepackage{algpseudocode}
\usepackage{enumitem}
\usepackage{geometry}
\def\TV{{\mathrm{\scriptscriptstyle TV} }}

\DeclareMathAlphabet\EuScriptBF{U}{eus}{b}{n}

\usepackage{authblk}

\allowdisplaybreaks
\begin{document}
	
	% \def\spacingset#1{\renewcommand{\baselinestretch}%
		% {#1}\small\normalsize} \spacingset{1}

	%\renewcommand{\baselinestretch}{1.0}
	
	%%%%%%%%%%%%%%%%%%%%%%%%%%%%%%%%%%%%%%%%%%%%%%%%%%%%%%%%%%%%%%%%%%%%%%%%%%%%%%
	
	\if1\blind
	{
		\title{\bf \hspace{.2cm}
			Bayesian Penalized Empirical Likelihood and Markov Chain Monte Carlo Sampling %\footnote{Address for correspondence: Jinyuan Chang, Joint Laboratory of Data Science and Business Intelligence, Southwestern University of Finance and Economics, Chengdu, Sichuan 611130, China. Email: changjinyuan@swufe.edu.cn}%\footnote{Add some connection with Shi (2016) and add some comparison between algorithms and optimization}
			\\}
		
		\author[a,b]{Jinyuan Chang }
		\author[c]{Cheng Yong Tang }
		\author[a]{Yuanzheng Zhu }
		\affil[a]{\it \small Joint Laboratory of Data Science and Business
			Intelligence, Southwestern University of Finance and Economics, Chengdu, China}
           % \affil[b]{\it \small Big Data Laboratory on Financial Security and Behavior (MOE Philosophy and Social Sciences Laboratory), Southwestern University of Finance and Economics, Chengdu, Sichuan, China}
		\affil[b]{\it \small Academy of Mathematics and Systems Science, Chinese Academy of Sciences, Beijing, China}
		\affil[c]{\it \small Department of Statistics, Operations, and Data Science, Temple University, Philadelphia, PA, USA
		}		\date{}
		
		\maketitle
	} \fi
	
	\if0\blind
	{
		\bigskip
		\bigskip
		\bigskip
		\begin{center}
			{\LARGE\bf On the Bayesian Empirical Likelihood for High-Dimensional Estimating Equations }
		\end{center}
		%\medskip
	} \fi
	
	\begin{abstract}
		
		In this study, we introduce a novel methodological framework called Bayesian Penalized Empirical Likelihood (BPEL), designed to address the computational challenges inherent in empirical likelihood (EL) approaches. Our approach has two primary objectives: (i) to enhance the inherent flexibility of EL in accommodating diverse model conditions, and (ii) to facilitate the use of well-established Markov Chain Monte Carlo (MCMC) sampling schemes as a convenient alternative to the complex optimization typically required for statistical inference using EL. To achieve the first objective, we propose a penalized approach that regularizes the Lagrange multipliers, significantly reducing the dimensionality of the problem while accommodating a comprehensive set of model conditions. For the second objective, our study designs and thoroughly investigates two popular sampling schemes within the BPEL context. We demonstrate that the BPEL framework is highly flexible and efficient, enhancing the adaptability and practicality of EL methods. Our study highlights the practical advantages of using sampling techniques over traditional optimization methods for EL problems, showing rapid convergence to the global optima of posterior distributions and ensuring the effective resolution of complex statistical inference challenges.

		%This framework provides a valuable tool for researchers and analysts grappling with complex problems.

	\end{abstract}

	\bigskip 
	\noindent%
	
	%\bigskip
	%\begin{quote}
	%\noindent
	%{\sl MSC2010 subject classifications}: Primary 62G99; secondary 62F40
	%\end{quote}
	
	%\thispagestyle{empty}
	%\pagenumbering{gobble}
	
	%\newpage
	%\tableofcontents
	
	%\newpage
	%\pagenumbering{arabic}
	%\setcounter{page}{1}
	
	%\linenumbers
	
	\bigskip
	\baselineskip 24pt

	%\newpage
	%\spacingset{2} % DON'T change the spacing!
	%\onehalfspacing

	\section{Introduction}
	%Empirical Likelihood (EL), Generalized Method of Moments (GMM), {\color{blue} Exponentially Tilted Empirical Likelihood (ETEL)}
	
	EL \citep{Owen_2001} is a versatile and flexible tool for statistical inference, providing a framework that accommodates broadly defined model conditions. Unlike traditional likelihood approaches, EL does not require the explicit specification of probability distributions governing the data generation process. This inherent flexibility offers numerous practical advantages, such as the ability to incorporate a wide range of model specifications and prior knowledge, making it highly adaptable for integrating information from multiple data sources. Additionally, EL retains key benefits of its parametric likelihood counterpart, including efficiency (in the semiparametric sense) and the convenience of conducting hypothesis tests and estimating confidence sets through the Wilks-type likelihood ratio framework.

	Recent developments in EL approaches have a focus on addressing the challenges posed by complex high-dimensional data. To handle the complexities  arising from various model conditions, researchers have explored regularization techniques applied to the Lagrange multipliers associated with EL or the empirical versions of moment conditions, aiming to achieve enhanced model parsimony.
	In \cite{Shi2016}, a two-step procedure is introduced. The first step involves employing a ``relaxed" EL that incorporates specific inequality constraints in its formulation. The second step includes moment selection and bias correction.  \cite{Chau2017} addresses a continuum of moment conditions where the numerical optimization problem becomes ill-conditioned. To resolve this, a penalty on the continuous version of the Lagrange multiplier's counterpart is proposed and investigated. 
	\cite{Changetal_2018} proposes a method to penalize the magnitudes of both the Lagrange multiplier and the model parameters, specifically to tackle high-dimensional model parameters under complex conditions. More recently, \cite{Chang_2021} explores the projection of high-dimensional moment conditions onto lower-dimensional spaces to facilitate statistical inference for specific components of model parameters and to assess model specification validity. 
	Besides addressing the challenge of handling many moment conditions, the development of EL approaches that incorporate penalties on model parameters to promote parsimonious structures can effectively manage high-dimensional problems, as discussed in \cite{TangLeng_2010_Bioka}, \cite{LengTang_2010}, \cite{ChangChenChen_2015}, and \cite{Chang2023}.

	The synergy of Bayesian methodologies with traditional likelihoods has consistently demonstrated its effectiveness. Leveraging advances in sampling techniques, Bayesian approaches have established their significance in tackling a wide array of challenges across various domains. This is particularly valuable when dealing with intricate statistical problems where maximizing or even computing the objective function becomes infeasible.
	The amalgamation of Bayesian principles with EL shows great promise in practical applications. 
	This integration enhances the adaptability and robustness of the Bayesian framework, 
	enabling the creation of statistical models that can accommodate a wide range of scenarios. Recent developments in the realm of Bayesian EL (BEL) methods are evident in a growing body of literature; see \cite{Lazar2003}, \cite{Rao2010}, \cite{Chaudhuri2011}, \cite{Yang2012}, \cite{Mengersen2013}, \cite{Chib2018}, \cite{Cheng2019}, \cite{Zhao2020},  \cite{Tang2022}, and \cite{Yu2023}. 
	
	The class of EL approaches often encounters significant challenges due to substantial computational 
	complexity, which frequently presents barriers in practice. These difficulties primarily arise from the nonconvex nature of the objective function and the potential nonconvexity of its support. As the complexity of the model increases with additional parameters and conditions, these computational obstacles become more severe. Thus, developing computationally efficient strategies is crucial to address these challenges.
	Indeed, as demonstrated in \cite{Chau2017} and related works, solving the associated optimization problem of penalized EL (PEL) can be  a dauntingly difficult task. In our study, we demonstrate that, when combined with the Bayesian framework, sampling schemes offer promising alternatives. Once successfully drawn, samples from the posterior distribution can be used to develop the estimator.

	In recent research, sampling techniques, often perceived as computationally demanding alternatives to optimization methods, demonstrate remarkable efficiency in approximating target distributions,  outperforming optimization alternatives in handling nonconvex problems; see  \cite{Ma2019}.
	While sampling techniques offer a promising approach within the framework of BEL, there exist numerous challenges associated with devising these computational schemes.
	On one hand, EL has the potential to leverage information from various model conditions, leading to more precise estimates of unknown model parameters. However, the inclusion of a large number of these conditions introduces additional complexities, both in theory and practical implementation. Indeed, the dimensionality of the problem remains a central obstacle in EL approaches, as elaborated in 
	\cite{Hjortetal_2008_AS}.
	Furthermore, the incorporation of an increasing number of moment conditions can substantially amplify the nonconvex nature of the associated optimization problems, making the development of an effective sampling scheme increasingly more challenging. As underscored in
	\cite{Chaudhuri2017}, 
	traditional MCMC techniques encounter significant hurdles when applied to BEL due to the intricate and nonconvex characteristics of the parameter space in which new samples are generated.

	%This combination of techniques  for surmounting obstacles and extending the practical utility of empirical likelihood in real-world scenarios.

	Our research aims to establish an innovative methodological framework, guided by two primary objectives: (i) our approach maintains the inherent flexibility and adaptability of EL, allowing for the incorporation of broad model conditions; and (ii) our framework provides convenient access to well-established MCMC computing schemes, streamlining practical implementations. 
	To address the first objective and mitigate challenges stemming from numerous model conditions, we propose a penalized approach. 
	By penalizing the magnitudes of the Lagrange multipliers used in evaluating EL at specific model parameter values, we create an effective mechanism similar to moment selection. This approach reduces the problem's dimensionality while still leveraging the potential efficiency gains from a comprehensive set of model conditions.
	For the second objective, our approach effectively overcomes the obstacles associated with devising sampling schemes for applying Bayesian approaches, thanks to the efficient dimensionality reduction achieved through PEL. In our study, we demonstrate the practicality of our framework using two well-established sampling methods: the popular Metropolis-Hastings sampling and the influential adaptive multiple importance sampling technique for approximate Bayesian computations.

	Our study makes several noteworthy contributions, in addition to the methodological advancement mentioned earlier. On a theoretical level, our analysis establishes the properties of the BPEL estimator, allowing for an exponentially increasing number of model conditions, thereby enabling unprecedented adaptability in practical applications.
	Furthermore, we develop theory that guarantees the convergence of the two showcased sampling schemes, thereby ensuring the validity of BPEL in statistical inference.
	Our study reinforces the observations made in a recent study by  \cite{Ma2019} that sampling techniques offer compelling alternatives to optimization methods in addressing computationally demanding problems. Our theoretical results and numerical studies demonstrate that sampling schemes converge rapidly to stationary distributions centered around the true global optimizer. % of the posterior distribution. 
	In contrast, optimization methods often require more time and can become trapped at local peaks, limiting their ability to locate the true optimum. %This finding underscores the practicality of employing our BPEL.

	The rest of this article are structured as follows. Section \ref{sec:method} delves into the framework of BPEL and introduces two MCMC algorithms. Numerical studies and real data analysis for an international trade dataset are presented in Sections \ref{sec:num_stu} and \ref{sec:real data}, respectively. Section \ref{sec:theoretical_analy} comprehensively develops the properties and theoretical guarantees of the proposed methods.  Some discussions are provided in Section \ref{sec:dis}, while all technical proofs are available in the supplementary material. The used real data and the code for implementing our proposed methods are available at the GitHub repository: {\tt https://github.com/JinyuanChang-Lab/BayesianPenalizedEL}.

	{\it Notation.} For any positive integer $q$, write $[q]=\{1,\ldots,q\}$ and let $\bfI_q$ be the $q \times q$ identity matrix. Denote by $I(\cdot)$ the indicator function. Let $\vech(\cdot)$ be an operator that stacks the columns of the lower triangular part of its argument square matrix. For a $q$-dimensional vector $\bfa = (a_1,\ldots,a_q)^\T$, we use $|\bfa|_2 = (\sum_{i=1}^q a_i^2)^{1/2}$ and ${\rm supp}(\bfa) = \{i \in [q]: a_i \neq 0\}$ to denote its $L_2$-norm and support, respectively. Let $\mathcal{U}(a,b)$ be the uniform distribution among $(a,b)$, and $\mathcal{N}(\bmu,\bSigma)$ be the Gaussian distribution with mean $\bmu$ and covariance matrix $\bSigma$. Denote by $\mathcal{T}_k(\bmu,\bSigma)$ the multivariate Student's distribution with $k$ degrees of freedom, mean $\bmu$, and covariance matrix $\bSigma$. For two positive real-valued sequences  $\{a_n\}$ and $\{b_n\}$, we write $a_n\lesssim b_n$  if $\limsup_{n\rightarrow\infty}a_n/b_n\leq c_0$ for some positive constant $c_0$, $a_n\asymp b_n$ if $a_n\lesssim b_n$ and $b_n\lesssim a_n$ hold simultaneously, and $a_n\ll b_n$  if $\limsup_{n\rightarrow\infty}a_n/b_n=0$.

	% compelling feature of Metropolican sampling, with with the help of the penalized empirical likelihood.
	% importance sampleing
	%role of penalty
	% bayesian methods for solving more chanllenging problems
	% empirical likelihood for solving more challenging problems
	% bayesian empirical likelihood and chanllenges
	% penalized empirical likelihood, relaxed empirical likelihood

	%	\cite{BubeckEldanLehec2015} suggest a projection step in unadjusted Langevin algorithm for sampling from a log-concave distribution with compact support. \cite{ MangoubiVishnoi2019} propose an MALA for a constrained convex body, where they used a similar step to constrain the Markov chain to stay at a given state space. \cite{Brosse2017} set the target distribution outside a convex set to zero in order to generate random samples within this set.

	%Denote by $I(\cdot)$ the indicator function.  %For a $q_1 \times q_2$ matrix $\bfB=(b_{i,j})_{q_1 \times q_2}$, let $\bfB^\T$ be its transpose. 

	\section{Methodology} \label{sec:method}
	
	\subsection{Penalized Empirical Likelihood}
	
	Let $\mathcal{X}_n = \{\mathbf{x}_1, \ldots, \mathbf{x}_n\}$ represent a set of $d$-dimensional independent and identically distributed observations, and let $\boldsymbol{\theta} = (\theta_1, \ldots, \theta_p)^\T \in \boldsymbol{\Theta}$ be a $p$-dimensional parameter. Here, the parameter space $\boldsymbol{\Theta} \subset \mathbb{R}^p$ is a compact set. 
	The information regarding the model parameter $\boldsymbol{\theta}$ is gathered through a set of unbiased moment conditions 
	$\mE\{\bfg(\bfx_{i}; \bthetazero)\}=\bzero$, 
	where  $\mathbf{g}(\cdot\,; \cdot) = \{g_1(\cdot\,; \cdot), \ldots, g_r(\cdot\,; \cdot)\}^\T\in{\mathbb R}^r$ is referred to as  the estimating function, and the true, yet unknown value $\boldsymbol{\theta}_0$ is situated within the interior of $\boldsymbol{\Theta}$.

	In existing studies,  it has been typically required  that $r\geq p$  for the identification of $\btheta_0$. When $p$ and $r$ are fixed constants,  the EL with the  estimating function $\bfg(\cdot\,;\cdot)$ considered in \cite{QinLawless_1994_AS} can be formulated as 
	%\begin{align*}
	%	{\rm EL}(\btheta)=\sup \bigg \{\prod_{i=1}^{n} \pi_{i}: \pi_{i}>0,\,\sum_{i=1}^{n} \pi_{i}=1,\,\sum_{i=1}^{n} \pi_{i} \bfg ( \bfX_{i}; \btheta)=\bzero \bigg\}\,.
	%\end{align*}
	%By employing standard optimization techniques for ${\rm EL}(\btheta)$, it can be shown that
	\begin{align}\label{eq:elnew}
		{\rm EL}(\btheta)=\exp\bigg[-n\log n- \max_{\blambda\in\hat{\Lambda}_n(\btheta)} \sum_{i=1}^{n} \log \{ 1+ \blambda^\T \bfg ( \bfx_{i}; \btheta)\}\bigg]\,,
	\end{align}
	where $\hat{\Lambda}_n(\btheta)=\{\blambda \in \R^{r}:\blambda^\T \bfg ( \bfx_{i}; \btheta) \in \mathcal{V}~\textrm{for any}~ i\in[n] \} $ with an open interval $\mathcal{V}$ containing zero. 
	The standard EL estimator for $\boldsymbol{\theta}_0$ is defined as $\tilde{\boldsymbol{\theta}}_n = \arg \max _{\boldsymbol{\theta} \in \boldsymbol{\Theta}}{\rm EL}(\boldsymbol{\theta})$, which is equivalent to solving the corresponding dual problem:
	\begin{align}\label{eq:mele}
		\tilde{\btheta}_n=\arg\min_{\btheta \in \bTheta} \max_{\blambda \in \hat{\Lambda}_n(\btheta)}\sum_{i=1}^{n} \log\{ 1+\blambda^\T \bfg ( \bfx_{i}; \btheta)\} \,.
	\end{align}
	The estimator $\tilde{\boldsymbol{\theta}}_n$ exhibits several desirable properties: (i) it is $\sqrt{n}$-consistent, (ii) it possesses asymptotic normality, and (iii) it attains the semiparametric efficiency bound of \cite{GodambeHeyde_1987_ISR}. However, in high-dimensional scenarios, the literature has highlighted the challenge of accommodating a diverging $r$. This issue is discussed in works such as \cite{Donald2003}, \cite{ChenPengQin_2008}, \cite{Hjortetal_2008_AS}, \cite{LengTang_2010}, and \cite{ChangChenChen_2015}. To elaborate, it is generally required that $r \ll n^{1/2}$ for the consistency and $r \ll n^{1/3}$ for the asymptotic normality of  $\tilde{\boldsymbol{\theta}}_n$. These constraints on the diverging rate of $r$ pose significant challenges when dealing with high-dimensional estimating equations.  %where $r\gg n$.

	To address scenarios where $r\gg n$ and $p$ remains fixed, we  investigate the PEL estimator for $\boldsymbol{\theta}_0$ as follows:
	\begin{align}\label{eq:thetan_hat}
		\hat{\btheta}_n=\arg\min_{\btheta \in \bTheta} \max_{\blambda \in \hat{\Lambda}_n(\btheta)}\bigg[\sum_{i=1}^{n} \log \{ 1+\blambda^\T \bfg ( \bfx_{i};\btheta) \} - n\sum_{j=1}^{r} P_{\nu}(|\lambda_j|)\bigg] \,,
	\end{align}
	where $\blambda = (\lambda_{1},\ldots,\lambda_{r})^\T$, and $P_{\nu}(\cdot)$ is a penalty function with the tuning parameter $\nu$. Given a penalty function $P_{\nu}(\cdot)$ with the tuning parameter $\nu$, we define $\rho(t;\nu)=\nu^{-1}P_\nu(t)$ for $t\in[0,\infty)$ and $\nu\in(0,\infty)$. %In the context of 
	For  $P_\nu(\cdot)$ in
	\eqref{eq:thetan_hat}, we %impose constraints on the selection of $P_\nu(\cdot)$, restricting it
	%to be a member of 
	consider the following class of  penalty functions:
	\begin{align} 
		\mathscr{P}=\big \{P_{\nu}(\cdot): &~\rho (t;\nu)\ \text{is increasing in} \ t \in [0,\infty)\ \text{and has continuous}\notag\\
		&~\text{derivative}\ \rho'(t;\nu)\ \text{for any}\ t \in (0,\infty)\ \text{with}\ \rho'(0^{+};\nu)\in (0,\infty),\label{eq:penalty}\\
		&~\text{where}\ \rho'(0^{+};\nu)\ \text{is independent of}\ \nu \big\}\,. \notag
	\end{align}
	The class $\mathscr{P}$ is broad and general, encompassing commonly used penalty functions. % such as the $L_1$ penalty, SCAD penalty \citep{FanLi_2001_JASA}, and MCP penalty \citep{zhang09}. %With appropriately chosen penalty function $P_\nu(\cdot)$ and tuning parameter $\nu$, the penalized EL estimator $\hat\btheta_n$ is associated with a sparse Lagrange multiplier $\blambda$. Since a sparse $\blambda$ effectively uses a subset of the estimating equations for the minimization of $\btheta$, $\hat\btheta_n$ can be seen as an estimator accompanied by automatic moment selection.
	Theorem \ref{th.norm} in Section \ref{sec:theory:1} demonstrates that the PEL estimator $\hat{\boldsymbol{\theta}}_n$ follows an asymptotically normal distribution and accommodates exponentially diverging $r$ with respect to $n$.

	To practically implement \eqref{eq:thetan_hat}, we encounter a two-layer optimization problem for $\boldsymbol{\theta} \in \boldsymbol{\Theta}$ and $\blambda \in \mathbb{R}^r$. 
	Let \begin{align}\label{eq:fn1}
		f_n(\blambda;\btheta)=\frac{1}{n}\sum_{i=1}^{n} \log \{ 1+\blambda^\T \bfg(\bfx_i;\btheta)\}-\sum_{j=1}^{r} P_{\nu}(|\lambda_j|)\,.
	\end{align} 
	Since $n^{-1}\sum_{i=1}^n\log\{1+\blambda^\T \mathbf{g}(\mathbf{x}_i;\boldsymbol{\theta})\}$ is concave in $\blambda$, the inner optimization layer of \eqref{eq:thetan_hat}, which seeks $\blambda$ given $\boldsymbol{\theta}$ by maximizing $f_n(\blambda;\boldsymbol{\theta})$, can be efficiently implemented even for large $r$ when $P_{\nu}(\cdot)$ is chosen as a convex function, such as the $L_1$ penalty.
	The main challenge is the outer optimization layer of \eqref{eq:thetan_hat}, which seeks the optimizer $\hat{\boldsymbol{\theta}}_n$. This is difficulty due to the nonconvex nature of the problem, making it NP-hard to find global minima  \citep{Jain2017}. As a result, this complexity often leads to computational inefficiency and a higher likelihood of converging to local optima.

	\subsection{Bayesian Penalized Empirical Likelihood}	
	
	We are motivated to explore an alternative approach using sampling techniques to solve the nonconvex problem associated with PEL.  Indeed, as an efficient alternative for addressing nonconvex optimization problems, \cite{Ma2019} has demonstrated that solving these issues with MCMC techniques can yield highly effective results.  Their findings indicate that the computational complexity of sampling algorithms exhibits linear scalability with the model dimension, in contrast to the exponential scaling of optimization algorithms in nonconvex settings.
	
	Applying sampling techniques to EL in conjunction with a Bayesian framework emerges as a compelling approach. 
	For ${\rm EL}(\boldsymbol{\theta})$ defined as \eqref{eq:elnew}, let $\pi_0(\cdot)$ represent a prior distribution for $\boldsymbol{\theta}$. Then, the posterior distribution $\pi(\boldsymbol{\theta}\,|\,\mathcal{X}_n)$ is proportional to $\pi_{0}(\boldsymbol{\theta})\times {\rm EL}(\boldsymbol{\theta})$.
	In cases where $r$ and $p$ are fixed constants, $\pi(\btheta\,|\,\mathcal{X}_n)$ converges to a Gaussian distribution with mean being the standard EL estimator $\tilde{\boldsymbol{\theta}}_n$ defined as \eqref{eq:mele}.
	Consequently, when samples are successfully drawn from the posterior distribution, their sample mean can serve as an estimator for $\boldsymbol{\theta}_0$.

	As the model's complexity increases, BEL faces challenges.
	In this study, we explore a scenario with high-dimensional model conditions ($r\gg n$), while keeping $p$ fixed. The flexibility by allowing large number $r$ also brings significant challenges.  For example, as demonstrated in \cite{Tsao_2004_AOS}, as $n\rightarrow\infty$, $\mathbb{P}\{{\rm EL}(\boldsymbol{\theta})=0\}\rightarrow1$ for any $\boldsymbol{\theta}$ in a small neighborhood of $\boldsymbol{\theta}_0$ if $r/n\geq0.5$. Such degeneration renders  ${\rm EL}(\boldsymbol{\theta})$ inapplicable in this scenario.
	To handle diverging $r$, we propose to replace ${\rm EL}(\btheta)$  by 
	\begin{align}\label{eq:pel}
		{\rm PEL}_\nu(\btheta)=
		\exp\bigg(-n\log n- \max_{\blambda\in\hat{\Lambda}_n(\btheta)} \bigg[\sum_{i=1}^{n} \log \{ 1+ \blambda^\T \bfg ( \bfx_{i}; \btheta)\} - n\sum_{j=1}^{r} P_{\nu}(|\lambda_j|)\bigg]\bigg)\,,
	\end{align}
	where $P_{\nu}(\cdot)$ is a penalty function with the tuning parameter $\nu$.  Since adding the penalty term $P_{\nu}(\cdot)$ encourages sparse Lagrange multiplier $\blambda$,  the PEL effectively performs a selection of the model conditions at each given $\btheta$. 
	%To achieve this, 
	We then consider the BPEL with a prior distribution $\pi_{0}(\cdot)$, which leads to a posterior distribution defined as
	\begin{align}\label{eq:postdis}
		\pi^{\dag}(\btheta\,|\,\mathcal{X}_n)\propto \pi_{0}(\btheta) \times		{\rm PEL}_\nu(\btheta) \times  I(\btheta \in \bTheta)\,. 
	\end{align}
	
	% which means that the standard posterior distribution $\pi(\btheta\,|\,\mathcal{X}_n)$ will be not informative around $\btheta_0$ if $r/n\geq0.5$. Hence, dimension reduction in the number of estimating function is necessary when $r\gg n$. Adding a penalty term on the Lagrange multiplier $\blambda$ actually serves for such purpose. For each given $\btheta$, we will obtain a sparse Lagrange multiplier, which is equivalent to a selection among available estimating equations for the model parameters to tackle the scenario with $r\gg n$. 
	%
	%While penalized empirical likelihood effectively handles situations with diverging $r$, the optimization of ${\rm PEL}_\nu(\boldsymbol{\theta})$ remains a challenge due to the inherently nonconvex nature of the problem.  
	%
	
	Our BPEL connects with and differs from the so-called Gibbs posterior in the literature of Bayesian methods \citep{Bissiri_2016, Tang2022, Frazier2023}. On one hand, they share a common foundation with the Gibbs posterior in that both are built upon generic loss functions. The key difference lies in the device each utilizes: EL employs an appropriate multinomial likelihood, \((p_1, \dots, p_n)\) with \(p_i \ge 0\) and \(\sum^n_{i=1} p_i = 1\), subject to a broad class of model conditions. In contrast, the Gibbs posterior uses a “pseudo-likelihood” proportional to the exponential loss. Furthermore, the inclusion of the penalty on the Lagrange multiplier helps achieve substantial dimension reduction of the problem, which is key in handling high-dimensional problems with many moment conditions. As shown in our numerical studies in Section \ref{sec:num_stu} and Section \ref{sec:add_fs} of the supplementary material, MCMC schemes developed from the proposed BPEL demonstrate compelling performance in their finite sample accuracy in approximating the posterior distributions.
	
	Our theory, as elaborated in Section \ref{sec:tv}, establishes the fundamental properties of BPEL.  Theorem \ref{th.TV} in Section \ref{sec:tv} demonstrates that the posterior distribution $\pi^\dag(\btheta\,|\,\mathcal{X}_n)$ defined as \eqref{eq:postdis} exhibits a Gaussian limiting distribution centered around the PEL estimator $\hbthetan$ as defined in \eqref{eq:thetan_hat}.
	Additionally, we define the expected value as
	\begin{align} \label{eq:exp_pi}
		\mE_{\btheta \sim \pi^\dag}(\btheta) = \int_{\R^p} \btheta \pi^\dag(\btheta\,|\,\mathcal{X}_n) \, {\rm d}\btheta\,. 
	\end{align}
	Corollary \ref{cor.pos_mean} in Section \ref{sec:tv} suggests that $\hat{\btheta}_n$ can be effectively approximated by $\mathbb{E}_{\btheta\sim\pi^\dag}(\btheta)$ with an approximation error that diminishes faster than $n^{-1/2}$. 
	This validates the approach to obtain
	$\hat{\btheta}_n$: generating samples from the posterior distribution $\pi^\dag(\btheta\,|\,\mathcal{X}_n)$ and then using the associated sample mean to approximate $\hat{\btheta}_n$. In Section \ref{sec:comp}, we will introduce two algorithms designed for implementing BPEL.

	The impact of prior specification on the properties of resulting estimators is a notable area of research. For instance, \cite{Vexetal_2014_Bioka} explores this in the context of EL. In various scenarios, the choice of prior can enhance desirable properties of the estimator derived from the posterior distribution, such as sparsity, as discussed in \cite{NarisettyHe2014}, \cite{Castillo_2015}, and \cite{OuyangBondell_2023}.
	Given the two primary goals of our study -- developing BPEL and investigating it with MCMC -- we use a non-informative prior in our numerical demonstrations. As detailed in Section \ref{sec:add_prior} of the supplementary material, we examined the effects of different prior specifications. The overall finding is intuitive: when the prior is specified closer to the true value, the resulting estimator performs better compared to using a non-informative prior. Conversely, if the prior is specified further from the true value, the performance of the estimator deteriorates and becomes less competitive.
	% see Table \ref{tab_prior} in the Supplementary Material for more details.

	\subsection{MCMC Algorithms} \label{sec:comp}
	\subsubsection{Algorithm 1}
	
	%Building upon the insights from \cite{Ma2019}, we explore the use of sampling techniques as an alternative to optimizing ${\rm PEL}_\nu(\boldsymbol{\theta})$. 
	In recent decades, MCMC sampling methods have achieved significant success and have garnered influential applications across diverse fields.  For an extensive overview of this body of work, we refer to the monograph by \cite{Brooks2011} and reference therein.
	The Metropolis-Hastings (M-H) algorithm family plays a central role in the practical implementation of MCMC techniques, serving as a cornerstone in the toolbox of statisticians and data scientists.
	
	Our first algorithm explores the utilization of the M-H algorithm for BPEL. To accomplish this, we begin by specifying a proposal distribution with a density function denoted as $\phi(\cdot\,|\,\bfx)$, where $\bfx \in \mathbb{R}^p$. Subsequently, we employ the M-H algorithm to generate samples from the posterior distribution $\pi^\dag(\btheta\,|\,\mathcal{X}_n)$, as defined in \eqref{eq:postdis}. The specific steps for this process are detailed in Algorithm \ref{alg3}.

	\begin{algorithm}[!ht]
		\setstretch{0.9}
		\caption{M-H algorithm to generate samples from $\pi^\dag(\btheta\,|\,\mathcal{X}_n) $}
		\label{alg3}
		{\footnotesize
			\hspace*{0.02in} {\bf Input:}
			the proposal distribution with density $\phi(\cdot\,|\,\cdot )$, the number of iteration $K$, an initial  point $\btheta^{0} \in \bTheta$.
			\begin{algorithmic}
				\For {$k=0, 1, \ldots, K-1$}
				\State
				{\bf Proposal step:} 
				\State
				~~~~~~generate $\bvartheta^{k+1}$ from the proposal distribution with density $\phi(\bvartheta\,|\,\btheta^k)$. 
				\State
				{\bf Accept-reject step:}
				\State
				~~~~~~compute 
				\begin{align*}
					\alpha^{k+1} = \left \{
					\begin{aligned}
						& \min\bigg\{1,  \frac {\pi^\dag(\bvartheta^{k+1}\,|\,\mathcal{X}_n)\phi(\btheta^k\,|\,\bvartheta^{k+1})} {\pi^\dag(\btheta^k\,|\,\mathcal{X}_n) \phi(\bvartheta^{k+1}\,|\,\btheta^k)} \bigg\} \,, && \text{ if~} \bvartheta^{k+1} \in \bTheta \text{~with~} \pi^\dag(\btheta^k\,|\,\mathcal{X}_n) \phi(\bvartheta^{k+1}\,|\,\btheta^k) \neq 0 \,,\\
						&~~~~~~~~~~~~~~~~~~~~~~1 \,, && \text{ if~} \bvartheta^{k+1} \in \bTheta \text{~with~} \pi^\dag(\btheta^k\,|\,\mathcal{X}_n) \phi(\bvartheta^{k+1}\,|\,\btheta^k) =0 \,, \\
						&~~~~~~~~~~~~~~~~~~~~~~0 \,, && \text{ if~} \bvartheta^{k+1} \notin \bTheta \,.
					\end{aligned}
					\right.
				\end{align*}
				\State
				~~~~~~generate $u \sim \mathcal{U}(0,1)$. 
				\State
				~~~~~~{\bf if} $u \leq \alpha^{k+1}$, {\bf then} $\btheta^{k+1} \leftarrow \bvartheta^{k+1}$, 
				~{\bf else} $\btheta^{k+1} \leftarrow \btheta^k$.
				\EndFor
			\end{algorithmic}
			\hspace*{0.02in} {\bf Output:} $\btheta^1, \ldots, \btheta^K$.
		}
	\end{algorithm}
	
	At each iteration $k$, Algorithm \ref{alg3} begins with a state $\btheta^k \in \bTheta$. In the proposal step, it generates a new parameter $\bvartheta^{k+1}$ from the proposal distribution centered at $\btheta^k$, denoted by $\phi(\cdot\,|\,\btheta^k)$. Following this, in the accept-reject step, Algorithm \ref{alg3} decides whether to accept $\bvartheta^{k+1}$ with a probability denoted as $\alpha^{k+1}$. This crucial step ensures that the Markov chain, guided by Algorithm \ref{alg3}, remains within the valid parameter space $\bTheta$. Consequently, it expedites the convergence of the resulting chain towards its stationary distribution, which is the posterior distribution $\pi^\dag(\btheta\,|\,\mathcal{X}_n)$.
	%
	%
	%This crucial step ensures that the Markov chain, guided by Algorithm \ref{alg3}, always remains within the feasible parameter space $\bTheta$;  speeding up the process that the resulting chain converging to its stationary distribution.  %Consequently, the resulting sequence of draws
	%spends a significant amount of time exploring the high-probability regions of the posterior distribution $\pi^\dag(\btheta\,|\,\mathcal{X}_n)$ while allocating less time to the lower-probability regions.
	%
	There exist various approaches for selecting the proposal distribution with density $\phi(\cdot\,|\,\cdot)$, including methods like the symmetric Metropolis algorithm, random walk M-H, and the independence sampler, as detailed by \cite{Roberts2004}.

	%\footnote{Need to add some discussion on the robust of the initial value.} %Furthermore, 

	\subsubsection{Algorithm 2}
	
	Another widely-used MCMC technique is Importance Sampling \citep{Ripley2006, Hesterberg1995}. This method involves generating samples from a proposal distribution and then applying importance weights to these samples to account for the disparities between the proposal distribution and the target distribution. 
	In practical applications, recycling successive samples often proves to be an effective strategy \citep{Marin2019},  particularly  when the computation of importance weights is computationally intensive. In this context, \cite{CORNUET2012} introduces the Adaptive Multiple Importance Sampling (AMIS) algorithm, which combines various importance sampling methods with adaptive techniques.  
	The integration of the AMIS approach with EL, as shown in \cite{Mengersen2013}, is particularly compelling. 
	To ensure the consistency of AMIS, \cite{Marin2019} introduces a modified variant called Modified AMIS (MAMIS) with a simpler recycling strategy compared to AMIS.

	We present and investigate an MAMIS algorithm, as outlined in Algorithm \ref{alg2}, specifically designed for computing BPEL. This algorithm operates in a scenario where a density function $\varphi(\cdot\,; \bzeta)$ is defined, with $\bzeta$ representing a parameter in $\R^s$, and where an explicit function $\bfh: \R^p \mapsto \R^s$ is known. This configuration allows us to generate weighted samples that effectively capture the characteristics of the posterior distribution $\pi^\dag(\btheta\,|\,\mathcal{X}_n)$, as defined in \eqref{eq:postdis}.

	\begin{algorithm}[!ht]
		\setstretch{0.9}
		\caption{An MAMIS algorithm to generate the weighted samples with respect to  $\pi^\dag(\btheta\,|\,\mathcal{X}_n)$}
		\label{alg2}
		{\footnotesize
			\hspace*{0.02in} {\bf Input:}
			the proposal distribution admits density $\varphi(\cdot\,; \bzeta)$ with the parameter $\bzeta \in \R^s$, an initial parameter $\hat{\bzeta}_1$, an 
			
			~~~~~~~~~~~  explicitly known function $\bfh: \R^p \mapsto \R^s$, the number of iteration $K$ and the increasing sampling sizes 
			
			~~~~~~~~~~~ $\{N_1, \ldots, N_K\}$.
			\begin{algorithmic}
				\For {$k\in[K]$}
				\For {$i\in[N_k]$}
				\State
				{\bf Proposal step:} 
				\State
				generate $\btheta^k_i$ from the proposal distribution with density $\varphi(\btheta\,; \hat{\bzeta}_{k})$.
				\State
				compute the importance weight 
				$\omega^k_i = \pi^\dag(\btheta^k_i\,|\,\mathcal{X}_n) / \varphi(\btheta^k_i\,;\hat{\bzeta}_{k})$. 
				%if $\btheta^k_i \in \bTheta$ and let $\omega^k_i = 0$ if $\btheta^k_i \notin \bTheta$.
				\EndFor
				\State
				update the parameter of the proposal distribution:  $\hat{\bzeta}_{k+1} = N_k^{-1} \sum_{i=1}^{N_k} \omega^k_i \bfh(\btheta^k_i)$.
				\EndFor
				\For {$k\in[K]$}
				\For {$i\in[N_k]$}
				\State
				{\bf Recycling process:} 
				\State
				update the importance weight 
				$\omega^k_i = \pi^\dag(\btheta^k_i\,|\,\mathcal{X}_n) /\{S_K^{-1} \sum_{l=1}^K N_l \varphi(\btheta^k_i\,;\hat{\bzeta}_l)\}$ with $S_K=N_1+\cdots+N_K$
				\State if $\btheta^k_i \in \bTheta$.
				\EndFor
				\EndFor
			\end{algorithmic}
			\hspace*{0.02in} {\bf Output:} 
			the weighted samples $(\btheta^1_1, \omega^1_1), \ldots, (\btheta^1_{N_1}, \omega^1_{N_1}),\ldots,(\btheta^K_1, \omega^K_1),\ldots, (\btheta^K_{N_K}, \omega^K_{N_K})$.
		}
	\end{algorithm}
	
	Algorithm \ref{alg2} generates a sequence of samples while progressively adjusting the parameter $\bzeta \in \R^s$ involved in the proposal distribution. At each iteration $k$ of Algorithm \ref{alg2}, the new value for the parameter $\bzeta$ of the proposal distribution is determined based on the most recent $N_k$ samples drawn. This represents the primary distinction between the MAMIS algorithm by \cite{Marin2019} and the AMIS algorithm by \cite{CORNUET2012}. Specifically, MAMIS updates the proposal distribution parameter using only the last $N_k$ samples at iteration $k$, while AMIS updates this parameter by considering all past $\sum_{j=1}^{k}N_j$ samples. The end product output of Algorithm \ref{alg2} is generated by updating the importance weights for all samples produced during the recycling process.

	%Denote by $\mE_{\pi^\dag}(\cdot)$ the expectation operator with respect to the posterior distribution $\pi^\dag(\btheta\,|\,\mathcal{X}_n)$ defined as \eqref{eq:postdis}. 

	\subsection{Sampling vs Optimizations}

	We advocate the utilization of sampling techniques as a practical and efficient alternative to optimization methods for addressing computationally challenging PEL problems.
	Specifically for obtaining the estimator $\hat{\boldsymbol{\theta}}_n$ as defined in \eqref{eq:thetan_hat}, 
	we can rely on samples $\btheta^1,\ldots,\btheta^K$ generated from the M-H algorithm (see Algorithm \ref{alg3}),  estimating  $\mathbb{E}_{\btheta\sim\pi^\dag}(\btheta)$, as defined in \eqref{eq:exp_pi}, by computing the sample mean, i.e., $K^{-1}\sum_{k=1}^K\btheta^k$.
	When employing the MAMIS algorithm (see Algorithm \ref{alg2}) and completing $K$ iterations, the estimator for $\mathbb{E}_{\btheta\sim\pi^\dag}(\btheta)$ is determined as a weighted average:
	\begin{align} \label{eq:esi_Alg2}
		\widehat{\mE}_{\pi^\dag,K}(\btheta) = \frac{1}{S_K} \sum_{k=1}^{K} \sum_{i=1}^{N_k} \omega_i^k \btheta^k_i \,,
	\end{align}
	where $S_K=N_1+\cdots+N_K$.

	Our theory in Section \ref{sec:tv} supports the use of sampling algorithms as efficient alternatives.
	For the M-H algorithm, 
	Theorem \ref{th.MH_alg} in Section \ref{sec:tv} demonstrates that, conditional on $\mathcal{X}_n$, the average $K^{-1}\sum_{k=1}^K\btheta^k$ converges almost surely to $\mathbb{E}_{\btheta\sim\pi^\dag}(\btheta)$ as $K\rightarrow\infty$.  For the MAMIS algorithm, Theorem \ref{th.MAMIS_alg} in Section \ref{sec:tv} establishes that, conditional on $\mathcal{X}_n$, $\widehat{\mE}_{\pi^\dag,K} (\btheta)$ in \eqref{eq:esi_Alg2} converges almost surely to $\mathbb{E}_{\btheta\sim\pi^\dag}(\btheta)$ as $K\rightarrow\infty$.
	%Theorem \ref{th.TV}
	These results, combined with Corollary \ref{cor.pos_mean} in Section \ref{sec:tv}, validate the properties of BPEL estimators obtained through these established sampling techniques. Another consideration in Algorithms \ref{alg3} and \ref{alg2} is the choice of the initial point, denoted, respectively, as $\btheta^{0}$ and $\hat{\bzeta}_1$. Our theoretical analyses only require $\btheta^0 \in \bTheta$ satisfying  $\pi^\dag(\btheta^0\,|\,\mathcal{X}_n)>0$ and do not impose any restriction on $\hat{\bzeta}_1$;  see Theorems  
	\ref{th.MH_alg} and \ref{th.MAMIS_alg} in Section \ref{sec:tv} for details. 
	Our empirical simulation studies in Section \ref{sec:num_stu} consistently demonstrate the proposed algorithms' robust performance, irrespective of the initial value chosen. Notice that the performance of the optimization methods for the nonconvex optimization problems usually depends crucially on the choice of the initial point. The combination of theoretical analysis and empirical evidence underscores that, in comparison to competing optimization methods, these sampling-based approaches offer significant advantages in terms of convergence speed, stability across replications, and resilience to variations in initial values. This reaffirms the benefits of incorporating BPEL into the methodology.

	The M-H and MAMIS algorithms each have their strengths.  M-H is easy to implement, but high rejection rates can reduce its efficiency, especially with a poorly tuned proposal distribution. MAMIS, while requiring more effort -- particularly in computing importance weights -- offers improved sampling efficiency and is less sensitive to the proposal distribution, making it ideal for complex posterior distributions.
	Choosing between these algorithms depends on the specific problem and the balance between implementation ease and sampling efficiency.

	\section{Numerical Studies} \label{sec:num_stu}
	\subsection{Data Generation Process} \label{sec.dgp}
	
	%\subsubsection{Data generation model} \label{sec: IV model}
	We conduct simulation studies to empirically assess the performance of our proposed methods. For the data generation process (DGP), we adopt the structural equation
	$y_i = \hbar(\bfu_i^\T \bthetazero) + e_i^{(0)}$, $i \in [n]$,
	where $\hbar: \R \mapsto \R$ is a continuous function, $e_i^{(0)}$ is the error, and $\bfu_i=(u_{i,1},u_{i,2})^\T$ represents two endogenous variables. The set of all instrumental variables (IVs) is denoted as $\bfz_{i}=(z_{i,1}, \ldots, z_{i,r})^\T$ for $i \in [n]$. The true reduced-form equations for the endogenous variables are specified as   $u_{i,1}=0.5z_{i,1}+0.5z_{i,2}+e_i^{(1)}$ and $ u_{i,2}=0.5z_{i,3}+0.5z_{i,4}+e_i^{(2)}$, where $(e_i^{(1)}, e_i^{(2)})$  represents the random errors.  
	Essentially, each of the two endogenous variables is influenced by  only two IVs.
	All IVs are selected orthogonal to the error term 
	$e_i^{(0)}$. Hence, we have 
	$
	\mE \{y_i - \hbar(\bfu_i^\T \bthetazero)\,|\, \bfz_{i}\} =\bzero 
	$, which implies that $\btheta_0$ can be identified by the $r$ unbiased moment conditions 
	$\mathbb{E}\{\bfg(\bfx_i;\btheta_0)\}=\bzero$, 
	where $\bfg(\bfx_i;\btheta) = \{y_i - \hbar(\bfu_i^\T \btheta) \}\bfz_i $ with $\bfx_i= (y_i, \bfu_i^\T, \bfz_i^\T)^\T$. In the DGP, we generate $\bfz_{i}\sim\mathcal{N}(\bzero,\bfI_r)$, and
	\begin{align*}
		\left(
		\begin{array}{c}
			e_i^{(0)} \\
			e_i^{(1)} \\
			e_i^{(2)}
		\end{array}
		\right)
		\sim \mathcal{N}
		\left(
		\left(
		\begin{array}{c}
			0 \\
			0 \\
			0
		\end{array}
		\right) ,
		\left(
		\begin{array}{ccc}
			0.43 & 0.3 & 0.3 \\
			0.3 & 0.34 & 0.09 \\
			0.3 & 0.09 & 0.34
		\end{array}
		\right)
		\right) .
	\end{align*}
	We set $\btheta_0=(0.5,0.5)^\T$ and consider two selections for the link function $\hbar(\cdot)$: (i) the linear case with $\hbar(v)=v$, and (ii) the nonlinear case with $\hbar(v)=\sin v$. 
	
	\subsection{Sampling Efficiency and Stability}
	
	We begin by demonstrating the improvement in sampling efficiency achieved through the use of PEL. In this context, we generate data following the DGP with linear link function $\hbar(\cdot)$ by setting $n=120$ and varying $r$ in the range $[50,1000]$. We aim to sample from two posterior distributions $\pi_{0}(\btheta) \times {\rm EL}(\btheta)$ and $\pi_{0}(\btheta) \times {\rm PEL}_{\nu}(\btheta)$, where ${\rm EL}(\btheta)$ and ${\rm PEL}_{\nu}(\btheta)$ are, respectively, given in \eqref{eq:elnew} and \eqref{eq:pel}. Evaluating ${\rm PEL}_{\nu}(\btheta)$ involves an optimization problem that solves for $\blambda$ by maximizing the objective function $f_n(\blambda;\btheta)$ defined as \eqref{eq:fn1} at given $\btheta$. To ensure the attainment of a sparse Lagrange multiplier and maintain the convexity of the objective function, we select $P_{\nu}(\cdot)$ as the $L_1$ penalty function.
	%Subsequently, we employ the interior-point method proposed by \cite{Koh2007a, Koh2007} to efficiently address the optimization problem involved in evaluating ${\rm PEL}_{\nu}(\btheta)$. This approach transforms the optimization problem, which incorporates an $L_1$ penalty, into a constrained optimization framework. It exhibits significant computational efficiency benefits, especially in high-dimensional settings. 
	In practice, since the prior information about the true parameter $\btheta_0$ is typically unavailable, we select $\pi_0(\cdot)$ as the improper uniform prior. We implement Algorithm \ref{alg3} to sample from both posterior distributions using identical settings, employing a proposal distribution $\mathcal{N}(\btheta, \sigma^2\bfI_p)$ with $\sigma^2 = 10^{-4}$ and initializing from $\btheta^0 = (0.3, 0.3)^\T$.  In the case of PEL, we set the tuning parameter $\nu=0.03$ involved in ${\rm PEL}_{\nu}(\boldsymbol{\theta})$.
	
\begin{figure}[!ht] 
		\centering
		\includegraphics[scale=0.55]{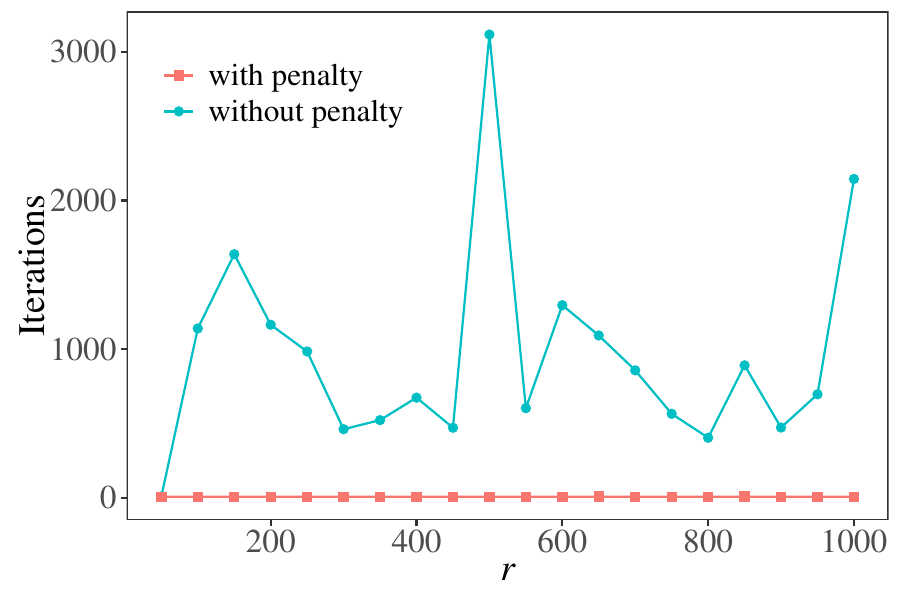}
		\caption{The average number of iterations over $500$ runs required to obtain $5$ valid samples. }
		\label{fig:1}
	\end{figure}

	To compare efficiency, we measure the number of iterations required to obtain the same number of accepted samples. Figure \ref{fig:1} illustrates the average number of iterations needed over 500 runs to accept 5 samples for different values of $r$, thereby providing a comparison between using EL and PEL within a Bayesian framework. 		
	The sampling efficiency of Algorithm \ref{alg3} when using ${\rm PEL}_{\nu}(\btheta)$ is notably superior to that achieved with ${\rm EL}(\btheta)$. The selection of $\sigma^2$ within the proposal distribution closely influences the acceptance rate in each step of the M-H algorithm. With our small choice of $\sigma^2$ in the simulation, the M-H algorithm should efficiently generate valid samples.  
	It is worth highlighting that the acceptance rate remains consistently high and stable when using PEL  across all $r$ settings. In contrast, when employing EL without any penalty, it may require thousands more iterations to achieve the same number of accepted samples. Additionally, it is evident that the M-H algorithm with EL becomes increasingly unstable as $r$ increases.

	\subsection{Comparison with the Optimization Methods}\label{sec:comp1}
	
	As we suggested in Section \ref{sec:comp}, the computation of the PEL estimator $\hat{\btheta}_n$ defined as \eqref{eq:thetan_hat} can be implemented using Algorithm \ref{alg3} (referred to as M-H) and Algorithm \ref{alg2} (referred to as MAMIS). In this part, we compare their performance with two optimization methods: (a) {\tt optim}: A versatile R function for general-purpose optimization of objective functions, supporting various optimization algorithms like Nelder-Mead, quasi-Newton, and conjugate-gradient; and (b) {\tt nlm}: An R function specialized in non-linear optimization, particularly designed for finding minima of objective functions using Newton-type algorithms.

	The choice of the proposal distribution plays a crucial role in achieving efficient sampling with BPEL. Within the context of the M-H algorithm, one commonly used scheme is the random walk M-H, where the proposal distribution takes the form of a Gaussian distribution $\mathcal{N}(\btheta, \sigma^2 \bfI_p)$ with the current state denoted as $\btheta$. It is essential to carefully select an appropriate value for $\sigma^2$. A small value for $\sigma^2$ can result in slow exploration of the state space, while a large value can lead to decreased acceptance rates, subsequently slowing down the algorithm.
	To strike a balance between exploration and acceptance rates, we can monitor the acceptance rate of the algorithm. In the simulation for M-H, we set $\sigma^2=C(n\log r)^{-1}$ with some constant $C>0$. We adjust the value of $C$ until the acceptance rate closely matches the desired rate, typically aiming for approximately $0.234$, as suggested in \cite{Gelman1997}. It is known that the M-H algorithm requires some time to converge to its stationary distribution, especially when the initial point $\btheta^0 \in \bTheta$ is situated in the tails of the posterior distribution $\pi^\dag(\btheta\,|\,\mathcal{X}_n)$.  
	Considering this, we set a burn-in period of $500$ iterations. For the MAMIS algorithm, we adhere to recommendations from \cite{CORNUET2012} and \cite{Mengersen2013} that advocate for the adoption of $\mathcal{T}_3(\bmu,\bSigma)$ as the proposal distribution.  During each iteration $k$ of MAMIS, we calculate the updated value  $\hat\bzeta_{k+1} = \{\hat{\bmu}_{k+1}^\T, \vech(\widehat{\bSigma}_{k+1})^\T\}^\T$ for the parameter vector $\bzeta = \{\bmu^\T, \vech(\bSigma)^\T\}^\T$ involved in the proposal distribution $\mathcal{T}_3(\bmu,\bSigma)$ as 
	$\hat{\bmu}_{k+1} = N_k^{-1} \sum_{i=1}^{N_k} \omega^k_i \btheta^k_i$ and  
	$\vech(\widehat{\bSigma}_{k+1}) = N_k^{-1} \sum_{i=1}^{N_k} \omega^k_i \vech\{(\btheta^k_i - \hat{\bmu}_{k+1})(\btheta^k_i - \hat{\bmu}_{k+1})^\T\}$, 
	where $\omega^k_i$ represents the corresponding importance weights, as outlined in Algorithm \ref{alg2}. In our simulations, we initialize $\widehat{\bSigma}_{1} = \bfI_p$, and the selection of $\hat{\bmu}_{1}$ is described in the next paragraph.

	We conduct 200 replications following the DGP and explore various combinations of dimensionalities. Specifically, we consider $n\in\{120, 240\}$ and $r\in\{80, 160, 320, 640\}$. 
	To assess the robustness of these methods with respect to initial points, we select $49$ equally spaced grid points on the plane within the range of $[-3, 4] \times [-3, 4]$ as our chosen initial points. In the case of MAMIS, which is not an iterative algorithm, we set these initial points as the initial means $\hat{\bmu}_{1}$ for its proposal distribution $\mathcal{T}_3(\bmu,\bSigma)$ to facilitate comparison. In our simulations, we identify the true global minima $\hat{\btheta}_n$ defined as \eqref{eq:thetan_hat} through exhaustive search. To achieve this, in each replication of the simulation (indexed by $k$), we generate a grid of $10201$ equally spaced points within the range $[-0.5, 1.5]\times[-0.5, 1.5]$. We then compute the posterior probabilities for these points and selected $\btheta^{\rm mode}_{k}$ as the point with the highest probability. Since $\pi_0(\cdot)$ is selected as the improper uniform prior, $\btheta^{\rm mode}_k$ is actually the required true global minima in the $k$-th replication. We repeat this process for $k=1$ to $200$, and compare the outcomes obtained from both optimization and sampling methods by calculating the measure 
	$$
	{\rm MSE}_1 = \frac{1}{200 \times 49} \sum_{k=1}^{200} \sum_{l=1}^{49} |\check{\btheta}_{k}(l) - \btheta^{\rm mode}_{k}|_2^2\,.$$ 
	Here, $\check{\btheta}_{k}(l)$ represents the related outcome in the $k$-th replication initiated from the $l$-th initial point.

	In the context of BPEL sampling, we explore three scenarios with varying sample sizes of $1500$, $2500$, and $3500$, which we label as (M-H-1, M-H-2, M-H-3) and (MAMIS-1, MAMIS-2, MAMIS-3), respectively, for Algorithms 1 and 2. 
	Additionally, we conduct an investigation into the influence of different values for the tuning parameter $\nu$.
	Table \ref{tab1} presents the simulation results. %Several observations can be made from our findings:
	The overall performance of the sampling approaches surpasses that of the optimization methods. Notably, for the nonlinear model, the optimization using the R function {\tt nlm} is proven to be unreliable, resulting in highly unstable results. As the size of the generated samples increases, the performance of BPEL improves. Both M-H and MAMIS exhibit promising performance in both linear and nonlinear cases. For the nonlinear models, MAMIS  significantly outperforms M-H, possibly owing to the advantages gained from employing importance weights for parameter estimation. The role of the tuning parameter $\nu$ is pivotal, underscoring the merits from using the PEL approach in achieving more parsimonious models by effectively selecting most useful model conditions within the constraints of the available data information.  When using very small values of $\nu$, such as 0.01,  the performance of the methods becomes less satisfactory. Overall, the BPEL performs satisfactorily for a reasonable range of choices for $\nu$.
	
	%Qualitatively similar findings can be drawn from Table \ref{tab2} for the results in the setting of $n=240$.

	\begin{sidewaystable}
		\setlength\tabcolsep{3pt}
		\centering
		%\rotatebox[origin=c]{90}{%
			%  \begin{varwidth}{\textheight}
				\caption{Comparison of BPEL and optimization methods} \label{tab1}
				\footnotesize
				\begin{spacing}{1}
					\begin{tabular} 
						{cl|cccc|cccc|cccc|cccc} \hline 
						& & \multicolumn{4}{c|}{{$\hbar(v) = v, n=120$} } & \multicolumn{4}{c|}{{$\hbar(v)=\sin v, n=120$}} & \multicolumn{4}{c|}{{$\hbar(v) = v, n=240$ }} & \multicolumn{4}{c}{{$\hbar(v)=\sin v, n=240$}} \\ 
						{$\nu$} & {Methods} & {$r=80$} & {$r=160$} & {$r=320$} & {$r=640$} & {$r=80$} & {$r=160$} & {$r=320$} & {$r=640$}  & {$r=80$} & {$r=160$} & {$r=320$} & {$r=640$} & {$r=80$} & {$r=160$} & {$r=320$} & {$r=640$} \\
						\hline
						0.01 
						& {MAMIS-1}  & 0.0606    & 0.0432    & 0.0371    & 0.0336  
						& 8.8598    & 7.0955    & 6.7810    & 6.7583  & 0.2600    & 0.0467    & 0.0297    & 0.0250    & 9.7398    & 7.7113    & 6.9972    & 6.7786 \\
						& {MAMIS-2}  & 0.0062    & 0.0020    & 0.0016    & 0.0015   & 8.2760    & 6.4363    & 6.0974    & 6.0576 & 0.1139    & 0.0047    & 0.0004    & 0.0004    & 9.1874    & 7.1095    & 6.4394    & 6.1002  \\
						& {MAMIS-3}  & 0.0023    & 0.0014    & 0.0014    & 0.0013   & 7.8634    & 6.0010    & 5.6087    & 5.6073 & 0.0911    & 0.0018    & 0.0002    & 0.0003    & 8.8130    & 6.6946    & 6.0186    & 5.6339  \\
						& {M-H-1}     & 0.0457    & 0.0015    & 0.0016    & 0.0015     & 12.7310    & 12.2970    & 12.2665    & 12.3473  & 0.4498    & 0.0371    & 0.0082    & 0.0004    & 13.2222    & 12.2080    & 12.0538    & 11.9004 \\
						& {M-H-2}     & 0.0411    & 0.0015    & 0.0015    & 0.0015   & 12.6824    & 12.2486    & 12.1999    & 12.3289  & 0.4279    & 0.0311    & 0.0067    & 0.0003    & 13.2205    & 12.1876    & 12.0544    & 11.8679 \\
						&{M-H-3}     & 0.0389    & 0.0014    & 0.0015    & 0.0014   & 12.6477    & 12.2179    & 12.1564    & 12.3042 & 0.4146    & 0.0277    & 0.0053    & 0.0003    & 13.2281    & 12.1745    & 12.0461    & 11.8451  \\
						& {optim}    & 0.2822    & 0.0720    & 0.0133    & 0.0132   & 12.2067    & 12.0219    & 11.9113    & 11.9760  & 1.0834    & 0.2247    & 0.0582    & 0.0160    & 12.8270    & 12.2083    & 12.2197    & 12.1037 \\
						& {nlm}      & 0.0956    & 0.0341    & 0.0222    & 0.0151   & 117934.2    & 115037.8    & 85081.4    & 83013.3  & 6.2781    & 0.0674    & 0.0163    & 0.0076    & 121272.7    & 113263.9    & 104125.3    & 87304.6 \\
						\hline 
						0.03
						& {MAMIS-1}  & 0.0465    & 0.0363    & 0.0317    & 0.0311   & 7.3502    & 6.2407    & 5.8879    & 6.2055  & 0.0707    & 0.0345    & 0.0279    & 0.0264    & 7.2303    & 6.6040    & 6.5472    & 6.3108  \\
						& {MAMIS-2}  & 0.0021    & 0.0011    & 0.0008    & 0.0009   & 6.5728    & 5.4791    & 5.0858    & 5.3622 & 0.0033    & 0.0012    & 0.0004    & 0.0004    & 6.3727    & 5.8076    & 5.8222    & 5.6557  \\
						& {MAMIS-3}  & 0.0008    & 0.0008    & 0.0007    & 0.0007   & 6.0018    & 4.9604    & 4.5459    & 4.7549 & 0.0007    & 0.0009    & 0.0002    & 0.0002    & 5.7517    & 5.2972    & 5.2865    & 5.1854  \\
						& {M-H-1}     & 0.0135    & 0.0009    & 0.0007    & 0.0008     & 12.5367    & 12.0546    & 12.1103    & 12.2611  & 0.0522    & 0.0034    & 0.0003    & 0.0002    & 12.8677    & 12.1031    & 12.1453    & 12.1239 \\
						&{M-H-2}     & 0.0116    & 0.0009    & 0.0007    & 0.0008   & 12.4304    & 11.9548    & 12.0239    & 12.1413   & 0.0450    & 0.0034    & 0.0003    & 0.0002    & 12.7587    & 12.0285    & 12.0758    & 12.0806 \\
						&{M-H-3}     & 0.0104    & 0.0008    & 0.0007    & 0.0007   & 12.3589    & 11.8816    & 11.9509    & 12.0655  & 0.0388    & 0.0034    & 0.0002    & 0.0002    & 12.6799    & 11.9762    & 12.0297    & 12.0417  \\
						& {optim}    & 0.0587    & 0.0109    & 0.0014    & 0.0014   & 12.8445    & 12.5361    & 12.3325    & 12.5015   & 0.2977    & 0.0454    & 0.0125    & 0.0005    & 13.2494    & 12.7951    & 12.7153    & 12.8332 \\
						& {nlm}      & 0.0385    & 0.0033    & 0.0015    & 0.0058   & 79382.0    & 73390.0    & 76096.0   & 94794.3  & 0.1321    & 0.0127    &  0.0057    & 0.0047    & 90321.8    & 70994.2    & 90414.5    & 79523.2 \\
						\hline   
						0.05
						& {MAMIS-1}  & 0.0451    & 0.0359    & 0.0331    & 0.0295   & 6.0750    & 5.5727    & 5.3221    & 5.3884 & 0.0598    & 0.0387    & 0.0305    & 0.0261    & 5.8614    & 5.5769    & 5.6081    & 5.8757  \\
						& {MAMIS-2}  & 0.0025    & 0.0018    & 0.0011    & 0.0008   & 5.0407    & 4.6182    & 4.3750    & 4.3938 & 0.0043    & 0.0024    & 0.0010    & 0.0007    & 4.7997    & 4.6707    & 4.7641    & 4.9901  \\
						& {MAMIS-3}  & 0.0017    & 0.0016    & 0.0008    & 0.0007   & 4.4061    & 3.9666    & 3.7920    & 3.7718 & 0.0027    & 0.0019    & 0.0009    & 0.0005    & 4.0542    & 4.0482    & 4.1549    & 4.3903  \\ 
						& {M-H-1}     & 0.0076    & 0.0015    & 0.0010    & 0.0008   & 12.0832    & 11.7687    & 11.8703    & 11.8866  & 0.0063    & 0.0018    & 0.0010    & 0.0006    & 12.2347    & 11.8718    & 12.0358    & 12.0665 \\
						&{M-H-2}     & 0.0074    & 0.0014    & 0.0009    & 0.0007   & 11.8869    & 11.5660    & 11.6859    & 11.7368  & 0.0059    & 0.0016    & 0.0009    & 0.0005    & 12.0296    & 11.7100    & 11.9241    & 11.9669 \\
						& {M-H-3}     & 0.0074    & 0.0013    & 0.0009    & 0.0007   & 11.7189    & 11.4244    & 11.5409    & 11.6131   & 0.0056    & 0.0016    & 0.0009    & 0.0005    & 11.8578    & 11.5979    & 11.8305    & 11.8745 \\
						& {optim}    & 0.0209    & 0.0010    & 0.0001    & 0.0000   & 13.2051    & 12.8147    & 12.6047    & 12.7401  & 0.0827    & 0.0130    & 0.0073    & 0.0000   & 13.5813    & 13.0596    & 13.0502    & 13.0375 \\
						& {nlm}      & 0.0167    & 0.0002    & 0.0004    & 0.0009   & 69229.7    & 58220.3    & 61362.5    & 63215.1  & 0.0387    & 0.0023    & 0.0032    & 0.0052    & 48458.3    & 69014.4    & 62235.7    & 80570.4 \\
						\hline
						0.07
						& {MAMIS-1}  & 0.0454    & 0.0381    & 0.0323    & 0.0293   & 4.6585    & 4.7192    & 4.4856    & 4.5785 & 0.0534    & 0.0381    & 0.0341    & 0.0278    & 4.5237    & 4.5635    & 4.9180    & 5.0988  \\
						& {MAMIS-2}  & 0.0049    & 0.0031    & 0.0023    & 0.0015   & 3.5217    & 3.6260    & 3.5231    & 3.5504  & 0.0071    & 0.0058    & 0.0036    & 0.0023    & 3.2789    & 3.5005    & 3.8945    & 4.1681  \\
						& {MAMIS-3}  & 0.0042    & 0.0029    & 0.0019    & 0.0014   & 2.8438    & 2.9326    & 2.9355    & 2.9075 & 0.0066    & 0.0054    & 0.0033    & 0.0021    & 2.5653    & 2.8253    & 3.2818    & 3.5420  \\
						& {M-H-1}     & 0.0065    & 0.0031    & 0.0020    & 0.0016   & 11.6003    & 11.3310    & 11.3530    & 11.5029  & 0.0069    & 0.0054    & 0.0036    & 0.0022    & 11.7248    & 11.5753    & 11.6807    & 12.0016 \\
						& {M-H-2}     & 0.0063    & 0.0030    & 0.0020    & 0.0015   & 11.2135    & 10.9535    & 11.1190    & 11.2268 & 0.0067    & 0.0053    & 0.0035    & 0.0021    & 11.3384    & 11.2954    & 11.4157    & 11.8458  \\
						& {M-H-3}     & 0.0062    & 0.0029    & 0.0019    & 0.0015   & 10.9140    & 10.7093    & 10.8781    & 11.0160  & 0.0067    & 0.0053    & 0.0035    & 0.0021    & 11.0287    & 11.0815    & 11.2360    & 11.7241  \\
						& {optim}    & 0.0117    & 0.0000    & 0.0007    & 0.0000   & 13.2631    & 13.0115    & 12.8281    & 12.8492 & 0.0169    & 0.0033    & 0.0007    & 0.0000    & 13.5938    & 13.3118    & 13.2545    & 13.2196  \\
						& {nlm}      & 0.0119    & 0.0000    & 0.0017    & 0.0010   & 49184.4    & 62755.1    & 63796.5    & 71037.4  & 0.0076    & 0.0015    & 0.0061    & 0.0006    & 72324.7    & 59709.3    & 80154.4    & 62439.4 \\
						\hline    
					\end{tabular}
				\end{spacing}
				%\end{varwidth}}
			\end{sidewaystable}

		\subsection{Comparison with Competing Methods}\label{sec:comp2}
		
		In this part, we compare the PEL estimator $\hat{\btheta}_n$ defined as \eqref{eq:thetan_hat} with two other estimators: the standard EL estimator $\tilde{\btheta}_n$ defined as \eqref{eq:mele} and the relaxed EL (REL) estimator introduced by \cite{Shi2016}.
		The REL is  tailored for high-dimensional estimating equations, making it resilient to minor deviations from the equality constraints. 
		%All these methods are designed for the estimation of the true structural parameter $\btheta_0$. 
		Notice that the standard EL can only work for low-dimensional estimating equations. In line with our model specifications, where the two endogenous variables $u_{i,1}$ and $u_{i,2}$ are linked to IVs ($z_{i,1}, z_{i,2}$ and $z_{i,3}, z_{i,4}$, respectively) for each $i\in[n]$, we only use the first four moment conditions, that are related to the IVs $z_{i,1}, z_{i,2}, z_{i,3}$ and $z_{i,4}$, to produce the standard EL estimator $\tilde{\btheta}_n$. The computation of $\tilde{\btheta}_n$ can be implemented by the function {\tt gel} in the R-package {\tt gmm}. For both our PEL estimator $\hat{\btheta}_n$ and the REL estimator, we use all the $r$ moment conditions.

		For the selection of the tuning parameter in the REL estimator, we follow the recommendation in \cite{Shi2016}, using a consistent tuning parameter $0.5n^{-1/2}(\log r)^{1/2}$ throughout the simulations.
		Regarding the tuning parameter $\nu$ in our BPEL, we employ the Bayesian Information Criterion (BIC) defined as 
		\begin{align} \label{eq:BIC}
			{\rm BIC}(\nu) = \log \bigg\{\frac{1}{r} \bigg|\frac{1}{n} \sum_{i=1}^{n} \bfg(\bfx_{i};\hbthetan^{(\nu)})\bigg|_2^2 \bigg\} + |\mathcal{R}_n^{(\nu)}| n^{-1} \log n  
		\end{align}
		for its selection, where $\hat{\btheta}_{n}^{(\nu)}$ is the associated PEL estimator with tuning parameter $\nu$ calculated by our sampling algorithm, and $\mathcal{R}_n^{(\nu)}={\rm supp}\{\hat{\blambda}(\hbthetan^{(\nu)})\}$ with $\hat{\blambda}(\hbthetan^{(\nu)})=(\hat\lambda_1^{(\nu)},\ldots,\hat\lambda_r^{(\nu)})^\T=\arg\max_{\blambda\in\hat{\Lambda}_n(\hbthetan^{(\nu)})}f_n(\blambda;\hbthetan^{(\nu)})$ with $f_n(\blambda;\btheta)$ defined as \eqref{eq:fn1}. In practice, we set $\mathcal{R}_n^{(\nu)} = \{j \in [r]: |\hat\lambda_j^{(\nu)}|>10^{-6}\}$ and restrict $\nu$ in the interval $[0.05n^{-1/2}(\log r)^{1/2}, 0.75n^{-1/2}(\log r)^{1/2}]$.
		
		For the same $49$ initial points of the $200$ replications mentioned in Section \ref{sec:comp1}, we calculate the measure 
		$$
		{\rm MSE}_2 = \frac{1}{200 \times 49} \sum_{k=1}^{200} \sum_{l=1}^{49} |\check{\btheta}_{k}(l) - \bthetazero|_2^2
		$$  to evaluate the performance of different estimators, 	where $\check{\btheta}_{k}(l)$ is the related estimator in the $k$-th replication initiated from the $l$-th initial point. Table \ref{tab3} compares the measure ${\rm MSE}_2$ for the three estimators: the PEL estimator (MAMIS, M-H), the standard EL estimator, and the REL estimator. The results for M-H and MAMIS are derived based on the generated samples of size $3500$. 
		It becomes clear that BPEL demonstrates substantial performance improvements, clearly establishing its superiority over the other estimation methods. Particularly noteworthy is the effectiveness of MAMIS in addressing the challenges posed by nonlinear estimating equations, showing its promising performance.  %through its tuning parameter selection.

		\begin{table}[!htbp]
			\setlength\tabcolsep{3pt}
			\centering
			\caption{Comparison of BPEL and other estimators} \label{tab3} 
			\footnotesize
			\begin{spacing}{0.95}
				\begin{tabular}{cl|cccc|cccccccc} \hline   
					& & \multicolumn{4}{c|}{$\hbar(v) = v$} & \multicolumn{4}{c}{$\hbar(v)=\sin v$} \\ 
					$n$ & Methods & $r=80$ & $r=160$ & $r=320$ & $r=640$ & $r=80$ & $r=160$ & $r=320$ & $r=640$  \\
					\hline
					120 
					& MAMIS  & 0.0080    & 0.0096    & 0.0096    & 0.0114    & 0.0963    & 0.0829    & 0.0661    & 0.0620   \\
					& M-H     & 0.0086    & 0.0108    & 0.0118    & 0.0140    & 6.8664    & 7.4009    & 6.8831    & 7.2457   \\
					& EL  & 59.8762    & 58.7258    & 60.5130    & 60.0313    & 13.9942    & 14.0453    & 14.2512    & 14.4454   \\
					& REL  & 8.6108    & 8.8342    & 8.8781    & 9.2086    & 18.1845    & 18.5454    & 18.5858    & 18.9122   \\
					
					\hline  
					240 
					& MAMIS  & 0.0036    & 0.0044    & 0.0047    & 0.0055    & 0.1417    & 0.1200    & 0.1069    & 0.1136   \\
					& M-H     & 0.0039    & 0.0048    & 0.0051    & 0.0061    & 13.1962    & 13.6146    & 12.9027    & 13.0548   \\
					& EL  & 57.9585    & 57.3146    & 57.5111    & 57.6992    & 14.0103    & 13.8869    & 14.0533    & 14.3296   \\
					& REL  & 8.2303    & 8.1680    & 8.1674    & 7.8542    & 19.3646    & 19.6221    & 19.9443    & 20.0887   \\
					\hline       
				\end{tabular}
			\end{spacing}
		\end{table}

		\subsection{Additional Numerical Studies}

		We provide additional simulation studies in the supplementary material: Section  \ref{sec:add_prior} examines the impact of prior specification, Section \ref{sec:dgp-t}  evaluates the performance of our method using an alternative data generation process with data from a Student's $t$-distribution instead of a normal distribution, Section \ref{sec:add_fs} assesses the finite sample accuracy of the MCMC algorithms in approximating the posterior distribution,  Section \ref{sec:add_truth} compares the posterior distributions resulting from different Bayesian EL formulations, and Section \ref{sec:ABC_BSL} presents the comparison between our method and two competing methods: approximate Bayesian computation and Bayesian synthetic likelihood. Overall, our findings confirm the highly competitive performance of the proposed BPEL with the MCMC framework in terms of finite sample performance and accuracy in approximating posterior distributions.

		% adding a few lines of elaboration here.
		
		\section{Real Data Analysis} \label{sec:real data}
		
		%\subsection{Data and Model}

		International trade refers to the cross-border exchange of capital, commodities, and services between nations or regions. This type of trade typically constitutes a substantial portion of a country's gross domestic product (GDP). \cite{Eaton2011}, hereafter referred to as EKK, combined an empirical model with microeconomic principles to analyze France's international trade patterns. Additionally, Shi (2016) utilized EKK's microeconomic model to derive parameter estimates for Chinese exporting companies. In this section, we reexamine the dataset previously examined in \cite{Shi2016},  employing the proposed BPEL approach.

		The model proposed by EKK comprises five parameters denoted as $\boldsymbol{\theta} = (\theta_1, \ldots, \theta_5)^\T \in \boldsymbol{\Theta}$. The first component, $\theta_1$, characterizes the distribution of production efficiency among firms, with a higher $\theta_1$ indicating a larger proportion of manufacturers with lower efficiency. The second component, $\theta_2$, quantifies the cost associated with accessing a fraction of potential buyers, where a higher $\theta_2$ corresponds to lower costs. Parameters $\theta_3$, $\theta_4$ and $\theta_5$ represent the standard deviation of the demand shock, the standard deviation of the entry cost shock, and the correlation coefficient between these two shocks, respectively. Each firm is identified by the index $i \in [n]$, while countries are represented by the index $j \in \{0\}\cup[r]$, with $j = 0$ denoting the home country.

		According to the EKK's model, the sales of firm $i$ in country $j$ is
		$Z_{i,j}(\btheta; e^{(1)}_{i,j}, e^{(2)}_{i,j}, e^{(3)}_{i}) = \kappa \bar{Z}_j (1-\tau_{i,j})^{\theta_2/\theta_1} \tau_{i,j}^{-1/\theta_1} a^{(1)}_{i,j}/a^{(2)}_{i,j}$,
		where $a^{(1)}_{i,j}= \exp\{\theta_3(1-\theta_5^2)^{1/2}e^{(1)}_{i,j} + \theta_3\theta_5 e^{(2)}_{i,j}\}$, $a^{(2)}_{i,j} = \exp\{\theta_4e^{(2)}_{i,j}\}$, $\tau_{i,j} = \min\{1, e^{(3)}_{i} \bar{u}_{i} / \bar{u}_{i,j}\}$ and  
		\begin{equation*} 
			\kappa = \bigg(\frac{\theta_1}{\theta_1 -1} - \frac{\theta_1}{\theta_1 + \theta_2 -1}\bigg) \exp\bigg\{\frac{1}{2}\big(\theta_3 - \theta_1^2\theta_4^2 \big) + \theta_3\theta_4\theta_5(\theta_1-1) + \frac{1}{2} \theta_4(\theta_1-1)^2 \bigg\}  
			%a^{(1)}_{i,j}&= \exp\{\theta_3(1-\theta_5^2)^{1/2}e^{(1)}_{i,j} + \theta_3\theta_5 e^{(2)}_{i,j}\} \,,
			%~~~a^{(2)}_{i,j} = \exp\{\theta_4e^{(2)}_{i,j}\} \,, ~~~\tau_{i,j} = \min\{1, e^{(3)}_{i} \bar{u}_{j} / \bar{u}_{i,j}\} 
		\end{equation*}
		with $\bar{u}_{i,j} = (a^{(2)}_{i,j})^{\theta_1} N_j$ and $
		\bar{u}_{i} = \min\{\bar{u}_{i,0}, \max_{j\in[r]}\bar{u}_{i,j}\}$, and $(\bar{Z}_j, N_j)_{j\in\{0\}\cup[r]}$ are known constants. 
		Here $e^{(1)}_{i,j} \sim \mathcal{N}(0,1)$, $e^{(2)}_{i,j} \sim \mathcal{N}(0,1)$ and  $e^{(1)}_{i} \sim \mathcal{U}(0,1)$ are mutually independent. 
		Furthermore, $Z_{i,j}(\btheta; e^{(1)}_{i,j}, e^{(2)}_{i,j}, e^{(3)}_{i}) = 0$ means that the firm $i$ is kept outside of the country $j$. As a pertinent economic indicator of our interest,  the mean sale of all firms in country $j$ is
		$\mu_j(\btheta) = \mathbb{E}\{Z_{i,j}(\btheta; e^{(1)}_{i,j}, e^{(2)}_{i,j}, e^{(3)}_{i})\}$, 
		where the expectation is taken respect to the random variables $\{e^{(1)}_{i,j}, e^{(2)}_{i,j}, e^{(3)}_{i}\}$. The dataset is sourced from the Chinese administrative databases, encompassing a total of $n=6754$ firms and their export data to $r=126$ foreign destination countries in 2006. Leveraging this dataset, we obtain the $r$-dimensional estimating function $\bfg(\bfx_{i}; \btheta) =\{{g}_{1} (\bfx_{i}; \btheta),\ldots,{g}_{r} (\bfx_{i}; \btheta)\}^\T $, $i\in[n]$, with $\bfx_{i} = (x_{i,1},\ldots,x_{i,r})^\T$ and
		$g_j(\bfx_{i}; \btheta) = x_{i,j} - \mu_j(\btheta)$
		for any $j\in[r]$ and $\btheta \in \bTheta$, where $x_{i,j}$ is the sale of firm $i$ in country $j$ from this dataset ($j=0$ is not considered in this dataset).
		
\begin{figure}[!ht] 
			\centering
			\includegraphics[scale=0.35]{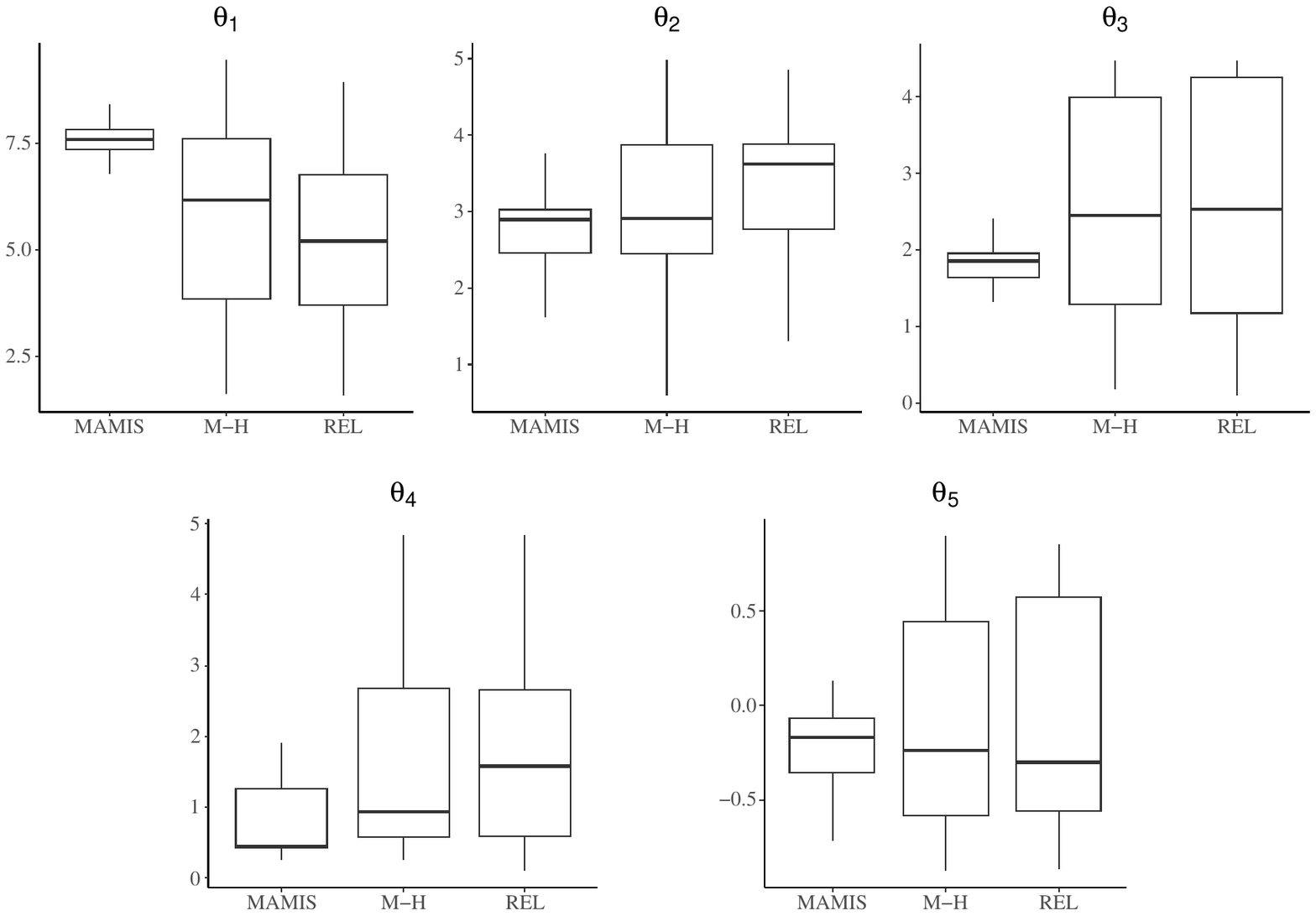}
			\caption{The box-plots of the estimated points.}
			\label{fig:2}
		\end{figure}

		Since the model is highly nonlinear with respect to $\btheta \in \bTheta$, resulting in no closed-form expression for $\mu_j(\btheta)$, we  approximate it via numerical simulation \citep{Eaton2011, Shi2016}. Specifically, in the estimation, we utilize the ``artificial data" for another $5n = 33770$ firms from the dataset. This involves simulating the entry decisions and sales across various countries for each of these artificial firms.  Subsequently, we calculate sample means to approximate $\mu_j(\btheta)$ for any $j\in \{0\}\cup[r]$ and $\btheta \in \bTheta$. We generated samples of size 3500 from the posterior distribution for 
		the BPEL. To select the tuning parameter $\nu$, we  employed the BIC as defined in \eqref{eq:BIC}. %We considered 10 equally spaced points within the interval $0.05 (6754^{-1}\log 126)^{1/2} \leq \nu \leq 0.75 (6754^{-1}\log 126)^{1/2}$ for this purpose. 
		For the parameter space $\boldsymbol{\Theta}$, we adopted a compact range of values, specifically $\boldsymbol{\Theta} = [1.5, 10] \times [0.5, 5] \times [0.1, 5] \times [0.1, 5] \times [-0.9, 0.9]$, which is consistent with the economic context and aligns with the study of \cite{Shi2016}. To initiate the analysis, we selected 15 samples uniformly distributed within the parameter space $\boldsymbol{\Theta}$. Figure \ref{fig:2} presents the box-plots of the corresponding 15 estimates  obtained by M-H and MAMIS from these initial values. The results for the REL with the same initial values are also included for comparative evaluation.
		
		It is evident that for all five parameters, MAMIS exhibits the smallest variations in the resulting estimates, whereas the variations of M-H and REL are relatively similar. This consistency with the findings in Sections \ref{sec:comp1} and \ref{sec:comp2} reaffirms the robustness of MAMIS when considering different initial points. Such robustness is desirable for conducting more in-depth analyses.
		For instance, let us take $\theta_5$ into consideration
		which represents the correlation coefficient between the demand shock and the entry cost shock. The sign of its estimate carries the key implication. 
		The 15 estimates of $\theta_5$ obtained by REL and M-H, from different initial values, fall within the ranges of $(-0.8738, 0.8507)$ and $(-0.8774, 0.8996)$, respectively. In contrast, the estimates of $\theta_5$ by MAMIS range in $(-0.7978, 0.1329)$, with the majority being negative, signaling a more assuring result.

		We then proceed to examine the specific moments selected by the respective methods. For REL, we employ the greedy algorithm outlined in Section 3.2 of \cite{Shi2016}. To assess the effectiveness of moment selection, we validate whether or not the top 10 trading partners of China in terms of export volume in this dataset, including the USA, Japan, Germany, etc., are either selected or partially selected. We find that, although REL selects at least some of these countries for 10 out of the 15 initial values, the number of selected countries does not exceed 3. In contrast, for 13 out of the 15 initial values, M-H identifies at least some of these countries, with 9 of them including more than 3. In the case of MAMIS, 13 out of the 15 initial values result in the identification of some of these countries, and all of them include more than 3 countries. %This observation underscores the advantageous moment selection performance of BPEL. 
		Additionally, the robustness of MAMIS with respect to the initial points provides enhanced reliability in this context.

		%Consistent with the findings in Section \ref{sec: IV model}, MAMIS exhibits the most stable performance among the methods, as evidenced by its minimal variations. The performances of M-H and REL are relatively similar.
		
		\section{Theoretical Analysis} \label{sec:theoretical_analy}

		We introduce some additional notation first. For simplicity, write  $\mathbb{E}_n(\cdot)=n^{-1}\sum_{i=1}^{n}\cdot$. % For any positive integer $q$, we write $[q]=\{1,\ldots,q\}$. 
		For a $q \times q$ symmetric matrix $\bfA$, denote by $\lambda_{\min}(\bfA)$ and $\lambda_{\max}(\bfA)$ the smallest and largest eigenvalues of $\bfA$, respectively. For a $q_1 \times q_2$ matrix $\bfB=(b_{i,j})_{q_1 \times q_2}$, let %$\bfB^\T$ be its transpose, 
		$|\bfB|_{\infty} = \max_{i \in [q_1], j \in [q_2]} |b_{i,j}|$ be the super-norm. % and $\|\bfB\|_2=\lambda^{1/2}_{\max}(\bfB^{ \otimes2})$ be the spectral norm with $\bfB^{\otimes2}=\bfB \bfB^\T$. 
		%Specifically, if $q_2=1$, we use $|\bfB|_{\infty}=\max_{i \in [q_1]} |b_{i,1}|$,  $|\bfB|_1=\sum_{i=1}^{q_1}|b_{i,1}|$ and $|\bfB|_2=(\sum_{i=1}^{q_1}b^2_{i,1})^{1/2}$ to denote the $L_{\infty}$-norm, $L_1$-norm and $L_2$-norm of the $q_1$-dimensional vector $\bfB$, respectively. 
		For the $r$-dimensional estimating function $\bfg(\cdot\,; \cdot)=\{g_1(\cdot\,;\cdot),\ldots,g_r(\cdot\,;\cdot)\}^\T$ and $p$-dimensional parameter $\btheta=(\theta_1,\ldots,\theta_p)^\T$, let $\nabla_{\btheta} \bfg (\cdot\,; \btheta)=\{\partial g_j(\cdot\,;\btheta)/\partial \theta_k\}_{j\in[r],k\in[p]}$, an $r\times p$ matrix, be the first-order partial derivative of $\bfg (\cdot\,; \btheta)$ with respect to $\btheta$. %Denote by $\nabla^2_{\btheta} g_j (\cdot\,; \btheta)=\{\partial^2 g_j(\cdot\,;\btheta)/\partial\theta_{k_1}\partial\theta_{k_2}\}_{k_1,k_2\in[p]}$, a $p\times p$ matrix, the second-order derivative of $g_j (\cdot\,; \btheta)$ with respect to $\btheta$. 
		Let $\bfV (\btheta)= \mE \{ \bfg (\bfx_{i}; \btheta)^{\otimes2} \}$ and $\bGamma(\btheta) =\mE\{\nabla_{\btheta} \bfg(\bfx_{i}; \btheta) \}$ for any $\btheta\in\bTheta$. For a given index set $\mathcal{F}$, let $|\mathcal{F}|$ be its cardinality. Denote by $\bfg_{\mathcal{F}}(\cdot\,; \cdot)$ the subvector of $\bfg(\cdot\,; \cdot)$ collecting the components indexed by $\mathcal{F}$. Let $\bfV_{\mathcal{F}} (\btheta)= \mE \{ \bfg_{\mathcal{F}} ( \bfx_{i}; \btheta)^{\otimes2} \}$ and $\bGamma_{\mathcal{F}}(\btheta) =\mE \{\nabla_{\btheta} \bfg_{\mathcal{F}}(\bfx_{i}; \btheta) \}$. Analogously, we also write $\bfa_{\mathcal{F}}$ as the corresponding subvector of vector $\bfa$. For any two probability measures $\mu$	and $\nu$, denote by $\mathcal{D}_\TV(\mu, \nu)$ the total variation distance between $\mu$ and $\nu$. 
		
		\subsection{Properties of the Penalized Empirical Likelihood Estimator }
		\label{sec:theory:1}
		
		%We commence our theoretical investigation by examining the asymptotic properties of the penalized empirical likelihood estimator $\hbthetan$ defined as \eqref{eq:thetan_hat}. %Subsequently, in Section \ref{sec:tv}, we establish the Bernstein-von Mises theorem for the posterior distribution $\pi^\dag(\btheta\,|\,\mathcal{X}_n)$, as defined in \eqref{eq:postdis}. Additionally, we provide the theoretical assurances for the performance of Algorithms \ref{alg3} and \ref{alg2} presented in Section \ref{sec:comp}.
		To investigate the asymptotic properties of  $\hat{\btheta}_n$ in \eqref{eq:thetan_hat},  
		we assume some regularity conditions.
		
		\begin{condition}\label{A.GIC}
			%  (global identification condition)
			%	$\bthetazero \in \bTheta$ is the unique
			%	solution to
			%	$
			%	\mE\{\bfg(\bfX_{i}; \btheta)\}=\bzero \,,
			%	$,
			%	Assume that
			For any $\varepsilon>0$, it holds that
			$$
			\inf_{\btheta \in \bTheta:\, |\btheta-\bthetazero|_\infty>\varepsilon} |\mE\{\bfg ( \bfx_{i}; \btheta)\}|_{\infty} \geq \Delta(\varepsilon)\,,$$ where $\Delta(\cdot)$ is a nonnegative function satisfying $\lim \inf _{\varepsilon \rightarrow 0^{+}} \varepsilon^{-1} \Delta(\varepsilon) \geq K_{1}$ for some universal constant $K_{1}>0$.
		\end{condition}
		
		\begin{condition}\label{A.ee}
			{\rm(a)} There exist universal constants $K_2>0$ and $\gamma>4$ such that $$
			\max_{j\in[r]}\mE\bigg\{ \sup_{\btheta \in \bTheta} |g_j ( \bfx_i; \btheta)|^{\gamma}\bigg\} \leq K_2$$ and $\sup_{\btheta \in \bTheta}\max_{j\in[r]} \mathbb{E}_n \{|g_j ( \bfx_{i}; \btheta)|^{\gamma}\} = O_{\rm p}(1)$. 
			{\rm(b)} There exist universal constants $0<K_3<K_4$ such that $K_3<\lambda_{\min}\{\bfV (\bthetazero)\} \leq \lambda_{\max}\{\bfV (\bthetazero)\} < K_4$.
			{\rm(c)} For any $\bfx$ and $j\in[r]$, $g_j(\bfx;\btheta)$ is twice continuously differentiable with respect to $\btheta \in \bTheta$ satisfying 
			$$
			\sup_{\btheta \in \bTheta} \max_{j \in [r], k \in [p]}  \mathbb{E}_n \bigg\{\bigg|\frac{\partial g_j(\bfx_{i}; \btheta)}{\partial \theta_{k}}\bigg|^2\bigg\}=O_{\rm p}(1)=
			\sup_{\btheta \in \bTheta} \max_{j \in [r], k_1, k_2 \in  [p]} \mathbb{E}_n \bigg\{\bigg|\frac{\partial^2 g_j(\bfx_{i}; \btheta)}{\partial \theta_{k_1} \partial \theta_{k_2}}\bigg|^2\bigg\}\,.$$
		\end{condition}

		Detailed discussion on Conditions \ref{A.GIC} and \ref{A.ee} are given in Section \ref{sec:discussioncond} of the  supplementary material. For any $\btheta \in \bTheta$, define
		$$
		\mathcal{M}_{\btheta}^*=\{j \in [r]: |\mathbb{E}_n\{ g_j (\bfx_{i}; \btheta)\} | \geq  C_*\nu \rho'(0^{+}) \}
		$$
		for some $C_* \in (0,1)$. We assume the existence of a sequence $\ell_n\rightarrow\infty$ such that 
		$$\mathbb{P}\bigg(\sup _{\btheta \in \bTheta:\, |\btheta-\bthetazero|_2 \leq c_n}|\mathcal{M}_{\btheta}^*|\leq \ell_n\bigg)\rightarrow1$$ as $n\rightarrow\infty$, 
		with some $c_n\rightarrow 0$ satisfying $\nu c_n^{-1} \rightarrow 0$. % Conditions \ref{A.GIC} and \ref{A.ee} are commonly used assumptions in the literature.  Condition \ref{A.GIC} is the identification condition for the unknown true parameter $\btheta_0$. A similar condition can be found in \cite{Shi2016} and \cite{Changetal_2018}. Condition \ref{A.ee}(b) requires the covariance matrix of $\bfg(\bfX_i;\btheta_0)$ behaves reasonably well. Conditions \ref{A.ee}(a) and \ref{A.ee}(c) impose the moments requirements on each estimating function $g_j(\cdot\,;\cdot)$ and its derivatives. If there exist functions $B_l(\cdot)$ with $\mathbb{E}\{B_l(\bfX_i)\}<\infty$, $l=1,2,3$, such that $|g_j(\bfX;\btheta)|^\gamma\leq B_1(\bfX)$, $|\partial g_j(\bfX;\btheta)/\partial\theta_k|^2\leq B_2(\bfX)$ and $|\partial^2g_j(\bfX;\btheta)/\partial\theta_{k_1}\partial\theta_{k_2}|^2\leq B_3(\bfX)$ for any $j\in[r]$ and $\btheta\in\bTheta$, then the second requirement in  Condition \ref{A.ee}(a) and the two requirements in Condition \ref{A.ee}(c) hold automatically. More generally, if there exist functions $B_{l,j}(\cdot)$ such that $|g_j(\bfX;\btheta)|^\gamma\leq B_{1,j}(\bfX)$, $|\partial g_j(\bfX;\btheta)/\partial\theta_k|^2\leq B_{2,j}(\bfX)$ and $|\partial^2g_j(\bfX;\btheta)/\partial\theta_{k_1}\partial\theta_{k_2}|^2\leq B_{3,j}(\bfX)$ for any $j\in[r]$ and $\btheta\in\bTheta$, and  $\max_{j\in[r]}\mathbb{E}\{B_{l,j}^m(\bfX_i)\}\leq Km!H^{m-2}$ for any $m\geq 2$ and $l=1,2,3$ with two universal constants $K, H>0$, it follows from Theorem 2.8 of \cite{Petrov_1995} that the second requirement in  Condition \ref{A.ee}(a) and the two requirements in Condition \ref{A.ee}(c) hold automatically provided $\log(rp)=o(n)$. In fact, the order $O_{\rm p}(1)$ in Conditions \ref{A.ee}(a) and \ref{A.ee}(c) can be replaced by $O_{\rm p}(\varpi_n)$ with some diverging sequence $\varpi_n$, and our main results remain valid. We use $O_{\rm p}(1)$ here for ease of presentation. To establish the consistency of the penalized empirical likelihood estimator $\hat{\btheta}_n$, Conditions \ref{A.GIC}, \ref{A.ee}(a) and \ref{A.ee}(b) are needed. Condition \ref{A.ee}(c) is needed for establishing the asymptotic normality of $\hat{\btheta}_n$.
		Proposition \ref{pro.cons} shows that $\hat{\btheta}_n$ is consistent to the true parameter $\btheta_0$, allowing $r$ growing exponentially with the sample size $n$.
		
		% Write $\alpha_n=n^{-1/2}(\log r)^{1/2}$. 
		
		\begin{proposition} \label{pro.cons}
			Let $P_{\nu}(\cdot) \in \mathscr{P}$ be a convex function for $\mathscr{P}$ defined as {\rm \eqref{eq:penalty}}. Under Conditions {\rm \ref{A.GIC}, \ref{A.ee}(a)} and {\rm\ref{A.ee}(b)}, if $\log r\ll n^{1/3}$ and $\ell_nn^{-1/2}(\log r)^{1/2}\ll\min \{\nu,n^{-1/\gamma}\}$, then the PEL estimator $\hbthetan$ defined as {\rm \eqref{eq:thetan_hat}} satisfies $|\hbthetan-\bthetazero|_{\infty}=O_{\rm p}(\nu)$.
		\end{proposition}
		
		Proposition \ref{pro.cons} establishes the consistency of the PEL estimator with diverging \(r\), incorporating the impact of the penalty function. In particular, the convergence rate of \(\hat{\btheta}_n\) is \(\nu\), provided that the tuning parameter \(\nu\) in \eqref{eq:thetan_hat} satisfies \(\nu \gg \ell_n n^{-1/2} (\log r)^{1/2}\). As a result, the convergence rate of \(\hat{\btheta}_n\) is slower than \(n^{-1/2}\), which can be viewed as the price paid for using the penalty in handling exponentially growing dimensionality \(r\).

		Recall $\rho(t;\nu)=\nu^{-1}P_\nu(t)$. For $P_\nu(\cdot)\in\mathscr{P}$ with $\mathscr{P}$ defined as \eqref{eq:penalty}, since $\rho'(0^{+};\nu)$ is independent of $\nu$, we write it as $\rho'(0^{+})$ for simplicity. Let $\mathcal{R}_n={\rm supp}\{\hat{\blambda}(\hbthetan)\}$ for the Lagrange multiplier  $\hat{\blambda}(\hbthetan)=(\hat\lambda_1,\ldots,\hat\lambda_r)^\T=\arg\max_{\blambda\in\hat{\Lambda}_n(\hbthetan)}f_n(\blambda;\hbthetan)$ with $f_n(\blambda;\btheta)$ defined as \eqref{eq:fn1}. Then $\hbthetan$ and $\hat{\blambda}(\hbthetan)$ satisfy the score equation: % $\nabla_{\blambda} f_n(\hat{\blambda};\hbthetan)=\bzero$, that is, 
		\begin{align} \label{eq:hat.eta}
			\bzero=\frac{1}{n} \sum_{i=1}^{n} \frac{\bfg(\bfx_{i};\hbthetan)}{1+\hat{\blambda}(\hbthetan)^\T \bfg(\bfx_{i};\hbthetan)}-\hat{\bseta} \,,
		\end{align}
		where $\hat{\bseta}=(\hat{\eta}_{1},\ldots ,\hat{\eta}_{r})^\T$ with $\hat{\eta}_j=\nu \rho' (|\hat{\lambda}_j|;\nu) \sgn(\hat{\lambda}_j)$ for $\hat{\lambda}_j\neq0$ and $\hat{\eta}_j \in [-\nu \rho'(0^{+}),\nu \rho'(0^{+})]$ for $\hat{\lambda}_j=0$. Here, an effective drastic dimension reduction is achieved with the associated sparse \(\hat{\blambda}(\hbthetan)\). The use of the penalty function \(P_\nu(\cdot)\) leads to \(\hat{\bseta}\) in \eqref{eq:hat.eta}, an extra term compared to that of the conventional EL. While \(P_\nu(\cdot)\) ensures the consistency of \(\hat{\btheta}_n\) as shown in Proposition \ref{pro.cons}, as we will show in Theorem \ref{th.norm} later, \(\hat{\bseta}\) leads to a bias of the PEL estimator \(\hat{\btheta}_n\).

		We further remark that while penalizing the Lagrange multiplier in our PEL does effectively achieve the selection of moments, its properties in terms of the validity of the selected moments remain an interesting research question. On one hand, it is reasonable to expect that under appropriate conditions and with a suitably chosen tuning parameter, our PEL may correctly select the set of valid moments. On the other hand, the major challenge lies in the ambiguity of defining valid moments when the corresponding moment functions are evaluated at broad candidate values of the model parameters rather than the truth. This consideration opens the door to a research question of its own interest in the context of moment selection that we are interested in investigating in our future research.

		To study the asymptotic distribution of $\hat{\btheta}_n$, we need the following regularity conditions.
		
		\begin{condition}\label{A.ee2}
			Let $\bfQ_{\mF}={\bGamma_{\mF}(\bthetazero)}^{\T, \otimes2}$ for any $\mathcal{F}\subset[r]$. There exist universal constants $0<K_5<K_6$ such that $K_5<\lambda_{\min}(\bfQ_{\mF}) \leq \lambda_{\max}(\bfQ_{\mF}) < K_6$ for any $\mathcal{F} $ with $p\leq |\mathcal{F}| \leq \ell_n$.
		\end{condition}		
		\begin{condition}\label{A.g_subgra.1}
			{\rm(a)}
			For the PEL estimator $\hbthetan$ defined as \eqref{eq:thetan_hat}, there exists a constant $\tilde{c} \in (C_*,1)$ such that
			$$
			\P\bigg[\bigcup_{j \in [r]}\{\tilde{c} \nu\rho'(0^+)\leq |\mathbb{E}_n\{g_j (\bfx_{i}; \hbthetan) \}| <\nu\rho'(0^+)\}\bigg] \rightarrow 0$$ as $n \rightarrow \infty$. 
			{\rm(b)} It holds that
			$$
			\P\bigg[\bigcup_{j \in \mathcal{R}_n^\c} \{|\hat{\eta}_j|=\nu \rho'(0^+)\}\bigg] \rightarrow 0
			$$
			as $n \rightarrow \infty$.
		\end{condition}
		
		Discussion of Conditions \ref{A.ee2} and \ref{A.g_subgra.1} are given in Section \ref{sec:discussioncond} of the supplementary material. %Condition \ref{A.ee2} is standard  in the literature. Due to the penalty imposed on the Lagrange multiplier $\blambda$ involved in the optimization \eqref{eq:thetan_hat}, the standard theoretical analysis of empirical likelihood cannot be applied here. Condition \ref{A.g_subgra.1}(a) is a technical assumption used to derive the convergence rate of the Lagrange multiplier $\hat{\blambda}(\hbthetan)=\arg\max_{\blambda\in\hat{\Lambda}_n(\hbthetan)}f_n(\blambda;\hbthetan)$ associated with $\hat{\btheta}_n$;  see the proof of Lemma \ref{l.lam.thetan.hat} in the supplementary material for details. Condition \ref{A.g_subgra.1}(b) requires that each $\hat{\eta}_j$ $(j\in \mathcal{R}_n^\c)$ lies in the interior of $[-\nu \rho'(0^{+}),\nu \rho'(0^{+})]$ with probability approaching one, which is realistic in practice. If the distribution function of the random variable $\hat{\eta}_j$ is continuous at $\pm \nu\rho'(0^+)$, we then have $\mathbb{P}\{|\hat{\eta}_j|=\nu\rho'(0^+)\}=0$. Condition \ref{A.g_subgra.1} makes sure that $\hat\blambda(\btheta)=\arg\max_{\blambda\in\hat{\Lambda}_n(\btheta)}f_n(\blambda;\btheta)$ is continuously differentiable at $\hbthetan$ with probability approaching one;  see Lemma \ref{l.envlope} in the supplementary material for details. 	
		Write $\widehat{\bfV}_{\Rn}(\hbthetan)=\mathbb{E}_n\{\bfg_{\Rn} ( \bfx_{i}; \hbthetan)^{\otimes2}\} $ and  $\widehat{\bGamma}_{\Rn}(\hbthetan) =\mathbb{E}_n\{ \nabla_{\btheta} \bfg_{\Rn}(\bfx_{i}; \hbthetan)\}$. Define
		\begin{align}
			& \widehat{\bfH}_\Rn=\{ \widehat{\bGamma}_\Rn(\hbthetan)^\T \widehat{\bfV}^{-1/2}_\Rn (\hbthetan)\}^{\otimes2} ~~\text{and}~~ \hat{\bpsi}_\Rn= \widehat{\bfH}_\Rn^{-1}\widehat{\bGamma}_\Rn(\hbthetan)^\T \widehat{\bfV}^{-1}_\Rn (\hbthetan) \hat{\bseta}_\Rn\,, \label{eq:bias}
		\end{align}
		where $\hat{\bseta}=(\hat{\eta}_{1},\ldots ,\hat{\eta}_{r})^\T$ is specified in \eqref{eq:hat.eta}. We assume $(r,\ell_n,\nu)$ satisfy the following restrictions: 
		\begin{align}\label{eq:rest1}
			&\log r\ll \min\{n^{1/3},n^{(\gamma-2)/(2\gamma)}\}\,,~~\ell_n\ll \min\{n^{(\gamma-2)/(3\gamma)}(\log r)^{-2/3},n^{1/5}(\log r)^{-2/5}\}\notag\\
			&~~~~~~~~~~~~~~~~~~~~~~~~\textrm{and}~~
			\ell_n n^{-1/2} (\log r)^{1/2} \ll \nu \ll \ell_n^{-1/4}n^{-1/4}\,. \end{align}
		The asymptotic distribution of $\hat{\btheta}_n$ is stated in Theorem \ref{th.norm}, where the bias term $\hat{\bpsi}_\Rn$ comes from the penalty function $P_{\nu}(\cdot)$ imposed on the Lagrange multiplier $\blambda$ in \eqref{eq:thetan_hat}.

		\begin{theorem}\label{th.norm}
			Let $P_{\nu}(\cdot) \in \mathscr{P}$ be convex with bounded second-order derivative around $0$, where $\mathscr{P}$ is defined as \eqref{eq:penalty}. Assume Conditions {\rm \ref{A.GIC}--\ref{A.g_subgra.1}} hold with $(r,\ell_n,\nu)$ satisfying \eqref{eq:rest1}. For any $\bft \in \R^{p}$ with $|\bft|_2=1$, the PEL estimator $\hbthetan$ defined as \eqref{eq:thetan_hat} satisfies
			$
			n^{1/2} \bft^\T \widehat{\bfH}_\Rn^{1/2}(\hbthetan-\bthetazero-\hat{\bpsi}_\Rn) \rightarrow \mathcal{N}(0,1)
			$ in distribution as $n\rightarrow\infty$, where $\widehat{\bfH}_\Rn$ and $\hat{\bpsi}_\Rn$ are defined in \eqref{eq:bias}.
		\end{theorem}
		
		Here, the estimated bias \(\hat{\bpsi}_\Rn\) can be easily calculated based on \eqref{eq:bias}. %; as mentioned earlier, this bias is due to the penalty on the Lagrange multiplier. 
		Theorem \ref{th.norm} indicates that, upon correcting the bias by subtracting it from $\hbthetan$, the resulting estimator $\hbthetan-\hat{\bpsi}_\Rn$ is \(n^{1/2}\)-consistent and asymptotically normal.

		\subsection{Properties of the Posterior 
			Distribution and Algorithms}
		\label{sec:tv}
		For the proposed BPEL, we establish the Bernstein-von Mises theorem for the posterior distribution $\pi^\dag(\btheta\,|\,\mathcal{X}_n)$, as defined in \eqref{eq:postdis}. Furthermore, we provide theoretical assurances for the performance of Algorithms \ref{alg3} and \ref{alg2} in Section \ref{sec:comp}.

		For any $\btheta \in \bTheta$, write $\mathcal{R}(\btheta)={\rm supp}\{\hat{\blambda}(\btheta)\}$ with $$\hat{\blambda}(\btheta)=\{\hat{\lambda}_1(\btheta),\ldots,\hat{\lambda}_r(\btheta)\}^\T=\arg\max_{\blambda\in\hat{\Lambda}_n(\btheta)}f_n(\blambda;\btheta)\,,$$ where $f_n(\blambda;\btheta)$ is defined as \eqref{eq:fn1}. Then $\btheta$ and $\hat{\blambda}(\btheta)$ satisfy the score equation:
		\begin{align} \label{eq:eta.hat}
			\bzero=\frac{1}{n} \sum_{i=1}^{n} \frac{\bfg(\bfx_{i};\btheta)}{1+\hat{\blambda}(\btheta)^\T \bfg(\bfx_{i};\btheta)}-\hat{\bseta}(\btheta) \,,
		\end{align}
		where $\hat{\bseta}(\btheta)=\{\hat{\eta}_{1}(\btheta),\ldots ,\hat{\eta}_{r}(\btheta)\}^\T$ with $\hat{\eta}_j(\btheta)=\nu \rho' \{|\hat{\lambda}_j(\btheta)|;\nu\} \sgn\{\hat{\lambda}_j(\btheta)\}$ for $\hat{\lambda}_j(\btheta) \neq 0$ and $\hat{\eta}_j(\btheta) \in [-\nu \rho'(0^{+}),\nu \rho'(0^{+})]$ for $\hat{\lambda}_j(\btheta)=0$. By the definition of the PEL estimator $\hat{\btheta}_n$, we have $f_n\{\hat{\blambda}(\btheta);\btheta\} \geq f_n\{\hat{\blambda}(\hat{\btheta}_n);\hat{\btheta}_n\}$ for any $\btheta\in\bTheta$. % with $f_n(\blambda;\btheta)$ defined as \eqref{eq:fn1}.
		To investigate the asymptotic properties of the posterior distribution $\pi^\dag(\btheta\,|\,\mathcal{X}_n)$ defined as  \eqref{eq:postdis}, we need to first study the asymptotic behavior of 
		$
		\aleph_n(\btheta)=f_n\{\hat{\blambda}(\btheta);\btheta\}-f_n\{\hat{\blambda}(\hat{\btheta}_n);\hat{\btheta}_n\}$ for $\btheta\in\bTheta$. 
		Given $\alpha_n=n^{-1/2}(\log r)^{1/2}$ and some $\beta_n$ satisfying $\ell_n^{1/2} \nu \ll \beta_{n} \ll \min\{ \ell_n^{-1}n^{-1/\gamma}, \nu^{2/3}\ell_n^{-2/3} n^{-1/(3\gamma)}\}$, we split the whole parameter space $\bTheta$ into three regions: 
		$\mathcal{C}_1=\{\btheta\in\bTheta:|\btheta-\hat{\btheta}_n|_2\leq \alpha_n\}$, $\mathcal{C}_2=\{\btheta\in\bTheta:\alpha_n<|\btheta-\hat{\btheta}_n|_2\leq \beta_n\}$ and $\mathcal{C}_3=\{\btheta\in\bTheta:|\btheta-\hat{\btheta}_n|_2>\beta_n\}$. Proposition \ref{pro.expan} in the supplementary material shows that the asymptotic behavior of $\aleph_n(\btheta)$ for $\btheta$ in these three regions are different.

		Investigating the asymptotic behavior of $\aleph_n(\btheta)$ calls some new technical arguments. Write 
		\begin{equation}\label{eq:tildef}
			\tilde{f}_n(\blambda;\btheta)=\frac{1}{n}\sum_{i=1}^{n} \log \{ 1+\blambda^\T \bfg(\bfx_i;\btheta)\}~~\textrm{and}~~\tilde{\blambda}(\btheta)=\arg\max_{\blambda\in\hat{\Lambda}_n(\btheta)}\tilde{f}_n(\blambda;\btheta)\,.   
		\end{equation}
		When $r$ is a fixed constant, we know $2n\tilde{f}_n\{\tilde{\blambda}(\btheta);\btheta\}$ is the conventional log-EL ratio in the literature. The asymptotic behavior of $2n\tilde{f}_n\{\tilde{\blambda}(\btheta);\btheta\}$ depends on the magnitude of $\mathbb{E}\{\bfg(\bfx_i;\btheta)\}$. More specifically, under some mild conditions, it holds that (i) $2n\tilde{f}_n\{\tilde{\blambda}(\btheta);\btheta\}$ is asymptotically chi-square distributed with degree of freedom $r$ if $|\mathbb{E}\{\bfg(\bfx_i;\btheta)\}|_2\ll n^{-1/2}$, (ii) $2n\tilde{f}_n\{\tilde{\blambda}(\btheta);\btheta\}$ converges to a noncentral chi-square distribution if $|\mathbb{E}\{\bfg(\bfx_i;\btheta)\}|_2\asymp n^{-1/2}$, and (iii) $2n\tilde{f}_n\{\tilde{\blambda}(\btheta);\btheta\}$ diverges to $\infty$ in probability if $|\mathbb{E}\{\bfg(\bfx_i;\btheta)\}|_2\gg n^{-1/2}$. See, for example, Proposition 1 and Theorem 1 of \cite{Changetal_2013_AOS} for such results with $r=1$. In comparison to $\tilde{f}_n(\blambda;\btheta)$ defined in \eqref{eq:tildef}, $f_n(\blambda;\btheta)$ involved in $\aleph_n(\btheta)$ includes a penalty term imposed on the Lagrange multiplier $\blambda$. This makes the standard technique for analyzing the conventional log-EL ratio inapplicable. To further establish the Bernstein-von Mises theorem for the posterior distribution $\pi^\dag(\btheta\,|\,\mathcal{X}_n)$ defined as \eqref{eq:postdis}, we assume the following regularity conditions.

		\begin{condition} \label{A.Pro2}
			{\rm (a)} 
			There exists a constant $\bar{c} \in (0,1)$ such that
			$$
			\P\bigg\{\sup_{\btheta \in \mathcal{C}_1}\max_{j \in \mathcal{R}(\btheta)^\c} |\hat{\eta}_j(\btheta)|\leq \bar{c} \nu \rho'(0^+)\bigg\} \rightarrow 1$$ as $n \rightarrow \infty$, 
			where $\hat \eta_j(\btheta)$ is specified in \eqref{eq:eta.hat}.
			{\rm (b)} There exists $\kappa_n$ satisfying $\max\{\ell_n^{1/2}n^{-1/2}(\log r)^{1/2},$ 
            $ \ell_n \beta_{n}^{3/2} n^{1/(2\gamma)}\} \ll \kappa_n \ll \nu$ such that  
			$$
			\P\bigg[\bigcup_{\btheta \in \mathcal{C}_2}\bigcup_{j \in \mathcal{R}_n} \{\nu\rho'(0^+)-\kappa_n < |\mathbb{E}_n\{g_j (\bfx_{i}; \btheta)\}| < \nu \rho'(0^+)+\kappa_n\}\bigg] \rightarrow 0$$ as $n \rightarrow \infty$.
			{\rm (c)} There exist universal constants $K_7, K_8>0$ such that 
			$$
			\P\bigg\{\inf_{\btheta\in\bTheta}\lambda_{\min}([\mathbb{E}_n\{\nabla_{\btheta}\bfg_{\mathcal{R}_n}(\bfx_i;\btheta)\}]^{\T,\otimes2})\geq K_7\bigg\} \rightarrow 1~~\textrm{and}~~\P\bigg[\sup_{\btheta \in \mathcal{C}_3}\lambda_{\max}\{\widehat\bfV_{\mathcal{R}_n}(\btheta)\} \leq K_8\bigg] \rightarrow 1$$
			as $n \rightarrow \infty$.
		\end{condition}
		
		\begin{condition}\label{A.prior}
			%	(prior)
			The prior density $\pi_{0}(\cdot)$ is continuously differentiable with bounded first-order derivatives and $\pi_{0}(\bthetazero)>0$. 
			%	(b) $-\log \pi_{0}(\btheta)$ is $L_{0}$-Lipschitz smooth and weakly convex in a neighborhood of $\bthetazero$.
		\end{condition}

		Discussion of Conditions \ref{A.Pro2} and \ref{A.prior} are given in Section \ref{sec:discussioncond} of the supplementary material.	Let $\Pi^{\dag}_n(\cdot)$ be the measure which admits the posterior distribution $\pi^{\dag}(\cdot\,|\,\mathcal{X}_n)$. Denote by  $\mathcal{N}_{\bmu,\bSigma}(\cdot)$ the Gaussian measure with mean $\bmu$ and covariance matrix $\bSigma$. To establish the Bernstein-von Mises theorem for the posterior distribution $\pi^\dag(\btheta\,|\,\mathcal{X}_n)$ as in Theorem \ref{th.TV}, we need to assume $(r,\ell_n,\nu)$ satisfy the following restrictions:
		\begin{align}\label{eq:restthm2}
			&\log r\ll n^{(\gamma-2)/(3\gamma)}\,,~~\ell_n \ll \min\{n^{(\gamma-2)/(9\gamma)}(\log r)^{-1/9},n^{1/3}(\log r)^{-1},n^{(\gamma-2)/(2\gamma)}(\log r)^{-3/2}\}\,,\notag\\
			&~~~~~~~~~~~~~~~~~~~~~~~~~\textrm{and}~~\ell_nn^{-1/2}(\log r)^{1/2} \ll \nu \ll \min\{\ell_n^{-7/2}n^{-1/\gamma},(\log r)^{-1}\}\,.
		\end{align}

		\begin{theorem}\label{th.TV}
			Let $P_{\nu}(\cdot) \in \mathscr{P}$ be convex and assume $\rho(t;\nu)=\nu^{-1}P_{\nu}(t)$ has bounded second-order derivative with respect to $t$ around $0$, where $\mathscr{P}$ is defined in {\rm \eqref{eq:penalty}}. Assume Conditions {\rm\ref{A.GIC}--\ref{A.prior}} hold with $(r,\ell_n,\nu)$ satisfying \eqref{eq:restthm2}. The posterior distribution $\pi^\dag(\btheta\,|\,\mathcal{X}_n)$ converges in total variation toward a Gaussian distribution $\mathcal{N}(\hat{\btheta}_n,n^{-1}\widehat{\bfH}_{\mathcal{R}_n}^{-1})$ in probability, that is,
			$\mathcal{D}_\TV(\Pi^{\dag}_n ,\, \mathcal{N}_{\hbthetan,n^{-1}\widehat{\bfH}_\Rn^{-1}})  \rightarrow 0$ in probability as $n\rightarrow \infty$,
			where $\hat{\btheta}_n$ is the PEL estimator in \eqref{eq:thetan_hat}, and $\widehat{\bfH}_\Rn$ is defined in \eqref{eq:bias}.	
		\end{theorem}
		
		Theorem \ref{th.TV} shows that $\pi^\dag(\btheta\,|\,\mathcal{X}_n)$ has a Gaussian limiting distribution and it concentrates on a $n^{-1/2}$-ball centered at the PEL estimator $\hbthetan$ of interest, which indicates that $\hat{\btheta}_n$  can be approximated by the mean of the posterior distribution $\pi^\dag(\btheta\,|\,\mathcal{X}_n)$. More specifically, as shown in Corollary \ref{cor.pos_mean}, the approximation error is of order smaller than $n^{-1/2}$. %\cite{ChibShinSimoni2018} establish a similar result for the Bayesian ETEL posterior in the low-dimensional setting. They show that the Bayesian ETEL posterior converges in total variation to a Gaussian distribution and concentrates on a $n^{-1/2}$-ball centered at the truth $\bthetazero$, which is different from ours. Such difference comes from the penalty term imposed on the Lagrange multiplier $\blambda$ when we define the posterior distribution $\pi^\dag(\btheta\,|\,\mathcal{X}_n)$ which is necessary for addressing the high-dimensional estimating function with $r\gg n$. Theorem \ref{th.TV} 

		%However, considering the consistency of $\hbthetan$ defined in \eqref{eq:thetan_hat}, the results of Theorem \ref{th.TV} are quite reasonable. %\cite{YangHe2012} consider the  Bayesian EL for quantile regression in the low-dimensional setting, where they show that the  Bayesian EL posterior density in a $n^{-1/2}$-neighborhood of the true parameter converges in probability to a Gaussian distribution centered at the maximized EL estimator. Their results are similar in spirit to ours.
		
		\begin{corollary} \label{cor.pos_mean}
			Under the conditions of Theorem {\rm \ref{th.TV}}, we have $|\mE_{\btheta \sim \pi^\dag}(\btheta) - \hbthetan|_\infty = o_{\rm p}(n^{-1/2})$, where $\hbthetan$ is the PEL estimator defined as {\rm \eqref{eq:thetan_hat}}, and $\mE_{\btheta \sim \pi^\dag}(\btheta)$ is defined in \eqref{eq:exp_pi}.  
		\end{corollary}

		%	\subsection{Theoretical guarantees for Algorithms \ref{alg3} and \ref{alg2} } \label{sec:ac}
		
		Theorems \ref{th.MH_alg} and \ref{th.MAMIS_alg} state the theoretical guarantees for Algorithms \ref{alg3} and \ref{alg2}, respectively.

		\begin{theorem}\label{th.MH_alg}
			For the density $\phi(\cdot\,|\,\cdot)$ of the proposal distribution in Algorithm {\rm \ref{alg3}}, we assume $\phi(\bvartheta\,|\,\btheta)$ is positive and continuous on $(\btheta, \bvartheta) \in \bTheta \times \bTheta$. Conditional on $\mathcal{X}_n$, for any $\btheta^0 \in \bTheta$ such that $\pi^\dag(\btheta^0\,|\,\mathcal{X}_n)>0$ with $\pi^\dag(\cdot\,|\,\mathcal{X}_n)$ defined as \eqref{eq:postdis}, it holds that $\mathcal{D}_\TV (\mathcal{T}^k_{\btheta^{0}},\, \Pi^{\dag}_n) \rightarrow 0$ as $k \rightarrow \infty$, where $\mathcal{T}^k_{\btheta^{0}}(\cdot)$ is the measure which admits the distribution of the Markov chain determined by Algorithm {\rm \ref{alg3}} at $k$-th step with initial point $\btheta^{0}$. Furthermore, conditional on $\mathcal{X}_n$, $|K^{-1}\sum_{k=1}^K \btheta^k - \mE_{\btheta \sim \pi^\dag}(\btheta)|_\infty \rightarrow 0$ almost surely as $K \rightarrow \infty$, where $\{\btheta^k\}_{k\geq1}$ are generated via Algorithm {\rm\ref{alg3}} with the initial point $\btheta^{0}$ satisfying $\pi^\dag(\btheta^0\,|\,\mathcal{X}_n)>0$.
		\end{theorem}
		
		%\subsection{Asymptotic properties for Algorithm } \label{sec:alg2}
		
		%Recall the MAMIS estimator $\widehat{\mE}_{\pi^\dag,K}(\btheta)$ defined in \eqref{eq:esi_Alg2}. In this part, we will establish the asymptotic properties of Algorithm \ref{alg2}. 

		\begin{theorem}\label{th.MAMIS_alg}
			For the density $\varphi(\cdot\,; \cdot)$ of the proposal distribution and the function $\bfh: \R^p \mapsto \R^s$ in Algorithm {\rm \ref{alg2}}, we assume $\varphi(\btheta\,; \bzeta)$ is positive and continuous on $(\btheta, \bzeta) \in \bTheta \times \R^s$ and $\sup_{\btheta \in \bTheta} |\bfh(\btheta)|_\infty \leq K_9$ for some universal constant $K_9>0$. Conditional on $\mathcal{X}_n$, if $\sum_{k=1}^{\infty}\exp(-CN_k) < \infty$ for any $C>0$, then   $|\widehat{\mE}_{\pi^\dag,K}(\btheta) - \mE_{\btheta \sim \pi^\dag}(\btheta)|_\infty \rightarrow 0$ almost surely as $K \rightarrow \infty$, where $\widehat{\mE}_{\pi^\dag,K}(\btheta)$ is the MAMIS estimator defined as \eqref{eq:esi_Alg2}.
		\end{theorem}

		\section{Discussion} \label{sec:dis}

		In this paper, we explore BPEL and demonstrate its promising performance using MCMC sampling as a competitive alternative to optimization in addressing EL problems. This framework has the potential for further advancements in several areas. To maintain focus and avoid digressions, we have confined our study to fixed-dimensional model parameters and exponentially growing moment conditions. However, there is significant interest in extending this approach to tackle variable and model selection using BPEL, which could accommodate high-dimensional sparse model parameters and potentially a continuum of moment conditions, as considered in \cite{Chau2017}. Incorporating specific priors in the context of concrete studies, particularly in high-dimensional problems, is another area of interest. Research in this direction presents additional challenges, especially in selecting appropriate priors, developing efficient sampling schemes, and conducting associated analyses.

		In the broader context of Bayesian methodology, approximate Bayesian computation (ABC) and Bayesian synthetic likelihood (BSL) are two competitive methods for handling situations where the likelihood is difficult to evaluate or intractable. ABC and BSL have been extensively compared in the literature. We demonstrate that the rationale of ABC integrates well with our BPEL method, achieving both accuracy and computational efficiency. Our Algorithm 2, inspired by ABC, uses importance weights for samples drawn from an alternative distribution to address challenging sampling situations. Empirical evidence shows promising performance, particularly in difficult cases. BSL leverages the limiting distribution, such as the normal distribution, to handle intractable probability distributions, with the advantage of easy sampling from the normal distribution. We view our BPEL as a compelling alternative to BSL: EL uses a multinomial likelihood that incorporates model information without requiring a fully specified parametric model, making it a competitive option when the full likelihood is intractable. 
		
		Furthermore, we foresee the use of more sophisticated sampling schemes in conjunction with PEL as highly valuable for addressing complex problems with specific considerations. Examples include the Hamiltonian MCMC method examined in \cite{Chaudhuri2017} and the variational Bayesian approach explored in \cite{Yu2023}. These avenues of research are part of our plans for future projects.
		
		% \bigskip
		
		% \noindent {\bf Acknowledgments}\\
		% The authors are grateful to the Co-Editor,  an Associate Editor and two referees for their helpful suggestions.

        % \bigskip

        % \noindent {\bf Supplementary material}\\
        % Supplementary material are available at Journal of the Royal Statistical Society: Series B.

% \bigskip 

% \noindent
% {\it Conflict of interest}:
% There are no relevant financial or nonfinancial competing interests to report.

% \bigskip

% \noindent
% {\bf Funding} \\
% Jinyuan Chang and Yuanzheng Zhu are supported in part by the National Natural Science Foundation grants (Grant No. 72125008, 72495122 and 71991472). 

% \bigskip

		\newpage
		
		\setcounter{equation}0
		\setcounter{section}0
		\setcounter{table}0
		\setcounter{figure}0
		
		\numberwithin{equation}{section}
		\renewcommand{\thesection}{\Alph{section}}
		\setcounter{page}{1}
		\renewcommand{\thepage}{S\arabic{page}}
		\renewcommand{\thetable}{S\arabic{table}}
		\renewcommand{\thefigure}{S\arabic{figure}}
		%\pagestyle{fancy}
		%\fancyhf{}
		\rhead{\bfseries\thepage}
		\lhead{\bfseries SUPPLEMENTARY MATERIAL}
		
		\begin{center}
			{\bf\Large Supplementary Material for ``Bayesian Penalized Empirical Likelihood and MCMC Sampling'' by Jinyuan Chang, Cheng Yong Tang and Yuanzheng Zhu}
		\end{center}
		
		In the sequel, we use the abbreviations ``w.p.a.1''  and ``w.r.t'' to denote, respectively, ``with probability approaching one'' and ``with respect to''. Let $C$, $\bar{C}$ and $\tilde{C}$ be generic positive finite constants that may be different in different uses. Let $\lfloor a \rfloor$ represent the largest integer not greater than $a \in \R$. For any positive integer $q$, we write $[q]=\{1,\ldots,q\}$. Denote by $I(\cdot)$ the indicator function. Let $\tr(\bfA)$ be the trace of a $q \times q$ matrix $\bfA=(a_{i,j})_{q \times q}$. For a $q \times q$ symmetric matrix $\bfA$, denote by $\lambda_{\min}(\bfA)$ and $\lambda_{\max}(\bfA)$ the smallest and largest eigenvalues of $\bfA$, respectively. For a $q_1 \times q_2$ matrix $\bfB=(b_{i,j})_{q_1 \times q_2}$, let  $\|\bfB\|_2=\lambda^{1/2}_{\max}(\bfB^{ \otimes2})$ be the spectral norm with $\bfB^{\otimes2}=\bfB \bfB^\T$. Specifically, if $q_2=1$, we use $|\bfB|_{\infty}=\max_{i \in [q_1]} |b_{i,1}|$,  $|\bfB|_1=\sum_{i=1}^{q_1}|b_{i,1}|$ and $|\bfB|_2=(\sum_{i=1}^{q_1}b^2_{i,1})^{1/2}$ to denote the $L_{\infty}$-norm, $L_1$-norm and $L_2$-norm of the $q_1$-dimensional vector $\bfB$, respectively. Given index sets $\mathcal{S}_1 \subset [q_1]$ and $\mathcal{S}_2 \subset [q_2]$, denote by $[\bfB]_{\mathcal{S}_1, \mathcal{S}_2}$ the $|\mathcal{S}_1| \times |\mathcal{S}_2|$ matrix that is obtained by extracting the rows of a $q_1 \times q_2$ matrix $\bfB$ indexed by $\mathcal{S}_1$ and columns indexed by $\mathcal{S}_2$. For simplicity and when no confusion arises, we use the notation $\bfg_i(\btheta)=\{g_{i,1}(\btheta),\ldots,g_{i,r}(\btheta)\}^\T$ as the equivalence to $\bfg(\bfx_i;\btheta)$, and denote by $\mathbb{E}_n(\cdot)=n^{-1}\sum_{i=1}^{n}\cdot$. Let $\bar{\bfg}(\btheta)=\mathbb{E}_n\{\bfg_i(\btheta)\}$, and write its $j$-th component as $\bar{g}_j(\btheta)=\mathbb{E}_n\{g_{i,j}(\btheta)\} $. Denote by $\nabla^2_{\btheta} g_{i,j} ( \btheta)=\{\partial^2 g_{i,j}(\btheta)/\partial\theta_{k_1}\partial\theta_{k_2}\}_{k_1,k_2\in[p]}$, a $p\times p$ matrix, the second-order derivative of $g_{i,j} (\btheta)$ with respect to $\btheta$. Let $\widehat{\bGamma}(\btheta) =\nabla_{\btheta} \bar{\bfg}(\btheta)$ and $\widehat{\bfV}(\btheta)=\mathbb{E}_n\{\bfg_i(\btheta)^{\otimes2} \}$. For a given set $\mathcal{F} \subset[r]$, we
		denote by $\bfg_{i,\mathcal{F}}(\btheta)$ the subvector of $\bfg_i(\btheta)$ collecting the components indexed by $\mathcal{F}$. Analogously, let $\bar{\bfg}_{\mathcal{F}}(\btheta)=\mathbb{E}_n\{\bfg_{i,\mathcal{F}}(\btheta)\}$, $\widehat{\bGamma}_{\mF}(\btheta) =\nabla_{\btheta} \bar{\bfg}_{\mF}(\btheta)$ and $\widehat{\bfV}_{\mF}(\btheta)=\mathbb{E}_n\{\bfg_{i,\mF}(\btheta)^{\otimes2} \} $. We also write $\bfa_{\mathcal{F}}$ as the corresponding subvector of vector $\bfa$.
		Recall $f_n(\blambda;\btheta)=\mathbb{E}_n[\log \{ 1+\blambda^\T \bfg_i (\btheta)\}]-\sum_{j=1}^{r} P_{\nu}(|\lambda_j|)$ and 
		$\hat\blambda(\btheta)=\{\hat\lambda_1(\btheta),\ldots,\hat\lambda_r(\btheta)\}^\T=\arg\max_{\blambda \in \wLambadn(\btheta)} f_n(\blambda;\btheta)$. Write $\mathcal{R}(\btheta)={\rm supp}\{\hat{\blambda}(\btheta)\}$, $\mathcal{R}_n={\rm supp}\{\hat{\blambda}(\hbthetan)\}$, and  $\mathcal{M}_{\btheta}^*=\{j\in [r]: |\bar{g}_j(\btheta)| \geq   C_*\nu \rho'(0^{+}) \}$ for some $C_*\in(0,1)$. Define $\mathcal{M}_{\btheta}(c)=\{ j\in[r]:|\bar{g}_j(\btheta)| \geq c\nu\rho'(0^{+}) \}$ for $c \in (C_*,1)$. Recall $\alpha_n=n^{-1/2}(\log r)^{1/2}$. 
		
		\section{Additional Numerical Results} 
		
		\subsection{The Impact of the Prior $\pi_0(\btheta)$} \label{sec:add_prior}
		In this section, we investigate the impact of the prior distribution $\pi_0(\btheta)$ in \eqref{eq:postdis} on our proposed Bayesian penalized EL methods in estimating the true parameter $\btheta_0$. More specifically, we adopt the data generation process outlined in Section \ref{sec.dgp} with the true parameter $\btheta_0=(0.5,0.5)^\T$, and consider three choices for the prior: 
		\begin{itemize}
			\item[(a)] the prior distribution $\mathcal{N}\{(-1, -1)^\T, 0.5^2\bfI_2\}$, which contains no correct information about the truth. 
			\item[(b)] the prior $\mathcal{N}\{(0.6, 0.6)^\T, 0.5^2\bfI_2\}$, which concentrates around the true value. 
			\item[(c)] the improper uniform prior, which provides no information about $\btheta_0$. 
		\end{itemize}
		For the $49$ initial points mentioned in Section \ref{sec:comp1}, we calculate the measure ${\rm MSE}_2$ defined in Section \ref{sec:comp2} to evaluate the performance of these estimators. The results for M-H and MAMIS are derived based on the generated samples of size $3500$. Table \ref{tab_prior} summarizes the performance of our proposed methods with such selected three priors. In particular, we have observed that when the prior is specified ``closer" to the truth, the resulting estimator has better performance in comparison to the one using a non-informative prior. Conversely, if a prior is specified ``further away" from the truth, the performance of the resulting estimator deteriorates and becomes less competitive.
		
		\begin{table}[!ht]
			\setlength\tabcolsep{3pt}
			\small
			\centering
			\captionsetup{justification=centering}
			\caption{Comparison of Bayesian penalized empirical likelihood under various priors and other estimators. All the reported results are based on $200$ replications. } \label{tab_prior}
			\begin{spacing}{1.4}
				\begin{tabular}{cl|ccc|ccccccc} \hline   
					& & \multicolumn{3}{c|}{$\hbar(v) = v$} & \multicolumn{3}{c}{$\hbar(v)=\sin v$} \\ 
					$n$ & Methods & $r=25$ & $r=50$ & $r=100$  & $r=25$ & $r=50$ & $r=100$ \\
					\hline
					120 
					& MAMIS + prior (a)  & 0.0155    & 0.0153    & 0.0142    & 0.6431    & 0.2683    & 0.2025    \\
					& MAMIS + prior (b)  & 0.0074    & 0.0071    & 0.0079    & 0.0630    & 0.0605    & 0.0525    \\
					& MAMIS + prior (c)  & 0.0090    & 0.0088    & 0.0091    & 0.3006    & 0.2026    & 0.1881    \\
					& M-H + prior (a)     & 0.0152    & 0.0155    & 0.0139    & 6.7466    & 6.2702    & 7.2373    \\
					& M-H + prior (b)     & 0.0078    & 0.0081    & 0.0093    & 0.0587    & 0.0565    & 0.0495    \\
					& M-H + prior (c)     & 0.0089    & 0.0092    & 0.0099    & 6.4315    & 6.1490    & 7.1785    \\
					& EL  & 60.0889    & 59.6533    & 59.5774    & 14.0751    & 14.1910    & 14.1341    \\
					& REL  & 8.5188    & 8.4780    & 8.5875    & 17.9534    & 18.1076    & 18.0692    \\
					\hline  
					240 
					& MAMIS + prior (a)  & 0.0065    & 0.0064    & 0.0078    & 0.2435    & 0.2270    & 0.1724    \\
					& MAMIS + prior (b)   & 0.0038    & 0.0032    & 0.0037    & 0.0609    & 0.0565    & 0.0503    \\
					& MAMIS + prior (c)   & 0.0041    & 0.0035    & 0.0041    & 0.1521    & 0.1188    & 0.1566    \\
					& M-H + prior (a)      & 0.0063    & 0.0060    & 0.0074    & 10.0399    & 12.9733    & 11.9599    \\
					& M-H + prior (b)     & 0.0041    & 0.0036    & 0.0042    & 0.0686    & 0.0531    & 0.0468    \\
					& M-H + prior (c)      & 0.0044    & 0.0038    & 0.0044    & 9.9999    & 12.5173    & 11.4562    \\
					& EL  & 58.5087    & 56.9165    & 57.9832    & 14.1962    & 13.9493    & 14.2135    \\
					& REL  & 8.2888    & 8.1843    & 8.1181    & 19.0072    & 19.1025    & 19.4474    \\
					\hline       
				\end{tabular}
			\end{spacing}
		\end{table}

		\subsection{Non-Gaussian Data Generation Process}\label{sec:dgp-t}
		
		In this section, we further validate the efficacy of our proposed methods by conducting some additional simulation studies. For the simulation examples considered in Sections \ref{sec:comp1} and \ref{sec:comp2}, we let all instrumental variables (IVs) $z_{i,j}$ be independently and identically distributed following the Student's $t$-distribution with three degrees of freedoms. For the $49$ initial points mentioned in Section \ref{sec:comp1}, we calculate the measure ${\rm MSE}_1$ defined in Section \ref{sec:comp1} to evaluate the performance of these methods. The results for Algorithms \ref{alg3} and \ref{alg2} based on sample sizes of 1500, 2500, and 3500 are denoted by (M-H-1, M-H-2,
		M-H-3) and (MAMIS-1, MAMIS-2, MAMIS-3), respectively. The results for $n=120$ and $n=240$ are presented in Tables \ref{tab_nonG_1} and \ref{tab_nonG_2}, respectively. Furthermore, we also compare the penalized empirical likelihood (EL) against the standard EL and the relaxed EL introduced by \cite{Shi2016_s} for the Student's $t$ IVs. For the $49$ initial points mentioned in Section \ref{sec:comp1}, we calculate the measure ${\rm MSE}_2$ defined in Section \ref{sec:comp2} to evaluate the performance of these estimators. The results are presented in Table \ref{tab_nonG_3}, where the results for M-H
		and MAMIS are derived based on the generated samples of size $3500$.
		
		Overall, these simulation results for the Student's $t$ IVs align closely with those listed in Sections \ref{sec:comp1} and \ref{sec:comp2}. These findings further validate the robustness and effectiveness of our proposed methods. 
		
		\begin{table}[!ht]
			\setlength\tabcolsep{3pt}
			\small
			\centering
			\captionsetup{justification=centering} 
			\caption{Comparison of Bayesian penalized empirical likelihood and optimization methods for Student's $t$ IVs. All the reported results are based on $200$ replications. ($n=120$)} \label{tab_nonG_1}
			\begin{spacing}{1.32}
				\begin{tabular}{cl|cccc|cccccccc} \hline   
					& & \multicolumn{4}{c|}{$\hbar(v) = v$} & \multicolumn{4}{c}{$\hbar(v)=\sin v$} \\ 
					$\nu$ & Methods & $r=80$ & $r=160$ & $r=320$ & $r=640$ & $r=80$ & $r=160$ & $r=320$ & $r=640$  \\
					\hline
					0.01 
					& MAMIS-1  & 0.0481    & 0.0376    & 0.0329    & 0.0316  
					& 9.8048    & 8.3530    & 7.7217    & 7.6243   \\
					& MAMIS-2  & 0.0019    & 0.0018    & 0.0038    & 0.0023  
					& 9.4126    & 7.6983    & 7.1222    & 7.0555   \\
					& MAMIS-3  & 0.0006    & 0.0016    & 0.0037    & 0.0022  
					& 9.0769    & 7.2673    & 6.6556    & 6.6122   \\
					& M-H-1     & 0.0034    & 0.0019    & 0.0039    & 0.0024  
					& 12.3905    & 12.0482    & 12.0274    & 12.0953   \\
					& M-H-2     & 0.0027    & 0.0018    & 0.0038    & 0.0023  
					& 12.4015    & 12.0261    & 12.0292    & 12.0976   \\
					& M-H-3     & 0.0026    & 0.0018    & 0.0038    & 0.0023  
					& 12.4149    & 12.0170    & 12.0302    & 12.0955   \\
					& optim    & 0.0930    & 0.0302    & 0.0152    & 0.0296  
					& 12.4895    & 12.4557    & 12.4459    & 12.5140   \\
					& nlm      & 0.0353    & 0.0272    & 0.0319    & 0.0471  
					& 108362.1    & 84673.0    & 78970.4    & 60638.1   \\
					\hline  
					0.03 
					& MAMIS-1  & 0.0351    & 0.0317    & 0.0306    & 0.0283  
					& 8.7835    & 7.5944    & 7.1610    & 7.1158   \\
					& MAMIS-2  & 0.0007    & 0.0012    & 0.0010    & 0.0013  
					& 8.1241    & 6.9087    & 6.4016    & 6.3359   \\
					& MAMIS-3  & 0.0004    & 0.0009    & 0.0009    & 0.0011  
					& 7.6302    & 6.4374    & 5.8968    & 5.7721   \\
					& M-H-1     & 0.0005    & 0.0010    & 0.0011    & 0.0012  
					& 12.5441    & 12.2105    & 12.0531    & 12.1171   \\
					& M-H-2     & 0.0005    & 0.0010    & 0.0010    & 0.0012  
					& 12.5179    & 12.1756    & 12.0277    & 12.1045   \\
					& M-H-3     & 0.0005    & 0.0009    & 0.0010    & 0.0012  
					& 12.5011    & 12.1475    & 11.9911    & 12.0974   \\
					& optim    & 0.0099    & 0.0025    & 0.0046    & 0.0040  
					& 12.5785    & 12.6160    & 12.5771    & 12.6170   \\
					& nlm      & 0.0014    & 0.0032    & 0.0061    & 0.0106  
					& 80729.2    & 80970.4    & 71142.2    & 83906.3   \\
					\hline    
					0.05 
					& MAMIS-1  & 0.0364    & 0.0327    & 0.0303    & 0.0251  
					& 7.7180    & 7.0629    & 6.5398    & 6.5053   \\
					& MAMIS-2  & 0.0009    & 0.0011    & 0.0010    & 0.0010  
					& 6.9023    & 6.1969    & 5.6669    & 5.6232   \\
					& MAMIS-3  & 0.0006    & 0.0006    & 0.0007    & 0.0009  
					& 6.3353    & 5.5782    & 5.0663    & 5.0143   \\
					& M-H-1     & 0.0006    & 0.0008    & 0.0007    & 0.0010  
					& 12.4400    & 12.1957    & 12.0535    & 12.1962   \\
					& M-H-2     & 0.0006    & 0.0007    & 0.0007    & 0.0009  
					& 12.3749    & 12.1346    & 11.9997    & 12.1664   \\
					& M-H-3     & 0.0006    & 0.0007    & 0.0006    & 0.0009  
					& 12.3236    & 12.0810    & 11.9597    & 12.1424   \\
					& optim    & 0.0028    & 0.0027    & 0.0069    & 0.0051  
					& 12.6724    & 12.6704    & 12.6800    & 12.7177   \\
					& nlm      & 0.0023    & 0.0036    & 0.0022    & 0.0137  
					& 58834.4    & 51792.2    & 55910.1    & 70604.6   \\
					\hline   
					0.07 
					& MAMIS-1  & 0.0345    & 0.0317    & 0.0278    & 0.0286  
					& 6.7844    & 6.4945    & 6.0224    & 6.0776   \\
					& MAMIS-2  & 0.0010    & 0.0008    & 0.0008    & 0.0008  
					& 5.7821    & 5.5112    & 5.0028    & 5.1639   \\
					& MAMIS-3  & 0.0007    & 0.0007    & 0.0007    & 0.0007  
					& 5.0577    & 4.7797    & 4.3840    & 4.5568   \\
					& M-H-1     & 0.0008    & 0.0008    & 0.0008    & 0.0008  
					& 12.3454    & 12.0507    & 11.9962    & 12.1738   \\
					& M-H-2     & 0.0008    & 0.0007    & 0.0007    & 0.0008  
					& 12.2338    & 11.9475    & 11.8991    & 12.0726   \\
					& M-H-3     & 0.0008    & 0.0007    & 0.0007    & 0.0008  
					& 12.1397    & 11.8704    & 11.8239    & 12.0118   \\
					& optim    & 0.0002    & 0.0002    & 0.0053    & 0.0006  
					& 12.7150    & 12.7020    & 12.7081    & 12.7720   \\
					& nlm      & 0.0002    & 0.0024    & 0.0013    & 0.0035  
					& 51010.8    & 61869.2    & 53782.9    & 90751.1   \\
					\hline    
				\end{tabular}
			\end{spacing}
		\end{table}
		
		\begin{table}[!ht]
			\setlength\tabcolsep{3pt}
			\small
			\centering
			\captionsetup{justification=centering} 
			\caption{Comparison of Bayesian penalized empirical likelihood and optimization methods for Student's $t$ IVs. All the reported results are based on $200$ replications. ($n=240$)} \label{tab_nonG_2}
			\begin{spacing}{1.32}
				\begin{tabular}{cl|cccc|cccccccc} \hline   
					& & \multicolumn{4}{c|}{$\hbar(v) = v$} & \multicolumn{4}{c}{$\hbar(v)=\sin v$} \\ 
					$\nu$ & Methods & $r=80$ & $r=160$ & $r=320$ & $r=640$ & $r=80$ & $r=160$ & $r=320$ & $r=640$  \\
					\hline
					0.01 
					& MAMIS-1  & 0.0616    & 0.0346    & 0.0316    & 0.0287  
					& 11.0028    & 8.9620    & 7.8281    & 7.5163   \\
					& MAMIS-2  & 0.0120    & 0.0003    & 0.0007    & 0.0006  
					& 10.7205    & 8.5181    & 7.2842    & 7.0322   \\
					& MAMIS-3  & 0.0062    & 0.0003    & 0.0007    & 0.0006  
					& 10.5070    & 8.1612    & 6.8805    & 6.6269   \\
					& M-H-1     & 0.1072    & 0.0004    & 0.0008    & 0.0007  
					& 12.4824    & 12.0179    & 11.9517    & 11.8182   \\
					& M-H-2     & 0.0988    & 0.0003    & 0.0008    & 0.0007  
					& 12.5009    & 12.0155    & 11.9503    & 11.8152   \\
					& M-H-3     & 0.0953    & 0.0003    & 0.0008    & 0.0007  
					& 12.5199    & 12.0181    & 11.9459    & 11.8169   \\
					& optim    & 0.4827    & 0.0582    & 0.0089    & 0.0047  
					& 12.6498    & 12.4680    & 12.4434    & 12.4251   \\
					& nlm      & 0.1528    & 0.0212    & 0.0178    & 0.0171  
					& 118679.2    & 138444.6    & 101989.6    & 90386.49   \\
					\hline  
					0.03 
					& MAMIS-1  & 0.0567    & 0.0295    & 0.0263    & 0.0192  
					& 8.9734    & 8.4032    & 7.4988    & 7.2192   \\
					& MAMIS-2  & 0.0027    & 0.0003    & 0.0005    & 0.0006  
					& 8.3184    & 7.8377    & 6.8258    & 6.5900   \\
					& MAMIS-3  & 0.0013    & 0.0002    &  0.0002    & 0.0005  
					& 7.8137    & 7.3651    & 6.3576    & 6.1302   \\
					& M-H-1     & 0.0187    & 0.0003    &  0.0003    & 0.0006  
					& 12.3586    & 12.0990    & 11.9399    & 11.8383   \\
					& M-H-2     & 0.0182    &0.0003     &  0.0003    & 0.0006  
					& 12.3348    & 12.1100    & 11.9372    & 11.8260   \\
					& M-H-3     & 0.0174    & 0.0002    &  0.0003    & 0.0006  
					& 12.3190    & 12.1200    & 11.9401    & 11.8233   \\
					& optim    & 0.1150    & 0.0038    & 0.0049    & 0.0143  
					& 12.5624    & 12.5882    & 12.5742    & 12.6203   \\
					& nlm      & 0.0349    & 0.0017    & 0.0065    & 0.0110  
					& 65201.5    & 79473.0    & 78277.7    & 89595.9   \\
					\hline    
					0.05 
					& MAMIS-1  & 0.0440    & 0.0306    & 0.0259    & 0.0197  
					& 7.6532    & 7.6072    & 7.4896    & 6.7806   \\
					& MAMIS-2  & 0.0013    & 0.0005    & 0.0005    & 0.0003  
					& 6.7688    & 6.8467    & 6.7353    & 6.0273   \\
					& MAMIS-3  & 0.0006    & 0.0002    &  0.0002    & 0.0002  
					& 6.1065    & 6.3622    & 6.2512    & 5.4572   \\
					& M-H-1     & 0.0004    &  0.0004    & 0.0003    & 0.0003  
					& 12.3166    & 12.1641    & 12.0266    & 12.0732   \\
					& M-H-2     & 0.0004    &  0.0003    & 0.0003    & 0.0003  
					& 12.2894    & 12.1465    & 12.0041    & 12.0479   \\
					& M-H-3     & 0.0003    &  0.0003    & 0.0003    & 0.0002  
					& 12.2555    & 12.1339    & 11.9867    & 12.0341   \\
					& optim    & 0.0340    & 0.0001    & 0.0011    & 0.0098  
					& 12.5571    & 12.6183    & 12.6591    & 12.7411   \\
					& nlm      & 0.0120    & 0.0007    & 0.0038    & 0.0228  
					& 73643.9    & 70545.1    & 62023.9    & 71602.1   \\
					\hline   
					0.07 
					& MAMIS-1  & 0.0385    & 0.0296    & 0.0243    & 0.0227  
					& 6.6666    & 7.0442    & 6.6153    & 6.3235   \\
					& MAMIS-2  & 0.0010    & 0.0011    & 0.0005    & 0.0004  
					& 5.5197    & 6.0508    & 5.7399    & 5.4336   \\
					& MAMIS-3  & 0.0006    & 0.0005    & 0.0003    & 0.0003  
					& 4.7946    & 5.4071    & 5.2108    & 4.8304   \\
					& M-H-1     & 0.0007    & 0.0006    & 0.0004    & 0.0004  
					& 12.3694    & 12.3079    & 12.1525    & 12.1026   \\
					& M-H-2     & 0.0007    & 0.0006    & 0.0004    & 0.0003  
					& 12.2878    & 12.2706    & 12.1128    & 12.0599   \\
					& M-H-3     & 0.0006    & 0.0006    & 0.0004    & 0.0003  
					& 12.2231    & 12.2467    & 12.0801    & 12.0354   \\
					& optim    & 0.0106    & 0.0001    & 0.0001    & 0.0015  
					& 12.5953    & 12.6723    & 12.7323    & 12.8134   \\
					& nlm      & 0.0001    & 0.0019    & 0.0079    & 0.0065  
					& 53267.2    & 58029.2    & 72518.9    & 71135.01   \\
					\hline    
				\end{tabular}
			\end{spacing}
		\end{table}
		
		\begin{table}[!ht]
			\setlength\tabcolsep{3pt}
			\small
			\centering
			\captionsetup{justification=centering}
			\caption{Comparison of Bayesian penalized empirical likelihood and other estimators for Student's $t$ IVs. All the reported results are based on $200$ replications.} \label{tab_nonG_3} 
			\begin{spacing}{1.4}
				\begin{tabular}{cl|cccc|cccccccc} \hline   
					& & \multicolumn{4}{c|}{$\hbar(v) = v$} & \multicolumn{4}{c}{$\hbar(v)=\sin v$} \\ 
					$n$ & Methods & $r=80$ & $r=160$ & $r=320$ & $r=640$ & $r=80$ & $r=160$ & $r=320$ & $r=640$  \\
					\hline
					120 
					& MAMIS  & 0.0034    & 0.0055    & 0.0071    & 0.0090    & 0.0346    & 0.0444    & 0.0374    & 0.0326   \\
					& M-H     & 0.0036    & 0.0058    & 0.0074    & 0.0093    & 10.7060    & 10.7426    & 10.0286    & 10.1199   \\
					& EL  & 62.7633    & 63.2831    & 62.7032    & 62.0324    & 11.6170    & 11.6186    & 11.5749    & 11.5220   \\
					& REL  & 9.7379    & 11.0635    & 12.3910    & 14.5793    & 20.3810    &  20.1763    & 20.2257    & 20.5747   \\
					
					\hline  
					240 
					& MAMIS  & 0.0017    & 0.0022    & 0.0028    & 0.0040    & 0.2323    & 0.2094    & 0.2222    & 0.2299   \\
					& M-H     & 0.0018    & 0.0023    & 0.0030    & 0.0041    & 10.6483    & 10.1344    & 10.3399    & 10.6642   \\
					& EL  & 60.9124    & 62.6785    & 60.5122    & 61.2036    & 11.3735    & 11.3767    & 11.3812    & 11.3883   \\
					& REL  & 7.9803    & 8.5840    & 9.5500    & 10.1348    & 22.3937    & 22.3504    & 22.3768    & 22.5482   \\
					\hline       
				\end{tabular}
			\end{spacing}
		\end{table}
		
		\subsection{The Normal Approximation in Finite Samples} \label{sec:add_fs}
		
		In this section, we conduct several numerical simulations to examine the performance of the normal approximation stated in Theorem \ref{th.TV} to the posterior distribution $\pi^{\dag}(\btheta\,|\,\mathcal{X}_n)$ defined as \eqref{eq:postdis} in finite samples. More specifically, we adopt the data generation process outlined in Section \ref{sec.dgp} with linear link function $\hbar(v)=v$. As described in Section \ref{sec:comp1}, we identify the true global minima $\hat{\btheta}_n$ defined as \eqref{eq:thetan_hat} through exhaustive search. Subsequently, we calculate its asymptotic covariance matrix $n^{-1}\widehat{\bfH}^{-1}_{\mathcal{R}_n}$ with $\widehat{\bfH}_{\mathcal{R}_n}$ defined as \eqref{eq:bias}. We then generate $5000$ samples, respectively, from the Gaussian distribution $\mathcal{N}(\hat{\btheta}_n, n^{-1}\widehat{\bfH}^{-1}_{\mathcal{R}_n})$ and the posterior distribution $\pi^{\dag}(\btheta\,|\,\mathcal{X}_n)$ defined as \eqref{eq:postdis}. To generate samples from $\mathcal{N}(\hat{\btheta}_n, n^{-1}\widehat{\bfH}^{-1}_{\mathcal{R}_n})$, we use the function {\tt mvrnorm} in the R-package {\tt MASS}. To generate samples from $\pi^{\dag}(\btheta\,|\,\mathcal{X}_n)$, we use Algorithm \ref{alg3} with the burn-in period of $1000$ iterations. Based on these samples, we compute the Wasserstein distance between the two distributions $\mathcal{N}(\hat{\btheta}_n, n^{-1}\widehat{\bfH}^{-1}_{\mathcal{R}_n})$ and $\pi^{\dag}(\btheta\,|\,\mathcal{X}_n)$ using the function {\tt wasserstein} in the R-package {\tt transport}. Figure \ref{fig:fs} below illustrates the average Wasserstein distance between the two distributions across different sample sizes $n$ under $500$ replications. It can be observed that, the two distributions exhibit a relatively large differences at smaller sample sizes, with this distance diminishing notably as the sample size increases. 
		\begin{figure}[!ht] 
			\begin{minipage}[t]{0.5\linewidth}
				\centering
				\includegraphics[scale=0.45]{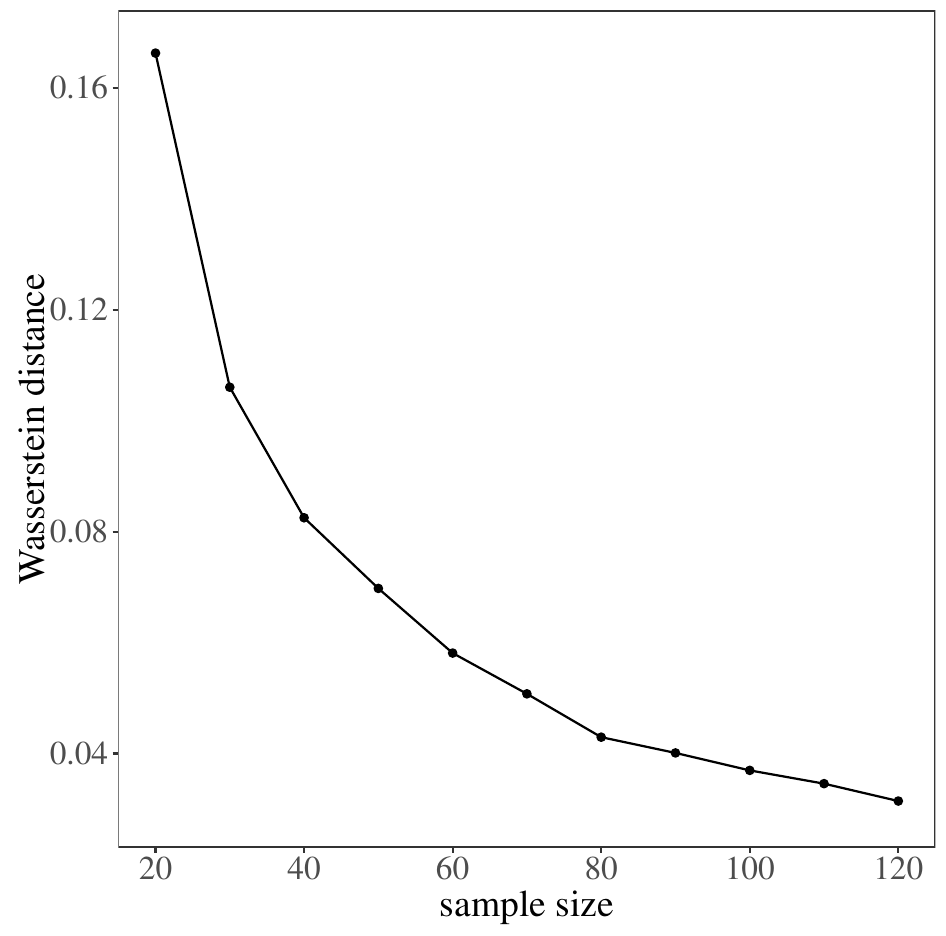}
				\centerline{(a)}
			\end{minipage}%
			\begin{minipage}[t]{0.5\linewidth}
				\centering
				\includegraphics[scale=0.45]{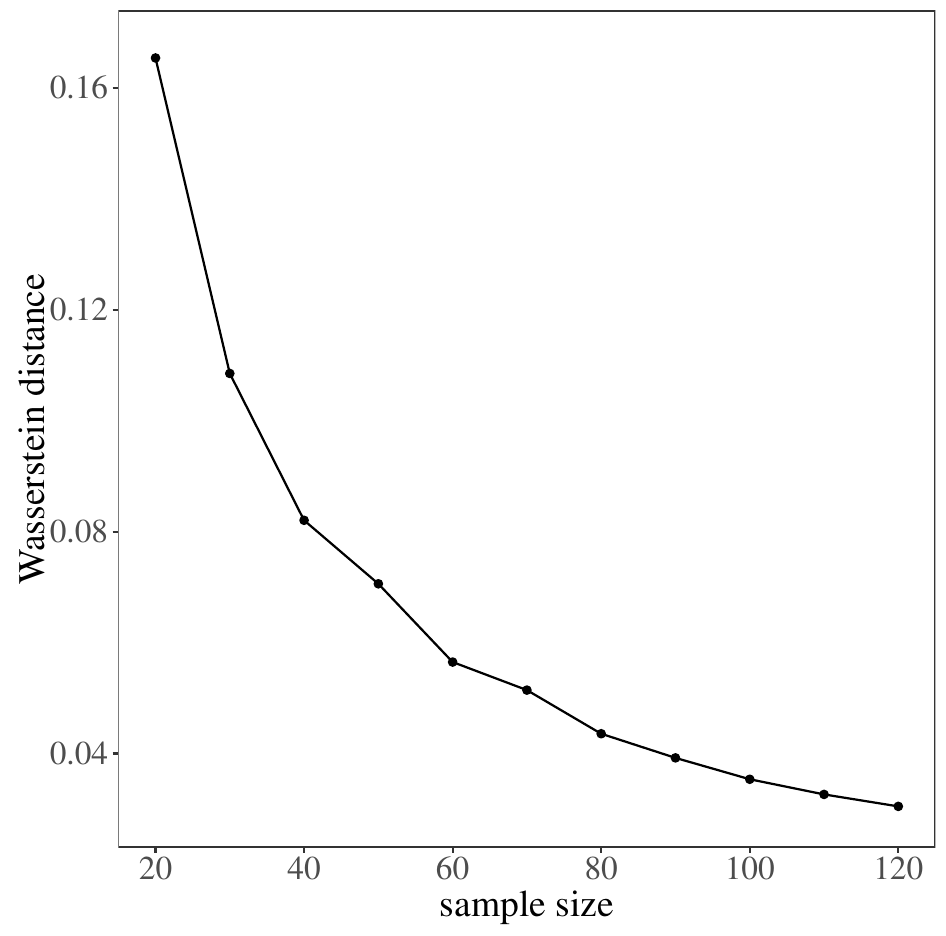}
				\centerline{(b)}
			\end{minipage}
			\caption{The average Wasserstein distance between the Gaussian distribution and the posterior under $500$ replications. (a) $r=80$. (b) $r=160$.}
			\label{fig:fs}
		\end{figure}
		
		We further validate the efficacy of our proposed methods in approximating the true global minima $\hat{\btheta}_n$ defined as \eqref{eq:thetan_hat}. Table \ref{tab_fs} below presents the measure ${\rm MSE} = \frac{1}{500}\sum_{k=1}^{500} |\check{\btheta}_{k} - \hat{\btheta}_n|_2^2$ across different sample sizes $n$, where $\check{\btheta}_{k}$ is the mean of the $5000$ samples drawn from the posterior distribution $\pi^{\dag}(\btheta\,|\,\mathcal{X}_n)$ in the $k$-th replication. It is evident that with small sample size $n$, although the normal approximation stated in Theorem \ref{th.TV} to the posterior distribution $\pi^{\dag}(\btheta\,|\,\mathcal{X}_n)$ may not work very well, our method can still effectively approximate the true global minimum $\hat{\btheta}_n$. As the sample size $n$ increases, the accuracy of the approximation exhibits a substantial improvement.  
		\begin{table}[!ht]
			\setlength\tabcolsep{3pt}
			\small
			\centering
			\captionsetup{justification=centering}
			\caption{The results of Bayesian penalized empirical likelihood based on $500$ replications.} \label{tab_fs}
			\begin{spacing}{1.4}
				\begin{tabular}{c|cccccc} \hline   
					$r$ & $n=20$ & $n=40$  & $n=60$ & $n=80$ & $n=100$ & $n=120$ \\
					\hline
					80 
					& 0.0221 & 0.0058 & 0.0031 & 0.0015 & 0.0011 & 0.0007    \\
					160 
					& 0.0216 & 0.0054 & 0.0028 & 0.0016 & 0.0010 & 0.0007   \\
					\hline       
				\end{tabular}
			\end{spacing}
		\end{table}

		\subsection{Comparison of the Performance of Posteriors Derived by Different Methods} \label{sec:add_truth}
		In this section, we conduct some additional numerical studies to further compare the performance of posteriors derived by different methods. Assume the observations $\bfx_1,\ldots,\bfx_n$ are drawn independently from the distribution $F(\btheta_0)$ with some unknown parameter $\btheta_0$.  The likelihood function  admits the form $L(\btheta)=\prod_{i=1}^n f(\bfx_i;\btheta)$ where $f(\cdot\,;\btheta)$ is the density function of $F_n(\btheta)$. Write $\mathcal{X}_n=\{\bfx_1,\ldots,\bfx_n\}$. Let $\pi_0(\cdot)$ represent a prior distribution for $\btheta$. Then the traditional posterior is given by $\pi^{\rm L}(\btheta\,|\,\mathcal{X}_n)\propto \pi_{0}(\btheta) \times L(\btheta)$. To estimate the unknown parameter $\btheta_0$, we can also identify it by $\mathbb{E}\{\mathbf{g}(\bfx_i;\btheta_0)\}=\bzero$ with some $r$-dimensional estimating function $\mathbf{g}(\cdot\,;\cdot)$.
		For given estimating function $\mathbf{g}(\cdot\,; \cdot)$, we can define the empirical likelihood ${\rm EL}(\btheta)$ and penalized empirical likelihood ${\rm PEL}_\nu(\btheta)$, respectively, as \eqref{eq:elnew} and \eqref{eq:pel} in the main paper. Hence, the associated EL-based posterior distribution $\pi^{\rm EL}(\btheta\,|\,\mathcal{X}_n)$ and PEL-based posterior distribution $\pi^{\dag}(\btheta\,|\,\mathcal{X}_n)$ satisfy $\pi^{\rm EL}(\btheta\,|\,\mathcal{X}_n)\propto \pi_{0}(\btheta) \times {\rm EL}(\btheta) $ and $\pi^{\dag}(\btheta\,|\,\mathcal{X}_n)\propto \pi_{0}(\btheta) \times {\rm PEL}_\nu(\btheta)$. We select $\pi_0(\cdot)$ as the improper uniform prior and compare these three posteriors via the following two models:
		\begin{itemize}
			\item Model ${\rm I}$: Let $x_1, \ldots, x_n$ be independent and identically distributed observations from the normal distribution $\mathcal{N}(0,\theta_0^2)$ with $\theta_0 = 1$. We can select $\mathbf{g}(\cdot\,; \cdot) = \{g_1(\cdot\,; \cdot), \ldots, g_r(\cdot\,; \cdot)\}^\T\in{\mathbb R}^r$ with $g_j(x_{i}; \theta) = x_i^{2j} - (2j-1)!! \theta^{2j}$.  
			
			\item Model ${\rm II}$: Consider the linear regression model $y_i = z_i \theta_0 + e_i, ~i \in [n]$, where $\theta_0 = 0.5$, $z_i \sim \mathcal{N}(0,1)$ is the covariate variable and $e_i \sim \mathcal{N}(0,0.9)$ is the error orthogonal to $z_i$. We can select $\mathbf{g}(\cdot\,; \cdot) = \{g_1(\cdot\,; \cdot), \ldots, g_r(\cdot\,; \cdot)\}^\T\in{\mathbb R}^r$ with $g_j(\bfx_{i}; \theta) = (y_i - z_i \theta) z_i^j$ and $\bfx_{i} = (y_i, z_i)^\T$.

				\item Model ${\rm III}$: Consider the generalized linear  model where the covariates $z_i, i \in [n]$, are drawn independently from the gamma distribution with shape parameter $2$ and rate parameter $1$. The response variables $y_i, i \in [n]$, are generated from the Bernoulli distribution such that $\mathbb{P}(y_i = 1\,|\,z_i) = \exp(z_i \theta_0)/\{1+\exp(z_i \theta_0)\}$ with the true parameter $\theta_0 = 0.2$. We can select $\mathbf{g}(\cdot\,; \cdot) = \{g_1(\cdot\,; \cdot), \ldots, g_r(\cdot\,; \cdot)\}^\T\in{\mathbb R}^r$ with $g_j(\bfx_{i}; \theta) = [y_i - \exp(z_i \theta)/\{1+\exp(z_i \theta)\}] z_i^j$ and $\bfx_{i} = (y_i, z_i)^\T$.
		\end{itemize}
		
\begin{figure}[!ht] 
			\centering 
			\caption*{~~~~~~~\small{$n=20$}~~~~~~~~~~~~~~~~~~~~~~~~~~~~~~~~~~~~\small{$n=60$}~~~~~~~~~~~~~~~~~~~~~~~~~~~~~~~~~~~\small{$n=100$}}
			\vspace{-0.35cm} 
			\subfigtopskip=2pt 
			\subfigbottomskip=2pt 
			\subfigcapskip=-5pt 
			\subfigure{ 
				\rotatebox{90}{~~~~~~~~~~~~~~\footnotesize{Model ${\rm I}$}}
				\includegraphics[width=0.29\linewidth]{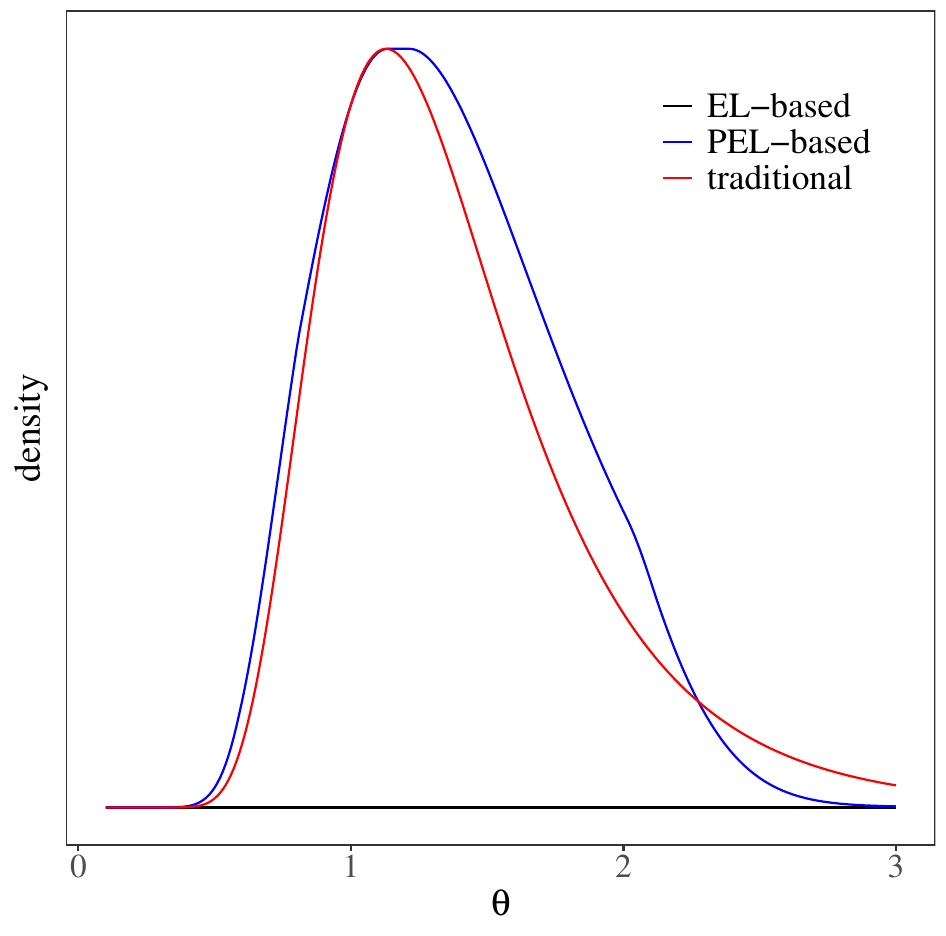}}
			\quad 
			\subfigure{
				\includegraphics[width=0.29\linewidth]{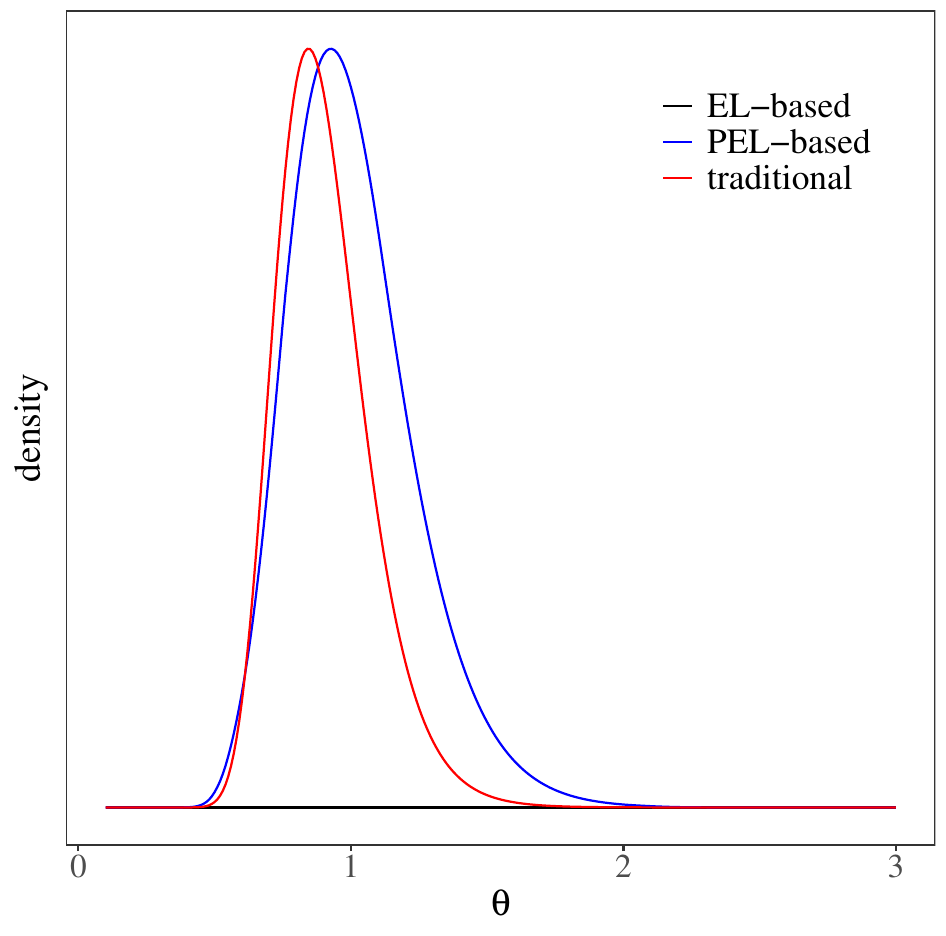}}
			\quad 
			\subfigure{
				\includegraphics[width=0.29\linewidth]{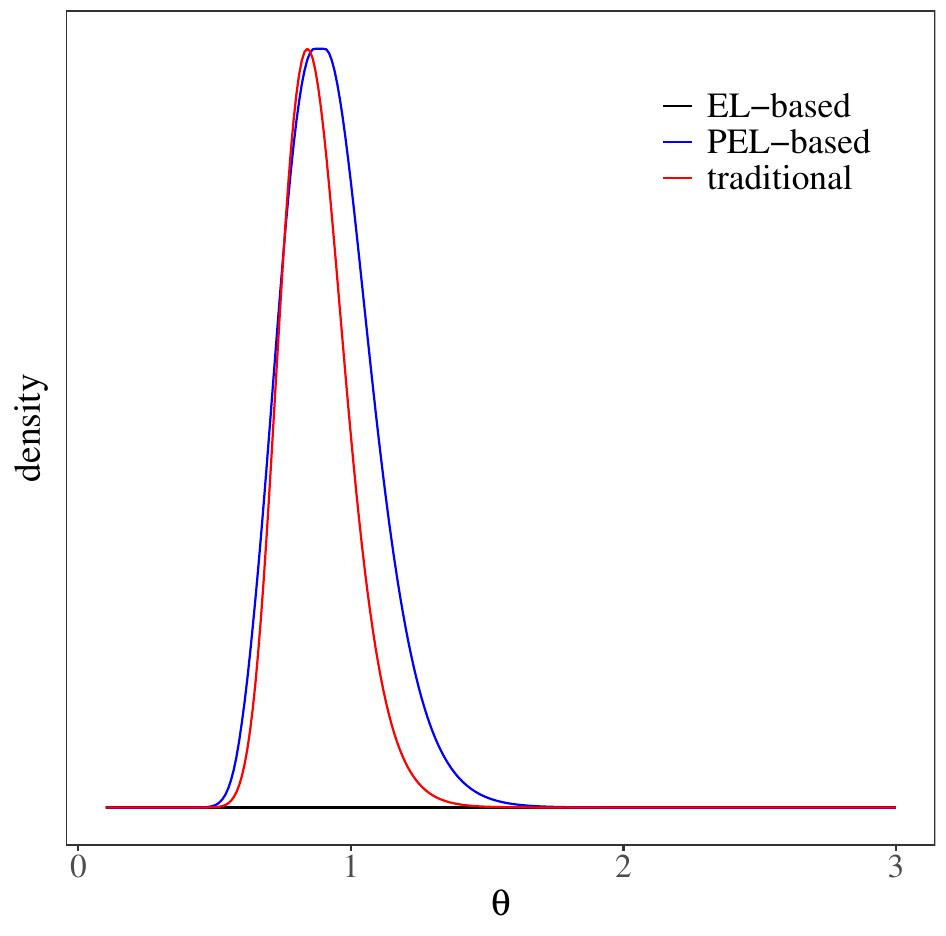}}
			\subfigure{
				\rotatebox{90}{~~~~~~~~~~~~~~\footnotesize{Model ${\rm II}$}}
				\includegraphics[width=0.29\linewidth]{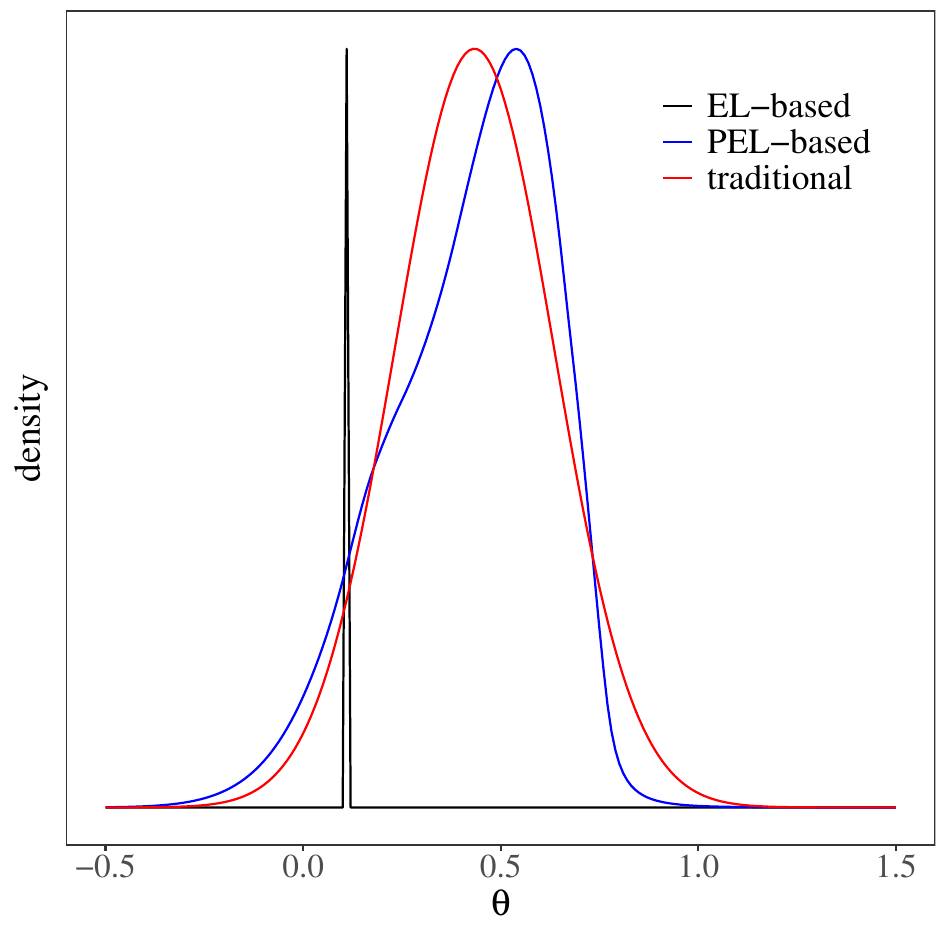}}
			\quad 
			\subfigure{
				\includegraphics[width=0.29\linewidth]{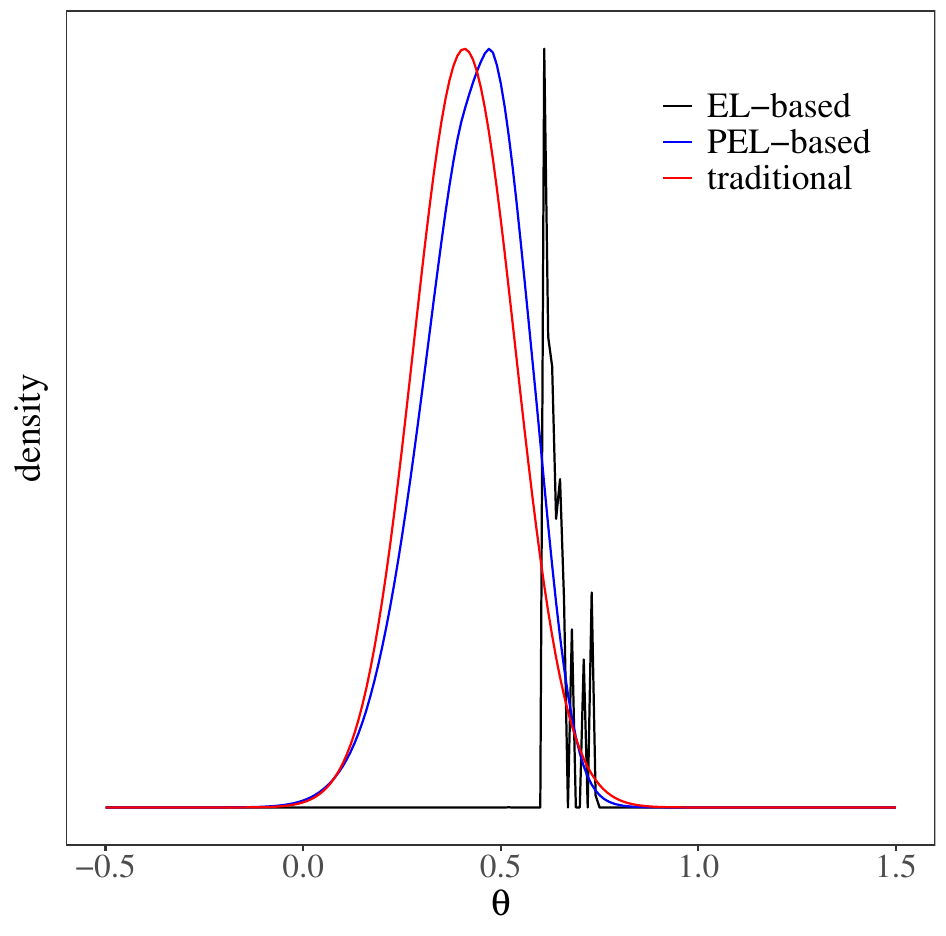}}
			\quad 
			\subfigure{
				\includegraphics[width=0.29\linewidth]{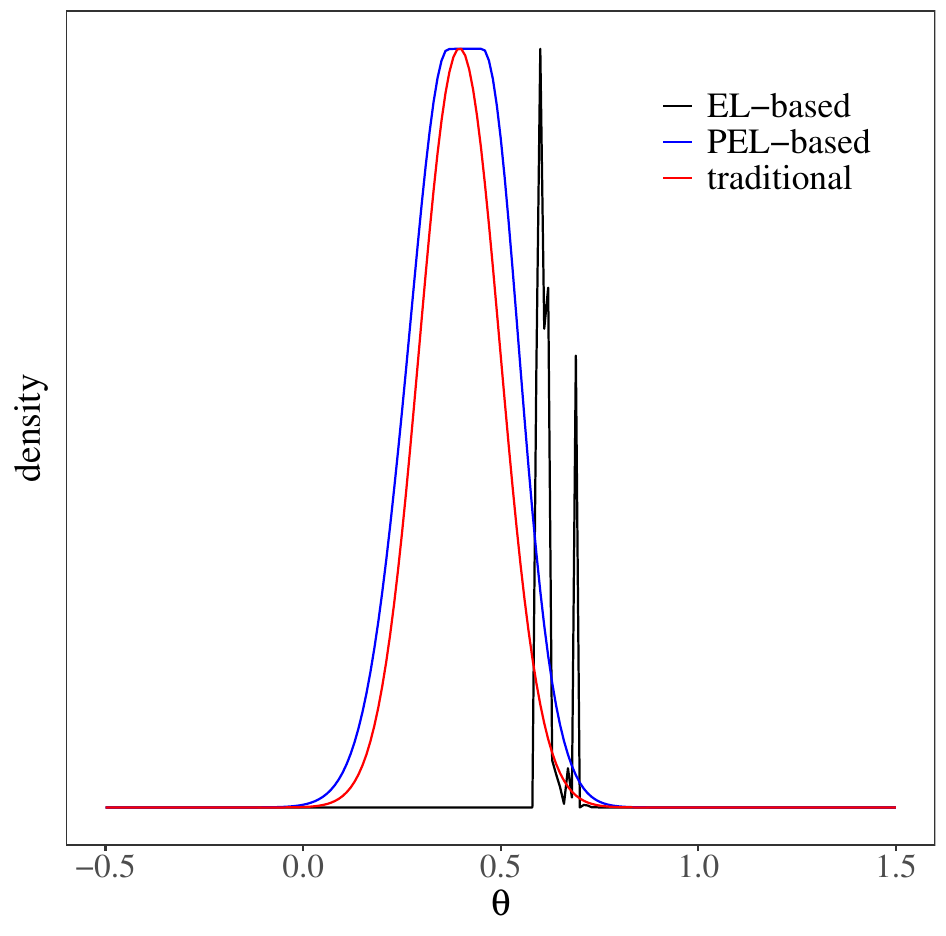}}
			\subfigure{
				\rotatebox{90}{~~~~~~~~~~~~~~\footnotesize{Model ${\rm III}$}}
				\includegraphics[width=0.29\linewidth]{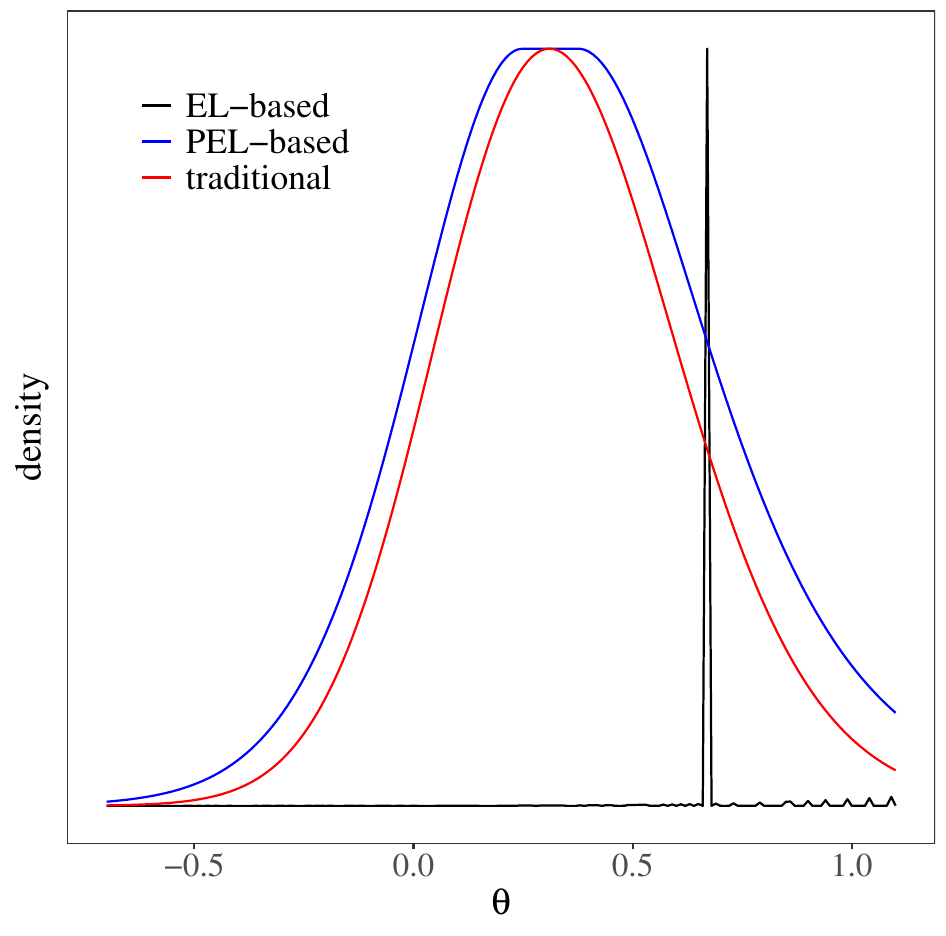}}
			\quad 
			\subfigure{
				\includegraphics[width=0.29\linewidth]{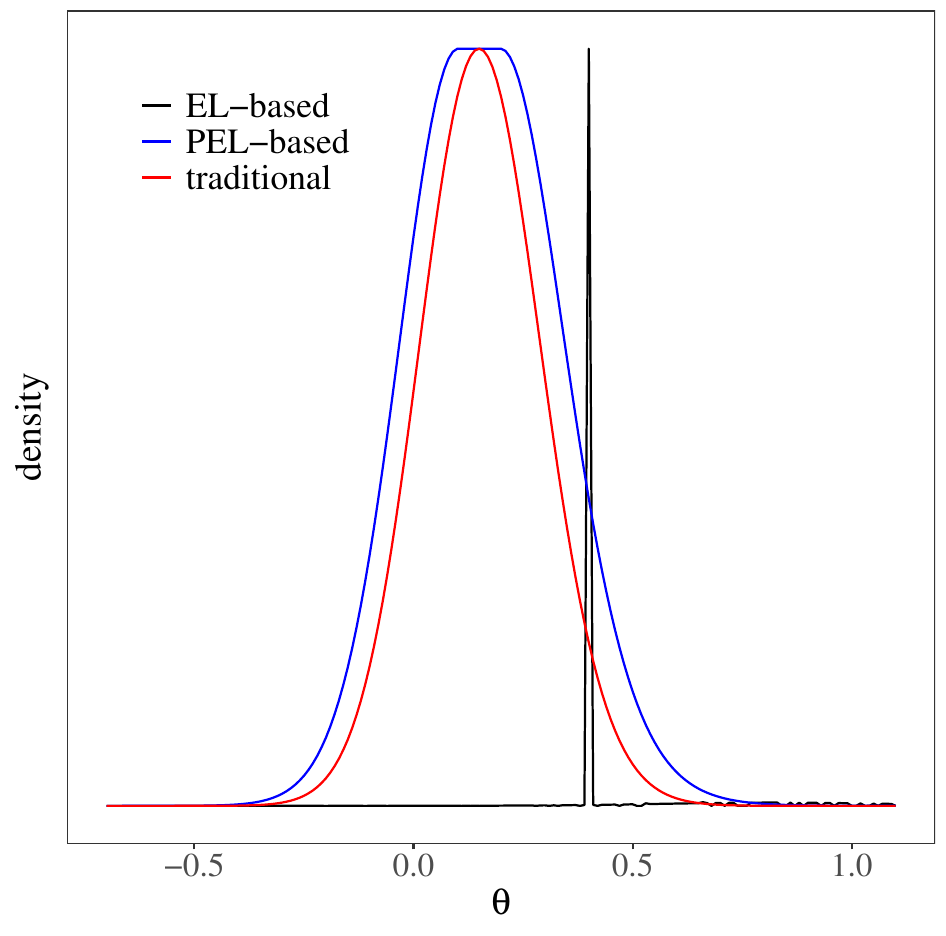}}
			\quad 
			\subfigure{
				\includegraphics[width=0.29\linewidth]{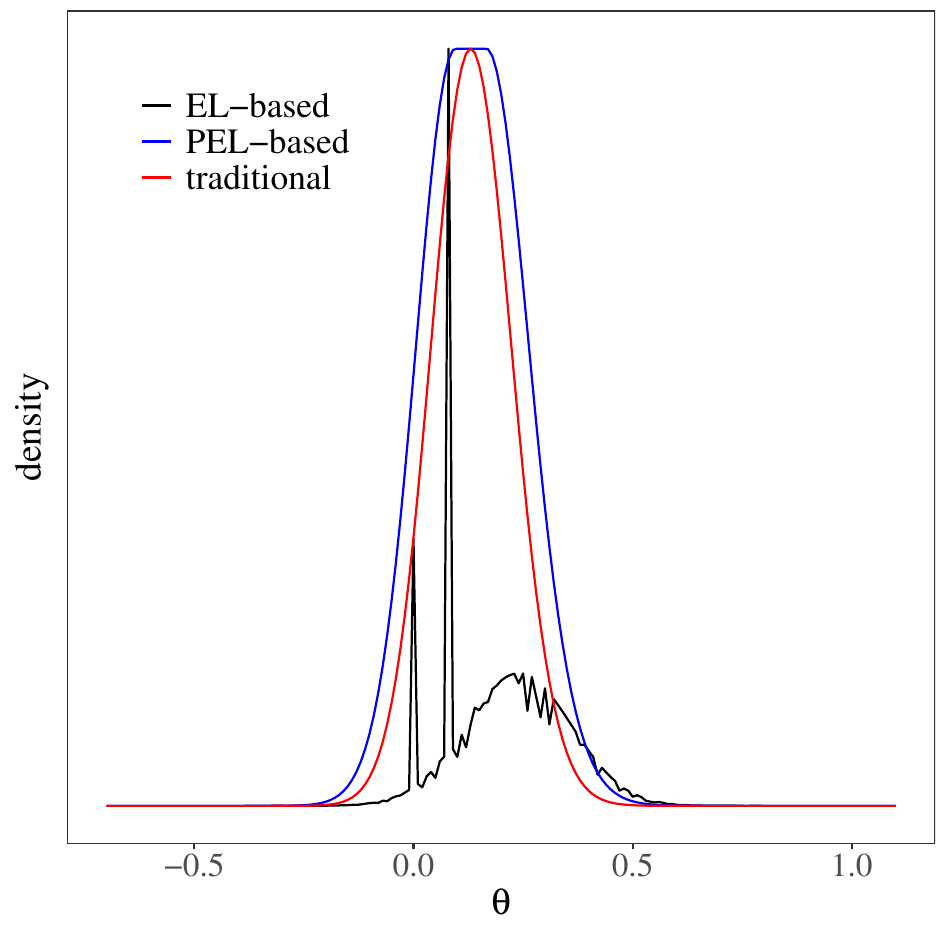}}
			\caption{Comparison of the density functions of the posterior distributions derived by different methods with $r=50$.}
			\label{fig:S3}
		\end{figure}

\begin{figure}[!ht] 
			\centering 
			\caption*{~~~~~~~\small{$n=20$}~~~~~~~~~~~~~~~~~~~~~~~~~~~~~~~~~~~~\small{$n=60$}~~~~~~~~~~~~~~~~~~~~~~~~~~~~~~~~~~~\small{$n=100$}}
			\vspace{-0.35cm} 
			\subfigtopskip=2pt 
			\subfigbottomskip=2pt 
			\subfigcapskip=-5pt 
			\subfigure{ 
				\rotatebox{90}{~~~~~~~~~~~~~~\footnotesize{Model ${\rm I}$}}
				\includegraphics[width=0.29\linewidth]{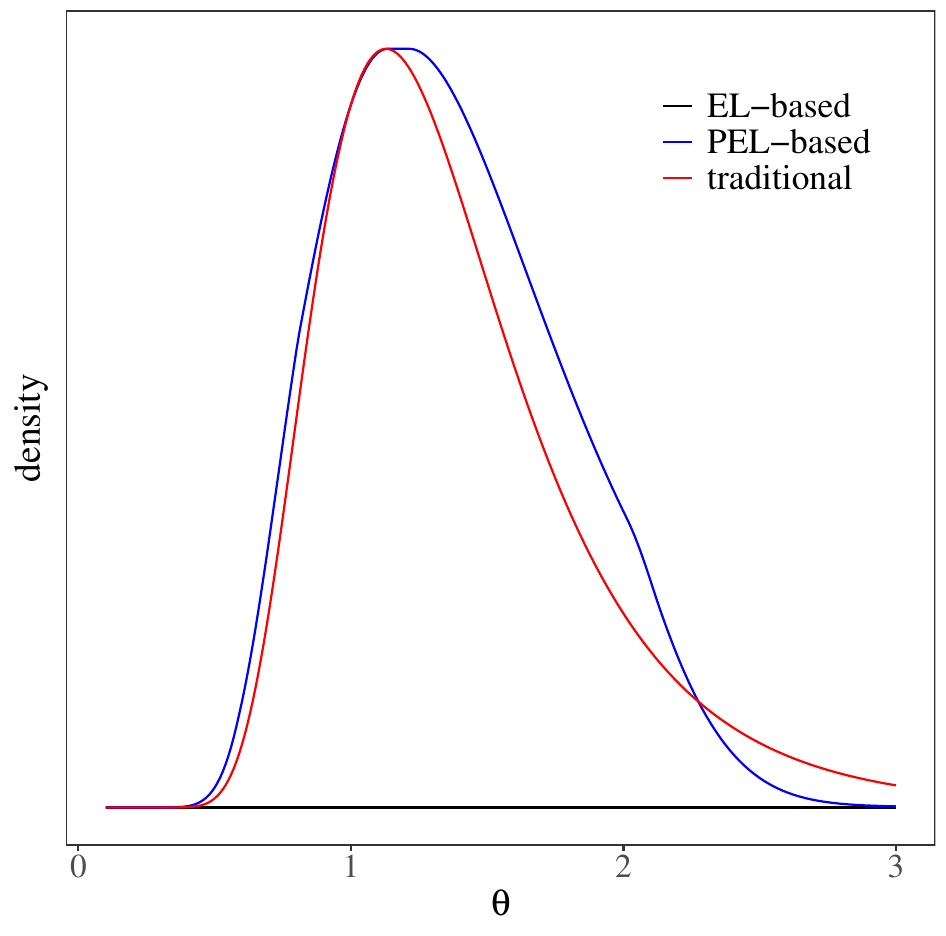}}
			\quad 
			\subfigure{
				\includegraphics[width=0.29\linewidth]{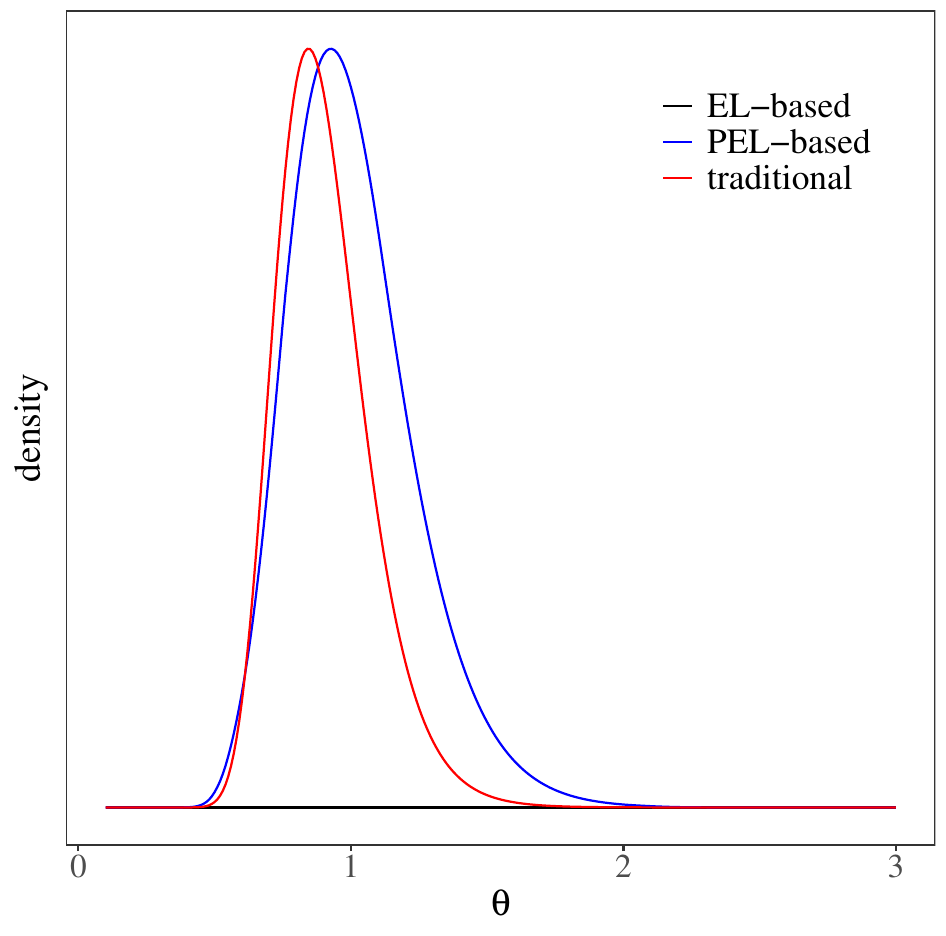}}
			\quad 
			\subfigure{
				\includegraphics[width=0.29\linewidth]{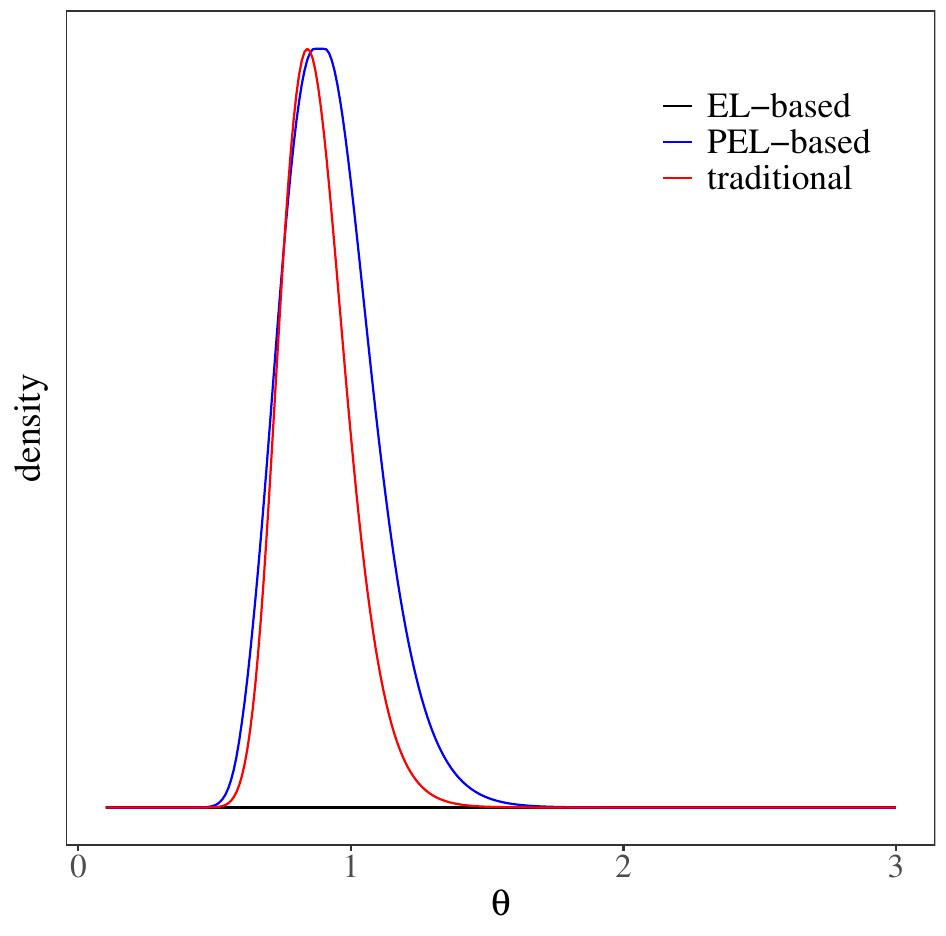}}
			\subfigure{
				\rotatebox{90}{~~~~~~~~~~~~~~\footnotesize{Model ${\rm II}$}}
				\includegraphics[width=0.29\linewidth]{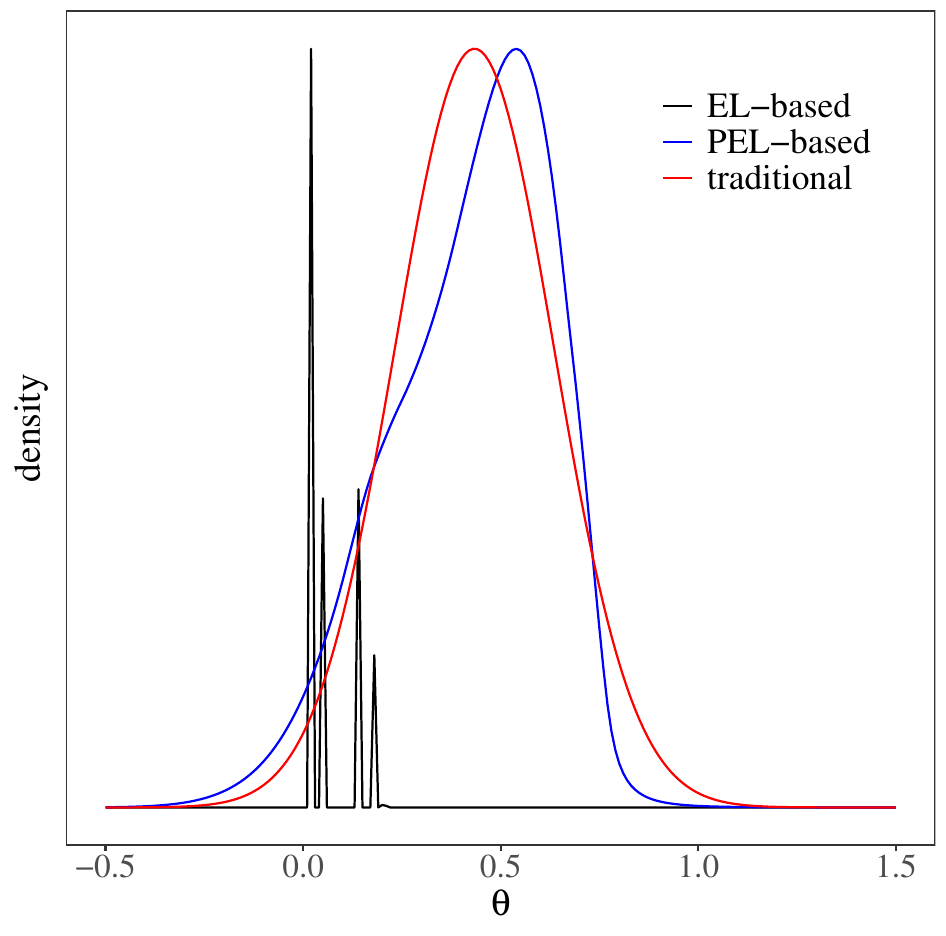}}
			\quad 
			\subfigure{
				\includegraphics[width=0.29\linewidth]{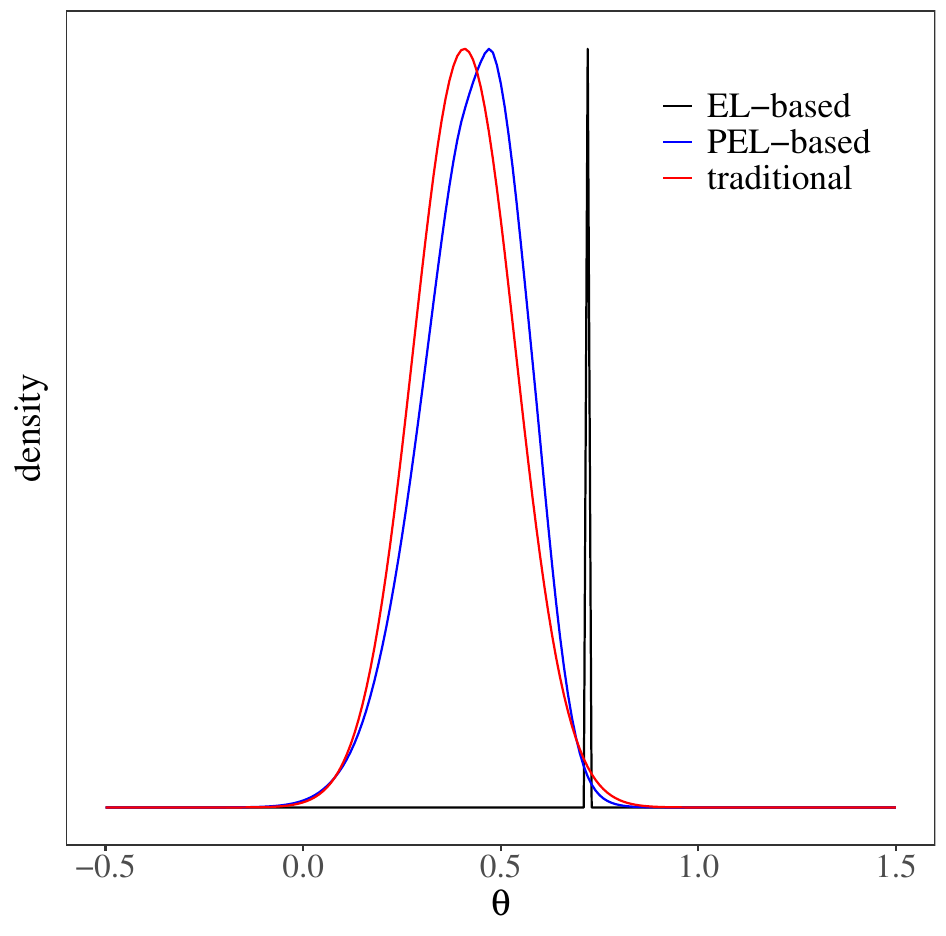}}
			\quad 
			\subfigure{
				\includegraphics[width=0.29\linewidth]{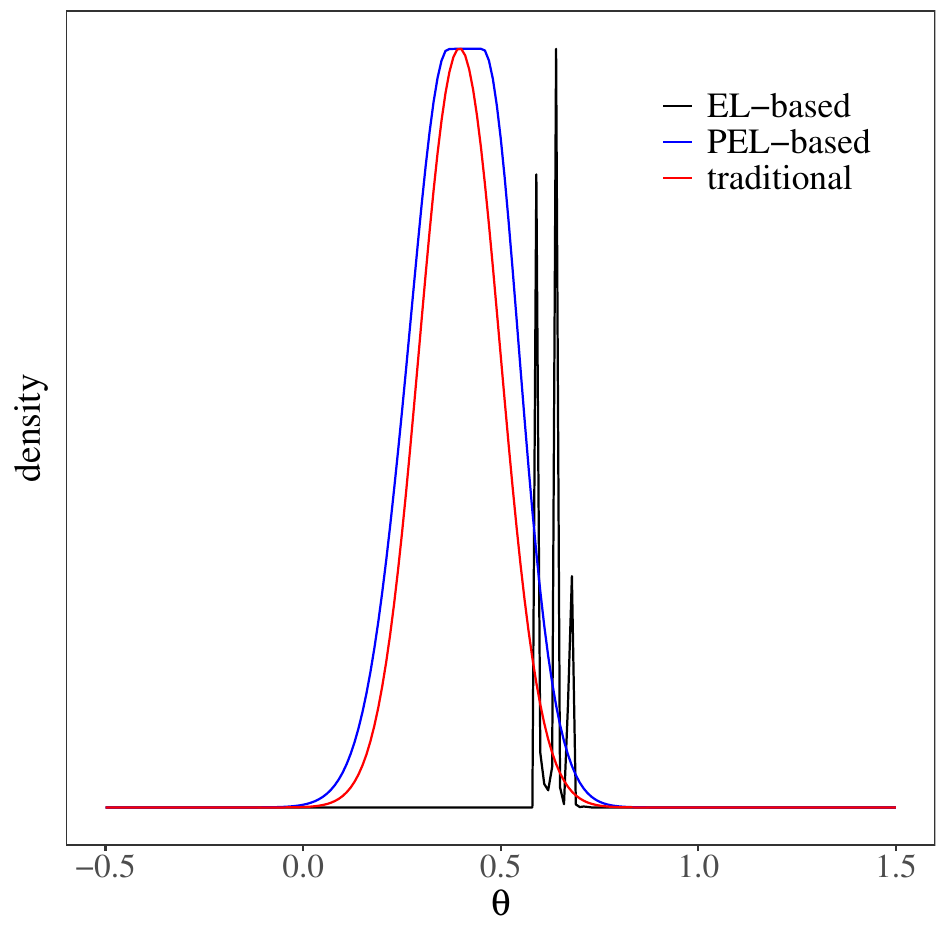}}
			\subfigure{
				\rotatebox{90}{~~~~~~~~~~~~~~\footnotesize{Model ${\rm III}$}}
				\includegraphics[width=0.29\linewidth]{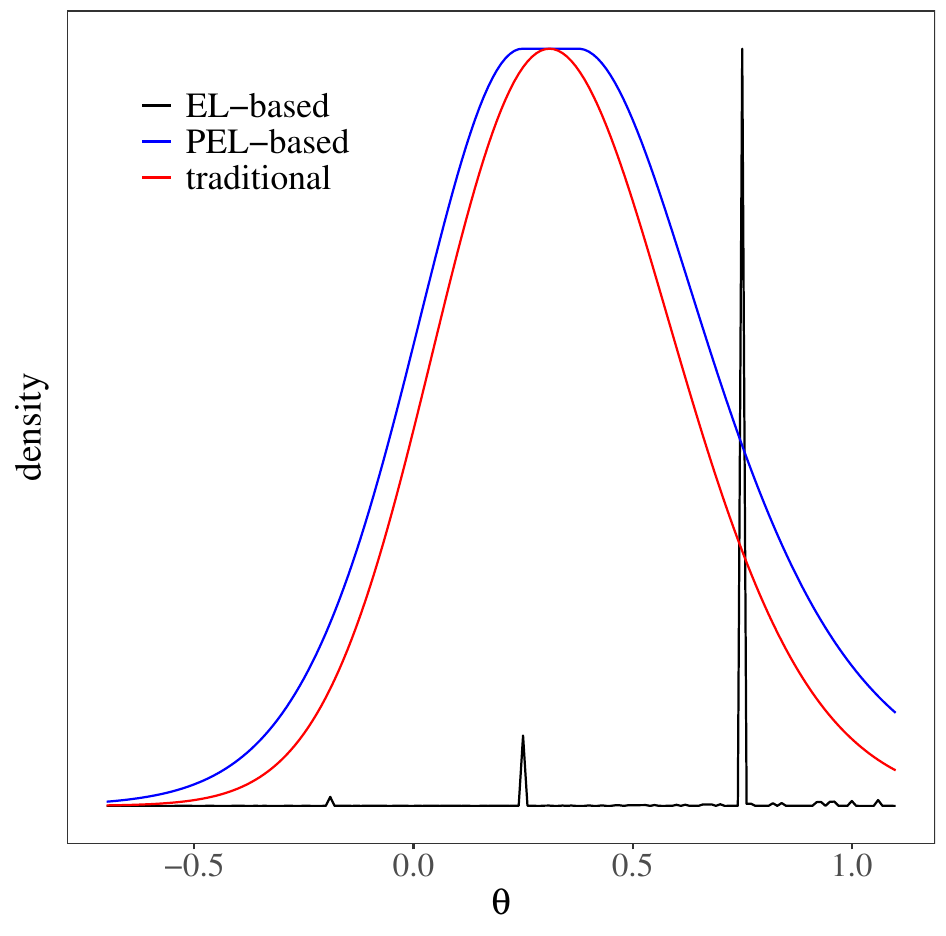}}
			\quad 
			\subfigure{
				\includegraphics[width=0.29\linewidth]{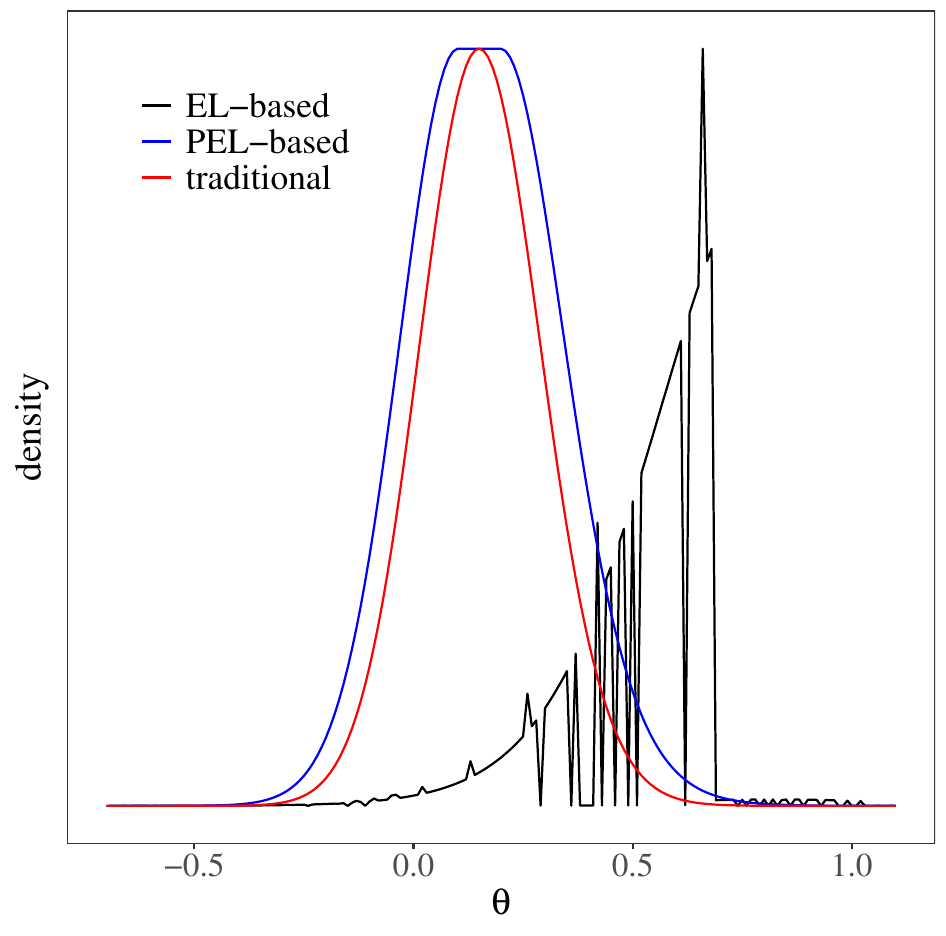}}
			\quad 
			\subfigure{
				\includegraphics[width=0.29\linewidth]{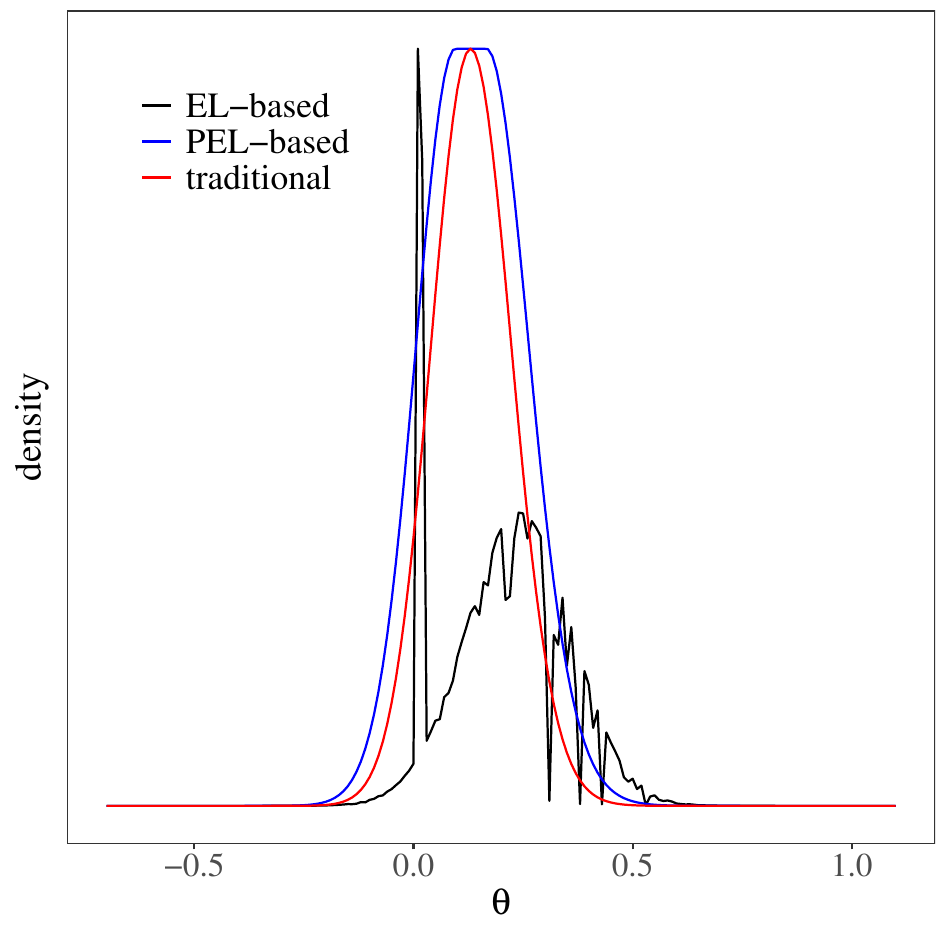}}
			\caption{Comparison of the density functions of the posterior distributions derived by different methods with $r=70$.}
			\label{fig:S4}
		\end{figure}

		We generate $5000$ samples from each of these posterior distributions via the M-H algorithm with the burn-in period of $2000$ iterations. Based on these samples, we compute the Wasserstein distances between $\pi^{\rm L}(\theta\,|\,\mathcal{X}_n)$ and $\pi^{\rm EL}(\theta\,|\,\mathcal{X}_n)$, as well as between $\pi^{\rm L}(\theta\,|\,\mathcal{X}_n)$ and $\pi^{\dag}(\theta\,|\,\mathcal{X}_n)$, using the function {\tt wasserstein} in the R-package {\tt transport}. Figure \ref{fig:S2} below illustrates the average Wasserstein distances between these posterior distributions across different sample sizes $n$ under $500$ replications. Figures \ref{fig:S3} and \ref{fig:S4} show the density functions of $\pi^{\rm L}(\theta\,|\,\mathcal{X}_n)$, $\pi^{\rm EL}(\theta\,|\,\mathcal{X}_n)$ and $\pi^{\dag}(\theta\,|\,\mathcal{X}_n)$ across different sample sizes $n$ in one replication. 
		
		\begin{figure}[!ht] 
			\centering  
			\vspace{-0.35cm} 
			\subfigtopskip=2pt 
			\subfigbottomskip=2pt 
			\subfigcapskip=-5pt 
			\subfigure[Model ${\rm I}$ with $r=50$]{
				\includegraphics[width=0.4\linewidth]{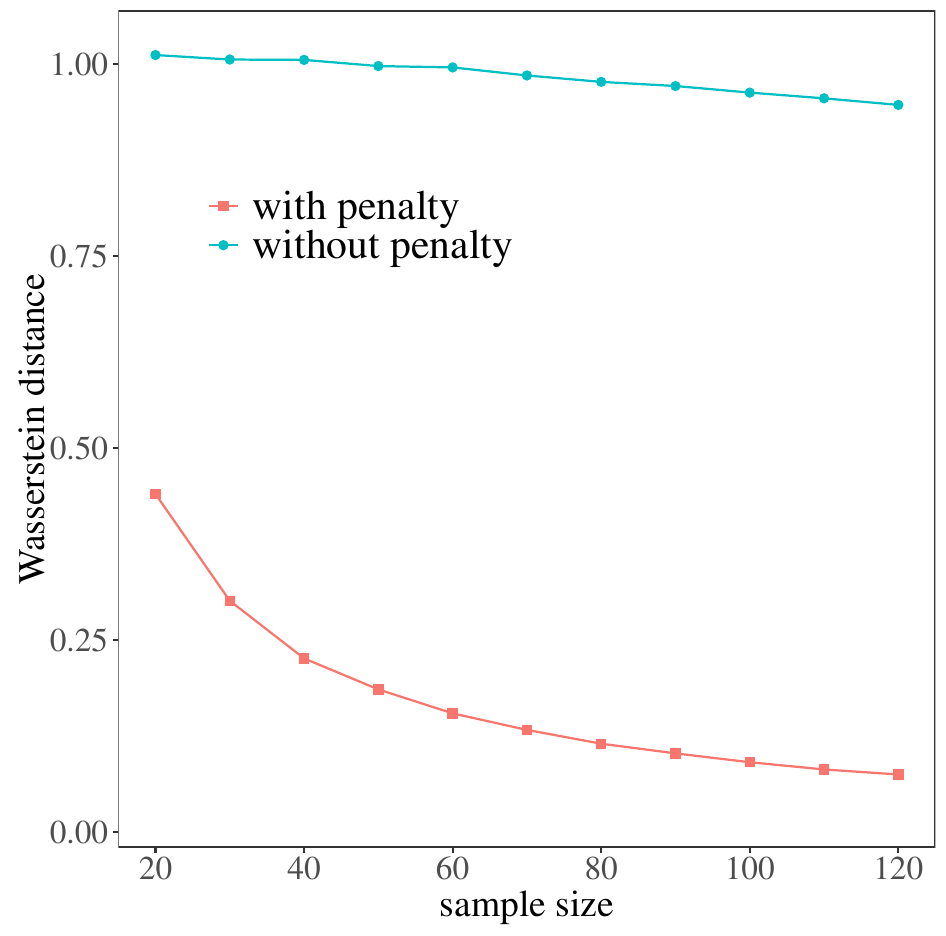}}
			\quad 
			\subfigure[Model ${\rm I}$ with $r=70$]{
				\includegraphics[width=0.4\linewidth]{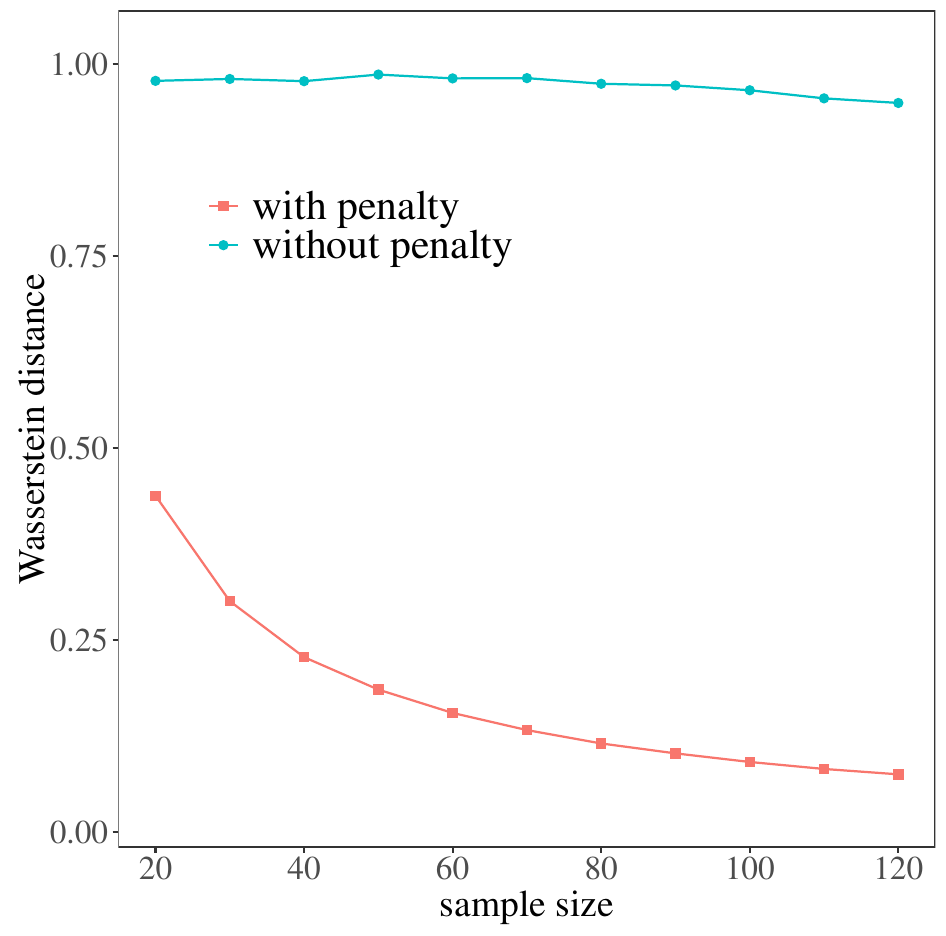}}
			\subfigure[Model ${\rm II}$ with $r=50$]{
				\includegraphics[width=0.4\linewidth]{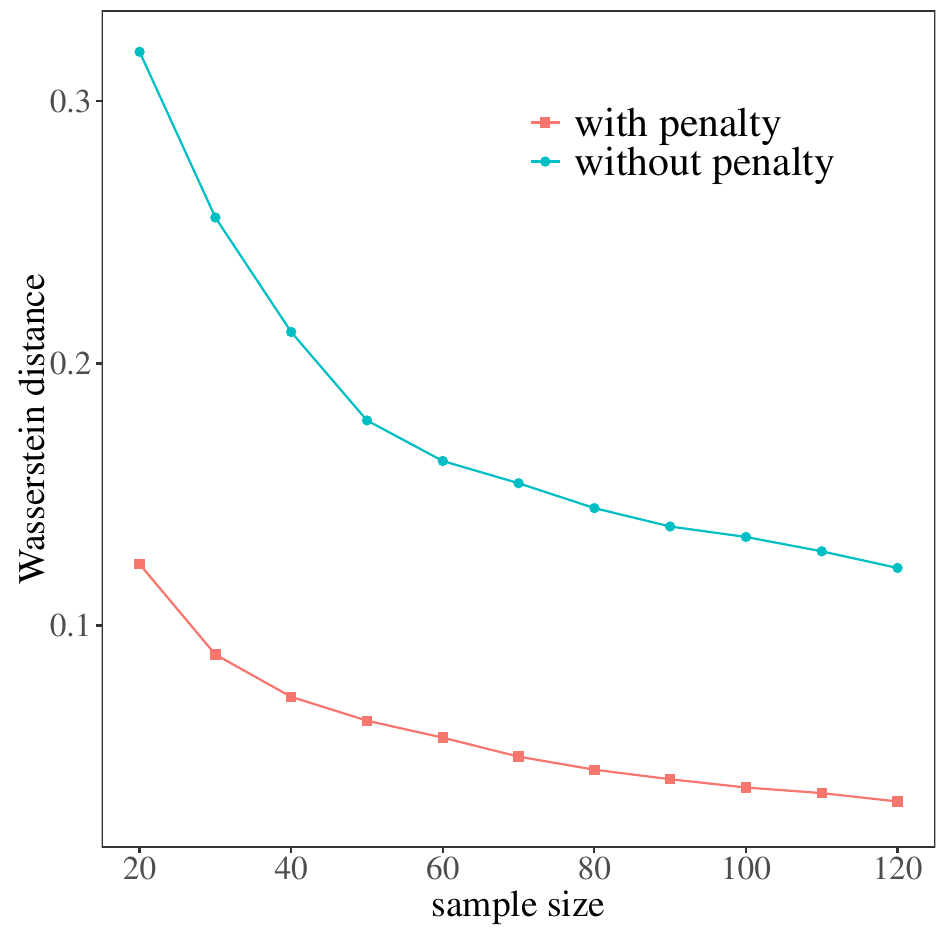}}
			\quad
			\subfigure[Model ${\rm II}$ with $r=70$]{
				\includegraphics[width=0.4\linewidth]{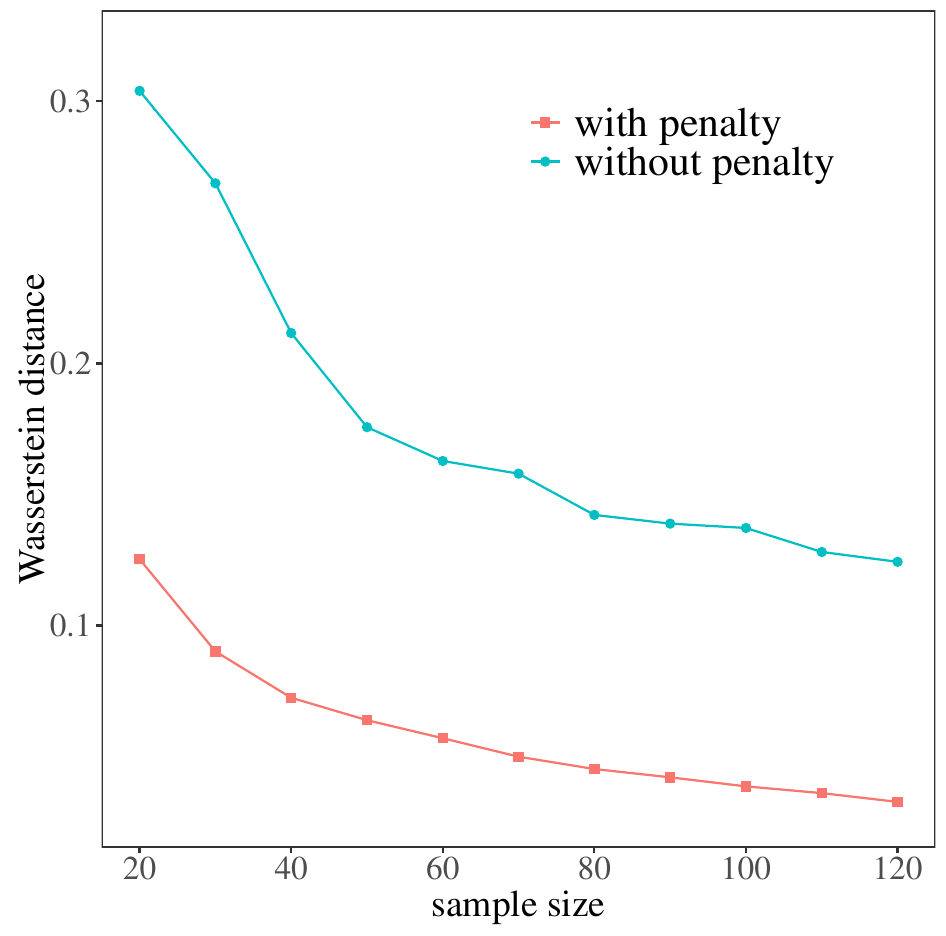}}
			\quad
			\subfigure[Model ${\rm III}$ with $r=50$]{
				\includegraphics[width=0.4\linewidth]{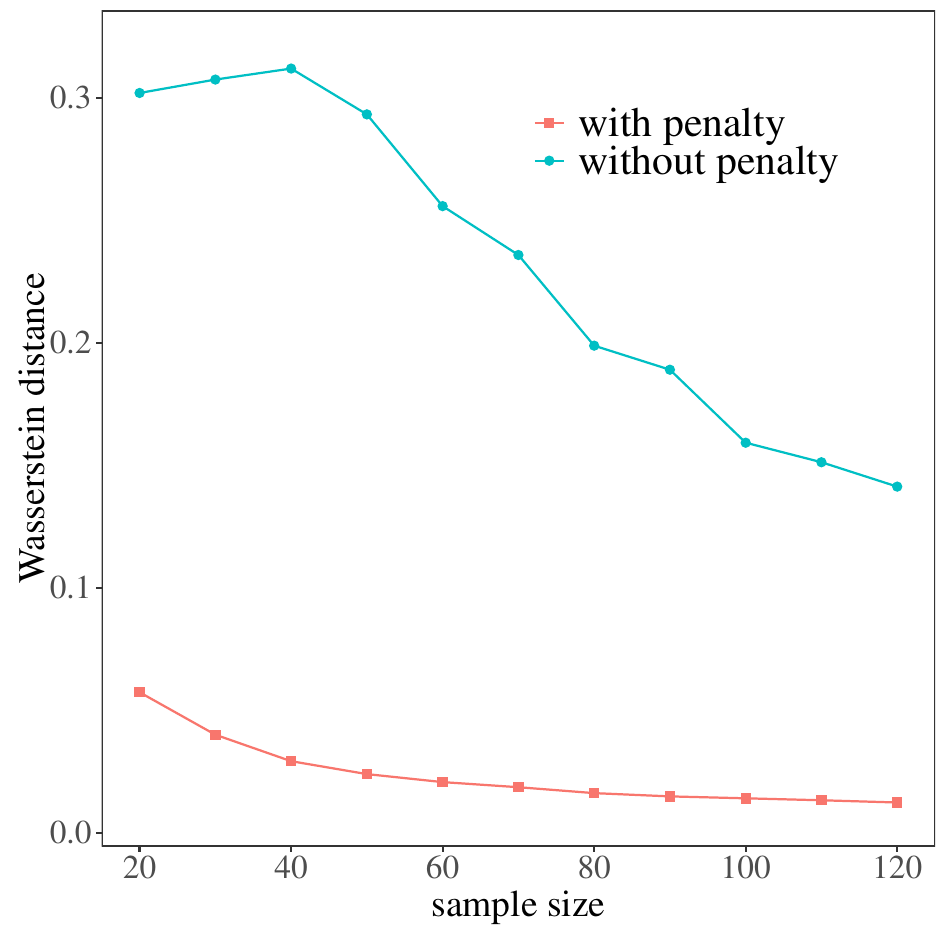}}
			\quad
			\subfigure[Model ${\rm III}$ with $r=70$]{
				\includegraphics[width=0.4\linewidth]{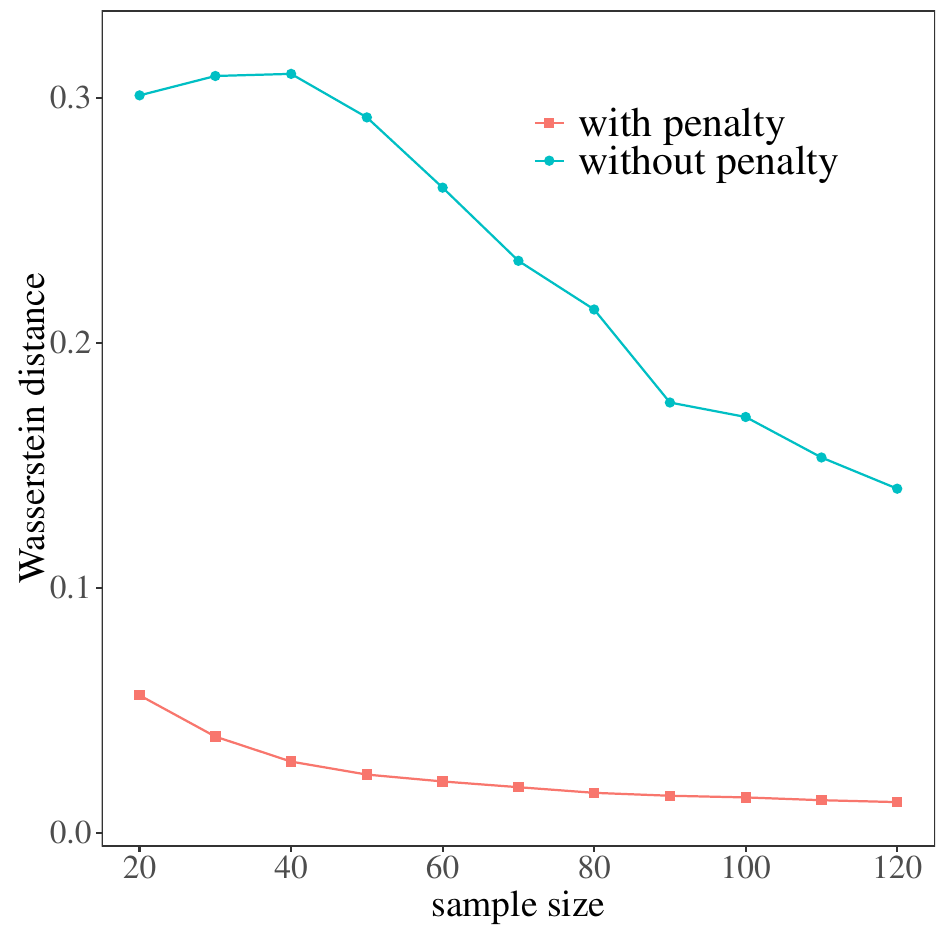}}
			\caption{Comparison of the two average Wasserstein distances under 500 replications. }
			\label{fig:S2}
		\end{figure}

		Overall, the numerical results indicate that the posterior distributions constructed the (un-penalized) empirical likelihood without the penalty on the Lagrange multiplier exhibit significant discrepancies from those derived from the likelihood function. However, this disparity can be markedly reduced through the introduction of a penalty term on the Lagrange multipliers in the empirical likelihood.

		\subsection{Comparison to ABC and BSL} \label{sec:ABC_BSL}
		
		In this section, we compare the performance of our proposed methods with the approximate Bayesian computation (ABC) and Bayesian synthetic likelihood (BSL) methods, implemented as described below.

		 \begin{itemize}
		 	\item {\tt abc}: The R function performs parameter estimation using the approximate Bayesian computation (ABC) algorithm in the R-package {\tt abc}.
		 	
		 	\item {\tt bsl}: The R function for performing Bayesian synthetic likelihood (BSL) to sample from the approximate posterior distribution in the R-package {\tt BSL}. 
		 \end{itemize}

We conduct the comparisons using the same three models as described in Section \ref{sec:add_truth}.
Both the ABC and BSL methods require the selection of summary statistics. In our numerical experiments, for demonstration purposes, we chose sufficient statistics for the parameters of interest, thereby favoring the ABC and BSL methods.  Specifically, for Model \({\rm I}\), we selected the sample standard deviation as the summary statistic. For Models \({\rm II}\) and \({\rm III}\), we selected the maximum likelihood estimator as the summary statistic.

When implementing {\tt abc}, we set the tolerance levels to \(0.05\), \(0.1\), and \(0.2\) to obtain \(5000\) valid approximate samples of the traditional posterior distributions for the three models. These tolerance levels correspond to \(100{,}000\), \(50{,}000\), and \(25{,}000\) MCMC iterations, respectively. The results are labeled as ({\tt abc}-0.05, {\tt abc}-0.1, {\tt abc}-0.2). For {\tt bsl}, we ran the MCMC sampler for \(7000\) iterations, discarding the first \(2000\) iterations for burn-in.

To evaluate the performance of {\tt abc} and {\tt bsl}, we used the {\tt wasserstein} function from the R package {\tt transport} to compute the Wasserstein distances between the approximate samples and those obtained directly via the Metropolis-Hastings (M-H) algorithm from the traditional posterior distribution. For our PEL method, we report results for the case with \(r=50\) and also evaluate the corresponding Wasserstein distances relative to those generated by the M-H algorithm. Figure \ref{fig:Wdis_three} below presents the results, showing the average Wasserstein distances calculated for different sample sizes \(n\) under \(500\) replications.

        \begin{figure}[!ht] 
			\centering
			\subfigure[Model ${\rm I}$]{\includegraphics[scale=0.32]{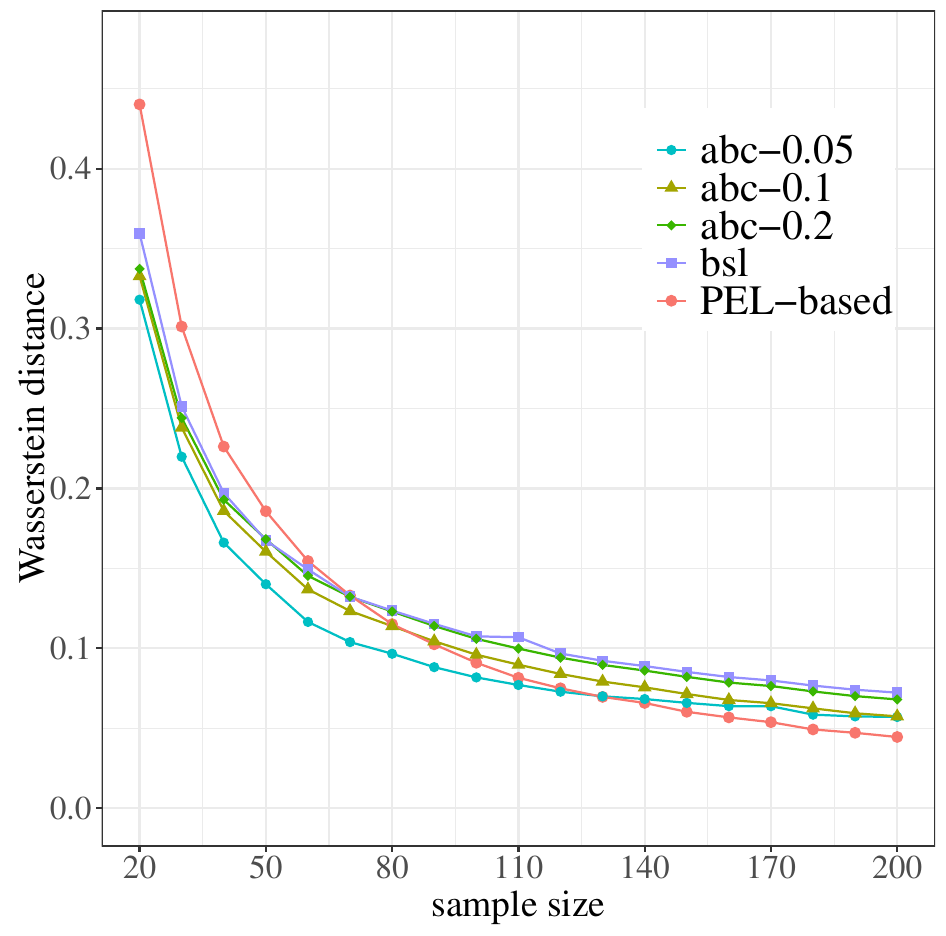}}
			\subfigure[Model ${\rm II}$]{\includegraphics[scale=0.32]{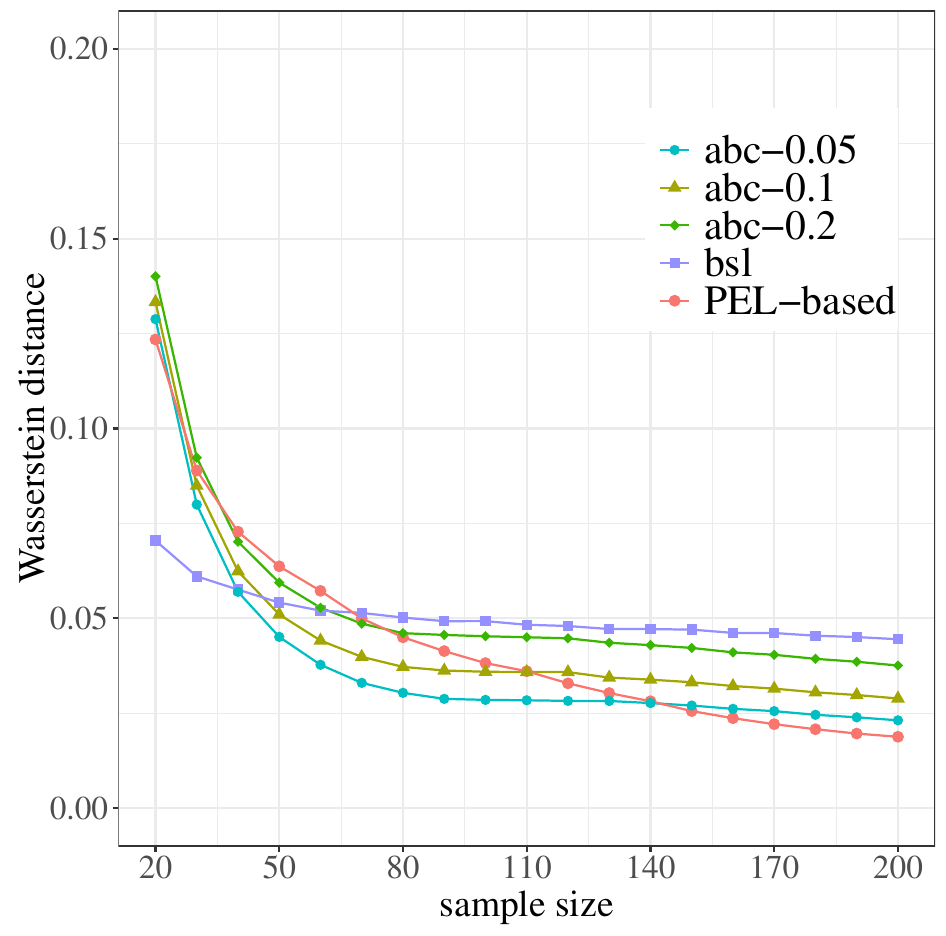}}
			\subfigure[Model ${\rm III}$]{\includegraphics[scale=0.32]{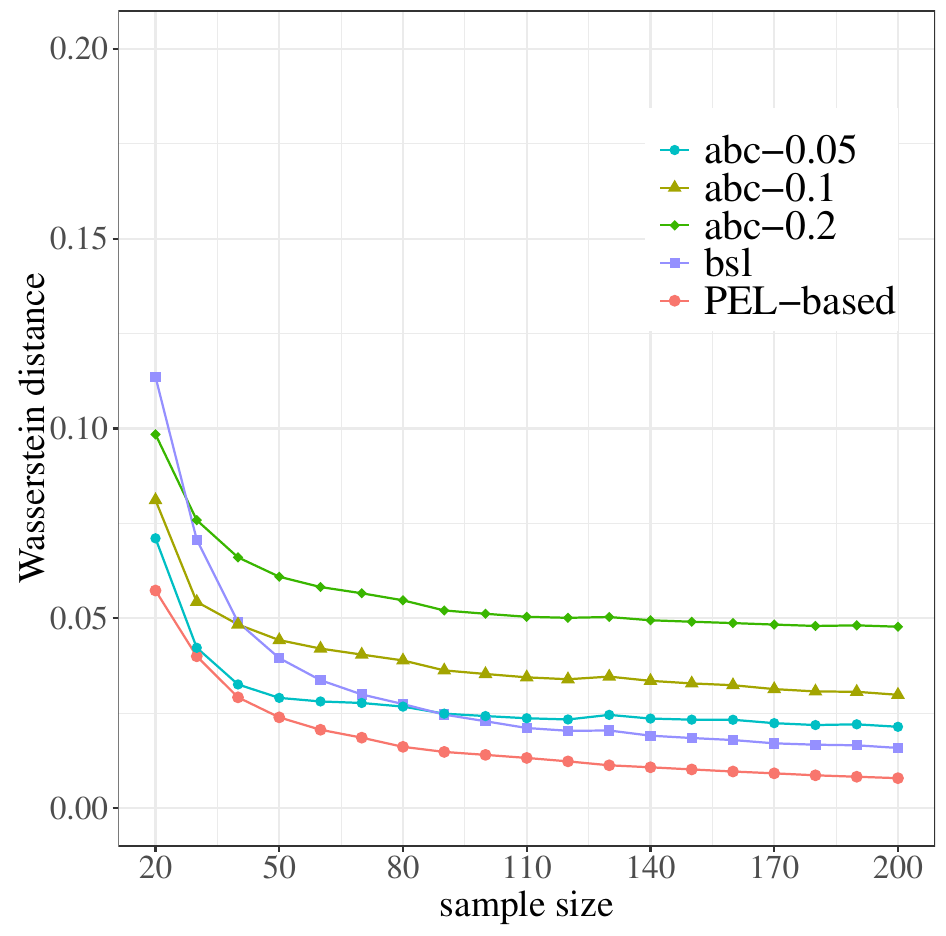}}
			\caption{Comparisons of the proposed method, ABC, and BSL. The average Wasserstein distances based on \(500\) replications are reported for the three models.}
			\label{fig:Wdis_three}
		\end{figure}

         	Overall, the numerical results indicate that our proposed method, along with {\tt abc} and {\tt bsl}, effectively approximates the posterior distribution, with the accuracy of the approximation improving as the sample size $n$ increases. For smaller sample sizes, {\tt abc} and {\tt bsl} exhibit  better accuracy. However, as the sample size $n$ grows, our proposed method demonstrates  substantial improvements,  achieving satisfactory approximation accuracy. Additionally, it is notable that the ABC method often requires significantly higher computational costs to achieve comparable approximation accuracy. 
		 
		%Overall, the numerical results indicate that  our proposed method, along with {\tt abc} and {\tt bsl}, all effectively approximate the traditional posterior distribution, with the accuracy of the approximation improving as the sample size increases. For smaller sample sizes, {\tt abc} and {\tt bsl} exhibit slightly higher accuracy. However, as the sample size increases, our proposed method demonstrates more substantial improvements, ultimately achieving superior approximation accuracy. Meanwhile, {\tt abc} often requires a considerably higher computational costs to achieve comparable approximation accuracy.

\begin{figure}[!ht] 
			\centering
			\subfigure[]
            {\includegraphics[scale=0.4]{Nsig_n_200.pdf}}
			\hspace{1cm}
			\subfigure[]
            {\includegraphics[scale=0.4]{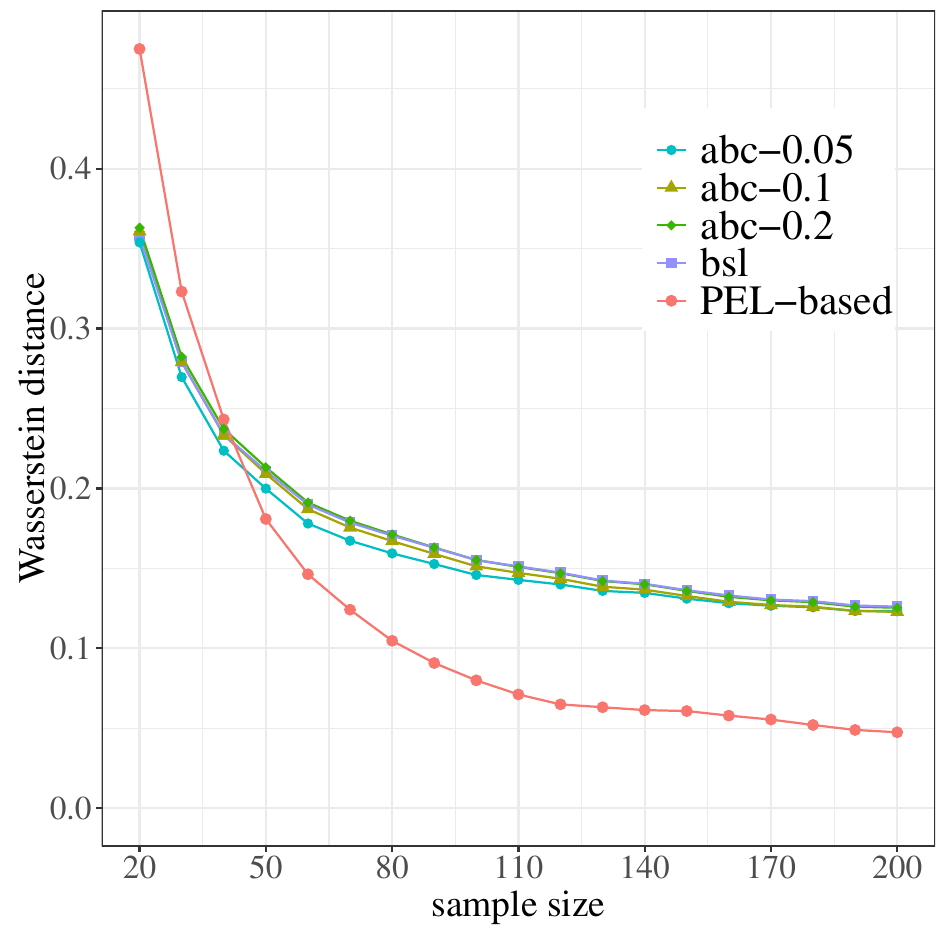}}
			\caption{Estimating the standard deviation parameter with: (a) normal distribution, (b) Student's $t$-distribution }
			\label{fig:Wdis_cor_mis}
		\end{figure}

Here, we also note that the settings favor the ABC and BSL methods in the choice of summary statistics, yet our method performs very competitively. Figure \ref{fig:Wdis_cor_mis}(b) presents results from a slightly modified setting of Model I, where the data generation process is changed from a normal distribution to a Student's $t$-distribution with $10$ degrees of freedom, and we are still interested in estimating the standard
deviation parameter. In this case, the summary statistic based on the sample standard deviation for ABC and BSL is no longer sufficient. For side-by-side comparisons, the corresponding case with results from the normal distribution is shown in Figure \ref{fig:Wdis_cor_mis}(a). In this scenario, the performance of the PEL-based method is superior, demonstrating the compelling performance of our approach, owing to the merits of using empirical likelihood.

		%Here, we also note that the settings favor the ABC and BSL methods in the choice of summary statistics, yet our method performs very competitively. Figure \ref{fig:Wdis_cor_mis}(b) presents results from a slightly modified setting of Model I, where the data-generating process is changed from a normal distribution to a $t$-distribution with $10$ degrees of freedom, representing a misspecified model case. The corresponding correctly specified model results are shown in Figure \ref{fig:Wdis_cor_mis}(a). In this scenario, the performance of the PEL-based method is superior, demonstrating the robustness of our approach, owing to the merits of using empirical likelihood.

		\section{Discussion of the Technical Conditions}\label{sec:discussioncond}
		
		Conditions \ref{A.GIC} and \ref{A.ee} are commonly used assumptions in the literature.  Condition \ref{A.GIC} is the identification condition for the unknown true parameter $\btheta_0$. A similar condition can be found in \cite{Shi2016} and \cite{Changetal_2018}. Condition \ref{A.ee}(b) requires the covariance matrix of $\bfg(\bfx_i;\btheta_0)$ behaves reasonably well. Conditions \ref{A.ee}(a) and \ref{A.ee}(c) impose the moments requirements on each estimating function $g_j(\cdot\,;\cdot)$ and its derivatives. If there exist functions $B_l(\cdot)$ with $\mathbb{E}\{B_l(\bfx_i)\}<\infty$, $l=1,2,3$, such that $|g_j(\bfx;\btheta)|^\gamma\leq B_1(\bfx)$, $|\partial g_j(\bfx;\btheta)/\partial\theta_k|^2\leq B_2(\bfx)$ and $|\partial^2g_j(\bfx;\btheta)/\partial\theta_{k_1}\partial\theta_{k_2}|^2\leq B_3(\bfx)$ for any $j\in[r]$ and $\btheta\in\bTheta$, then the second requirement in  Condition \ref{A.ee}(a) and the two requirements in Condition \ref{A.ee}(c) hold automatically. More generally, if there exist functions $B_{l,j}(\cdot)$ such that $|g_j(\bfx;\btheta)|^\gamma\leq B_{1,j}(\bfx)$, $|\partial g_j(\bfx;\btheta)/\partial\theta_k|^2\leq B_{2,j}(\bfx)$ and $|\partial^2g_j(\bfx;\btheta)/\partial\theta_{k_1}\partial\theta_{k_2}|^2\leq B_{3,j}(\bfx)$ for any $j\in[r]$ and $\btheta\in\bTheta$, and  $\max_{j\in[r]}\mathbb{E}\{B_{l,j}^m(\bfx_i)\}\leq Km!H^{m-2}$ for any $m\geq 2$ and $l=1,2,3$ with two universal constants $K, H>0$, it follows from Theorem 2.8 of \cite{Petrov_1995} that the second requirement in  Condition \ref{A.ee}(a) and the two requirements in Condition \ref{A.ee}(c) hold automatically provided $\log(rp)=o(n)$. In fact, the order $O_{\rm p}(1)$ in Conditions \ref{A.ee}(a) and \ref{A.ee}(c) can be replaced by $O_{\rm p}(\varpi_n)$ with some diverging sequence $\varpi_n$, and our main results remain valid. We use $O_{\rm p}(1)$ here for ease of presentation. To establish the consistency of the penalized empirical likelihood estimator $\hat{\btheta}_n$, Conditions \ref{A.GIC}, \ref{A.ee}(a) and \ref{A.ee}(b) are needed. Condition \ref{A.ee}(c) is needed for establishing the asymptotic normality of $\hat{\btheta}_n$. 
		
		Condition \ref{A.ee2} is standard  in the literature. Due to the penalty imposed on the Lagrange multiplier $\blambda$ involved in the optimization \eqref{eq:thetan_hat}, the standard theoretical analysis of empirical likelihood cannot be applied here. Condition \ref{A.g_subgra.1}(a) is a technical assumption used to derive the convergence rate of the Lagrange multiplier $\hat{\blambda}(\hbthetan)=\arg\max_{\blambda\in\hat{\Lambda}_n(\hbthetan)}f_n(\blambda;\hbthetan)$ associated with $\hat{\btheta}_n$;  see the proof of Lemma \ref{l.lam.thetan.hat} in the supplementary material for details. Condition \ref{A.g_subgra.1}(b) requires that each $\hat{\eta}_j$ $(j\in \mathcal{R}_n^\c)$ lies in the interior of $[-\nu \rho'(0^{+}),\nu \rho'(0^{+})]$ with probability approaching one, which is realistic in practice. If the distribution function of the random variable $\hat{\eta}_j$ is continuous at $\pm \nu\rho'(0^+)$, we then have $\mathbb{P}\{|\hat{\eta}_j|=\nu\rho'(0^+)\}=0$. Condition \ref{A.g_subgra.1} makes sure that $\hat\blambda(\btheta)=\arg\max_{\blambda\in\hat{\Lambda}_n(\btheta)}f_n(\blambda;\btheta)$ is continuously differentiable at $\hbthetan$ with probability approaching one;  see Lemma \ref{l.envlope} in Section \ref{sec:pfThm1}. 
		
		Condition \ref{A.Pro2}(a) guarantees that the Lagrange multiplier $\hat{\blambda}(\btheta)$ for $\btheta \in \mathcal{C}_1$ satisfies two properties: (i) $\hat{\blambda}(\btheta)$ is continuously differentiable on $\mathcal{C}_1$ with probability approaching one, and (ii) $\mathcal{R}(\btheta) = \mathcal{R}(\hbthetan)$ for any $\btheta \in \mathcal{C}_1$ with probability approaching one; see Lemmas \ref{l.diff.lam.hat} and \ref{l.supp} in Section \ref{sec.pf_pro2}. When $\btheta\notin\mathcal{C}_1$, characterizing the asymptotic property of $\hat{\blambda}(\btheta)$ is quite challenging. Due to $f_n\{\hat{\blambda}(\btheta);\btheta\}\geq f_n(\blambda;\btheta)$ for any $\blambda\in\hat{\Lambda}_n(\btheta)$, a feasible strategy to construct the lower bound of $f_n\{\hat{\blambda}(\btheta);\btheta\}$ with $\btheta\notin\mathcal{C}_1$ is to find a specific $\blambda_*(\btheta)\in\hat{\Lambda}_n(\btheta)$ and then derive the lower bound of $f_n\{\blambda_*(\btheta);\btheta\}$ directly, where the asymptotic behavior of $\blambda_*(\btheta)$ can be well characterized even if $\btheta\notin\mathcal{C}_1$. Such strategy has been also used in \cite{Changetal2013, Changetal_2016_AOS} to study the diverging rate of the conventional empirical likelihood ratio evaluated at a value not near the truth. In current setting, Condition \ref{A.Pro2}(b) is applied to derive the lower bound of $f_n\{\blambda_*(\btheta);\btheta\}$ for $\btheta\in\mathcal{C}_2$;  see Section \ref{sec:pfpn2b} for details.
		Condition \ref{A.Pro2}(c) says that the sample covariance matrix of the estimating function and the gradient of the estimating function should behave reasonably well, which will be used to obtain the lower bound of $f_n\{\blambda_*(\btheta);\btheta\}$ for $\btheta\in\mathcal{C}_3$. See Section \ref{sec:pfpn2c} for details. Condition \ref{A.prior} is a standard assumption concerning the prior distribution.

		\section{Proof of Proposition  \ref{pro.cons}}
		
		%\begin{proposition} \label{pro.cons}
		%	Let $P_{\nu}(\cdot) \in \mathscr{P}$ be a convex function for $\mathscr{P}$ defined in {\rm \eqref{eq:penalty}}. Assume Conditions {\rm \ref{A.GIC} and \ref{A.ee}} hold, if $\log r=o(n^{1/3})$ and $\ell_n^2 n^{-1} \log r=o[\min \{\nu^2,n^{-2/\gamma}\}]$, then the penalized EL estimator $\hbthetan$ defined in {\rm \eqref{eq:thetan_hat}} satisfies $|\hbthetan-\bthetazero|_{\infty}=O_{\rm p}(\nu)$.
		%\end{proposition}
		
		Write $\mathcal{R}_{0}=\mathcal{R}(\bthetazero)$. Then %we have
		\begin{align*}
			\max_{\blambda \in \wLambadn(\bthetazero)}f_n(\blambda;\bthetazero)=&~\max_{\bseta \in \hat{\Lambda}^{\dag}_n(\bthetazero)} \bigg\{\mathbb{E}_n\big[\log \{ 1+\bseta^\T \bfg_{i,\mathcal{R}_{0}} (\bthetazero)\}\big]-\sum_{j=1}^{|\mathcal{R}_{0}|} P_{\nu}(|\eta_j|) \bigg\} \\
			\leq&~ \max_{\bseta \in\hat{\Lambda}^{\dag}_n(\bthetazero)} \mathbb{E}_n\big[\log \{ 1+\bseta^\T \bfg_{i,\mathcal{R}_{0}} (\bthetazero)\}\big]\,,
		\end{align*}
		where $\hat{\Lambda}^{\dag}_n(\bthetazero)=\{\bseta=(\eta_{1},\ldots, \eta_{|\mathcal{R}_{0}|})^\T \in \R^{|\mathcal{R}_{0}|}:\bseta^\T \bfg_{i,\mathcal{R}_{0}} (\bthetazero) \in \mathcal{V}~\textrm{for any}~i\in[n]\}$ for some open interval $\mathcal{V}$ containing zero. Our first step is to show
		$\max_{\bseta \in\hat{\Lambda}^{\dag}_n(\bthetazero)} \mathbb{E}_n[\log \{ 1+\bseta^\T \bfg_{i,\mathcal{R}_{0}} (\bthetazero)\}]=O_{\rm p}(\ell_n \alpha_n^2)$. 
		To do this, we need the following two lemmas whose proofs are given in Sections \ref{sec:pflemma1} and \ref{sec:pflemma2}, respectively.
		\begin{lemma} \label{l.V.hat}
			Let $\mathscr{F}=\{\mathcal{F} \subset [r]:|\mathcal{F}|\leq \ell_n\}$ and $\mathcal{B}_{\infty,{\rm p}}(\btheta_0,\varphi_n)=\{\btheta \in \bTheta: |\btheta-\bthetazero|_{\infty}\leq O_{\rm p}(\varphi_n)\}$ for $\varphi_n=o(\ell_n^{-1/2})$. Under Condition {\rm \ref{A.ee}}, if $\log r=o(n^{1/3})$ and $\ell_n\alpha_n=o(1)$, then
			$$
			\sup_{\btheta \in \mathcal{B}_{\infty,{\rm p}}(\btheta_0,\varphi_n)} \sup_{\mathcal{F} \in \mathscr{F}} \| \widehat{\bfV}_{\mF} (\btheta) - \bfV_{\mF} (\bthetazero)\|_2
			= O_{\rm p}(\ell_n^{1/2} \varphi_n)+O_{\rm p}(\ell_n \alpha_n)\,.$$ 
		\end{lemma}
		
		\begin{lemma}\label{l.lam.theta0}
			Let $\log r=o(n^{1/3})$, $\ell_n \alpha_n=o[\min \{\nu,n^{-1/\gamma}\}]$, and $P_{\nu}(\cdot) \in \mathscr{P}$ be a convex function for $\mathscr{P}$ defined in {\rm \eqref{eq:penalty}}. Assume Conditions {\rm \ref{A.ee}(a)} and {\rm \ref{A.ee}(b)} hold. For any $c \in (C_*,1)$, the global maximizer $\hat{\blambda}(\bthetazero)$ for $f_n(\blambda;\bthetazero)$ w.r.t $\blambda$ satisfies ${\rm supp}\{\hat{\blambda}(\bthetazero)\} \subset \mathcal{M}_{\bthetazero}(c)$ w.p.a.1.
		\end{lemma}
		
		Define $\tilde{\bseta}=\arg \max _{\bseta \in \hat{\Lambda}^{\dag}_n(\bthetazero)} A_n(\bthetazero,\bseta)$ with $A_n(\btheta,\bseta)=\mathbb{E}_n[\log \{ 1+\bseta^\T \bfg_{i,\mathcal{R}_{0}} (\btheta) \}]$. %Recall $\mathbb{P}(\max _{\btheta \in \bTheta:\,|\btheta-\bthetazero|_2 \leq c_n}|\mathcal{M}_{\btheta}^*|\leq\ell_n)\rightarrow1$ for some $c_n\rightarrow 0$ satisfying $\nu c_n^{-1} \rightarrow 0$. 
		By Lemma \ref{l.lam.theta0}, we have $|\mathcal{R}_{0}| \leq \ell_n$ w.p.a.1. Pick $\delta_n$ satisfying $\delta_n=o(\ell_n^{-1/2} n^{-1/\gamma})$ and $\ell_n^{1/2} \alpha_n=o(\delta_n) $ for $\gamma$ defined in Condition \ref{A.ee}(a), which can be guaranteed by $\ell_n\alpha_n=o(n^{-1/\gamma})$. Let $\Lambda_{0}=\{\bseta \in \R^{|\mathcal{R}_{0}|}:|\bseta|_2\leq \delta_n \}$ and $\check{\bseta}=\arg \max _{\bseta \in \Lambda_{0}} A_n(\bthetazero,\bseta)$. Condition \ref{A.ee}(a) implies $\max _{i \in [n], \bseta \in \Lambda_{0}} |\bseta^\T \bfg_{i,\mathcal{R}_{0}} (\bthetazero)|
		= o_{\rm p}(1)$. Then, by the Taylor expansion, we have
		\begin{align*}
			0=A_n(\bthetazero,\bzero)
			\leq A_n(\bthetazero,\check{\bseta})
			= \check{\bseta}^\T \bar{\bfg}_{\mathcal{R}_{0}}(\bthetazero)- \frac{1}{2n}\sum _{i=1}^{n}\frac{\check{\bseta}^\T \bfg_{i,\mathcal{R}_{0}} (\bthetazero)^{\otimes2} \check{\bseta}}{\{1+C\check{\bseta}^\T \bfg_{i,\mathcal{R}_{0}} (\bthetazero)\}^2} \notag 
		\end{align*}
		for some $C\in(0,1)$. By Condition \ref{A.ee}(b) and the same arguments for deriving Lemma \rm\ref{l.V.hat}, if $\log r=o(n^{1/3})$ and $\ell_n\alpha_n=o(1)$, we have $\lambda_{\min}\{\widehat{\bfV}_{\mathcal{M}_{\bthetazero}} (\bthetazero)\}$ is uniformly bounded away from zero w.p.a.1. Thus $0\leq |\check{\bseta}|_2 |\bar{\bfg}_{\mathcal{R}_{0}}(\bthetazero)|_2 - 4^{-1}K_3|\check{\bseta}|_2^2 \{1+o_{\rm p}(1)\}$ w.p.a.1, where $K_3$ is specified in Condition \ref{A.ee}(b). By the moderate deviation of self-normalized sums \citep{JingShaoWang2003}, we have 
		$
		|\bar{\bfg}(\bthetazero)|_{\infty}
		=O_{\rm p}(\alpha_n)
		$, which implies $|\bar{\bfg}_{\mathcal{R}_{0}}(\bthetazero)|_2= O_{\rm p}(\ell_n^{1/2} \alpha_n)$ and $|\check{\bseta}|_2=O_{\rm p}(\ell_n^{1/2} \alpha_n)=o_{\rm p}(\delta_n)$. Hence, $\check{\bseta} \in {\rm int} (\Lambda_{0}) $ w.p.a.1. Since $\Lambda_{0} \subset \hat{\Lambda}^{\dag}_n(\bthetazero)$ w.p.a.1, we have $\tilde{\bseta}=\check{\bseta}$ w.p.a.1 by the concavity of $A_n(\bthetazero,\bseta)$ w.r.t $\bseta$. Then $
		\max_{\bseta \in\hat{\Lambda}^{\dag}_n(\bthetazero)} A_n(\btheta_0,\bseta)=O_{\rm p}(\ell_n \alpha_n^2)$. Let $b_n^2=\ell_n \alpha_n^2$ and $F_n(\btheta)=\max_{\blambda \in \wLambadn(\btheta)}f_n(\blambda;\btheta) $ for any $\btheta \in \bTheta$. Due to $\hbthetan=\arg \min _{\btheta \in \bTheta}F_n(\btheta) $, we have $F_n(\hbthetan) \leq F_n(\bthetazero)=O_{\rm p}(b_n^2)$.
		
		Our second step is to show that for any $\epsilon_n \rightarrow \infty$ satisfying $b_n^2\epsilon_n^2 n^{2/\gamma}=o(1)$, there exists a universal constant $K>0$ independent of $\btheta$ such that $\P\{F_n(\btheta)>K b_n^2 \epsilon_n^2\} \rightarrow 1$ as $n \rightarrow \infty$ for any $\btheta \in \bTheta$ satisfying $|\btheta-\bthetazero|_{\infty} > \epsilon_n
		\nu$. Thus $|\hbthetan-\bthetazero|_\infty=O_{\rm p}(\epsilon_n\nu)$. Due to $b_n^2=o(n^{-2/\gamma})$, we can select arbitrary slowly diverging $\epsilon_n$. We then have $|\hbthetan-\bthetazero|_\infty=O_{\rm p}(\nu)$ by a standard result from probability theory. For any $\btheta \in \bTheta$ satisfying $|\btheta-\bthetazero|_{\infty} > \epsilon_n \nu$, let $j_{0}=\arg \max _{j \in [r]}|\mE\{g_{i,j}(\btheta)\}|$ and $\mu_{j_{0}}=\mE\{g_{i,j_{0}}(\btheta)\} $. Define $\tilde{\blambda}=\tau b_n \epsilon_n \bfe_{j_{0}}$, where $\tau>0$ is a constant to be determined later and $\bfe_{j_{0}}$ is an $r$-dimensional vector with the $j_{0}$-th component being $1$ and other components being $0$. Without loss of generality, we assume $\mu_{j_{0}}>0$. Condition \ref{A.ee}(a) implies $\max_{i\in[n]} |\tilde{\blambda}^\T \bfg_{i}(\btheta)|= O_{\rm p}(b_n \epsilon_n n^{1/\gamma})=o_{\rm p}(1)$. Then $\tilde{\blambda} \in \wLambadn(\btheta)$ w.p.a.1. Write $\tilde{\blambda}=(\tilde{\lambda}_{1},\ldots, \tilde{\lambda}_{r})^\T$. By  the Taylor expansion, it holds w.p.a.1 that
		\begin{align*}
			F_n(\btheta)
			\geq \frac{1}{n} \sum_{i=1}^{n} \log \{ 1+\tilde{\blambda}^\T \bfg_i (\btheta) \} - P_{\nu}(|\tilde{\lambda}_{j_{0}}|)
			& \geq \tilde{\lambda}_{j_{0}} \bar{g}_{j_{0}}(\btheta) - \frac{1}{2n} \sum _{i=1}^{n}\frac{\{\tilde{\lambda}_{j_{0}} g_{i,j_{0}}(\btheta)\}^2 }{\{1+\bar{C}\tilde{\lambda}_{j_{0}} g_{i,j_{0}}(\btheta)\}^2}-C\nu \tilde{\lambda}_{j_{0}} \\
			& \geq \tilde{\lambda}_{j_{0}} \bar{g}_{j_{0}}(\btheta) - \tilde{\lambda}_{j_{0}}^2 \mathbb{E}_n\{g_{i,j_{0}}^2(\btheta)\}-C\nu \tilde{\lambda}_{j_{0}}
		\end{align*}
		for some $\bar{C}\in(0,1)$, which implies 
		$$
		\P\big\{F_n(\btheta)
		\leq Kb_n^2 \epsilon_n^2 \big\}
		\leq \P\big\{\bar{g}_{j_{0}}(\btheta)-\mu_{j_{0}} \leq b_n \epsilon_n[ K\tau^{-1} + \tau \mathbb{E}_n\{g^2_{i,j_{0}}(\btheta)\}]+C\nu-\mu_{j_{0}}\big\}+o(1)\,.$$ 
		From Condition \ref{A.ee}(a) and Jensen's inequality, there exists a universal constant $L>0$ independent of $\btheta$ such that $\P[\mathbb{E}_n\{g^2_{i,j_{0}}(\btheta)\} >L] \rightarrow 0$. Taking $\tau=(KL^{-1})^{1/2}$, we obtain
		$
		\P\{F_n(\btheta)
		\leq K b_n^2 \epsilon_n^2 \}
		\leq \P \{\bar{g}_{j_{0}}(\btheta)-\mu_{j_{0}} \leq 2 b_n\epsilon_n (KL)^{1/2} +C\nu -\mu_{j_{0}}\}+o(1)$. 
		By Condition \ref{A.GIC}, $\mu_{j_{0}} \geq \Delta(\epsilon_n \nu) \geq K_{1} \epsilon_n \nu/2$ with $K_1$ defined in Condition \ref{A.GIC} for sufficiently large $n$. We select sufficiently small $K>0$. Due to $b_n=o(\nu)$, when $n$ is sufficiently large, $2 b_n \epsilon_n (KL)^{1/2}+C\nu-\mu_{j_{0}} \leq -\check{C}\mu_{j_{0}}$ for some $\check{C}\in(0,1)$. Hence, we have
		$
		\sqrt{n}\{2 b_n\epsilon_n (KL)^{1/2}+C\nu-\mu_{j_{0}}\}
		\leq -\check{C}\sqrt{n} \mu_{j_{0}} \rightarrow -\infty$. 
		By the Central Limit Theorem, $\sqrt{n}\{\bar{g}_{j_{0}}( \btheta)- \mu_{j_{0}}\}\rightarrow\mathcal{N}(0, \sigma^2)$ in distribution for some $\sigma>0$, which implies that $\P\{F_n(\btheta) \leq Kb_n^2 \epsilon_n^2\} \rightarrow 0$ as $n \rightarrow \infty$ for any $\btheta \in \bTheta$ satisfying $|\btheta-\bthetazero|_{\infty} > \epsilon_n\nu$. We complete the proof of Proposition \ref{pro.cons}.
		\hfill $\Box$
		
		\section{Proof of Theorem \ref{th.norm}}\label{sec:pfThm1}
		
		Recall $\hat{\blambda}(\btheta)=\arg \max _{\blambda \in \wLambadn(\btheta)} f_n(\blambda;\btheta)$ and $\mathcal{R}_n={\rm supp}\{\hat{\blambda}(\hbthetan)\}$. We first present two lemmas whose proofs are given in Sections \ref{sec:pflemma3} and \ref{sec:pflemma4}, respectively.
		\begin{lemma}\label{l.lam.thetan.hat}
			Let $P_{\nu}(\cdot) \in \mathscr{P}$ be a convex function with bounded second-order derivative around $0$, where $\mathscr{P}$ is defined in {\rm \eqref{eq:penalty}}. Under the conditions of Proposition {\rm \ref{pro.cons}} and Conditions {\rm \ref{A.ee}(c)} and {\rm \ref{A.g_subgra.1}(a)}, if $\ell_n\nu^2=o(1)$, it holds w.p.a.1 that the global maximizer $\hat{\blambda}(\hbthetan)=(\hat{\lambda}_{1},\ldots,\hat{\lambda}_{r})^\T$ for $f_n(\blambda;\hbthetan)$ w.r.t $\blambda$ satisfies: ${\rm (i)}$ $|\hat{\blambda}(\hbthetan)|_2=O_{\rm p}(\ell_n^{1/2} \alpha_n)$, ${\rm (ii)}$ $\mathcal{R}_n \subset \mathcal{M}_{\hbthetan}(\tilde{c})$ with $\tilde{c}$ given in Condition {\rm \ref{A.g_subgra.1}(a)}, and ${\rm (iii)}$ $\sgn(\hat{\lambda}_j) =\sgn\{\bar{g}_j(\hbthetan)\}$ for any $j \in \mathcal{M}_{\hbthetan}(\tilde{c})$ with $\hat{\lambda}_j \neq 0$.
		\end{lemma}
		
		\begin{lemma}\label{l.envlope}
			%	Let $P_{\nu}(\cdot) \in \mathcal{P}$ be a convex function for $\mathcal{P}$ defined in {\rm \eqref{eq:penalty}} with bounded second derivative around 0. Under the conditions of Proposition {\rm \ref{pro.cons}} and Condition {\rm \ref{A.g_subgra.1}}, if $\ell_n\nu^2=o(1)$,
			Under the conditions of Lemma {\rm \ref{l.lam.thetan.hat}} and Condition {\rm \ref{A.g_subgra.1}(b)}, it holds w.p.a.1 that the global maximizer $\hat{\blambda}(\btheta)$ for $f_n(\blambda;\btheta)$ w.r.t $\blambda$ is continuously differentiable at $\hbthetan$ and $[\nabla_{\btheta} \hat{\blambda}(\hbthetan)]_{\mathcal{R}_n^\c,[p]}=\bzero$.
		\end{lemma}
		For simplicity, we write $\hat{\blambda}(\hbthetan)$ as $\hat{\blambda}=(\hat\lambda_1,\ldots,\hat\lambda_r)^\T$. Then we have %$\nabla_{\blambda}f_n(\hat{\blambda};\hbthetan)=\bzero$, i.e.
		\begin{align}\label{eq:partial1}
			\bzero=\frac{1}{n} \sum_{i=1}^{n} \frac{\bfg_{i}(\hbthetan)}{1+\hat{\blambda}^\T \bfg_{i}(\hbthetan)}-\hat{\bseta} \,,
		\end{align}
		where $\hat{\bseta}=(\hat{\eta}_{1},\ldots, \hat{\eta}_{r})^\T$ with $\hat{\eta}_j=\nu \rho' (|\hat{\lambda}_j|;\nu) \sgn(\hat{\lambda}_j)$ for $\hat{\lambda}_j\neq0$ and $\hat{\eta}_j \in [-\nu \rho'(0^{+}),\nu \rho'(0^{+})]$ for $\hat{\lambda}_j=0$. By the Taylor expansion, we know that
		\begin{align*}
			\bzero
			=\bar{\bfg}_{\Rn}(\hbthetan) - \frac{1}{n}\sum _{i=1}^{n}\frac{\bfg_{i,\Rn}(\hbthetan)^{\otimes2} \hat{\blambda}_\Rn}{\{1+C \hat{\blambda}_\Rn^\T \bfg_{i,\Rn}(\hbthetan)\}^2} - \hat{\bseta}_\Rn =:\, \bar{\bfg}_\Rn(\hbthetan) - \bfA(\hbthetan)\hat{\blambda}_\Rn- \hat{\bseta}_\Rn
		\end{align*}
		for some $C\in(0,1)$, which implies $\hat{\blambda}_\Rn= \bfA^{-1}(\hbthetan)\{\bar{\bfg}_\Rn(\hbthetan)- \hat{\bseta}_\Rn\}$.
		Since $\hbthetan=\arg\min_{\btheta \in \bTheta} f_n\{\hat{\blambda}(\btheta);\btheta\}$, we have $\bzero
		=\nabla_{\btheta} f_n\{\hat{\blambda}(\btheta);\btheta\}|_{\btheta=\hbthetan}$. Notice that
		\begin{align*}
			\nabla_{\btheta} f_n\{\hat{\blambda}(\btheta);\btheta\}|_{\btheta=\hbthetan}
			=\bigg\{\frac{\partial f_n(\hat{\blambda} ; \hbthetan)}{\partial \blambda_\Rn^\T}  [\nabla_{\btheta} \hat{\blambda}(\hbthetan)]_{\Rn,[p]} + \frac{\partial f_n(\hat{\blambda} ; \hbthetan)}{\partial\blambda_{\mathcal{R}_n^\c}^\T} [\nabla_{\btheta} \hat{\blambda}(\hbthetan)]_{\mathcal{R}_n^\c,[p]}\bigg\}^\T + \frac{\partial f_n(\hat{\blambda} ; \hbthetan)}{\partial \btheta} \,.
		\end{align*}
		By Lemma \ref{l.envlope} and \eqref{eq:partial1}, we have $[\nabla_{\btheta} \hat{\blambda}(\hbthetan)]_{\mathcal{R}_n^\c,[p]}=\bzero$ w.p.a.1 and $\partial f_n(\hat{\blambda};\hat{\btheta}_n)/\partial\blambda_{\Rn} = \bzero$. Thus, it holds w.p.a.1 that
		\begin{align*}
			\bzero = \frac{\partial f_n(\hat{\blambda} ; \hbthetan)}{\partial \btheta}
			=\bigg\{\frac{1}{n} \sum_{i=1}^{n} \frac{\nabla_{\btheta} \bfg_{i, \Rn} ( \hbthetan)}{1+\hat{\blambda}^\T_\Rn\bfg _{i, \Rn}(\hbthetan)} \bigg\}^\T \hat{\blambda}_\Rn=: \bfB(\hbthetan)^\T\hat{\blambda}_\Rn\,.
		\end{align*}
		We then obtain 
		$
		\bzero
		= \bfB(\hbthetan)^\T \bfA^{-1}(\hbthetan) \{\bar{\bfg}_\Rn(\hbthetan)- \hat{\bseta}_\Rn\}$. %Recall $\widehat{\bGamma}_{\mF}(\btheta) =\nabla_{\btheta} \bar\bfg_{\mF}(\btheta)$ and $\widehat{\bfV}_{\mF}(\btheta)=\mathbb{E}_n\{\bfg_{i,\mF}(\btheta)^{\otimes2}\}$ for any $\btheta\in\bTheta$ and $\mF\subset [r]$. 
		To derive the limiting distribution of $\hbthetan$, we need the following lemmas whose proofs are given in Sections \ref{sec:pflemma5}, \ref{sec:pflemma6} and \ref{sec:pflemma7}, respectively.
		\begin{lemma}\label{l.ee}
			%Assume that the conditions of Proposition {\rm \ref{pro.cons}} hold. If $\ell_n\nu^2=o(1)$, then
			Under the conditions of Lemma {\rm \ref{l.lam.thetan.hat}},
			$
			\| \bfA(\hbthetan) - \widehat{\bfV}_\Rn (\hbthetan) \|_2 =O_{\rm p}(\ell_n n^{1/\gamma}\alpha_n)
			$,
			and
			$
			| \{\bfB(\hbthetan)- \widehat{\bGamma}_\Rn (\hbthetan) \}\bft |_2=|\bft|_2 \cdot O_{\rm p}(\ell_n\alpha_n)
			$
			holds uniformly over $\bft \in \R^p$.
		\end{lemma}
		
		\begin{lemma}\label{l.Gamma.hat}
			Assume that the conditions of Proposition {\rm \ref{pro.cons}} and Condition {\rm \ref{A.ee}(c)} hold. For $\mathscr{F}$ defined in Lemma {\rm\ref{l.V.hat}},
			$$
			\sup_{\mathcal{F} \in \mathscr{F}} | \{\widehat{\bGamma}_{\mF}(\hbthetan)- \bGamma_{\mF}(\bthetazero)\} \bft |_2= |\bft|_2 \cdot\{O_{\rm p}(\ell_n^{1/2} \nu)+O_{\rm p}(\ell_n^{1/2} \alpha_n)\}
			$$
			holds uniformly over $\bft \in \R^p$.
		\end{lemma}
		
		\begin{lemma}\label{l.norm}
			Let $\widehat{\bfH}_{\mF}=\{\widehat{\bGamma}_{\mF}(\hbthetan)^\T \widehat{\bfV}^{-1/2}_{\mF} (\hbthetan)\}^{\otimes2}$ for any $\mathcal{F} \in \mathscr{F}$, where $\mathscr{F}$ is defined in Lemma {\rm\ref{l.V.hat}}. Assume that the conditions of Proposition {\rm \ref{pro.cons}} and Conditions {\rm \ref{A.ee}(c)} and {\rm \ref{A.ee2}} hold. If $\ell_n^2 \nu^2\log r=o(1)$ and $\ell_n^{3}\alpha_n^2\log r=o(1)$, for any $\bft \in \R^p$ with $|\bft|_2=1$, we have 
			$$\sup_{\mathcal{F} \in \mathscr{F}} \sup_{u \in \R} |\P\{n^{1/2} \bft^\T \widehat{\bfH}_{\mF}^{-1/2} \widehat{\bGamma}_{\mF}(\hbthetan)^\T \widehat{\bfV}^{-1}_{\mF} (\hbthetan) \bar{\bfg}_{\mF} (\bthetazero) \leq u \} - \Phi(u) | \rightarrow 0$$ as $n \rightarrow \infty$, 
			where $\Phi(\cdot)$ is the cumulative distribution function of the standard Gaussian distribution.
		\end{lemma}
		
		For any $\bft \in \R^p$ with $|\bft|_2=1$, let $\bdelta=\widehat{\bfH}_\Rn^{-1/2} \bft$ and $\bfU = \widehat{\bfV}^{-1/2}_\Rn (\hbthetan) \widehat{\bGamma}_\Rn(\hbthetan)$. Then $\widehat{\bfH}_{\Rn} = \bfU^{\T, \otimes2}$ and 
		$
		|\widehat{\bGamma}_\Rn(\hbthetan) \bdelta|_2^2
		\leq \lambda_{\max} \{\widehat{\bfV}_\Rn (\hbthetan)\} |\bfU (\bfU^{\T, \otimes2})^{-1/2} \bft|_2^2= \lambda_{\max} \{\widehat{\bfV}_\Rn (\hbthetan)\}$. 
		By Condition \ref{A.ee}(b) and Lemma \ref{l.V.hat}, $|\widehat{\bGamma}_\Rn(\hbthetan) \bdelta|_2=O_{\rm p}(1)$.
		Under Conditions \ref{A.ee}(b) and \ref{A.ee2}, Lemmas \ref{l.V.hat} and \ref{l.Gamma.hat} imply $|\bdelta|_2^2 \leq \lambda_{\max} \{\widehat{\bfV}_\Rn (\hbthetan)\} \lambda_{\min}^{-1}\{\widehat{\bGamma}_\Rn(\hbthetan)^{\T,\otimes2} \} = O_{\rm p}(1)$.
		By Lemma \ref{l.lam.thetan.hat}, we have w.p.a.1 that $\mathcal{R}_n \subset \mathcal{M}_{\hbthetan}(\tilde{c})$ and $\sgn(\hat{\lambda}_j) =\sgn\{\bar{g}_j(\hbthetan)\}$ for any $j \in \mathcal{R}_n$. Since $P_{\nu}(\cdot) \in \mathscr{P}$ has bounded second-order derivative around $0$, by Lemma \ref{l.lam.thetan.hat}, it holds w.p.a.1 that
		\begin{align*}
			|\nu \rho'(0^+) \sgn \{\bar{\bfg}_\Rn(\hbthetan)\} - \hat{\bseta}_\Rn|_2^2=&\, \sum_{j \in \Rn} \{\nu \rho'(0^+) \sgn(\hat{\lambda}_j)-\nu \rho' (|\hat{\lambda}_j|;\nu) \sgn(\hat{\lambda}_j)\}^2 \\
			=&\, \sum_{j \in \Rn} \{\nu \rho'' (c_j|\hat{\lambda}_j|;\nu) |\hat{\lambda}_j| \}^2 \leq C|\hat{\blambda}|_2^2=O_{\rm p}(\ell_n\alpha_n^2)
		\end{align*}
		for some $c_j\in(0,1)$. As shown in the proof of Lemma \ref{l.lam.thetan.hat}, 
		$
		|\bar{\bfg}_{\mathcal{M}_{\hbthetan}(\tilde{c})}(\hbthetan)-\nu \rho'(0^+) \sgn\{\bar{\bfg}_{\mathcal{M}_{\hbthetan}(\tilde{c})}(\hbthetan)\}|_2=O_{\rm p}(\ell_n^{1/2} \alpha_n)$. Then $|\bar{\bfg}_\Rn(\hbthetan)-\hat{\bseta}_\Rn|_2 =O_{\rm p}(\ell_n^{1/2}\alpha_n)$. Due to $
		\bfB(\hbthetan)^\T \bfA^{-1}(\hbthetan) \{\bar{\bfg}_\Rn(\hbthetan)- \hat{\bseta}_\Rn\}=\bzero$, by the triangle inequality, %we have
		\begin{align*}
			|\bdelta^\T \widehat{\bGamma}_\Rn(\hbthetan)^\T \widehat{\bfV}_\Rn^{-1} (\hbthetan) \{\bar{\bfg}_\Rn(\hbthetan)-\hat{\bseta}_\Rn \}|\leq &\, \underbrace{|\bdelta^\T \widehat{\bGamma}_\Rn(\hbthetan)^\T \{\widehat{\bfV}_\Rn^{-1} (\hbthetan)-\bfA^{-1}(\hat{\btheta}_n) \}\{\bar{\bfg}_\Rn(\hbthetan)- \hat{\bseta}_\Rn\}|}_{T_1} \\
			&+\underbrace{|\bdelta^\T \{\widehat{\bGamma}_\Rn(\hbthetan)-\bfB(\hat{\btheta}_n)\}^\T \bfA^{-1}(\hbthetan) \{\bar{\bfg}_\Rn(\hbthetan)- \hat{\bseta}_\Rn\} |}_{T_2} \,.
		\end{align*}
		By Lemma \ref{l.ee}, %we have
		\begin{align*}
			&T_1\leq |\widehat{\bGamma}_\Rn(\hbthetan) \bdelta|_2 \|\widehat{\bfV}_\Rn^{-1} (\hbthetan)-\bfA^{-1}(\hbthetan)\|_2|\bar{\bfg}_\Rn(\hbthetan)- \hat{\bseta}_\Rn|_2
			=O_{\rm p}(\ell_n^{3/2} n^{1/\gamma}\alpha_n^2)\,,\\ 
			&~~~T_2
			\leq |\{\widehat{\bGamma}_\Rn(\hbthetan)- \bfB(\hbthetan)\}\bdelta|_2\|\bfA^{-1}(\hbthetan)\|_2 |\bar{\bfg}_\Rn(\hbthetan)- \hat{\bseta}_\Rn|_2
			= O_{\rm p}(\ell_n^{3/2}\alpha_n^2)\,.
		\end{align*}
		Hence, $\bdelta^\T \widehat{\bGamma}_\Rn(\hbthetan)^\T \widehat{\bfV}_\Rn^{-1} (\hbthetan) \{\bar{\bfg}_\Rn(\hbthetan)-\hat{\bseta}_\Rn \}= O_{\rm p}(\ell_n^{3/2} n^{1/\gamma} \alpha_n^2)$. By the Taylor expansion, we have
		\begin{align} \label{eq:asp_normal}
			&\bdelta^\T \widehat{\bGamma}_\Rn(\hbthetan)^\T \widehat{\bfV}_\Rn^{-1} (\hbthetan) \{\widehat{\bGamma}_\Rn(\tbtheta) (\hbthetan-\bthetazero) -\hat{\bseta}_\Rn \}\notag\\
			&~~~~~~~~~~~~~~~=-\bdelta^\T \widehat{\bGamma}_\Rn(\hbthetan)^\T \widehat{\bfV}_\Rn^{-1} (\hbthetan) \bar{\bfg}_\Rn(\bthetazero)+ O_{\rm p}(\ell_n^{3/2} n^{1/\gamma} \alpha_n^2) \,,
		\end{align}
		where $\tbtheta$ is on the line joining $\bthetazero$ and $\hbthetan$. Write $\hbthetan=(\hat{\theta}_{n,1},\ldots,\hat{\theta}_{n,p})^\T$, $\bthetazero=(\theta_{0,1},\ldots,\theta_{0,p})^\T$ and $\tbtheta=(\tilde{\theta}_{1},\ldots,\tilde{\theta}_{p})^\T$. 
		By the Taylor expansion, Jensen's inequality and Cauchy-Schwarz inequality, it holds that
		\begin{align*}
			| \{\widehat{\bGamma}_\Rn(\tbtheta)- \widehat{\bGamma}_\Rn(\hbthetan)\} (\hbthetan-\bthetazero)|_2^2
			=&\,\sum _{j \in \Rn} \bigg[ \frac{1}{n} \sum_{i=1}^{n} \sum_{k=1}^p (\hat{\theta}_{n,k}-\theta_{0,k}) \sum_{l=1}^p \frac{\partial^2 g_{i,j}\{\dot{\btheta}^{(j,k)}\}}{\partial \theta_{k} \partial \theta_{l}}(\tilde{\theta}_{l}-\hat{\theta}_{n,l})\bigg]^2 \\
			%\leq&\, \sum _{j \in \Rn} \frac{1}{n} \sum_{i=1}^{n} \bigg\{\sum_{k=1}^p (\hat{\theta}_{n,k}-\theta_{0,k}) \sum_{l=1}^p \frac{\partial^2 g_{i,j}(\dot{\btheta})}{\partial \theta_{k} \partial \theta_{l}}(\tilde{\theta}_{l}-\hat{\theta}_{n,l}) \bigg\}^2\\
			%\leq&\, |\hbthetan-\bthetazero|^2_2 \cdot \frac{1}{n} \sum _{j \in \Rn} \sum_{i=1}^{n} \sum_{k=1}^p \bigg\{ \sum_{l=1}^p \frac{\partial^2 g_{i,j}(\dot{\btheta})}{\partial \theta_{k} \partial \theta_{l}}(\tilde{\theta}_{l}-\hat{\theta}_{n,l}) \bigg\}^2 \\
			%\leq&\, |\hbthetan-\bthetazero|^2_2 \cdot\frac{1}{n} \sum _{j \in \Rn} \sum_{i=1}^{n} \sum_{k=1}^p \sum_{l=1}^p \bigg|\frac{\partial^2 g_{i,j}(\dot{\btheta})}{\partial \theta_{k} \partial \theta_{l}}\bigg|^2 |\tbtheta-\hbthetan|_2^2 \\
			\leq&\, \frac{1}{n} \sum _{j \in \Rn} \sum_{i=1}^{n} \sum_{k=1}^p \sum_{l=1}^p \bigg|\frac{\partial^2 g_{i,j}\{\dot{\btheta}^{(j,k)}\}}{\partial \theta_{k} \partial \theta_{l}}\bigg|^2 \cdot|\hbthetan-\bthetazero|^4_2 \,,
		\end{align*}
		where $\dot{\btheta}^{(j,k)}$ lies on the jointing line between $\tbtheta$ and $\hbthetan$. Recall $p$ is fixed. 
		By Proposition \ref{pro.cons}, $|\hbthetan-\bthetazero|_2=O_{\rm p}(\nu)$. Together with Condition {\rm \ref{A.ee}(c)}, we have $| \{\widehat{\bGamma}_\Rn(\tbtheta)- \widehat{\bGamma}_\Rn(\hbthetan)\} (\hbthetan-\bthetazero)|_2= O_{\rm p}(\ell_n^{1/2} \nu^2)$. Recall $\hat{\bpsi}_\Rn= \widehat{\bfH}_\Rn^{-1} \widehat{\bGamma}_\Rn(\hbthetan)^\T \widehat{\bfV}^{-1}_\Rn (\hbthetan) \hat{\bseta}_\Rn$ and $\bdelta=\widehat{\bfH}_\Rn^{-1/2} \bft$. Then \eqref{eq:asp_normal} leads to
		\begin{align*}
			n^{1/2}\bft^\T \widehat{\bfH}_\Rn^{1/2}(\hbthetan-\bthetazero-\hat{\bpsi}_\Rn) 
			=&\, -n^{1/2}\bft^\T \widehat{\bfH}^{-1/2}_\Rn \widehat{\bGamma}_\Rn(\hbthetan)^\T \widehat{\bfV}_\Rn^{-1} (\hbthetan) \bar{\bfg}_\Rn(\bthetazero)\\
			&\,+ O_{\rm p}(\ell_n^{3/2} n^{1/2+1/\gamma} \alpha_n^2) +O_{\rm p}(\ell_n^{1/2} \nu^2 n^{1/2})\,.
		\end{align*}
		By Lemma \ref{l.norm}, we have %$n^{1/2} \bft^\T \widehat{\bfH}_{\Rn}^{-1/2} \widehat{\bGamma}_{\Rn}(\hbthetan)^\T \widehat{\bfV}^{-1}_{\Rn} (\hbthetan) \bar{\bfg}_{\Rn} (\bthetazero) \rightarrow \mathcal{N}(0,1)$ in distribution as $n \rightarrow \infty$. Then 
		$
		n^{1/2} \bft^\T \widehat{\bfH}_\Rn^{1/2}(\hbthetan-\bthetazero-\hat{\bpsi}_\Rn)\rightarrow \mathcal{N}(0,1)$ in distribution as $n \rightarrow \infty$. \hfill$\Box$
		%We complete the proof of Theorem \ref{th.norm}.
		%\hfill $\Box$

		\section{Proof of Theorem \ref{th.TV}} \label{sec.th2}
		
		Assume $(r,\ell_n,\nu)$ satisfy the following restrictions:
		\begin{equation}\label{eq:restriction2}
			\log r=o(n^{1/3})\,,~~\ell_n \ll n^{(\gamma-2)/(9\gamma)}(\log r)^{-1/9}~~\textrm{and}~~\ell_nn^{-1/2}(\log r)^{1/2} \ll \nu \ll \ell_n^{-7/2}n^{-1/\gamma}\,.
		\end{equation}	 
		To construct Theorem \ref{th.TV}, we need the following proposition whose proof is given in Section \ref{sec.pf_pro2}.
		\begin{proposition}\label{pro.expan}
			Let $P_{\nu}(\cdot) \in \mathscr{P}$ be convex and assume $\rho(t;\nu)=\nu^{-1}P_{\nu}(t)$ has bounded second-order derivative with respect to $t$ around $0$, where $\mathscr{P}$ is defined as {\rm \eqref{eq:penalty}}. Assume $(r,\ell_n,\nu)$ satisfy \eqref{eq:restriction2}. 
			
			{\rm (i)} Under Conditions {\rm \ref{A.GIC}--\ref{A.ee2}}, {\rm \ref{A.g_subgra.1}(a)} and {\rm \ref{A.Pro2}(a)}, then 
			$
			\aleph_n(\btheta)=  2^{-1} (\btheta-\hat\btheta_n)^\T \widehat\bfH_{\mathcal{R}_n} (\btheta-\hat\btheta_n) + |\btheta-\hat\btheta_n|_2^2 \cdot O_{\rm p}(\varpi_n)
			$
			holds uniformly over $\btheta \in \mathcal{C}_1$ with $\varpi_n=\max\{\ell_n^{3/2}\alpha_n, \nu, \ell_nn^{1/\gamma}\alpha_n\}$, where $\widehat\bfH_{\mathcal{R}_n}$ is defined in \eqref{eq:bias} and the term $O_{\rm p}(\varpi_n) $ holds uniformly over $\btheta \in \mathcal{C}_1$.
			
			{\rm (ii)} Under Conditions {\rm \ref{A.GIC}}, {\rm \ref{A.ee}}, {\rm \ref{A.g_subgra.1}(a)} and {\rm \ref{A.Pro2}(b)}, then 
			$
			\inf_{\btheta \in \mathcal{C}_2}  \aleph_n(\btheta)\geq (8K_4)^{-1}\kappa_{n}^2
			$
			with probability approaching one, where $K_4$ and $\kappa_n$ are specified in Conditions {\rm\ref{A.ee}(b)} and {\rm\ref{A.Pro2}(b)}, respectively.
			
			{\rm (iii)} Under Conditions {\rm \ref{A.GIC}}, {\rm \ref{A.ee}}, {\rm \ref{A.g_subgra.1}(a)} and {\rm \ref{A.Pro2}(c)}, then $\inf_{\btheta \in \mathcal{C}_3}  \aleph_n(\btheta) \geq  4^{-1}K_7^{1/2}  \xi_n\beta_{n}$ with probability approaching one for any $\xi_n$ satisfying $\beta_n^{-1}\ell_n\alpha_n^2 \ll \xi_n \ll \beta_n$, where $K_7$ is specified in Condition {\rm\ref{A.Pro2}(c)}.
			
		\end{proposition}

		Recall the posterior distribution $\pi^\dag(\btheta\,|\,\mathcal{X}_n) \propto  \pi_{0}(\btheta)\times \exp [-n\log n-n f_n\{\hat\blambda(\btheta);\btheta \}]I(\btheta \in \bTheta)$. 
		For any $\btheta \in \bTheta$, let $w_n(\btheta)=-n\log n-nf_n\{\hat\blambda(\btheta); \btheta\}$ and write $\bft = n^{1/2}(\btheta-\hat\btheta_n)$. Define $\mathcal{T}_n = \{\bft \in \R^p: \bft = n^{1/2}(\btheta-\hat\btheta_n), \btheta \in \bTheta\}$. Denote by $\pi_{0,\bft}(\cdot)$ and $\pi^{\dag}_{\bft}(\cdot\,|\,\mathcal{X}_n)$ the prior and the posterior distributions of $\bft$, respectively. Then,  $\pi_{0,\bft}(\bft)=n^{-p/2}\pi_{0}(\hbthetan+n^{-1/2}\bft)$ and
		\begin{align} \label{eq:pos_t}
			\pi^{\dag}_{\bft}(\bft\,|\,\mathcal{X}_n)
			&=\frac{\pi_{0}(\hbthetan+n^{-1/2}\bft) \exp \{ w_n(\hbthetan+n^{-1/2}\bft)-w_n(\hbthetan)\} I(\bft \in \mathcal{T}_n)} {\int_{\R^{p}} \pi_{0}(\hbthetan+n^{-1/2}\bfs) \exp \{ w_n(\hbthetan+n^{-1/2}\bfs)-w_n(\hbthetan)\}  I(\bfs \in \mathcal{T}_n) \,{\rm d}\bfs} \notag\\
			&=:C^{-1}_n \pi_{0}(\hbthetan+n^{-1/2}\bft) \exp \{ w_n(\hbthetan+n^{-1/2}\bft)-w_n(\hbthetan)\} I(\bft \in \mathcal{T}_n) \,.
		\end{align} 
		To prove Theorem \ref{th.TV}, it is equivalent to show
		\begin{align} \label{eq:Int.TV}
			&\int_{\R^{p}} \big |C^{-1}_n \pi_{0}(\hbthetan+n^{-1/2}\bft) \exp \{ w_n(\hbthetan+n^{-1/2}\bft)-w_n(\hbthetan)\} I(\bft \in \mathcal{T}_n) \notag \\ 
			&~~~~~~~~~~~~~~~~~~~~~- (2\pi)^{-p/2}|\widehat{\bfH}_\Rn|^{1/2} \exp(-\bft^\T \widehat{\bfH}_\Rn \bft /2 )   \big | \,{\rm d}\bft \rightarrow 0 
		\end{align}
		in probability. 
		It follows from the triangle inequality that
		\begin{align} \label{eq:Int.TV_tri}
			& \int_{\R^{p}} \big |C^{-1}_n \pi_{0}(\hbthetan+n^{-1/2}\bft) \exp \{ w_n(\hbthetan+n^{-1/2}\bft)-w_n(\hbthetan)\} I(\bft \in \mathcal{T}_n) \notag \\ 
			&~~~~~~~~~~~~~~~~~~~~~  - (2\pi)^{-p/2}|\widehat{\bfH}_\Rn|^{1/2} \exp( -\bft^\T \widehat{\bfH}_\Rn \bft/2 )   \big | \,{\rm d}\bft \notag \\
			&~~~~~\leq C^{-1}_n \int_{\R^{p}} \big | \pi_{0}(\hbthetan+n^{-1/2}\bft) \exp \{ w_n(\hbthetan+n^{-1/2}\bft)-w_n(\hbthetan)\} I(\bft \in \mathcal{T}_n) \notag \\
			&~~~~~~~~~~~~~~~\underbrace{~~~~~~~~~~~~~~~~~ - \pi_{0}(\hbthetan)\exp( -\bft^\T \widehat{\bfH}_\Rn \bft/2 ) \big | \,{\rm d}\bft~~~~~~~~~~~~~~~~~~~~~~~~}_{\rm I} \notag \\
			&~~~~~~~~ + C^{-1}_n  \underbrace{\int_{\R^{p}} \big| \pi_{0}(\hbthetan)\exp( -\bft^\T \widehat{\bfH}_\Rn \bft/2 )-C_n (2\pi)^{-p/2}|\widehat{\bfH}_\Rn|^{1/2} \exp( -\bft^\T \widehat{\bfH}_\Rn \bft/2 ) \big| \,{\rm d}\bft}_{\rm II} \,.
		\end{align}
		Notice that
		$
		{\rm I}
		\geq |C_n - (2\pi)^{p/2} \pi_{0}(\hbthetan) |\widehat{\bfH}_\Rn|^{-1/2} | 
		=
		{\rm II}
		$.
		%Notice that $\|\widehat{\bfH}_\Rn\|_2$ and $\|\widehat{\bfH}^{-1}_\Rn\|_2$ are bounded away from zero and infinity w.p.a.1 as shown in the proof of Lemma \ref{l.norm}, which implies $|\widehat{\bfH}_\Rn|^{\pm 1/2}=O_{\rm p}(1)$. 
		To show \eqref{eq:Int.TV}, it suffices to show $C_n^{-1} {\rm I} =o_{\rm p}(1)$. Recall $\widehat{\bfH}_{\Rn}=\{\widehat{\bGamma}_{\Rn}(\hat\btheta_n)^\T\widehat{\bfV}_{\Rn}^{-1/2}(\hat\btheta_n)\}^{\otimes2}$. Under Conditions \ref{A.ee}(b) and \ref{A.ee2}, by Proposition \ref{pro.cons}, Lemmas \ref{l.V.hat} and \ref{l.Gamma.hat}, if $\log r=o(n^{1/3})$, $\ell_n\alpha_n=o[\min\{\nu,n^{-1/\gamma}\}]$ and $\ell_n \nu^2=o(1)$, we know that the eigenvalues of $\widehat{\bfH}_\Rn$ are uniformly bounded away from zero and infinity w.p.a.1. Notice that $\widehat{\bfH}_{\Rn}$ is a $p \times p$ matrix with fixed $p$. Thus, $|\widehat{\bfH}_\Rn|^{-1/2}$ is uniformly bounded away from zero w.p.a.1. Since $\pi_{0}(\btheta)$ is bounded away from zero around $\bthetazero$ and $|\hbthetan - \bthetazero|_\infty=O_{\rm p}(\nu)$, we know $\pi_0(\hbthetan)$ is bounded away from zero w.p.a.1. If ${\rm I} =o_{\rm p}(1)$, then $| C_n - (2\pi)^{p/2} \pi_{0}(\hbthetan)  |\widehat{\bfH}_\Rn|^{-1/2} | =o_{\rm p}(1)$, which implies $C_n^{-1}=O_{\rm p}(1)$. Hence, to show \eqref{eq:Int.TV}, we only need to show ${\rm I} =o_{\rm p}(1)$.
		% i.e.
		%\begin{equation}\label{eq:total_integral}
		%	\int_{\R^{p}} \big | \pi_{0}(\hbthetan+\bft/\sqrt{n}) \exp \{ w_n(\hbthetan+\bft/\sqrt{n})-w_n(\hbthetan)\}- \pi_{0}(\hbthetan)\exp( -\bft^\T \widehat{\bfH}_\Rn \bft/2 ) \big | \,{\rm d}\bft \stackrel{P}{\longrightarrow} 0 \ .
		%\end{equation}
		Recall $\ell_n \ll \min\{n^{(\gamma-2)/(9\gamma)}(\log r)^{-1/9},n^{1/3}(\log r)^{-1},n^{(\gamma-2)/(2\gamma)}(\log r)^{-3/2}\}$ and $\ell_nn^{-1/2}(\log r)^{1/2} \ll \nu \ll \min\{\ell_n^{-7/2}n^{-1/\gamma},(\log r)^{-1}\}$. We break the domain of integration into four regions: 
		\begin{align} \label{eq:four_D}
			&\mathcal{D}_1= \{\bft\in \mathcal{T}_n: |\bft|_2 \leq n^{1/2} \alpha_n \}\,, ~~
			\mathcal{D}_2= \{\bft\in \mathcal{T}_n:  n^{1/2}\alpha_n < |\bft|_2 \leq n^{1/2}\beta_{n} \}\,, \notag \\
			&~~~~~~~~~~~~~~~~~\mathcal{D}_3= \{\bft\in \mathcal{T}_n:
			|\bft|_2 > n^{1/2} \beta_{n} \} \,, ~~ \mathcal{D}_4 = \mathcal{T}_n^\c \,.
		\end{align}
		Then 
		$
		{\rm I} = {\rm I}(1) + {\rm I}(2)+ {\rm I}(3)+{\rm I}(4)   
		$ with 
		\begin{align*}
			{\rm I}(k) = \int _{\mathcal{D}_k} \big | \pi_{0}(\hbthetan+ n^{-1/2}\bft) \exp \{w_n(\hbthetan+n^{-1/2}\bft)-w_n(\hbthetan)\} I(\bft \in \mathcal{T}_n) - \pi_{0}(\hbthetan)\exp( -\bft^\T \widehat{\bfH}_\Rn \bft/2 ) \big | \,{\rm d}\bft \,.
		\end{align*}
		In the sequel, we will show each ${\rm I}(k) = o_{\rm p}(1)$.  
		
		For ${\rm I}(3)$, by the triangle inequality, we have
		\begin{align*}
			{\rm I}(3) 
			\leq \int _{\mathcal{D}_3}  \pi_{0}(\hbthetan+n^{-1/2}\bft) \exp \{ w_n(\hbthetan+n^{-1/2}\bft)-w_n(\hbthetan)\} \,{\rm d}\bft
			+ \pi_{0}(\hbthetan) \int _{\mathcal{D}_3} \exp( -\bft^\T \widehat{\bfH}_\Rn \bft/2 ) \,{\rm d}\bft \,.
		\end{align*}
		Due to $\int_{\mathcal{D}_3}  \pi_{0}(\hbthetan+n^{-1/2}\bft) \,{\rm d}\bft \leq n^{p/2} \int_{\R^{p}} \pi_{0}(\btheta) \,{\rm d}\btheta \leq n^{p/2}$, Proposition \ref{pro.expan}(iii) implies that
		\begin{align*}
			&\int _{\mathcal{D}_3}  \pi_{0}(\hbthetan+n^{-1/2}\bft) \exp \{ w_n(\hbthetan+n^{-1/2}\bft)-w_n(\hbthetan)\} \,{\rm d}\bft \\
			&~~~~~\leq  \sup_{\bft \in \mathcal{D}_3}  \exp \{ w_n(\hbthetan +n^{-1/2}\bft)-w_n(\hbthetan)\} \cdot  \int_{\mathcal{D}_3}  \pi_{0} (\hbthetan+n^{-1/2}\bft) \,{\rm d}\bft
			\leq n^{p/2} \exp(- Cn\xi_n\beta_{n})
		\end{align*}
		w.p.a.1 for any $\beta_n^{-1}\ell_n \alpha_n^2 \ll \xi_n \ll \beta_n$. Since $r\gg n$, %$\ell_n \ll n^{1/9-2/(9\gamma)}(\log r)^{-1/9}$ and $\beta_n \gg \ell_n^{1/2} \nu \gg \ell_n^{3/2} n^{-1/2}(\log r)^{1/2}$, 
		we can select suitable $\xi_n$ satisfying $ n \beta_n \xi_n \gg  \log n $. Then 
		\begin{align*}
			\int _{\mathcal{D}_3}  \pi_{0}(\hbthetan+n^{-1/2}\bft) \exp \{ w_n(\hbthetan+n^{-1/2}\bft)-w_n(\hbthetan)\} \,{\rm d}\bft = o_{\rm p}(1) \,.
		\end{align*}
		%{\color{blue}Notice that $\bTheta\subset\mathbb{R}^p$ is a compact set including an Euclidean ball with radius $\iota>0$.} 
		Due to $n\beta_n^2 \rightarrow \infty$, Proposition 1.1 of \cite{Hsu2012} implies that 
		\begin{align*}
			\pi_{0}(\hbthetan)\int _{\mathcal{D}_3} \exp( -\bft^\T \widehat{\bfH}_\Rn \bft/2 ) \,{\rm d}\bft \leq (2\pi)^{p/2} \pi_{0}(\hbthetan) |\widehat{\bfH}_\Rn|^{-1/2} \exp(-\bar C n\beta_n^2) = o_{\rm p}(1) \,.
		\end{align*}
		Therefore, ${\rm I}(3) = o_{\rm p}(1)$.
		
		For ${\rm I}(2)$, it holds that
		\begin{align*}
			{\rm I}(2)
			\leq  \int_{\mathcal{D}_2}  \pi_{0}(\hbthetan+n^{-1/2}\bft) \exp \{ w_n(\hbthetan+n^{-1/2}\bft)-w_n(\hbthetan)\} \,{\rm d}\bft
			+ \pi_{0}(\hbthetan) \int_{\mathcal{D}_2} \exp( -\bft^\T \widehat{\bfH}_\Rn \bft/2 ) \,{\rm d}\bft \,.
		\end{align*}
		Since $n\alpha_n^2 \rightarrow \infty$, using the same arguments given above, we have $\pi_{0}(\hbthetan)\int _{\mathcal{D}_2} \exp( -\bft^\T \widehat{\bfH}_\Rn \bft/2 ) \,{\rm d}\bft = o_{\rm p}(1)$. By Proposition \ref{pro.expan}(ii), it then holds w.p.a.1 that
		\begin{align*}
			&\int _{\mathcal{D}_2}  \pi_{0}(\hbthetan+n^{-1/2}\bft) \exp \{ w_n(\hbthetan+n^{-1/2}\bft)-w_n(\hbthetan)\} \,{\rm d}\bft \\
			&~~~~~\leq \sup_{\bft \in \mathcal{D}_2} \exp \{ w_n(\hbthetan +n^{-1/2}\bft)-w_n(\hbthetan)\} \cdot \int_{\mathcal{D}_2}  \pi_{0}(\hbthetan+n^{-1/2}\bft) \,{\rm d}\bft
			\leq  n^{p/2} \exp(- Cn\kappa_{n}^2) \,.
		\end{align*}
		Since $\log n \ll n\kappa_{n}^2$, we have
		\begin{align*}
			\int _{\mathcal{D}_2}  \pi_{0}(\hbthetan+n^{-1/2}\bft) \exp \{ w_n(\hbthetan+n^{-1/2}\bft)-w_n(\hbthetan)\}\, {\rm d}\bft = o_{\rm p}(1) \,.
		\end{align*} 
		Therefore, ${\rm I}(2) = o_{\rm p}(1)$. 
		
		For ${\rm I}(1)$, by the triangle inequality, we have
		\begin{align*}
			{\rm I}(1) \leq& \int _{\mathcal{D}_{1}} \pi_{0}(\hbthetan+ n^{-1/2}\bft) \big |\exp \{w_n(\hbthetan+n^{-1/2}\bft)-w_n(\hbthetan)\} I(\bft \in \mathcal{T}_n) - \exp( -\bft^\T \widehat{\bfH}_\Rn \bft/2 ) I(\bft \in \mathcal{T}_n) \big | \,{\rm d}\bft \\
			&+ \int _{\mathcal{D}_{1}} \big |\pi_{0}(\hbthetan+n^{-1/2}\bft) I(\bft \in \mathcal{T}_n) - \pi_{0}(\hbthetan) \big | \exp( -\bft^\T \widehat{\bfH}_\Rn \bft/2 ) \,{\rm d}\bft  \,.
		\end{align*}
		Due to $\mathcal{D}_1 \subset \mathcal{T}_n$, by Proposition \ref{pro.expan}(i), it holds that
		\begin{align*}
			{\rm I}(1)
			\leq& \int _{\mathcal{D}_{1}} \pi_{0}(\hbthetan+n^{-1/2}\bft) \big | \exp\{-\bft^\T \widehat{\bfH}_\Rn \bft/2+ |\bft|_2^2\cdot O_{\rm p}(\varpi_n)\} -\exp( -\bft^\T \widehat{\bfH}_\Rn \bft/2 )\big |  \,{\rm d}\bft \\
			&+\int _{\mathcal{D}_{1}} \big |\pi_{0}(\hbthetan+n^{-1/2}\bft)-\pi_{0}(\hbthetan) \big | \exp( -\bft^\T \widehat{\bfH}_\Rn \bft/2 ) \,{\rm d}\bft \,,
		\end{align*}
		where $\varpi_n=\max \{\ell_n^{3/2} \alpha_n, \nu,  \ell_nn^{1/\gamma}\alpha_n\}$. By Condition \ref{A.prior}, we have $ \sup_{\bft \in \mathcal{D}_{1}} |\pi_{0}(\hbthetan+n^{-1/2}\bft)-\pi_{0}(\hbthetan)| =o_{\rm p}(1)$, which implies 
		\begin{align*}
			&\int _{\mathcal{D}_{1}}  |\pi_{0}(\hbthetan+n^{-1/2}\bft)-\pi_{0}(\hbthetan)| \exp( -\bft^\T \widehat{\bfH}_\Rn \bft/2 ) \,{\rm d}\bft \\
			&~~~~~~\leq (2\pi)^{p/2} |\widehat{\bfH}_\Rn|^{-1/2} \sup_{\bft \in \mathcal{D}_{1}} |\pi_{0}(\hbthetan+n^{-1/2}\bft)-\pi_{0}(\hbthetan)|=  o_{\rm p}(1) \,.
		\end{align*}
		Due to $\varpi_nn \alpha_n^2=o(1)$, then $\sup_{\bft \in \mathcal{D}_{1}}\{|\bft|_2^2\cdot O_{\rm p}(\varpi_n)\}=o_{\rm p}(1)$. Notice that $|e^{x}-1|\leq |x|e^{x}$ for any $x \in \R$. Then $\sup_{\bft \in \mathcal{D}_1}|\exp\{ |\bft|_2^2\cdot O_{\rm p}(\varpi_n)\}-1| = o_{\rm p}(1)$, which implies that
		\begin{align*}
			&\int _{\mathcal{D}_{1}} \pi_{0}(\hbthetan+n^{-1/2}\bft) \exp( -\bft^\T \widehat{\bfH}_\Rn \bft/2 ) |\exp\{ |\bft|_2^2\cdot O_{\rm p}(\varpi_n)\}-1| \,{\rm d}\bft \\
			&~~~~~\leq  o_{\rm p}(1) \cdot \sup_{\bft \in \mathcal{D}_{1}} \pi_{0}(\hbthetan+n^{-1/2}\bft) \int _{\mathcal{D}_{1}} \exp( -\bft^\T \widehat{\bfH}_\Rn \bft/2) \,{\rm d}\bft =o_{\rm p}(1) \,.
		\end{align*} 
		Therefore, ${\rm I}(1) = o_{\rm p}(1)$. 
		
		For ${\rm I}(4)$, due to $\mathcal{D}_4 \cap \mathcal{T}_n = \emptyset$, we have
		$
		{\rm I}(4)
		= \pi_{0}(\hbthetan) \int _{\mathcal{D}_4} \exp( -\bft^\T \widehat{\bfH}_\Rn \bft/2 ) \,{\rm d}\bft 
		$.
		Since $\bthetazero$ is an interior point of $\bTheta$, there exists a constant $\iota>0$ such that $\bTheta \supset \mathcal{B}_2(\bthetazero,\iota) := \{\btheta \in\R^p:\,|\btheta-\bthetazero|_2 \leq \iota \}$, which implies $\mathcal{D}_4 = \mathcal{T}_n^\c \subset \mathcal{T}_n^{*,\c}$ with $\mathcal{T}_n^* = \{\bft \in \R^p: \bft = n^{1/2}(\btheta-\hat\btheta_n), \btheta \in \mathcal{B}_2(\bthetazero,\iota)\}$. By Proposition \ref{pro.cons}, it holds w.p.a.1 that $n^{-1/2}|\bft|_2 \geq |n^{-1/2}\bft + \hat\btheta_n - \bthetazero|_2 - |\hat\btheta_n - \bthetazero|_2 \geq \iota/2$ for any $\bft \in \mathcal{D}_4$. Together with Proposition 1.1 of \cite{Hsu2012}, we have w.p.a.1 that
		\begin{align*}
			\pi_{0}(\hbthetan)\int _{\mathcal{D}_4} \exp( -\bft^\T \widehat{\bfH}_\Rn \bft/2 ) \,{\rm d}\bft \leq (2\pi)^{p/2} \pi_{0}(\hbthetan) |\widehat{\bfH}_\Rn|^{-1/2} \exp(-\tilde{C}n) = o_{\rm p}(1) \,.
		\end{align*}
		Therefore, ${\rm I}(4) = o_{\rm p}(1)$. 
		\hfill $\Box$

		\section{Proof of Proposition \ref{pro.expan}} \label{sec.pf_pro2}

		\subsection{Proof of part {(i)} of Proposition \ref{pro.expan}} 
		
		Recall $\mathcal{R}(\btheta)={\rm supp}\{\hat{\blambda}(\btheta)\}$ and $\mathcal{C}_1=\{\btheta\in\bTheta:|\btheta-\hat{\btheta}_n|_2\leq\alpha_n\}$ with $\alpha_n=n^{-1/2}(\log r)^{1/2}$. To prove part {\rm (i)} of Proposition \ref{pro.expan}, we need the following lemmas whose proofs are given in Sections \ref{sec:pflemma8}, \ref{sec:pflemma9} and \ref{sec:pflemma10}, respectively.
		
		\begin{lemma}\label{l.lam.theta}
			Let $c \in (\tilde{c},1)$ be some constant with $\tilde{c}$ given in Condition {\rm \ref{A.g_subgra.1}(a)}. Under the conditions of Lemma {\rm \ref{l.lam.thetan.hat}}, it holds w.p.a.1 that the global maximizer $\hat{\blambda}(\btheta)=\{\hat{\lambda}_{1}(\btheta),\ldots,\hat{\lambda}_{r}(\btheta)\}^\T$ for $f_n(\blambda;\btheta)$ w.r.t $\blambda$ satisfies the results: ${\rm (i)}$ $\sup_{\btheta \in \mathcal{C}_1}|\hat{\blambda}(\btheta)|_2=O_{\rm p}(\ell_n^{1/2} \alpha_n)$, ${\rm (ii)}$ $\mathcal{R}(\btheta) \subset \mathcal{M}_{\btheta}(c)$ for any $\btheta \in \mathcal{C}_1$, and ${\rm (iii)}$ $\sgn\{\hat{\lambda}_j(\btheta)\}=\sgn \{\bar{g}_j(\btheta)\}$ for any $\btheta \in \mathcal{C}_1$ and $j \in \mathcal{M}_{\btheta}(c)$ with $\hat{\lambda}_j(\btheta) \neq 0$.
		\end{lemma}
		
		\begin{lemma}\label{l.diff.lam.hat}
			Under the conditions of Lemma {\rm \ref{l.lam.thetan.hat}} and Condition {\rm\ref{A.Pro2}(a)}, it holds w.p.a.1 that the global maximizer $\hat{\blambda}(\btheta)$ for $f_n(\blambda;\btheta)$ w.r.t $\blambda$ is continuously differentiable in $\btheta\in\mathcal{C}_1$ with $[\nabla_{\btheta} \hat{\blambda}(\btheta)]_{\Rtheta^\c,[p]}=\bzero$ and
			\begin{align*}
				[\nabla_{\btheta} \hat{\blambda}(\btheta)]_{\Rtheta,[p]}
				=&\,\bigg(\frac{1}{n} \sum _{i=1}^{n}\frac{\bfg_{i,\Rtheta}(\btheta)^{\otimes2}}{\{1+\hat{\blambda}_\Rtheta(\btheta)^\T\bfg_{i,\Rtheta}(\btheta)\}^2} +\nu \diag[\rho''\{|\tilde\lambda_{1}(\btheta)|;\nu \},\ldots,\rho''\{|\tilde\lambda_{|\Rtheta|}(\btheta)|;\nu \}]\bigg)^{-1}\\
				\qquad &\times \bigg\{\frac{1}{n} \sum_{i=1}^{n} \frac{[\nabla_{\btheta} \bfg_{i}(\btheta)]_{\Rtheta,[p]}}{1+\hat{\blambda}_\Rtheta(\btheta)^\T\bfg_{i,\Rtheta}(\btheta)}-\frac{1}{n} \sum_{i=1}^{n} \frac{\bfg_{i,\Rtheta}(\btheta)\hat{\blambda}_\Rtheta(\btheta)^\T [\nabla_{\btheta} \bfg_{i}(\btheta)]_{\Rtheta,[p]}}{\{1+\hat{\blambda}_\Rtheta(\btheta)^\T\bfg_{i,\Rtheta}(\btheta)\}^2} \bigg\} \,,
			\end{align*}
			where $\hat\blambda_{\Rtheta}(\btheta)=\{\tilde{\lambda}_1(\btheta),\ldots,\tilde{\lambda}_{|\Rtheta|}(\btheta)\}^\T$.
		\end{lemma}
		
		\begin{lemma}\label{l.supp}
			Under the conditions of Lemma {\rm \ref{l.lam.thetan.hat}} and Condition {\rm\ref{A.Pro2}(a)}, it holds that $\mathcal{R}(\btheta)={\rm supp}\{\hat{\blambda}(\hbthetan)\}$ for any $\btheta \in \mathcal{C}_1$  w.p.a.1.
		\end{lemma}
		Notice that
		\begin{align*}
			\nabla_{\btheta} f_n\{\hat{\blambda}(\btheta);\btheta\}
			=&\,\bigg\{\frac{\partial f_n\{\hat{\blambda}(\btheta); \btheta\}}{\partial \blambda_{\mathcal{R}(\btheta)}^\T} [\nabla_{\btheta} \hat{\blambda}(\btheta)]_{\Rtheta,[p]} + \frac{\partial f_n\{\hat{\blambda}(\btheta); \btheta\}}{\partial\blambda_{\mathcal{R}(\btheta)^\c}^\T} [\nabla_{\btheta} \hat{\blambda}(\btheta)]_{\Rtheta^\c,[p]}\bigg\}^\T \\
			&\,+ \frac{\partial f_n(\blambda; \btheta)}{\partial \btheta} \bigg|_{\blambda=\hat\blambda(\btheta)}
		\end{align*}
		for any $\btheta \in \mathcal{C}_1$. Due to $\hat{\blambda}(\btheta)=\arg \max _{\blambda \in \wLambadn(\btheta)} f_n(\blambda;\btheta)$, then $\partial f_n\{\hat{\blambda}(\btheta) ; \btheta\} / \partial \blambda_{\mathcal{R}(\btheta)} = \bzero$. By Lemma \ref{l.diff.lam.hat}, we have $[\nabla_{\btheta} \hat{\blambda}(\btheta)]_{\Rtheta^\c,[p]} = \bzero$ for any $\btheta \in \mathcal{C}_1$ w.p.a.1. Thus, it holds w.p.a.1 that
		\begin{align*}
			\nabla_{\btheta}f_n\{\hat{\blambda}(\btheta);\btheta\}
			=\bigg\{\frac{1}{n} \sum_{i=1}^{n} \frac{\nabla_{\btheta} \bfg_{i}( \btheta)}{1+\hat{\blambda}(\btheta)^\T\bfg _{i}(\btheta)} \bigg\}^\T \hat{\blambda}(\btheta) 
		\end{align*}
		for any $\btheta \in \mathcal{C}_1$. By Lemma \ref{l.diff.lam.hat}, %we know $\hat{\blambda}(\btheta)$ is continuously differentiable on $\mathcal{C}_1$ w.p.a.1 and  $[\nabla_{\btheta}\hat{\blambda}(\btheta)]_{\Rtheta^\c,[p]}=\bzero$ for any $\btheta \in \mathcal{C}_1$ w.p.a.1. Then
		\begin{align}
			\nabla_{\btheta}^2 f_n\{\hat{\blambda}(\btheta);\btheta\}=&-\underbrace{\frac{1}{n}\sum_{i=1}^{n} \frac{\{[\nabla_{\btheta} \bfg_i (\btheta)]_{\Rtheta,[p]}^\T \hat{\blambda}_\Rtheta(\btheta) \}^{\otimes2}}{\{1+\hat{\blambda}_\Rtheta(\btheta)^\T\bfg_{i,\Rtheta} ( \btheta)\}^2}}_{T_{\btheta,1}}\notag \\
			&-\underbrace{\frac{1}{n} \sum_{i=1}^{n} \frac{[\nabla_{\btheta} \bfg_i (\btheta)]_{\Rtheta,[p]}^\T \hat{\blambda}_\Rtheta(\btheta) \bfg_{i,\Rtheta}(\btheta)^\T [\nabla_{\btheta} \hat{\blambda}(\btheta)]_{\Rtheta,[p]}}{\{1+\hat{\blambda}_\Rtheta(\btheta)^\T\bfg_{i,\Rtheta} ( \btheta)\}^2}}_{T_{\btheta,2}}\label{eq:fourterms}\\
			&+ \underbrace{\frac{1}{n} \sum_{i=1}^{n} \frac{\sum_{j \in \Rtheta} \hat{\lambda}_j(\btheta) \nabla^2_{\btheta}g_{i,j}(\btheta)  }{1+\hat{\blambda}_\Rtheta(\btheta)^\T\bfg_{i,\Rtheta} ( \btheta)}}_{T_{\btheta,3}}+ \underbrace{\frac{1}{n} \sum_{i=1}^{n} \frac{[\nabla_{\btheta} \bfg_i (\btheta)]_{\Rtheta,[p]}^\T [\nabla_{\btheta} \hat{\blambda}(\btheta)]_{\Rtheta,[p]}}{1+\hat{\blambda}_\Rtheta(\btheta)^\T\bfg_{i,\Rtheta} (\btheta)}}_{T_{\btheta,4}} \notag
		\end{align}
		for any $\btheta \in \mathcal{C}_1$. Lemma \ref{l.fourterms} specifies the leading term of $\bft^\T[\nabla_{\btheta}^2 f_n\{\hat{\blambda}(\btheta);\btheta\}]\bft$ for $\bft\in\mathbb{R}^p$, whose proof is given in Section \ref{sec.pflemma11}. 
		
		\begin{lemma}\label{l.fourterms}
			Let $P_{\nu}(\cdot) \in \mathscr{P}$ be convex and assume $\rho(t;\nu)=\nu^{-1}P_{\nu}(t)$ has bounded second-order derivative w.r.t $t$ around $0$, where $\mathscr{P}$ is defined in {\rm \eqref{eq:penalty}}. Under the conditions of Lemma {\rm \ref{l.lam.thetan.hat}} and Conditions {\rm \ref{A.ee2}} and {\rm\ref{A.Pro2}(a)}, then %$|\bft^\T T_{\btheta,1} \bft| =
			%	|\bft|_2^2 \cdot O_{\rm p}(\ell_n^2 n^{-1}\log r)$, $|\bft^\T T_{\btheta,2} \bft| = |\bft|_2^2 \cdot O_{\rm p}\{\ell_n^{3/2} n^{-1/2}(\log r)^{1/2}\} $, $|\bft^\T T_{\btheta,3} \bft|=|\bft|_2^2 \cdot O_{\rm p}\{\ell_n n^{-1/2}(\log r )^{1/2}\} $ and 
			$$\bft^\T\big[\nabla_{\btheta}^2 f_n\{\hat{\blambda}(\btheta);\btheta\}\big]\bft=\bft^\T \big\{\widehat{\bGamma}_\Rtheta(\btheta)^\T \widehat{\bfV}^{-1/2}_\Rtheta(\btheta)\big\}^{\otimes2} \bft + |\bft|_2^2 \cdot \{O_{\rm p}(\ell_n^{3/2} \alpha_n) + O_{\rm p}(\ell_nn^{1/\gamma}\alpha_n) + O_{\rm p}(\nu) \}$$ holds uniformly over $\btheta \in \mathcal{C}_1$ and $\bft \in \mathbb{R}^p$. 
		\end{lemma}
		Let $\bft =\btheta-\hbthetan$ for any $\btheta \in \mathcal{C}_1$. Since $\hbthetan=\arg\min_{\btheta \in \bTheta} f_n\{\hat{\blambda}(\btheta);\btheta\}$, then
		$\nabla_{\btheta}f_n\{\hat{\blambda}(\btheta);\btheta\}|_{\btheta=\hbthetan} =\bzero$.  
		By the Taylor expansion, it holds that
		\begin{align}\label{eq:ln_Taylor_expan}
			\aleph_n(\btheta)=&\,f_n\{\hat{\blambda}(\btheta);\btheta\}-f_n\{\hat{\blambda}(\hbthetan);\hbthetan\}\notag\\ 
			=&\, \big[\nabla_{\btheta}f_n\{\hat{\blambda}(\btheta);\btheta\}|_{\btheta=\hbthetan}\big]^\T  \bft + \frac{1}{2}\bft^\T \big[\nabla_{\btheta}^2 f_n\{\hat{\blambda}(\btheta);\btheta\} |_{\btheta=\tilde \btheta}\big] \bft =\frac{1}{2}\bft^\T \nabla_{\btheta}^2 f_n\{\hat{\blambda}(\tbtheta);\tbtheta\} \bft
		\end{align}
		for some $\tbtheta$ lying on the jointing line between $\hbthetan$ and $\btheta$. Let $\widehat{\bfH}_{\Rtheta} = \{\widehat{\bGamma}_\Rtheta(\btheta)^\T \widehat{\bfV}^{-1/2}_\Rtheta(\btheta)\}^{\otimes2}$ and recall $\widehat{\bfH}_\Rn=\{\widehat{\bGamma}_\Rn(\hbthetan)^\T \widehat{\bfV}^{-1/2}_\Rn(\hbthetan)\}^{\otimes2}$. By Lemma \ref{l.supp}, we know  $\mathcal{R}(\btheta)=\mathcal{R}_n$ for any $\btheta \in \mathcal{C}_1$ w.p.a.1. Under Conditions \ref{A.ee}(b) and \ref{A.ee2}, by Lemmas \ref{l.V.hat} and \ref{l.Gamma.hat}, if $\log r=o(n^{1/3})$, $\ell_n\nu^2=o(1)$ and $\ell_n\alpha_n=o[\min\{\nu,n^{-1/\gamma}\}]$, we have $\|\widehat{\bfV}_{\Rn} (\hbthetan)\|_2=O_{\rm p}(1)$, $\|\widehat{\bfV}^{-1}_{\Rn} (\hbthetan)\|_2=O_{\rm p}(1)$ and $\|\widehat{\bGamma}_\Rn(\hbthetan)\|_2=O_{\rm p}(1)$. Using the same arguments in the proof of Lemmas \ref{l.V.hat} and \ref{l.Gamma.hat}, if $\log r=o(n^{1/3})$, $\ell_n\alpha_n=o[\min\{\nu,n^{-1/\gamma}\}]$ and $\ell_n\nu^2=o(1)$, we have 
		\begin{align*}
			&~~~~~\sup_{\btheta \in \mathcal{C}_1}\|\widehat{\bfV}^{-1}_{\Rn} (\btheta) - \widehat{\bfV}^{-1}_{\Rn} (\hbthetan)\|_2=O_{\rm p}(\ell_n^{1/2} \alpha_n)\,,\\
			&\sup_{\btheta \in \mathcal{C}_1}|\{\widehat{\bGamma}_{\Rn} (\btheta) - \widehat{\bGamma}_{\Rn} (\hbthetan)\}\bft|_2= |\bft|_2 \cdot O_{\rm p}(\ell_n^{1/2}\alpha_n)\,,
		\end{align*}
		which implies 
		$
		\sup_{\btheta \in \mathcal{C}_1}\|\widehat{\bfH}_{\Rtheta}-\widehat{\bfH}_\Rn\|_2=O_{\rm p}(\ell_n^{1/2} \alpha_n)
		$.
		Together with Lemma \ref{l.fourterms}, \eqref{eq:ln_Taylor_expan} yields that 
		$
		\aleph_n(\btheta) - 2^{-1}(\btheta-\hat\btheta_n)^\T \widehat\bfH_{\mathcal{R}_n} (\btheta-\hat\btheta_n) 
		= |\btheta-\hat\btheta_n|_2^2 \cdot O_{\rm p}(\varpi_n)$ 
		with $\varpi_n=\max\{\ell_n^{3/2}\alpha_n, \nu, \ell_nn^{1/\gamma}\alpha_n\}$, where $O_{\rm p}(\varpi_n)$ holds uniformly over $\btheta \in \mathcal{C}_1$. \hfill $\Box$
		%We complete the proof of part {\rm (i)} of Proposition \ref{pro.expan}.
		
		\subsection{Proof of part {(ii)} of Proposition \ref{pro.expan}}\label{sec:pfpn2b}
		Recall $\mathcal{R}_n={\rm supp}\{\hat{\blambda}(\hbthetan)\}$ and $\mathcal{C}_2=\{\btheta\in\bTheta:\alpha_n<|\btheta-\hat{\btheta}_n|_2\leq\beta_n\}$. Select $\delta_n$ satisfying $\delta_n=o(\ell_n^{-1/2}n^{-1/\gamma})$ and $\ell_n^{1/2}\beta_{n}=o(\delta_n)$, which can be guaranteed by $\ell_n \beta_{n}=o(n^{-1/\gamma})$. For any $\btheta \in \mathcal{C}_2$, let $\tilde\blambda(\btheta) = \arg\max_{\blambda \in \tilde\Lambda}f_n(\blambda;\btheta)$ and $\tilde{\mathcal{R}}(\btheta)={\rm supp}\{\tilde{\blambda}(\btheta)\}$, where $\tilde\Lambda=\{\blambda\in \mathbb{R}^r:\, |\blambda_{\mathcal{R}_n}|_2\leq \delta_n, \blambda_{\mathcal{R}_n^\c}=\bzero\}$. Write $\tilde\blambda(\btheta)=\{\tilde\lambda_1(\btheta),\ldots,\tilde\lambda_r(\btheta)\}^\T$. By the Taylor expansion, we have
		\begin{align} \label{eq:bar.lam}
			0 = f_n(\bzero;\btheta)
			&\leq f_n\{\tilde\blambda(\btheta);\btheta\} \notag \\
			&= \tilde\blambda_{\tilde{\mathcal{R}}(\btheta)}(\btheta)^\T \bar\bfg_{\tilde{\mathcal{R}}(\btheta)}(\btheta) - \frac{1}{2n}\sum_{i=1}^n \frac{\tilde\blambda_{\tilde{\mathcal{R}}(\btheta)}(\btheta)^\T  \bfg_{i,\tilde{\mathcal{R}}(\btheta)}(\btheta)^{\otimes2} \tilde\blambda_{\tilde{\mathcal{R}}(\btheta)}(\btheta) }{\{1+C\tilde\blambda_{\tilde{\mathcal{R}}(\btheta)}(\btheta)^\T \bfg_{i,\tilde{\mathcal{R}}(\btheta)}(\btheta)\}^2} - \sum_{j \in \tilde{\mathcal{R}}(\btheta)} P_\nu\{|\tilde\lambda_j(\btheta)|\} \\
			&=: \tilde\blambda_{\tilde{\mathcal{R}}(\btheta)}(\btheta)^\T \bar\bfg_{\tilde{\mathcal{R}}(\btheta)}(\btheta) - \frac{1}{2} \tilde\blambda_{\tilde{\mathcal{R}}(\btheta)}(\btheta)^\T \tilde{\bfA}(\btheta) \tilde\blambda_{\tilde{\mathcal{R}}(\btheta)}(\btheta)- \sum_{j \in \tilde{\mathcal{R}}(\btheta)} P_\nu\{|\tilde\lambda_j(\btheta)|\} \notag
		\end{align}
		for some $C\in(0,1)$. By Lemma \ref{l.lam.thetan.hat}, we have $|\tilde{\mathcal{R}}(\btheta)|\leq |\mathcal{R}_n|\leq \ell_n$ w.p.a.1. Notice that $\nu=o(\beta_{n})$. By Proposition \ref{pro.cons}, Lemma \ref{l.V.hat} and Condition \ref{A.ee}(b), if $\log r=o(n^{1/3})$, $\ell_n\alpha_n=o[\min\{\nu,n^{-1/\gamma}\}]$ and $\ell_n\beta_{n}=o(n^{-1/\gamma})$, we have $\inf_{\btheta \in \mathcal{C}_2} \lambda_{\min}\{\tilde{\bfA}(\btheta)\}\geq K_3/2$ w.p.a.1, where $K_3$ is specified in Condition \ref{A.ee}(b). Recall $\rho(t;\nu)$ is convex w.r.t $t$. Thus $$0 \leq \tilde\blambda_{\tilde{\mathcal{R}}(\btheta)}(\btheta)^\T\big[\bar\bfg_{\tilde{\mathcal{R}}(\btheta)}(\btheta) - \nu\rho'(0^+)\sgn\{\tilde\blambda_{\tilde{\mathcal{R}}(\btheta)}(\btheta)\}\big] - 4^{-1}K_3|\tilde\blambda_{\tilde{\mathcal{R}}(\btheta)}(\btheta)|_2^2$$ w.p.a.1.  Then $|\tilde\blambda_{\tilde{\mathcal{R}}(\btheta)}(\btheta)|_2 \leq 4K_3^{-1}  |\bar\bfg_{\tilde{\mathcal{R}}(\btheta)}(\btheta) - \nu\rho'(0^+)\sgn\{\tilde\blambda_{\tilde{\mathcal{R}}(\btheta)}(\btheta)\}|_2$ w.p.a.1. By the Taylor expansion and Condition \ref{A.ee}(c),  $\sup_{\btheta \in \mathcal{C}_2} |\bar\bfg(\btheta) - \bar\bfg(\hat{\btheta}_n)|_\infty =  O_{\rm p}(\beta_{n})$. Together with the fact $|\tilde{\mathcal{R}}(\btheta)|\leq \ell_n$ w.p.a.1, we have $\sup_{\btheta \in \mathcal{C}_2} |\bar\bfg_{\tilde{\mathcal{R}}(\btheta)}(\btheta) - \bar\bfg_{\tilde{\mathcal{R}}(\btheta)}(\hat{\btheta}_n)|_2 = O_{\rm p}(\ell_n^{1/2} \beta_{n})$. By the triangle inequality, Lemma \ref{l.lam.thetan.hat} and \eqref{eq:g_bar}, since $\alpha_n=o(\nu)$, it holds that $|\bar\bfg_{\tilde{\mathcal{R}}(\btheta)}(\hat{\btheta}_n)|_2 = O_{\rm p}(\ell_n^{1/2} \nu)$.  Due to $\nu=o(\beta_{n})$, we then have $$\sup_{\btheta \in \mathcal{C}_2}|\bar\bfg_{\tilde{\mathcal{R}}(\btheta)}(\btheta) - \nu\rho'(0^+)\sgn\{\tilde\blambda_{\tilde{\mathcal{R}}(\btheta)}(\btheta)\}|_2 = O_{\rm p}(\ell_n^{1/2} \beta_{n})\,,$$ which implies $\sup_{\btheta \in \mathcal{C}_2}|\tilde\blambda_{\tilde{\mathcal{R}}(\btheta)}(\btheta)|_2 = O_{\rm p}(\ell_n^{1/2} \beta_{n}) = o_{\rm p}(\delta_n)$. Recall $\tilde\blambda(\btheta)=\arg\max_{\blambda \in \tilde\Lambda}f_n(\blambda;\btheta)$ and $\tilde\blambda(\btheta) \in {\rm int}(\tilde\Lambda)$ for any $\btheta \in \mathcal{C}_2$ w.p.a.1. Write $\tilde\blambda_{\mathcal{R}_n}(\btheta)=\{\dot\lambda_1(\btheta),\ldots,\dot\lambda_{|\mathcal{R}_n|}(\btheta)\}^\T$. Restricted on $\tilde\Lambda$, by the first-order condition, we have 
		$$
		\bzero=\frac{1}{n}\sum_{i=1}^{n} \frac{\bfg_{i,\mathcal{R}_n}(\btheta)}{1+\tilde{\blambda}_{\mathcal{R}_n}(\btheta)^\T\bfg_{i,\mathcal{R}_n}(\btheta)} - \tilde\bseta(\btheta)$$ 
		holds for any $\btheta \in \mathcal{C}_2$ w.p.a.1, where $\tilde\bseta(\btheta)=\{\tilde\eta_1(\btheta), \ldots,\tilde\eta_{|\mathcal{R}_n|}(\btheta)\}^\T$ with $\tilde\eta_j(\btheta)=\nu\rho'\{|\dot{\lambda}_j(\btheta)|;\nu\}\sgn\{\dot{\lambda}_j(\btheta)\}$ for $\dot{\lambda}_j(\btheta)\neq0$ and $\tilde\eta_j(\btheta) \in [-\nu\rho'(0^+), \nu\rho'(0^+)]$ for $\dot{\lambda}_j(\btheta)=0$. 
		By the Taylor expansion, we have
		\begin{align*}
			\bzero
			=\bar\bfg_{\tilde{\mathcal{R}}(\btheta)}(\btheta)-\frac{1}{n} \sum_{i=1}^{n} \frac{\bfg_{i,\tilde{\mathcal{R}}(\btheta)}(\btheta)^{\otimes2}\tilde{\blambda}_{\tilde{\mathcal{R}}(\btheta)}(\btheta)}{\{1+\tilde{C}\tilde{\blambda}_{\tilde{\mathcal{R}}(\btheta)}(\btheta)^\T\bfg_{i,\tilde{\mathcal{R}}(\btheta)}(\btheta)\}^2} - \tilde{\bseta}_{*}(\btheta)
			=:\bar\bfg_{\tilde{\mathcal{R}}(\btheta)}(\btheta) - \check\bfA(\btheta) \tilde{\blambda}_{\tilde{\mathcal{R}}(\btheta)}(\btheta) - \tilde\bseta_{*}(\btheta)
		\end{align*}
		for some $\tilde{C}\in(0,1)$, where $\tilde\bseta_{*}(\btheta)\in\mathbb{R}^{|\tilde{\mathcal{R}}(\btheta)|}$ includes all elements $\tilde\eta_j(\btheta)$'s in $\tilde\bseta(\btheta)$ such that the associated $\dot \lambda_j(\btheta) \neq0$. Hence, $\tilde{\blambda}_{\tilde{\mathcal{R}}(\btheta)}(\btheta)=\check\bfA^{-1}(\btheta)\{\bar\bfg_{\tilde{\mathcal{R}}(\btheta)}(\btheta)-\tilde\bseta_{*}(\btheta)\}$. 
		Using the same arguments in the proof of Lemma \ref{l.ee}, we have $\sup_{\btheta \in \mathcal{C}_2}\|\check\bfA(\btheta) - \widehat{\bfV}_{\tilde{\mathcal{R}}(\btheta)}(\btheta)\|_2=O_{\rm p}(\ell_n\beta_nn^{1/\gamma})=o_{\rm p}(1)$. Applying Proposition \ref{pro.cons}, Lemma \ref{l.V.hat} and Condition \ref{A.ee}(b), we know $\sup_{\btheta \in \mathcal{C}_2}\lambda_{\max}\{\check\bfA(\btheta)\}\leq 2K_4$ w.p.a.1 for $K_4$ specified in Condition \ref{A.ee}(b). For $\tilde{\bfA}(\btheta)$ specified in \eqref{eq:bar.lam}, we can show $\sup_{\btheta \in \mathcal{C}_2}\|\tilde\bfA(\btheta)- \check\bfA(\btheta)\|_2=O_{\rm p}(\ell_n \beta_{n}n^{1/\gamma})$. By \eqref{eq:bar.lam} and Condition \ref{A.Pro2}(b), we have
		\begin{align*}
			f_n\{\tilde\blambda(\btheta);\btheta\}  &=\frac{1}{2}\big[\bar\bfg_{\tilde{\mathcal{R}}(\btheta)}(\btheta)-\nu\rho'(0^+)\sgn\{\tilde\blambda_{\tilde{\mathcal{R}}(\btheta)}(\btheta)\}\big]^\T \check\bfA^{-1}(\btheta) \big[\bar\bfg_{\tilde{\mathcal{R}}(\btheta)}(\btheta)-\nu\rho'(0^+)\sgn\{\tilde\blambda_{\tilde{\mathcal{R}}(\btheta)}(\btheta)\}\big] \\
			&~~~+ O_{\rm p}(\ell_n^2\beta_{n}^3 n^{1/\gamma}) \\
			&\geq \frac{1}{4K_4}|\bar\bfg_{\tilde{\mathcal{R}}(\btheta)}(\btheta)-\nu\rho'(0^+)\sgn\{\tilde\blambda_{\tilde{\mathcal{R}}(\btheta)}(\btheta)\}|_2^2+ O_{\rm p}(\ell_n^2\beta_{n}^3 n^{1/\gamma})\geq\frac{\kappa^2_n}{4K_4} + O_{\rm p}(\ell_n^2\beta_{n}^3 n^{1/\gamma})
		\end{align*}
		holds uniformly over $\btheta \in \mathcal{C}_2$ w.p.a.1, where the term $O_{\rm p}(\ell_n^2\beta_{n}^3 n^{1/\gamma})$ holds uniformly over $\btheta \in \mathcal{C}_2$. 
		As we have shown in the proof of Proposition \ref{pro.cons},  $f_n\{\hat\blambda(\hbthetan);\hbthetan\}=O_{\rm p}(\ell_n \alpha_n^2)$. Notice that $f_n\{\hat{\blambda}(\btheta);\btheta\}\geq f_n\{\tilde{\blambda}(\btheta);\btheta\}$ for any $\btheta \in \mathcal{C}_2$. If $\max\{\ell_n \alpha_n^2, \ell_n^2\beta_{n}^3 n^{1/\gamma}\} = o(\kappa_n^2)$, then 
		\begin{align*}
			\P \bigg\{\inf_{\btheta \in \mathcal{C}_2}  \aleph_n(\btheta) \geq \frac{\kappa_{n}^2}{8K_4} \bigg\}\geq \P \bigg\{\frac{\kappa^2_n}{4K_4} + O_{\rm p}(\ell_n^2\beta_{n}^3 n^{1/\gamma}) - O_{\rm p}(\ell_n \alpha_n^2) \geq \frac{\kappa_{n}^2}{8K_4}\bigg\}-o(1)
			\rightarrow 1
		\end{align*}
		as $n \rightarrow \infty$. We complete the proof of part {\rm (ii)} of Proposition \ref{pro.expan}. \hfill $\Box$
		
		\subsection{Proof of part {(iii)} of Proposition \ref{pro.expan}}\label{sec:pfpn2c}
		Recall $\mathcal{R}_n={\rm supp}\{\hat{\blambda}(\hbthetan)\}$ and $\mathcal{C}_3=\{\btheta\in\bTheta:|\btheta-\hat{\btheta}_n|_2>\beta_n\}$. For any $\btheta \in \mathcal{C}_3$, we consider  $\check\blambda(\btheta)=\{\check\lambda_1(\btheta),\ldots,\check\lambda_r(\btheta)\}^\T\in \mathbb{R}^r$ with  $\check\blambda_{\mathcal{R}_n^\c}(\btheta) =\bzero$ and $$\check\blambda_{\mathcal{R}_n}(\btheta) = \frac{\xi_n\{\bar\bfg_{\mathcal{R}_n}(\btheta)-\bar\bfg_{\mathcal{R}_n}(\hbthetan)\}}{|\bar\bfg_{\mathcal{R}_n}(\btheta)-\bar\bfg_{\mathcal{R}_n}(\hbthetan)|_2}\,.$$ Due to $\ell_n^{1/2}\xi_n =o(n^{-1/\gamma})$, Condition \ref{A.ee}(a) yields $\sup_{\btheta \in \mathcal{C}_3} \max_{ i \in [n]} |\check\blambda(\btheta)^\T \bfg_i(\btheta)|=o_{\rm p}(1)$, which implies the event $\bigcap_{\btheta \in \mathcal{C}_3} \{\check\blambda(\btheta) \in \hat\Lambda_n(\btheta)\}$ holds w.p.a.1. Then
		$$
		\P \bigg[\inf_{\btheta \in \mathcal{C}_3}  f_n\{\hat\blambda(\btheta);\btheta\} \geq  \inf_{\btheta \in \mathcal{C}_3}  f_n\{\check\blambda(\btheta);\btheta\}\bigg] =1-o(1)\,.
		$$ Recall $\sup_{\btheta \in \mathcal{C}_3} \lambda_{\max}\{\widehat\bfV_{\mathcal{R}_n}(\btheta)\}\leq K_8$ w.p.a.1 for $K_8$ specified in Condition \ref{A.Pro2}(c). By the Taylor expansion, we have
		\begin{align*}
			f_n\{\check\blambda(\btheta);\btheta\} 
			=&~\check\blambda_{\mathcal{R}_n}(\btheta)^\T \bar\bfg_{\mathcal{R}_n}(\btheta) - \frac{1}{2n}\sum_{i=1}^n \frac{\check\blambda_{\mathcal{R}_n}(\btheta)^\T  \bfg_{i,\mathcal{R}_n}(\btheta)^{\otimes2} \check\blambda_{\mathcal{R}_n}(\btheta) }{\{1+\check C\check\blambda_{\mathcal{R}_n}(\btheta)^\T \bfg_{i,\mathcal{R}_n}(\btheta)\}^2}\\
			&- \sum_{j \in \Rn} P_\nu\{|\check\lambda_j(\btheta)|\} \\
			\geq&~ \check\blambda_{\mathcal{R}_n}(\btheta)^\T \big[\bar\bfg_{\mathcal{R}_n}(\btheta) - \nu\rho'(0^+) \sgn\{\check\blambda_{\mathcal{R}_n}(\btheta)\}\big]
			- \frac{K_8\xi_n^2}{2}\{1+o_{\rm p}(1)\} \\
			&- \frac{1}{2} \sum_{j \in \Rn} \nu\rho''\{c_j|\check\lambda_j(\btheta)|;\nu\}|\check\lambda_j(\btheta)|^2 \\
			\geq &~ \xi_n
			|\bar\bfg_{\mathcal{R}_n}(\btheta) - \bar\bfg_{\mathcal{R}_n}(\hbthetan)|_2 + \check\blambda_{\mathcal{R}_n}(\btheta)^\T\big[\bar\bfg_{\mathcal{R}_n}(\hbthetan) - \nu\rho'(0^+) \sgn\{\check\blambda_{\mathcal{R}_n}(\btheta)\}\big] \\
			&- C\xi_n^2\{1+o_{\rm p}(1)\} 
		\end{align*}
		holds uniformly over $\btheta \in \mathcal{C}_3$ w.p.a.1, where $\check C,c_j\in(0,1)$. By Condition \ref{A.Pro2}(c), we have $\inf_{\btheta \in \mathcal{C}_3}|\bar\bfg_{\mathcal{R}_n}(\btheta) - \bar\bfg_{\mathcal{R}_n}(\hbthetan)|_2 = \inf_{\btheta \in \mathcal{C}_3} |\{\nabla_{\btheta} \bar\bfg_{\mathcal{R}_n}(\tilde \btheta)\}^\T (\btheta-\hbthetan)|_2 \geq K_7^{1/2}\beta_{n}$ w.p.a.1. Since $\alpha_n=o(\nu)$, by Lemma \ref{l.lam.thetan.hat} and \eqref{eq:g_bar}, we have $|\bar\bfg_{\mathcal{R}_n}(\hbthetan)|_2=O_{\rm p}(\ell_n^{1/2}\nu)$. Due to $\ell_n^{1/2}\nu=o(\beta_{n})$ and $\xi_n=o(\beta_{n})$, it holds that $$\inf_{\btheta \in \mathcal{C}_3}  f_n\{\check\blambda(\btheta);\btheta\} \geq K_7^{1/2}\xi_n\beta_{n} + O_{\rm p}(\xi_n \ell_n^{1/2}\nu) + O_{\rm p}(\xi_n^2) \geq \frac{1}{2}K_7^{1/2} \xi_n\beta_{n}$$ w.p.a.1. Since $f_n\{\hat\blambda(\hbthetan);\hbthetan\}=O_{\rm p}(\ell_n \alpha_n^2)=o_{\rm p}(\xi_n\beta_{n})$, then 
		\begin{align*}
			\P \bigg\{\inf_{\btheta \in \mathcal{C}_3}  \aleph_n(\btheta) \geq \frac{K_7^{1/2} \xi_n\beta_{n}}{4} \bigg\}\geq\P \bigg[\inf_{\btheta \in \mathcal{C}_3}  f_n\{\check\blambda(\btheta);\btheta\} - f_n\{\hat\blambda(\hbthetan);\hbthetan\} \geq  \frac{K_7^{1/2} \xi_n \beta_{n}}{4}  \bigg] - o(1) \rightarrow 1
		\end{align*}
		as $n\rightarrow \infty$. We complete the proof of part {\rm (iii)} of Proposition \ref{pro.expan}. 
		\hfill $\Box$

		\section{Proof of Corollary \ref{cor.pos_mean}}
		
		%Recall the posterior distribution $\pi^\dag(\btheta\,|\,\mathcal{X}_n) \propto  \pi_{0}(\btheta)\times \exp [-n\log n-n f_n\{\hat\blambda(\btheta);\btheta \}]I(\btheta \in \bTheta)$. 
		%For any $\btheta \in \bTheta$, let $w_n(\btheta)=-n\log n-nf_n\{\hat\blambda(\btheta); \btheta\}$ and write $\bft = n^{1/2}(\btheta-\hat\btheta_n)$. Define $\mathcal{T}_n = \{\bft \in \R^p: \bft = n^{1/2}(\btheta-\hat\btheta_n), \btheta \in \bTheta\}$. Denote by $\pi_{0,\bft}(\cdot)$ and $\pi^{\dag}_{\bft}(\cdot\,|\,\mathcal{X}_n)$ the prior and the posterior distributions of $\bft$, respectively. Then,  $\pi_{0,\bft}(\bft)=n^{-p/2}\pi_{0}(\hbthetan+n^{-1/2}\bft)$ and
		%\begin{align*}
		%	\pi^{\dag}_{\bft}(\bft\,|\,\mathcal{X}_n)
		%	&=\frac{\pi_{0}(\hbthetan+n^{-1/2}\bft) \exp \{ w_n(\hbthetan+n^{-1/2}\bft)-w_n(\hbthetan)\} I(\bft \in \mathcal{T}_n)}{\int_{\R^{p}} \pi_{0}(\hbthetan+n^{-1/2}\bfs) \exp \{ w_n(\hbthetan+n^{-1/2}\bfs)-w_n(\hbthetan)\}I(\bfs \in \mathcal{T}_n) \,{\rm d}\bfs} \\
		%	&=:C^{-1}_n \pi_{0}(\hbthetan+n^{-1/2}\bft) \exp \{ w_n(\hbthetan+n^{-1/2}\bft)-w_n(\hbthetan)\}I(\bft \in \mathcal{T}_n) \,.
		%\end{align*} 
		Let $\mE_{\bft \sim \pi^{\dag}_{\bft}}(\bft) = \int_{\R^p} \bft \pi^{\dag}_{\bft}(\bft\,|\,\mathcal{X}_n) \, {\rm d}\bft$ for $\pi^{\dag}_{\bft}(\bft\,|\,\mathcal{X}_n)$ given in \eqref{eq:pos_t}. Notice that $\mE_{\bft \sim \pi^{\dag}_{\bft}}(\bft) = n^{1/2}\{\mE_{\btheta \sim \pi^\dag}(\btheta) - \hbthetan\}$ and $\mE_{\bft \sim \mathcal{N}(\bzero, \widehat{\bfH}_\Rn^{-1})}(\bft) = \bzero$. To prove Corollary \ref{cor.pos_mean}, it is equivalent to show $|\mE_{\bft \sim \pi^{\dag}_{\bft}}(\bft) - \mE_{\bft \sim \mathcal{N}(\bzero, \widehat{\bfH}_\Rn^{-1})}(\bft)|_\infty = o_{\rm p}(1)$. It follows from the triangle inequality that
		\begin{align*} 
			& |\mE_{\bft \sim \pi^{\dag}_{\bft}}(\bft) - \mE_{\bft \sim \mathcal{N}(\bzero, \widehat{\bfH}_\Rn^{-1})}(\bft)|_\infty \notag \\
			&~~~~\leq \int_{\R^{p}} |\bft|_\infty \big |C^{-1}_n \pi_{0}(\hbthetan+n^{-1/2}\bft) \exp \{ w_n(\hbthetan+n^{-1/2}\bft)-w_n(\hbthetan)\} I(\bft \in \mathcal{T}_n) \notag \\
			&~~~~~~~~~~~~~~~~~~~~~~~~~~~
			-(2\pi)^{-p/2}|\widehat{\bfH}_\Rn|^{1/2} \exp( -\bft^\T \widehat{\bfH}_\Rn \bft/2 )   \big | \,{\rm d}\bft \notag \\ 
			&~~~~\leq C^{-1}_n \int_{\R^{p}} |\bft|_\infty \big | \pi_{0}(\hbthetan+n^{-1/2}\bft) \exp \{ w_n(\hbthetan+n^{-1/2}\bft)-w_n(\hbthetan)\} I(\bft \in \mathcal{T}_n)  \notag \\
			&~~~~~~~~~~~~~\underbrace{~~~~~~~~~~~~~~~~~~~~
				-\pi_{0}(\hbthetan)\exp( -\bft^\T \widehat{\bfH}_\Rn \bft/2 ) \big | \,{\rm d}\bft~~~~~~~~~~~~~~~~~~~~~~~~~~~}_{\rm III} \notag \\
			&~~~~~~+ C^{-1}_n  \underbrace{\int_{\R^{p}} |\bft|_\infty \big| \pi_{0}(\hbthetan)\exp( -\bft^\T \widehat{\bfH}_\Rn \bft/2 )-C_n (2\pi)^{-p/2}|\widehat{\bfH}_\Rn|^{1/2} \exp( -\bft^\T \widehat{\bfH}_\Rn \bft/2 ) \big| \,{\rm d}\bft}_{\rm IV} \,.
		\end{align*}
		As shown in Section \ref{sec.th2}, we have $C_n^{-1} = O_{\rm p}(1)$. It suffices to show ${\rm III} = o_{\rm p}(1)$ and ${\rm IV}=o_{\rm p}(1)$.  Notice that
		\begin{align*}
			{\rm IV}
			= \big|(2\pi)^{p/2} \pi_{0}(\hbthetan) |\widehat{\bfH}_\Rn|^{-1/2} - C_n\big| \cdot \mE_{\bft \sim \mathcal{N}(\bzero, \widehat{\bfH}_\Rn^{-1})}(|\bft|_\infty)  \,.
		\end{align*}
		Recall $\widehat{\bfH}_{\Rn}=\{\widehat{\bGamma}_{\Rn}(\hat\btheta_n)^\T\widehat{\bfV}_{\Rn}^{-1/2}(\hat\btheta_n)\}^{\otimes2}$. Under Conditions \ref{A.ee}(b) and \ref{A.ee2}, by Proposition \ref{pro.cons}, Lemmas \ref{l.V.hat} and \ref{l.Gamma.hat}, if $\log r=o(n^{1/3})$, $\ell_n\alpha_n=o[\min\{\nu,n^{-1/\gamma}\}]$ and $\ell_n \nu^2=o(1)$, we know that the eigenvalues of $\widehat{\bfH}_\Rn$ are uniformly bounded away from zero and infinity w.p.a.1. Since $\widehat{\bfH}_{\Rn}$ is a $p \times p$ matrix with fixed $p$, then 
		\begin{align} \label{eq:exp_t}
			\mE_{\bft \sim \mathcal{N}(\bzero, \widehat{\bfH}_\Rn^{-1})}(|\bft|_\infty) \leq \mE_{\bft \sim \mathcal{N}(\bzero, \widehat{\bfH}_\Rn^{-1})}(|\bft|_1) = O_{\rm p}(1) \,.
		\end{align} 
		As shown in the proof of Theorem \ref{th.TV}, we have $|C_n - (2\pi)^{p/2} \pi_{0}(\hbthetan)  |\widehat{\bfH}_\Rn|^{-1/2} | \leq {\rm I} = o_{\rm p}(1)$ for ${\rm I}$ defined in \eqref{eq:Int.TV_tri}, which implies ${\rm IV} \leq {\rm I} \cdot \mE_{\bft \sim \mathcal{N}(\bzero, \widehat{\bfH}_\Rn^{-1})}(|\bft|_\infty)= o_{\rm p}(1)$. In the sequel, we will show that ${\rm III} =o_{\rm p}(1)$. Recall $\ell_n \ll \min\{n^{(\gamma-2)/(9\gamma)}(\log r)^{-1/9},n^{1/3}(\log r)^{-1},n^{(\gamma-2)/(2\gamma)}(\log r)^{-3/2}\}$ and $\ell_nn^{-1/2}(\log r)^{1/2} \ll \nu \ll \min\{\ell_n^{-7/2}n^{-1/\gamma},(\log r)^{-1}\}$. For $(\mathcal{D}_1, \mathcal{D}_2, \mathcal{D}_3, \mathcal{D}_4)$ defined as \eqref{eq:four_D}, it holds that ${\rm III} = {\rm III}(1) +  {\rm III}(2)+ {\rm III}(3) + {\rm III}(4)$ with
		\begin{align*}
			{\rm III}(k) 
			&= \int _{\mathcal{D}_k} |\bft|_\infty \big | \pi_{0}(\hbthetan+ n^{-1/2}\bft) \exp \{w_n(\hbthetan+n^{-1/2}\bft)-w_n(\hbthetan)\} I(\bft \in \mathcal{T}_n) \\
			&~~~~~~~~~~~~~~~~~~~- \pi_{0}(\hbthetan)\exp( -\bft^\T \widehat{\bfH}_\Rn \bft/2 ) \big | \,{\rm d}\bft \,.
		\end{align*}

		For ${\rm III}(3)$, by the triangle inequality, we have
		\begin{align*}
			{\rm III}(3)
			\leq&\, \int _{\mathcal{D}_3}  |\bft|_\infty \pi_{0}(\hbthetan+n^{-1/2}\bft) \exp \{ w_n(\hbthetan+n^{-1/2}\bft)-w_n(\hbthetan)\} \,{\rm d}\bft \\
			&+ \pi_{0}(\hbthetan) \int _{\mathcal{D}_3} |\bft|_\infty \exp( -\bft^\T \widehat{\bfH}_\Rn \bft/2 ) \,{\rm d}\bft \,.
		\end{align*}
		Since $\bTheta \subset \R^p$ is a compact set, then 
		\begin{align*}
			\int_{\mathcal{D}_3}  |\bft|_\infty \pi_{0}(\hbthetan+n^{-1/2}\bft) \,{\rm d}\bft \leq \tilde{C} n^{(p+1)/2} \int_{\R^{p}} \pi_{0}(\btheta) \,{\rm d}\btheta \leq \tilde{C} n^{(p+1)/2}\,.
		\end{align*}
		By Proposition \ref{pro.expan}(iii), 
		\begin{align*}
			&\int _{\mathcal{D}_3} |\bft|_\infty \pi_{0}(\hbthetan+n^{-1/2}\bft) \exp \{ w_n(\hbthetan+n^{-1/2}\bft)-w_n(\hbthetan)\} \,{\rm d}\bft \\
			&~~~~~~\leq \sup_{\bft \in \mathcal{D}_3}  \exp \{ w_n(\hbthetan +n^{-1/2}\bft)-w_n(\hbthetan)\} \cdot  \int_{\mathcal{D}_3} |\bft|_\infty \pi_{0}(\hbthetan+n^{-1/2}\bft) \,{\rm d}\bft \\
			&~~~~~~\leq \tilde{C} n^{(p+1)/2} \exp(- Cn\xi_n\beta_{n})
		\end{align*}
		w.p.a.1 for any $\beta_n^{-1}\ell_n \alpha_n^2 \ll \xi_n \ll \beta_n$. Since $r\gg n$, we can select suitable $\xi_n$ satisfying $ n \beta_n \xi_n \gg  \log n $. Then 
		\begin{align*}
			\int _{\mathcal{D}_3} |\bft|_\infty \pi_{0}(\hbthetan+n^{-1/2}\bft) \exp \{ w_n(\hbthetan+n^{-1/2}\bft)-w_n(\hbthetan)\} \,{\rm d}\bft = o_{\rm p}(1) \,.
		\end{align*}
		Recall that %$\bTheta\subset\mathbb{R}^p$ is a compact set including an Euclidean ball with radius $\iota>0$ and 
		the eigenvalues of $\widehat{\bfH}_\Rn$ are uniformly bounded away from zero and infinity w.p.a.1. Since $n\beta_n^2 \rightarrow \infty$ and $p$ is fixed, by the Cauchy-Schwarz inequality and Proposition 1.1 of \cite{Hsu2012}, we have 
		\begin{align*}
			\mE_{\bft \sim \mathcal{N}(\bzero, \widehat{\bfH}_\Rn^{-1})}\{|\bft|_\infty I(\bft \in \mathcal{D}_3)\}  
			\leq&~ \mE^{1/2}_{\bft \sim \mathcal{N}(\bzero, \widehat{\bfH}_\Rn^{-1})}(|\bft|^2_\infty) \mE^{1/2}_{\bft \sim \mathcal{N}(\bzero, \widehat{\bfH}_\Rn^{-1})}\{I^2(\bft \in \mathcal{D}_3)\} \\
			\leq&~ \mE^{1/2}_{\bft \sim \mathcal{N}(\bzero, \widehat{\bfH}_\Rn^{-1})}(|\bft|^2_2)  \exp(-\bar C n\beta_n^2) = o_{\rm p}(1)\,,
		\end{align*}
		which implies 
		\begin{align} \label{eq:exp_t_D3}
			\pi_{0}(\hbthetan) \int _{\mathcal{D}_3} |\bft|_\infty \exp( -\bft^\T \widehat{\bfH}_\Rn \bft/2 ) \,{\rm d}\bft 
			=&~ (2\pi)^{p/2} \pi_{0}(\hbthetan) |\widehat{\bfH}_\Rn|^{-1/2} \mE_{\bft \sim \mathcal{N}(\bzero, \widehat{\bfH}_\Rn^{-1})}\{|\bft|_\infty I(\bft \in \mathcal{D}_3)\} \notag \\
			=&~ o_{\rm p}(1) \,.
		\end{align} 
		Therefore, ${\rm III}(3) = o_{\rm p} (1)$. 
		%\begin{align*}
		%\int _{\mathcal{D}_3} |\bft|_\infty \exp( -\bft^\T \widehat{\bfH}_\Rn \bft/2 ) \,{\rm d}\bft 
		%&= (2\pi)^{p/2} |\widehat{\bfH}_\Rn|^{-1/2} \mE_{\bft \sim \mathcal{N}(\bzero, \widehat{\bfH}_\Rn^{-1})}\{|\bft|_\infty I(\bft \in \mathcal{D}_3)\} \\
		%&\leq (2\pi)^{p/2} |\widehat{\bfH}_\Rn|^{-1/2}  \mE^{1/2}_{\bft \sim \mathcal{N}(\bzero, \widehat{\bfH}_\Rn^{-1})}(|\bft|^2_\infty) \mE^{1/2}_{\bft \sim \mathcal{N}(\bzero, \widehat{\bfH}_\Rn^{-1})}\{I^2(\bft \in \mathcal{D}_3)\} \\
		%&\leq (2\pi)^{p/2} |\widehat{\bfH}_\Rn|^{-1/2}  \mE^{1/2}_{\bft \sim \mathcal{N}(\bzero, \widehat{\bfH}_\Rn^{-1})}(|\bft|^2_2)  \exp(-\bar C n\beta_n^2) = o_{\rm p}(1) \,,
		%\end{align*} 
		
		For ${\rm III}(2)$, it holds that
		\begin{align*}
			{\rm III}(2) 
			\leq &\, \int_{\mathcal{D}_2} |\bft|_\infty \pi_{0}(\hbthetan+n^{-1/2}\bft) \exp \{ w_n(\hbthetan+n^{-1/2}\bft)-w_n(\hbthetan)\} \,{\rm d}\bft \\
			&+ \pi_{0}(\hbthetan) \int_{\mathcal{D}_2} |\bft|_\infty \exp( -\bft^\T \widehat{\bfH}_\Rn \bft/2 ) \,{\rm d}\bft \,.
		\end{align*}
		Since $n\alpha_n^2 \rightarrow \infty$, using the same arguments for \eqref{eq:exp_t_D3}, we have 
		$$
		\pi_{0}(\hbthetan)\int _{\mathcal{D}_2} |\bft|_\infty \exp( -\bft^\T \widehat{\bfH}_\Rn \bft/2 ) \,{\rm d}\bft = o_{\rm p}(1) \,.
		$$
		Due to $\log n \ll n\kappa_{n}^2$, by Proposition \ref{pro.expan}(ii), it then holds w.p.a.1 that
		\begin{align*}
			&\int _{\mathcal{D}_2} |\bft|_\infty \pi_{0}(\hbthetan+n^{-1/2}\bft) \exp \{ w_n(\hbthetan+n^{-1/2}\bft)-w_n(\hbthetan)\} \,{\rm d}\bft \\
			&~~~~~\leq \sup_{\bft \in \mathcal{D}_2} \exp \{ w_n(\hbthetan +n^{-1/2}\bft)-w_n(\hbthetan)\} \cdot \int_{\mathcal{D}_2} |\bft|_\infty \pi_{0}(\hbthetan+n^{-1/2}\bft) \,{\rm d}\bft \\
			&~~~~~\leq \tilde{C} n^{(p+1)/2} \exp(- Cn\kappa_{n}^2) = o_{\rm p}(1) \,.
		\end{align*} 
		Therefore, ${\rm III}(2)=o_{\rm p}(1)$. 
		
		For ${\rm III}(1)$, by Proposition \ref{pro.expan}(i), we have
		\begin{align*}
			{\rm III}(1)
			\leq&\, \int _{\mathcal{D}_{1}} |\bft|_\infty \pi_{0}(\hbthetan+n^{-1/2}\bft) \big | \exp\{-\bft^\T \widehat{\bfH}_\Rn \bft/2+ |\bft|_2^2\cdot O_{\rm p}(\varpi_n)\} -\exp( -\bft^\T \widehat{\bfH}_\Rn \bft/2 )\big |  \,{\rm d}\bft \\
			&+\int _{\mathcal{D}_{1}} |\bft|_\infty \big |\pi_{0}(\hbthetan+n^{-1/2}\bft)-\pi_{0}(\hbthetan) \big | \exp( -\bft^\T \widehat{\bfH}_\Rn \bft/2 ) \,{\rm d}\bft \,,
		\end{align*}
		where $\varpi_n=\max \{\ell_n^{3/2} \alpha_n, \nu,  \ell_nn^{1/\gamma}\alpha_n\}$. Under Condition \ref{A.prior}, we know $ \sup_{\bft \in \mathcal{D}_{1}} |\pi_{0}(\hbthetan+n^{-1/2}\bft)-\pi_{0}(\hbthetan)| =o_{\rm p}(1)$. By \eqref{eq:exp_t}, we have 
		\begin{align*}
			&\int _{\mathcal{D}_{1}} |\bft|_\infty  |\pi_{0}(\hbthetan+n^{-1/2}\bft)-\pi_{0}(\hbthetan)| \exp( -\bft^\T \widehat{\bfH}_\Rn \bft/2 ) \,{\rm d}\bft \\
			&~~~~~~\leq \sup_{\bft \in \mathcal{D}_{1}} |\pi_{0}(\hbthetan+n^{-1/2}\bft)-\pi_{0}(\hbthetan)| \cdot (2\pi)^{p/2} |\widehat{\bfH}_\Rn|^{-1/2} \mE_{\bft \sim \mathcal{N}(\bzero, \widehat{\bfH}_\Rn^{-1})}(|\bft|_\infty) = o_{\rm p}(1) \,.
		\end{align*}
		Due to $\varpi_nn \alpha_n^2=o(1)$, then $\sup_{\bft \in \mathcal{D}_{1}}\{|\bft|_2^2\cdot O_{\rm p}(\varpi_n)\}=o_{\rm p}(1)$. Notice that $|e^{x}-1|\leq |x|e^{x}$ for any $x \in \R$. Then $\sup_{\bft \in \mathcal{D}_1}|\exp\{ |\bft|_2^2\cdot O_{\rm p}(\varpi_n)\}-1| = o_{\rm p}(1)$, which implies that
		\begin{align*}
			&\int _{\mathcal{D}_{1}} |\bft|_\infty \pi_{0}(\hbthetan+n^{-1/2}\bft) \exp( -\bft^\T \widehat{\bfH}_\Rn \bft/2 ) |\exp\{ |\bft|_2^2\cdot O_{\rm p}(\varpi_n)\}-1| \,{\rm d}\bft \\
			&~~~~~\leq  o_{\rm p}(1) \cdot \sup_{\bft \in \mathcal{D}_{1}} \pi_{0}(\hbthetan+n^{-1/2}\bft) \int _{\mathcal{D}_{1}} |\bft|_\infty \exp( -\bft^\T \widehat{\bfH}_\Rn \bft/2) \,{\rm d}\bft =o_{\rm p}(1) \,.
		\end{align*} 
		Therefore, ${\rm III}(1)= o_{\rm p}(1)$. 
		
		For ${\rm III}(4)$,  due to $\mathcal{D}_4 \cap \mathcal{T}_n = \emptyset$, we have 
		$
		{\rm III}(4) 
		= \pi_{0}(\hbthetan) \int _{\mathcal{D}_4} |\bft|_\infty \exp( -\bft^\T \widehat{\bfH}_\Rn \bft/2 ) \,{\rm d}\bft 
		$.
		As shown in Section \ref{sec.th2}, it holds w.p.a.1 that $n^{-1/2}|\bft|_2  \geq \iota/2$  for any $\bft \in \mathcal{D}_4$. Using the same arguments for \eqref{eq:exp_t_D3}, we have $\mE_{\bft \sim \mathcal{N}(\bzero, \widehat{\bfH}_\Rn^{-1})}\{|\bft|_\infty I(\bft \in \mathcal{D}_4)\} = o_{\rm p}(1)$, which implies
		\begin{align*}
			\pi_{0}(\hbthetan) \int _{\mathcal{D}_4} |\bft|_\infty \exp( -\bft^\T \widehat{\bfH}_\Rn \bft/2 ) \,{\rm d}\bft 
			=&~ (2\pi)^{p/2} \pi_{0}(\hbthetan) |\widehat{\bfH}_\Rn|^{-1/2} \mE_{\bft \sim \mathcal{N}(\bzero, \widehat{\bfH}_\Rn^{-1})}\{|\bft|_\infty I(\bft \in \mathcal{D}_4)\} \\
			=&~ o_{\rm p}(1) \,.
		\end{align*}
		Therefore, ${\rm III}(4) = o_{\rm p}(1)$. 
		\hfill $\Box$
		
		\section{Proof of Theorem \ref{th.MH_alg}} \label{sec.pfproMH_alg} 
		To prove Theorem \ref{th.MH_alg}, we first introduce the following two concepts.
		
		\begin{definition} [$\Pi$-irreducibility]
			\label{def.irredu}
			For a distribution $\Pi$ on $\mathcal{D}$, a Markov chain is called $\Pi$-irreducible if for each $A \in \mathscr{B}(\mathcal{D})$ with $\Pi(A)>0$ and $\bfx \in \mathcal{D}$, there exists $k \in \mathbb{N}$ such that $\Psi^k(\bfx, A)>0$, where $\mathscr{B}(\mathcal{D})$ is the Borel $\sigma$-algebra on $\mathcal{D}$, and $\Psi^k$ is the $k$-step transition probability defined recursively as $\Psi^{k}(\bfx, {\rm d}\bfy) = \int_{\bfz \in \mathcal{D} } \Psi^{k-1}(\bfx, {\rm d}\bfz)\Psi(\bfz, {\rm d}\bfy)$.
		\end{definition}
		
		\begin{definition} [Aperiodic]
			\label{def.ap}
			A Markov chain with stationary distribution $\Pi$ on $\mathcal{D}$ and transition probability $\Psi(\cdot\,, \cdot)$ is aperiodic if there do not exist $T \geq 2$  and disjoint subsets $\mathcal{D}_1, \ldots, \mathcal{D}_T \subset \mathcal{D}$ with each $\Pi(\mathcal{D}_i)>0$ such that (i)  $\Psi(\bfx\,, \mathcal{D}_{i+1}) = 1$ for all $\bfx \in \mathcal{D}_{i}$ and $i= 1, \ldots, T-1$, and (ii) $\Psi(\bfx\,, \mathcal{D}_{1}) = 1$ for all $\bfx \in \mathcal{D}_{T}$. 
		\end{definition}
		
		%Now we begin to prove Theorem \ref{th.MH_alg}. 
		Denote by $\Psi(\btheta, \cdot)$ the transition probability of the Markov chain determined by Algorithm \ref{alg3} at $\btheta \in \bTheta$. For given $\btheta \in \bTheta$, $\alpha_{\btheta}(\bvartheta) = \min \{1, R_{\btheta}(\bvartheta)\}$ is the acceptance probability at $\bvartheta \in \R^p$, where
		\begin{align*}
			R_{\btheta}(\bvartheta) = \left \{
			\begin{aligned}
				& \frac{\pi^\dag(\bvartheta\,|\,\mathcal{X}_n) \phi(\btheta\,|\,\bvartheta)}{\pi^\dag(\btheta\,|\,\mathcal{X}_n) \phi(\bvartheta\,|\,\btheta)} \,, && \text{ if~} \bvartheta \in \bTheta \text{~with~} \pi^\dag(\btheta\,|\,\mathcal{X}_n) \phi(\bvartheta\,|\,\btheta)\neq 0 \,,\\
				&~~~~~~~~~~~~1 \,, && \text{ if~} \bvartheta \in \bTheta \text{~with~} \pi^\dag(\btheta\,|\,\mathcal{X}_n) \phi(\bvartheta\,|\,\btheta)=0 \,, \\
				&~~~~~~~~~~~~0 \,, && \text{ if~} \bvartheta \notin \bTheta \,.
			\end{aligned}
			\right.
		\end{align*}
		Then the transition probability of the associated Markov chain at $\btheta \in \bTheta$ has a probability mass $\psi_{\btheta} = 1- \int_{\bTheta} \phi(\bvartheta\,|\,\btheta) \alpha_{\btheta}(\bvartheta) \,{\rm d}\bvartheta$. Define %the transition kernel
		$\psi(\btheta,\bvartheta) = \phi(\bvartheta\,|\,\btheta) \alpha_{\btheta}(\bvartheta)$ for any $\btheta,\bvartheta \in \bTheta$. 
		%Notice that $\pi^\dag(\btheta\,|\,\mathcal{X}_n)  \phi(\bvartheta\,|\,\btheta) \alpha_{\btheta}(\bvartheta) = \pi^\dag(\bvartheta\,|\,\mathcal{X}_n)  \phi(\btheta\,|\,\bvartheta) \alpha_{\bvartheta}(\btheta)$ for any $\btheta, \bvartheta \in \bTheta$, 
		We have $\pi^\dag(\btheta\,|\,\mathcal{X}_n) \psi(\btheta,\bvartheta) = \pi^\dag(\bvartheta\,|\,\mathcal{X}_n) \psi(\bvartheta,\btheta)$ for any $\btheta, \bvartheta \in \bTheta$. Since the Markov chain determined by Algorithm {\rm \ref{alg3}} always stays in $\bTheta$, its transition probability $\Psi(\cdot\,, \cdot): \bTheta \times \mathscr{B}(\bTheta) \mapsto \R_+$ satisfies $
		\Psi(\btheta, {\rm d}\bvartheta)
		= \psi_{\btheta} \delta_{\btheta}({\rm d}\bvartheta) + \psi(\btheta,\bvartheta) \, {\rm d}\bvartheta $, 
		where $\mathscr{B}(\bTheta)$ is the Borel $\sigma$-algebra on $\bTheta$, and $\delta_{\btheta}$ is the Dirac-delta function at $\btheta$ with $\delta_{\btheta}(A)=I(\btheta \in A)$. For any $A,B \in \mathscr{B}(\bTheta)$, we have
		\begin{align*}
			\int_A \pi^\dag(\btheta\,|\,\mathcal{X}_n) \Psi(\btheta, B) \,{\rm d}\btheta
			=& \int_{A\bigcap B} \pi^\dag(\btheta\,|\,\mathcal{X}_n) \psi_{\btheta} \,{\rm d}\btheta
			+ \int_{(\btheta,\bvartheta) \in A\times B} \pi^\dag(\btheta\,|\,\mathcal{X}_n) \psi(\btheta,\bvartheta) \,{\rm d}\btheta {\rm d}\bvartheta \\
			=& \int_B \pi^\dag(\btheta\,|\,\mathcal{X}_n) \psi_{\btheta } \delta_{\btheta}(A)\,{\rm d}\btheta + \int_{(\btheta,\bvartheta) \in A\times B} \pi^\dag(\bvartheta\,|\,\mathcal{X}_n) \psi(\bvartheta,\btheta) \,{\rm d}\btheta {\rm d}\bvartheta \\
			=& \int_B \pi^\dag(\btheta\,|\,\mathcal{X}_n) \Psi(\btheta, A) \,{\rm d}\btheta \,.
		\end{align*}
		%Recall that $\Pi^{\dag}_n(\cdot)$ is the measure which admits the posterior distribution $\pi^{\dag}(\cdot\,|\,\mathcal{X}_n)$. 
		Therefore, $\Pi^\dag_n(A) = \int_A \pi^\dag(\btheta\,|\,\mathcal{X}_n)\,{\rm d}\btheta = \int_A \pi^\dag(\btheta\,|\,\mathcal{X}_n) \Psi(\btheta, \bTheta) \,{\rm d}\btheta = \int_{\bTheta} \pi^\dag(\btheta\,|\,\mathcal{X}_n) \Psi(\btheta, A) \,{\rm d}\btheta$ for any $A \in \mathscr{B}(\bTheta)$, which implies that $\Pi^\dag_n$ is the stationary distribution of such Markov chain with transition probability $\Psi(\cdot\,,\cdot)$. 
		
		Denote by $\mathbb{L}(\cdot)$ the Lebesgue measure on $\R^p$. For any $A\in\mathscr{B}(\bTheta)$ with $\Pi_n^\dag(A)>0$, due to $\Pi_n^\dag(A)=\int_A \pi^\dag(\btheta\,|\,\mathcal{X}_n)\,{\rm d}\btheta$, we know $\mathbb{L}(A)>0$. Recall that $\bTheta \subset \R^p$ is a compact set. Since $\phi(\bvartheta\,|\,\btheta)$ is positive and continuous on $(\btheta, \bvartheta) \in \bTheta \times \bTheta$, there exists a constant $C>0$ such that $\inf_{\btheta, \bvartheta \in \bTheta}\phi(\btheta\,|\,\bvartheta) > C$. On one hand, for any $A\in\mathscr{B}(\bTheta)$ and $\btheta \in \bTheta$ such that $\Pi_n^\dag(A)>0$ and $\pi^\dag(\btheta\,|\,\mathcal{X}_n) = 0$, we have $\Psi(\btheta, A) \geq \int_A \psi(\btheta,\bvartheta) \, {\rm d}\bvartheta = \int_A \phi(\bvartheta\,|\,\btheta) \, {\rm d}\bvartheta >0$. On the other hand, for any $A\in\mathscr{B}(\bTheta)$ and $\btheta \in \bTheta$ such that $\Pi_n^\dag(A)>0$ and $\pi^\dag(\btheta\,|\,\mathcal{X}_n)>0$, we have
		\begin{align*}
			\Psi(\btheta, A) &= \psi_{\btheta} \delta_{\btheta}(A) + \int_A \psi(\btheta,\bvartheta) \, {\rm d}\bvartheta  
			\geq \int_A \phi(\bvartheta\,|\,\btheta) \min \bigg\{1, \frac{\pi^\dag(\bvartheta\,|\,\mathcal{X}_n) \phi(\btheta\,|\,\bvartheta)}{\pi^\dag(\btheta\,|\,\mathcal{X}_n) \phi(\bvartheta\,|\,\btheta)} \bigg\} \, {\rm d}\bvartheta  \\
			&= \int_{\bvartheta \in A:\, \pi^\dag(\bvartheta\,|\,\mathcal{X}_n) \geq  \pi^\dag(\btheta\,|\,\mathcal{X}_n)} \min \bigg\{\phi(\bvartheta\,|\,\btheta), \frac{\pi^\dag(\bvartheta\,|\,\mathcal{X}_n) }{\pi^\dag(\btheta\,|\,\mathcal{X}_n)} \phi(\btheta\,|\,\bvartheta)  \bigg\} \, {\rm d}\bvartheta \\
			&~~~~~~~+ \int_{\bvartheta \in A:\, \pi^\dag(\bvartheta\,|\,\mathcal{X}_n) <   \pi^\dag(\btheta\,|\,\mathcal{X}_n)} \frac{1}{\pi^\dag(\btheta\,|\,\mathcal{X}_n)} \min\{ \pi^\dag(\btheta\,|\,\mathcal{X}_n)\phi(\bvartheta\,|\,\btheta), \pi^\dag(\bvartheta\,|\,\mathcal{X}_n) \phi(\btheta\,|\,\bvartheta)\} \, {\rm d}\bvartheta  \\
			&\geq C \mathbb{L}(\{\bvartheta \in A: \pi^\dag(\bvartheta\,|\,\mathcal{X}_n) \geq  \pi^\dag(\btheta\,|\,\mathcal{X}_n)\}) +  \frac{C}{\pi^\dag(\btheta\,|\,\mathcal{X}_n)} \int_{\bvartheta \in A:\, \pi^\dag(\bvartheta\,|\,\mathcal{X}_n) < \pi^\dag(\btheta\,|\,\mathcal{X}_n)} \pi^\dag(\bvartheta\,|\,\mathcal{X}_n) \, {\rm d}\bvartheta \,.
		\end{align*}
		Since $\mathbb{L}(\{\bvartheta \in A: \pi^\dag(\bvartheta\,|\,\mathcal{X}_n) \geq  \pi^\dag(\btheta\,|\,\mathcal{X}_n)\})$ and $\int_{\bvartheta \in A:\, \pi^\dag(\bvartheta\,|\,\mathcal{X}_n) < \pi^\dag(\btheta\,|\,\mathcal{X}_n)} \pi^\dag(\bvartheta\,|\,\mathcal{X}_n) \, {\rm d}\bvartheta$ cannot be zero simultaneously for any $A\in\mathscr{B}(\bTheta)$ with $\Pi_n^\dag(A)>0$, then $\Psi(\btheta, A) >0$ for any $A\in\mathscr{B}(\bTheta)$ and $\btheta \in \bTheta$ such that $\Pi_n^\dag(A)>0$ and $\pi^\dag(\btheta\,|\,\mathcal{X}_n) > 0$. Therefore, it holds that $\Psi(\btheta, A) >0$ for any $\btheta \in \bTheta$ and $A\in\mathscr{B}(\bTheta)$ with $\Pi_n^\dag(A)>0$. By Definition \ref{def.irredu}, the Markov chain with transition probability $\Psi(\cdot\,,\cdot)$ is $\Pi^\dag_n$-irreducible. Furthermore, by Definition \ref{def.ap}, we know the Markov chain $\{\btheta^k\}_{k\geq1}$ with transition probability $\Psi(\cdot\,,\cdot)$ and initial point $\btheta^0$ is aperiodic. Notice that $\mathscr{B}(\bTheta)$ is a countably generated $\sigma$-algebra. Denote by $\mathcal{T}^k_{\btheta^{0}}(\cdot)$ the measure which admits the distribution of such Markov chain at $k$-th step with initial point $\btheta^{0}$. Conditional on $\mathcal{X}_n$, for any $\btheta^0 \in \bTheta$ such that $\pi^\dag(\btheta^0\,|\,\mathcal{X}_n)>0$, by Theorem 4 of \cite{RobertsRosenthal2004}, we have $\mathcal{D}_\TV(\mathcal{T}^k_{\btheta^{0}} ,\, \Pi^{\dag}_n) \rightarrow 0$ as $k \rightarrow \infty$. Furthermore, notice that $\bTheta \subset \R^p$ is a compact set with fixed $p$. Conditional on $\mathcal{X}_n$, for any $\btheta^0 \in \bTheta$ such that $\pi^\dag(\btheta^0\,|\,\mathcal{X}_n)>0$, it follows from Fact 5 of \cite{RobertsRosenthal2004} that $|K^{-1}\sum_{k=1}^K \btheta^k - \mE_{\btheta \sim \pi^\dag}(\btheta)|_\infty \leq |K^{-1}\sum_{k=1}^K \btheta^k - \mE_{\btheta \sim \pi^\dag}(\btheta)|_1 \rightarrow 0 $ almost surely as $K \rightarrow \infty$, where $\{\btheta^k\}_{k\geq 1}$ are generated via Algorithm \ref{alg3} with the initial $\btheta^0$ and $\mE_{\btheta \sim \pi^\dag}(\btheta)$ is defined in \eqref{eq:exp_pi}. We complete the proof of Theorem \ref{th.MH_alg}.
		\hfill $\Box$
		
		\section{Proof of Theorem \ref{th.MAMIS_alg}} \label{sec.pfproMAMIS_alg} 
		
		For the function $\bfh: \R^p \mapsto \R^s$ involved in Algorithm {\rm \ref{alg2}}, let $\bzeta^* = \mE_{\btheta \sim \pi^\dag}\{\bfh(\btheta)\}$.  
		%with $\pi^\dag(\btheta\,|\,\mathcal{X}_n)$ defined as \eqref{eq:postdis}. 
		%Notice that $\bTheta \subset \R^p$ is a compact set and $\sup_{\btheta \in \bTheta} |\bfh(\btheta)|_\infty \leq K_9$ for some universal constant $K_9>0$. Then $\bvartheta^* = \mE_{\pi^\dag}\{\bfh(\btheta)\}$ is well defined. 
		Define 
		\begin{align} \label{eq:esi_E_btheta}
			\widehat{\mE}^*_{\pi^\dag,K}(\btheta) = \frac{1}{S_K} \sum_{k=1}^{K} \sum_{i=1}^{N_k}  \frac{\pi^\dag(\btheta^k_i\,|\,\mathcal{X}_n)}{\varphi(\btheta^k_i\,;\bzeta^*) } \btheta^k_i 
		\end{align}
		with $S_K=N_1+\cdots+N_K$, where 
		$\{\btheta^1_1, \ldots,  \btheta^1_{N_1}, \ldots,\btheta^K_1,\ldots, \btheta^K_{N_K}\}$ are generated via Algorithm {\rm \ref{alg2}}. To construct Theorem \ref{th.MAMIS_alg}, we need the following two lemmas whose proofs are given in Sections \ref{sec:pflemma12} and \ref{sec:pflemma14}, respectively.
		
		\begin{lemma}\label{l.h.cons}
			Assume that the conditions of Theorem {\rm \ref{th.MAMIS_alg}} hold. Conditional on $\mathcal{X}_n$, $|\hat{\bzeta}_{k} - \bzeta^*|_\infty \rightarrow 0$ almost surely as $k \rightarrow \infty$, where $\hat{\bzeta}_{k}$ is defined in Algorithm {\rm \ref{alg2}}.
		\end{lemma}
		
		\begin{lemma}\label{l.three_res}
			Assume that the conditions of Theorem {\rm \ref{th.MAMIS_alg}} hold. Conditional on $\mathcal{X}_n$, $|\widehat{\mE}^*_{\pi^\dag,K}(\btheta) - \mE_{\btheta \sim \pi^\dag}(\btheta)|_\infty \rightarrow 0$ almost surely as $K \rightarrow \infty$, where $\widehat{\mE}^*_{\pi^\dag,K}(\btheta)$ is defined in \eqref{eq:esi_E_btheta}.  
		\end{lemma}
		
		Denote by $\P_{\mathcal{X}_n}(\cdot)$ the conditional probability given $\mathcal{X}_n$. For some sufficiently large $M>0$, by Lemma \ref{l.h.cons}, we have that for any $\epsilon>0$, there exists a sufficiently large integer $k_\epsilon$ such that $\mathbb{P}_{\mathcal{X}_n}(\mathcal{A}) \leq \epsilon$ with $\mathcal{A} = \bigcup_{t=k_\epsilon}^\infty \{|\hat{\bzeta}_{t} - \bzeta^*|_\infty > M\}$. 
		Define a compact set $\mathcal{B}=\{\bzeta \in \R^s: |\bzeta - \bzeta^*|_\infty \leq M \} $. Recall $\bTheta \subset \R^p$ is a compact set with fixed $p$. Since $\varphi(\btheta\,; \bzeta)$ is positive and continuous on $(\btheta, \bzeta) \in \bTheta \times \R^s$, then $\inf_{\btheta \in \bTheta, \bzeta \in \mathcal{B}} \varphi(\btheta\,; \bzeta) \geq C_M$ for some constant $C_M>0$ and $\varphi(\btheta\,; \bzeta)$ is uniformly continuous on $(\btheta\,; \bzeta) \in \bTheta \times \mathcal{B}$. For any $\varepsilon>0$, there exists $\delta(\varepsilon)>0$ such that $|\varphi(\btheta_1\,; \bzeta_1) - \varphi(\btheta_2\,; \bzeta_2)| < C_M (2+2C_M)^{-1}\varepsilon$ for any $(\btheta_1, \bzeta_1), (\btheta_2, \bzeta_2)  \in \bTheta \times \mathcal{B}$ satisfying $|\btheta_1 - \btheta_2|_\infty \leq \delta(\varepsilon)$ and $|\bzeta_1 - \bzeta_2|_\infty \leq \delta(\varepsilon)$. For any $K \geq k_\epsilon$,  it holds that
		$$
		\inf_{\btheta \in \bTheta}\frac{1}{S_K} \sum_{k=1}^K N_k \varphi(\btheta\,;\bzeta) \geq
		\inf_{\btheta \in \bTheta} \frac{1}{S_K} \sum_{k=k_\epsilon}^K N_k \varphi(\btheta\,;\bzeta) \geq
		\frac{C_M}{S_K}\bigg(S_K - \sum_{k=1}^{k_\epsilon-1} N_k\bigg)
		$$
		for any $\bzeta \in \mathcal{B}$. Notice that $S_K \to \infty$ as $K \to \infty$. Given $k_\epsilon$, there exists a sufficiently large integer $K^*$ such that $S_K / (S_K - \sum_{k=1}^{k_\epsilon-1} N_k) \leq 1+C_M$ for all $K > K^*$. Recall $N_k\leq N_{k+1}$ for any $k\geq 1$ and $N_k\rightarrow\infty$ as $k\rightarrow\infty$. Since $\sup_{\btheta\in\bTheta,\bzeta\in\mathbb{R}^s}\varphi(\btheta\,;\bzeta)<\infty$, we have
		\[
		\frac{1}{S_t}\sum_{k=1}^{\lfloor\sqrt{t}\rfloor}N_k\sup_{\btheta\in\bTheta}|\varphi(\btheta\,;\bzeta^*)-\varphi(\btheta\,;\hat{\bzeta}_k)|\lesssim \frac{1}{S_t}\sum_{k=1}^{\lfloor\sqrt{t}\rfloor}N_k\rightarrow0
		\]
		as $t\rightarrow\infty$. For given $\varepsilon>0$, there exists some sufficiently large integer $\tilde{K}$ such that 
		\[
		\frac{1}{S_t}\sum_{k=1}^{\lfloor\sqrt{t}\rfloor}N_k\sup_{\btheta\in\bTheta}|\varphi(\btheta\,;\bzeta^*)-\varphi(\btheta\,;\hat{\bzeta}_k)|\leq \frac{C_M\varepsilon}{2(1+C_M)}
		\]
		for any $t\geq \tilde{K}$. 
		It holds that 
		\begin{align*}
			&\bigg\{\sup_{\btheta \in \bTheta}\bigg| \frac{\varphi(\btheta\,; \bzeta^*)}{S_t^{-1} \sum_{k=1}^t N_k \varphi(\btheta \,;\hat{\bzeta}_k)} - 1\bigg| > \varepsilon, \mathcal{A}^\c \bigg\} \\
			&~~~~~~~~~~\subset  \bigg\{ \frac{1}{S_t} \sum_{k=1}^t N_k \sup_{\btheta \in \bTheta}|\varphi(\btheta \,;\bzeta^*) - \varphi(\btheta \,;\hat{\bzeta}_k)| > \frac{C_M\varepsilon}{1+C_M}, \mathcal{A}^\c \bigg\}\\
			&~~~~~~~~~~\subset \bigg\{ \frac{1}{S_t} \sum_{k=\lfloor \sqrt{t}\rfloor+1}^t N_k \sup_{\btheta \in \bTheta}|\varphi(\btheta \,;\bzeta^*) - \varphi(\btheta \,;\hat{\bzeta}_k)| > \frac{C_M\varepsilon}{2+2C_M}, \mathcal{A}^\c \bigg\}\\
			&~~~~~~~~~~\subset \bigcup_{k=\lfloor \sqrt{t}\rfloor +1}^t \{ |\hat{\bzeta}_{k} - \bzeta^*|_\infty >  \delta(\varepsilon), \mathcal{A}^\c\}\subset \bigcup_{k=\lfloor \sqrt{t}\rfloor +1}^t \{ |\hat{\bzeta}_{k} - \bzeta^*|_\infty >  \delta(\varepsilon)\} 
		\end{align*}
		for any $t > \max(K^*, k_\epsilon,\tilde{K})$. We then have
		\begin{align*}
			&\limsup_{m \to \infty} \P_{\mathcal{X}_n}\bigg[\bigcup_{t=m}^\infty \bigg\{\sup_{\btheta \in \bTheta}\bigg| \frac{\varphi(\btheta\,; \bzeta^*)}{S_t^{-1} \sum_{k=1}^t N_k \varphi(\btheta \,;\hat{\bzeta}_k)} - 1\bigg| > \varepsilon\bigg\} \bigg] \\
			&~~~~~~\leq \P_{\mathcal{X}_n}(\mathcal{A}) + \limsup_{m \to \infty}\ \P_{\mathcal{X}_n}\bigg[\bigcup_{k=\lfloor m\rfloor+1}^\infty \{|\hat{\bzeta}_{k} - \bzeta^*|_\infty >  \delta(\varepsilon)  \} \bigg]=\mathbb{P}_{\mathcal{X}_n}(\mathcal{A}) \leq \epsilon \,,
		\end{align*}
		where the second step is due to the fact that conditional on $\mathcal{X}_n$ we have $|\hat{\bzeta}_{k} - \bzeta^*|_\infty \rightarrow 0$ almost surely as $k \rightarrow \infty$. Letting $\epsilon \to 0$, we know that conditional on $\mathcal{X}_n$, 
		\begin{align} \label{eq:sup_bound}
			\sup_{\btheta \in \bTheta} \bigg|\frac{\varphi(\btheta\,;\bzeta^*)}{S_K^{-1} \sum_{k=1}^K N_k \varphi(\btheta\,;\hat{\bzeta}_k)} - 1 \bigg| \rightarrow 0
		\end{align}
		almost surely as $K \rightarrow \infty$. 
		
		Define 
		$$
		\widehat{\mE}^*_{\pi^\dag,K}(|\btheta|_\infty) = \frac{1}{S_K} \sum_{k=1}^{K} \sum_{i=1}^{N_k} \frac{\pi^\dag(\btheta^k_i\,|\,\mathcal{X}_n)}{\varphi(\btheta^k_i\,;\bzeta^*)}|\btheta^k_i|_\infty  ~~\textrm{and}~~ \mE_{\btheta \sim \pi^\dag}(|\btheta|_\infty) = \int_{\R^p} |\btheta|_\infty \pi^\dag(\btheta\,|\,\mathcal{X}_n) \, {\rm d}\btheta \,,
		$$
		where 
		$\{\btheta^1_1, \ldots,  \btheta^1_{N_1}, \ldots,\btheta^K_1,\ldots, \btheta^K_{N_K}\}$ are generated via Algorithm {\rm \ref{alg2}}. For  $\widehat{\mE}_{\pi^\dag,K}(\btheta)$ defined in \eqref{eq:esi_Alg2} and $\widehat{\mE}^*_{\pi^\dag,K}(\btheta)$ defined in \eqref{eq:esi_E_btheta}, since $\pi^\dag(\btheta\,|\,\mathcal{X}_n) |\btheta|_\infty / \varphi(\btheta\,;\bzeta^*) = 0$ for any $\btheta \notin \bTheta$, we then have
		\begin{align*}
			|\widehat{\mE}_{\pi^\dag,K}(\btheta)- \widehat{\mE}^*_{\pi^\dag,K}(\btheta) |_\infty 
			&\leq \frac{1}{S_K} \sum_{k=1}^{K} \sum_{i=1}^{N_k} \frac{\pi^\dag(\btheta^k_i\,|\,\mathcal{X}_n)|\btheta^k_i|_\infty}{\varphi(\btheta^k_i\,;\bzeta^*)}  \bigg|\frac{\varphi(\btheta^k_i\,;\bzeta^*)}{S_K^{-1} \sum_{l=1}^K N_l \varphi(\btheta^k_i\,;\hat{\bzeta}_l)} - 1 \bigg| \\
			&\leq \widehat{\mE}^*_{\pi^\dag,K}(|\btheta|_\infty) \sup_{\btheta \in \bTheta} \bigg|\frac{\varphi(\btheta\,;\bzeta^*)}{S_K^{-1} \sum_{k=1}^K N_k \varphi(\btheta\,;\hat{\bzeta}_k)} - 1 \bigg| \\
			&\leq \mE_{\btheta \sim \pi^\dag}(|\btheta|_\infty) \sup_{\btheta \in \bTheta} \bigg|\frac{\varphi(\btheta\,;\bzeta^*)}{S_K^{-1} \sum_{k=1}^K N_k \varphi(\btheta\,;\hat{\bzeta}_k)} - 1 \bigg| \\
			&~~~+ \big|\widehat{\mE}^*_{\pi^\dag,K}(|\btheta|_\infty) -  \mE_{\btheta \sim \pi^\dag}(|\btheta|_\infty)\big| \sup_{\btheta \in \bTheta} \bigg|\frac{\varphi(\btheta\,;\bzeta^*)}{S_K^{-1} \sum_{k=1}^K N_k \varphi(\btheta\,;\hat{\bzeta}_k)} - 1 \bigg|
		\end{align*}
		Using the same arguments for the proof of Lemma \ref{l.three_res} in Section \ref{sec:pflemma14}, it holds that conditional on $\mathcal{X}_n$ we have $|\widehat{\mE}^*_{\pi^\dag,K}(|\btheta|_\infty) -  \mE_{\btheta \sim \pi^\dag}(|\btheta|_\infty)| \rightarrow 0$ almost surely as $K \rightarrow \infty$. Notice that $\bTheta \subset \R^p$ is a compact set with fixed $p$. Then $\mE_{\btheta \sim\pi^\dag}(|\btheta|_\infty) < \infty$. Together with \eqref{eq:sup_bound}, it holds that conditional on $\mathcal{X}_n$ we have $|\widehat{\mE}_{\pi^\dag,K}(\btheta)- \widehat{\mE}^*_{\pi^\dag,K}(\btheta) |_\infty \rightarrow 0$ almost surely as $K \rightarrow \infty$. By the triangle inequality and Lemma \ref{l.three_res}, conditional on $\mathcal{X}_n$, $|\widehat{\mE}_{\pi^\dag,K}(\btheta)- \mE_{\btheta \sim \pi^\dag}(\btheta)|_\infty \leq  |\widehat{\mE}_{\pi^\dag,K}(\btheta)- \widehat{\mE}^*_{\pi^\dag,K}(\btheta)|_\infty + |\widehat{\mE}^*_{\pi^\dag,K}(\btheta)- \mE_{\btheta \sim \pi^\dag}(\btheta)|_\infty \rightarrow 0$ almost surely as $K \rightarrow \infty$. We complete the proof of Theorem \ref{th.MAMIS_alg}.
		\hfill $\Box$
		
		\section{Proofs of auxiliary lemmas} \label{PF of lemmas}
		
		\subsection{Proof of Lemma \ref{l.V.hat}}\label{sec:pflemma1}
		
		The proof is almost identical to that of Lemma 1 in \cite{Changetal2018}. Recall $p$ is fixed. We only need to replace $\{\varrho_n,\omega_n,\xi_n,b_n^{1/(2\beta)},s\}$ appeared in the proof of Lemma 1 in \cite{Changetal2018} by $(1,1,1,\varphi_n,p)$ and all the arguments still hold. \hfill $\Box$
		
		\subsection{Proof of Lemma \ref{l.lam.theta0}}\label{sec:pflemma2}
		
		Due to the convexity of $P_{\nu}(\cdot)$, $f_n(\blambda;\bthetazero)$ is concave w.r.t $\blambda$. We only need to show that there exists a local maximizer $\hat{\blambda}(\bthetazero)$ satisfying the results stated in the lemma. Recall $\mathcal{M}_{\bthetazero}^*=\{j\in[r]: |\bar{g}_j(\bthetazero)| \geq  C_*\nu \rho'(0^{+}) \}$ for some $C_* \in (0,1)$, and $\mathbb{P}(\max _{\btheta \in \bTheta:\,|\btheta-\bthetazero|_2 \leq c_n}|\mathcal{M}_{\btheta}^*|\leq\ell_n)\rightarrow1$ for some $c_n\rightarrow 0$ satisfying $\nu c_n^{-1} \rightarrow 0$. For any given $c\in(C_*,1)$, write $\mathcal{M}_{\bthetazero}:=\mathcal{M}_{\bthetazero}(c)=\{ j\in[r]:|\bar{g}_j(\bthetazero)| \geq c\nu\rho'(0^{+}) \}$. Then $\ell_n \geq |\mathcal{M}_{\bthetazero}^*| \geq |\mathcal{M}_{\bthetazero}|$ w.p.a.1. To prove Lemma \ref{l.lam.theta0}, we establish its validity separately with Case 1: $\mathcal{M}_{\bthetazero} \neq \emptyset$  and Case 2: $\mathcal{M}_{\bthetazero} = \emptyset$.
		
		\subsubsection{Case 1: $ \mathcal{M}_{\bthetazero} \neq \emptyset$}

		Restricted on $\mathcal{M}_{\bthetazero}$, we select $\delta_n$ satisfying $\delta_n=o(\ell_n^{-1/2} n^{-1/\gamma})$ and $\ell_n^{1/2} \alpha_n=o(\delta_n) $, which can be guaranteed by $\ell_n\alpha_n=o(n^{-1/\gamma})$. Let $\Lambda_{0}=\{\blambda\in \R^{r}:|\blambda_{\mathcal{M}_{\bthetazero}}|_2\leq \delta_n {\rm \ and \ }\blambda_{\mathcal{M}^\c_{\bthetazero}}=\bzero \}$ and $\tilde{\blambda}_{0}=\arg \max _{\blambda \in \Lambda_{0}} f_n(\blambda; \bthetazero)$. By Condition \ref{A.ee}(a), we have $\max_{i\in[n],j\in[r]}|g_{i,j} (\bthetazero)| = O_{\rm p}(n^{1/\gamma})$, which implies $\max_{i\in[n]} |\bfg_{i,\mathcal{M}_{\bthetazero}} (\bthetazero)|_2 = O_{\rm p}(\ell_n^{1/2} n^{1/\gamma})$. Then $\max_{i\in[n]} |\tilde{\blambda}^\T_{0} \bfg_{i}(\bthetazero)|=o_{\rm p}(1)$. Write $\tilde{\blambda}_{0}=(\tilde{\lambda}_{0,1},\ldots,\tilde{\lambda}_{0,r})^\T$. By the Taylor expansion, we have
		\begin{align*}
			0=f_n(\bzero;\bthetazero) \leq f_n(\tilde{\blambda}_{0};\bthetazero)
			=\tilde{\blambda}^\T_{0} \bar{\bfg}(\bthetazero)- \frac{1}{2n} \sum_{i=1}^{n} \frac{\tilde{\blambda}^\T_{0} \bfg_{i} (\bthetazero)^{\otimes2} \tilde{\blambda}_{0}}{\{1+C\tilde{\blambda}^\T_{0} \bfg_{i} (\bthetazero)\}^2} - \sum_{j=1}^{r} P_{\nu}(|\tilde{\lambda}_{0,j}|)
		\end{align*}
		for some $C \in (0,1)$. By Condition \ref{A.ee}(b) and the same arguments for deriving Lemma \rm\ref{l.V.hat}, if $\log r=o(n^{1/3})$ and $\ell_n\alpha_n=o(1)$, we have $\lambda_{\min}\{\widehat{\bfV}_{\mathcal{M}_{\bthetazero}} (\bthetazero)\}$ is uniformly bounded away from zero w.p.a.1. Thus 
		$0 \leq|\tilde{\blambda}_{0,\mathcal{M}_{\bthetazero}}|_2 |\bar{\bfg}_{\mathcal{M}_{\bthetazero}}(\bthetazero)|_2 - 4^{-1}K_3 |\tilde{\blambda}_{0,\mathcal{M}_{\bthetazero}}|_2^2$ w.p.a.1, where $K_3$ is specified in Condition \ref{A.ee}(b).  By the moderate deviation of self-normalized sums \citep{JingShaoWang2003}, %we have
		$
		|\bar{\bfg}(\bthetazero)|_{\infty}
		=O_{\rm p}(\alpha_n)$. Then
		$
		|\bar{\bfg}_{\mathcal{M}_{\bthetazero}}(\bthetazero)|_2= O_{\rm p}(\ell_n^{1/2} \alpha_n)
		$ and 
		$
		|\tilde{\blambda}_{0,\mathcal{M}_{\bthetazero}}|_2=O_{\rm p}(\ell_n^{1/2} \alpha_n)=o_{\rm p}(\delta_n)$. 
		Write $\tilde{\blambda}_{0,\mathcal{M}_{\bthetazero}}=(\tilde{\lambda}_{1},\ldots,\tilde{\lambda}_{|\mathcal{M}_{\bthetazero}|})^\T$. We then have w.p.a.1 that
		\begin{align} \label{eq:M_0}
			\bzero= \frac{1}{n} \sum_{i=1}^{n} \frac{\bfg_{i,\mathcal{M}_{\bthetazero}}(\bthetazero)}{1+\tilde{\blambda}_{0,\mathcal{M}_{\bthetazero}}^\T \bfg_{i,\mathcal{M}_{\bthetazero}}(\bthetazero)} - \tilde{\bseta} \,,
		\end{align}
		where $\tilde{\bseta}=(\tilde{\eta}_1,\ldots,\tilde{\eta}_{|\mathcal{M}_{\bthetazero}|})^\T$ with $\tilde{\eta}_j=\nu\rho'(|\tilde\lambda_{j}|;\nu)\sgn(\tilde{\lambda}_{j})$ for $\tilde\lambda_{j}\neq0$ and $\tilde\eta_j\in[-\nu\rho'(0^+),\nu\rho'(0^+)]$ for $\tilde\lambda_{j}=0$.
		In the sequel, we will show that $\tilde{\blambda}_{0}$ is a local maximizer for $f_n(\blambda;\bthetazero)$ w.p.a.1.
		
		Firstly, define $\Lambda_{0}^* = \{\blambda\in \R^r:|\blambda_{\mathcal{M}_{\bthetazero}^*}|_2 \leq \varepsilon,\blambda_{\mathcal{M}^{*,\c}_{\bthetazero}}=\bzero\}$ for some sufficiently small constant $\varepsilon>0$. For $\tilde{\blambda}_{0}$ defined before, we will prove $\tilde\blambda_{0}=\arg \max_{\blambda\in\Lambda_{0}^*} f_n(\blambda;\bthetazero)$ w.p.a.1. Since $\tilde{\blambda}_{0} \in\Lambda_{0}$ and $\mathcal{M}_{\bthetazero} \subset \mathcal{M}_{\bthetazero}^*$, we know $\tilde{\blambda}_{0} \in \Lambda_{0}^*$ for sufficiently large $n$. Restricted on $\blambda \in \Lambda_{0}^*$, by the concavity of $f_n(\blambda;\bthetazero)$ w.r.t $\blambda$, it suffices to show that $\bfw = \tilde{\blambda}_{0,\mathcal{M}^*_{\bthetazero}}=: (w_{1},\ldots,w_{|\mathcal{M}^*_{\bthetazero}|})^\T \in \R^{|\mathcal{M}^*_{\bthetazero}|}$ satisfies the equation
		\begin{align} \label{eq:1.order.con.1}
			\bzero= \frac{1}{n} \sum_{i=1}^{n} \frac{\bfg_{i,\mathcal{M}^*_{\bthetazero}}(\bthetazero)}{1+\bfw ^\T \bfg_{i,\mathcal{M}^*_{\bthetazero}}(\bthetazero)} - \tilde{\bseta}^*
		\end{align}
		w.p.a.1, where $\tilde{\bseta}^*=(\tilde\eta^*_1,\ldots,\tilde\eta^*_{|\mathcal{M}_{\bthetazero}^*|})^\T$ with $\tilde{\eta}^*_j=\nu\rho'(|w_j|;\nu)\sgn(w_j)$ for $w_j\neq0$ and $\tilde{\eta}^*_j \in [-\nu\rho'(0^+),\nu\rho'(0^+)]$ for $w_j=0$. By \eqref{eq:M_0}, we know  $0= n^{-1}\sum_{i=1}^{n} g_{i,j}(\bthetazero)/\{1+\bfw ^\T \bfg_{i,\mathcal{M}_{\bthetazero}^*}(\bthetazero)\} - \tilde{\eta}^*_j$ holds for any $j\in\mathcal{M}_{\bthetazero}$. For any $j \in \mathcal{M}_{\bthetazero}^*\backslash \mathcal{M}_{\bthetazero}$, since $\max_{i\in[n]} |\bfw^\T \bfg_{i,\mathcal{M}_{\bthetazero}^*}(\bthetazero)|=\max_{i\in[n]} |\tilde\blambda^\T_{0} \bfg_{i}(\bthetazero)|=o_{\rm p}(1)$, it holds that
		\begin{align}\label{eq:rjlemma2}
			\frac{1}{n}\sum_{i=1}^n\frac{g_{i,j}(\bthetazero)}{1+\bfw^\T \bfg_{i,\mathcal{M}_{\bthetazero}^*}(\bthetazero)}=\bar g_j(\bthetazero)+R_j
		\end{align}
		with
		\begin{align}
			|R_j|^2=\bigg |\frac{1}{n} \sum_{i=1}^{n} \frac{\bfw^\T \bfg_{i,\mathcal{M}_{\bthetazero}^*}(\bthetazero) g_{i,j}(\bthetazero)}{1+\bfw^\T \bfg_{i,\mathcal{M}_{\bthetazero}^*}(\bthetazero)} \bigg |^2 
			& \leq\max_{j \in [r]} \bigg \{\frac{1}{n} \sum_{i=1}^{n} |\bfw^\T \bfg_{i,\mathcal{M}_{\bthetazero}^*}(\bthetazero)| |g_{i,j}(\bthetazero)|\bigg\}^2 \cdot\{1+o_{\rm p}(1)\} \notag\\
			& \leq\bfw^\T \widehat{\bfV}_{\mathcal{M}_{\bthetazero}^*} (\bthetazero) \bfw \cdot \max_{j \in [r]} \mathbb{E}_n\{ |g_{i,j}(\bthetazero)|^2\} \cdot\{1+o_{\rm p}(1)\} \,.\label{eq:rjbd}
		\end{align}
		Due to $|\bfw|_2 = |\tilde{\blambda}_{0,\mathcal{M}_{\bthetazero}}|_2 = O_{\rm p}(\ell_n^{1/2} \alpha_n)$, by Conditions \ref{A.ee}(a) and \ref{A.ee}(b), $\max_{j\in[r]}|R_j|=O_{\rm p}(|\bfw|_2)=O_{\rm p}(\ell_n^{1/2} \alpha_n)$. Notice that $C_* \nu\rho'(0^+) \leq |\bar{g}_j(\bthetazero)| < c \nu\rho'(0^+)$ for any $j\in\mathcal{M}_{\bthetazero}^*\backslash \mathcal{M}_{\bthetazero}$, and $\ell_n^{1/2} \alpha_n=o(\nu)$. Then 
		\[
		\max_{j\in\mathcal{M}_{\bthetazero}^*\backslash \mathcal{M}_{\bthetazero}}\bigg|\frac{1}{n}\sum_{i=1}^n\frac{g_{i,j}(\bthetazero)}{1+\bfw^\T \bfg_{i,\mathcal{M}_{\bthetazero}^*}(\bthetazero)}\bigg|\leq \nu\rho'(0^+)
		\]
		w.p.a.1, which implies \eqref{eq:1.order.con.1} holds. Thus $\tilde{\blambda}_{0}=\arg\max_{\blambda\in\Lambda_{0}^*}f_n(\blambda;\bthetazero)$ w.p.a.1.
		
		Secondly, define $\tilde\Lambda_{0}=\{\blambda\in \R^r: |\blambda_{\mathcal{M}_{\bthetazero}^*}-\tilde\blambda_{0,\mathcal{M}_{\bthetazero}^*}|_2 \leq O(\ell_n^{1/2} \alpha_n), |\blambda_{\mathcal{M}^{*,\c}_{\bthetazero}}|_1\leq O(\ell_n \alpha_n) \}$. We will show $\tilde{\blambda}_{0}=\arg \max_{\blambda\in \tilde\Lambda_{0}} f_n(\blambda;\bthetazero)$ w.p.a.1.  Recall $\max_{i\in[n],j\in[r]} |g_{i,j} (\bthetazero)| = O_{\rm p}(n^{1/\gamma})$ and $|\tilde\blambda_{0}|_2=O_{\rm p}(\ell_n^{1/2} \alpha_n)$. Since $\ell_n\alpha_n=o(n^{-1/\gamma})$, we have 
		\begin{align*}
			\sup_{i\in[n],\blambda \in \tilde\Lambda_{0}}|\blambda^\T\bfg_{i}(\bthetazero)|
			\leq&\,\sup_{i\in[n],\blambda \in \tilde\Lambda_{0}}|\blambda^\T_{\mathcal{M}_{\bthetazero}^*}\bfg_{i,\mathcal{M}_{\bthetazero}^*}(\bthetazero)|+\sup_{i\in[n],\blambda \in \tilde\Lambda_{0}}|\blambda^\T_{\mathcal{M}^{*,\c}_{\bthetazero}}\bfg_{i,\mathcal{M}^{*,\c}_{\bthetazero}}(\bthetazero)| \\
			\leq&\,\sup_{i\in[n],\blambda \in \tilde\Lambda_{0}} |\blambda _{\mathcal{M}_{\bthetazero}^*}|_2|\bfg_{i,\mathcal{M}_{\bthetazero}^*}(\bthetazero)|_2 + \sup_{i\in[n],\blambda \in \tilde\Lambda_{0}} \max_{j \in [r]}|g_{i,j}(\bthetazero)| |\blambda^\T_{\mathcal{M}^{*,\c}_{\bthetazero}}|_{1} =o_{\rm p}(1) \,.
		\end{align*}
		For any $\blambda \in \tilde\Lambda_{0}$, denote by $\mathring\blambda=(\blambda^\T_{\mathcal{M}_{\bthetazero}^*},\bzero^\T)^\T$ the projection of $\blambda=(\blambda^\T_{\mathcal{M}_{\bthetazero}^*},\blambda^\T_{\mathcal{M}^{*,\c}_{\bthetazero}})^\T$ onto $\Lambda_{0}^*$. Write $\blambda=(\lambda_1,\ldots,\lambda_r)^\T$. By the Taylor expansion, it holds that
		\[
		f_n(\blambda;\bthetazero)-f_n(\mathring\blambda;\bthetazero)=\frac{1}{n}\sum_{i=1}^n\frac{\bfg_{i}(\bthetazero)^\T (\blambda-\mathring\blambda)}{1+\blambda_*^\T\bfg_{i}(\bthetazero)}-\sum_{j\in\mathcal{M}^{*,\c}_{\bthetazero}}P_{\nu}(|\lambda_j|)\,,
		\]
		where $\blambda_*$ is on the jointing line between $\blambda$ and $\mathring\blambda$. Let $\check{\blambda}_{0}=\arg \max_{\blambda\in \tilde\Lambda_{0}} f_n(\blambda;\bthetazero)$. Due to $\tilde\blambda_{0} \in {\rm int} (\tilde\Lambda_{0})$, then $f_n(\tilde{\blambda}_0;\btheta_0)\leq f_n(\check{\blambda}_0;\btheta_0)$.  For any $\blambda\in\tilde\Lambda_{0}$, due to $\sum_{j\in\mathcal{M}^{*,\c}_{\bthetazero}}P_{\nu}(|\lambda_j|) \geq \nu\rho'(0^+) |\blambda_{\mathcal{M}^{*,\c}_{\bthetazero}}|_1$ and 
		\begin{align*}
			\bigg|\frac{1}{n}\sum_{i=1}^n\frac{\bfg_{i}(\bthetazero)^\T (\blambda-\mathring\blambda)}{1+\blambda_*^\T\bfg_{i}(\bthetazero)}\bigg|
			=&\,\bigg|\blambda_{\mathcal{M}^{*,\c}_{\bthetazero}}^\T\bar\bfg_{\mathcal{M}^{*,\c}_{\bthetazero}}(\bthetazero)-\frac{1}{n}\sum_{i=1}^n \frac{\blambda_*^\T\bfg_{i}(\bthetazero)\bfg_{i,\mathcal{M}^{*,\c}_{\bthetazero}}(\bthetazero)^\T\blambda_{\mathcal{M}^{*,\c}_{\bthetazero}}}{1+\blambda_*^\T\bfg_{i}(\bthetazero)} \bigg|\\
			\leq&\, |\bar\bfg_{\mathcal{M}^{*,\c}_{\bthetazero}}(\bthetazero)|_\infty|\blambda_{\mathcal{M}^{*,\c}_{\bthetazero}}|_1+  \frac{1}{n}\sum_{i=1}^n\sum_{j=1}^r\sum_{k\in\mathcal{M}^{*,\c}_{\bthetazero}}| \lambda_{*,j}g_{i,j}(\bthetazero)\lambda_kg_{i,k}(\bthetazero) |\{1+o_{\rm p}(1)\} \\
			\leq&\, C_*\nu\rho'(0^+) |\blambda_{\mathcal{M}^{*,\c}_{\bthetazero}}|_1+|\blambda_{\mathcal{M}^{*,\c}_{\bthetazero}}|_1|\blambda_*|_1 \{1+o_{\rm p}(1)\}\max_{j \in [r]}\mathbb{E}_n\{|g_{i,j}(\bthetazero)|^2\} \\
			\leq&\, \{C_*\nu\rho'(0^+)  + O_{\rm p}(\ell_n\alpha_n)\}|\blambda_{\mathcal{M}^{*,\c}_{\bthetazero}}|_1\,,
		\end{align*}
		then  
		$
		f_n(\blambda;\bthetazero)-f_n(\mathring\blambda;\bthetazero)  \leq  \{-(1-C_*)\nu\rho'(0^+)+ O_{\rm p}(\ell_n\alpha_n)\}|\blambda_{\mathcal{M}^{*,\c}_{\bthetazero}}|_1$ for any $\blambda\in\tilde{\Lambda}_0$, where the term $O_{\rm p}(\ell_n\alpha_n)$ holds uniformly over $\blambda\in\tilde{\Lambda}_0$. Since $\ell_n \alpha_n=o(\nu)$, we have $\check{\blambda}_{0,\mathcal{M}_{\btheta_0}^{*,\c}}=\bzero$ w.p.a.1, which implies $\check{\blambda}_0\in{\rm int}(\Lambda_0^*)$ w.p.a.1.  
		Recall $\tilde \blambda_{0}=\arg\max_{\blambda\in\Lambda_{0}^*}f_n(\blambda;\bthetazero)$ w.p.a.1. Then $f_n(\tilde{\blambda}_0;\btheta_0)\geq f_n(\check{\blambda}_0;\btheta_0)$ w.p.a.1. Therefore, $f_n(\tilde{\blambda}_0;\btheta_0)= f_n(\check{\blambda}_0;\btheta_0)$ w.p.a.1. By the concavity of $f_n(\blambda;\bthetazero)$ w.r.t $\blambda$, we have $\check{\blambda}_0=\tilde\blambda_{0}$ w.p.a.1, which indicates that $\tilde{\blambda}_0$ is a local maximizer for $f_n(\blambda;\btheta_0)$ w.p.a.1. Then $\hat{\blambda}(\bthetazero)=\tilde\blambda_{0}$ and ${\rm supp}\{\hat{\blambda}(\bthetazero)\}\subset \mathcal{M}_{\bthetazero}$ w.p.a.1. 
		\hfill $\Box$

		\subsubsection{Case 2: $ \mathcal{M}_{\bthetazero} = \emptyset$}
		In this case, we will show $\bzero \in \mathbb{R}^r$ is a local maximizer for $f_n(\blambda;\btheta_0)$ w.p.a.1. Due to the concavity of $f_n(\blambda;\btheta_0)$ w.r.t $\blambda$, we then have $\hat{\blambda}(\bthetazero) = \bzero$ w.p.a.1, which implies ${\rm supp}\{\hat{\blambda}(\bthetazero)\} \subset \mathcal{M}_{\bthetazero}$ w.p.a.1. Let $\breve{\blambda}_0 = \arg\max_{\blambda \in \breve{\Lambda}_0} f_n(\blambda;\btheta_0)$, where $\breve{\Lambda}_0 =\{ \blambda\in\mathbb{R}^r: |\lambda_{j_0}|\leq (\log n)^{-1}n^{-1/\gamma}, \blambda_{[r]\setminus \{j_0\}}=\bzero \}$ with $j_0 = \arg\max_{j\in[r]}\mathbb{E}\{g^2_{i,j}(\bthetazero)\}$. It follows from Condition \ref{A.ee}(a) that $\max_{i\in[n]}|g_{i,j_0} (\bthetazero)| = O_{\rm p}(n^{1/\gamma})$. Hence, $\max_{i\in[n]} |\breve{\blambda}_0^\T \bfg_{i} (\bthetazero)|_2 = O_{\rm p}\{(\log n)^{-1}\} = o_{\rm p}(1)$. Write $\breve{\blambda}_0 =(\breve{\lambda}_{0,1},\ldots,\breve{\lambda}_{0,r})^{\T}$. By the Taylor expansion, we have 
		\begin{align*}
			0 = f_n(\bzero;\bthetazero) 
			\leq f_n(\breve{\blambda}_{0};\bthetazero)
			=&~\breve{\blambda}^\T_{0} \bar{\bfg}(\bthetazero)- \frac{1}{2n} \sum_{i=1}^{n} \frac{\breve{\blambda}^\T_{0} \bfg_{i} (\bthetazero)^{\otimes2} \breve{\blambda}_{0}}{\{1+C\breve{\blambda}^\T_{0} \bfg_{i} (\bthetazero)\}^2} - \sum_{j=1}^{r} P_{\nu}(|\breve{\lambda}_{0,j}|) \\
			\leq&~|\breve{\lambda}_{0,j_0}||\bar{g}_{j_0}(\btheta_0)|-2^{-1} |\breve{\lambda}_{0,j_0}|^2 \mathbb{E}_n\{g_{i,j_0}^2(\btheta_0)\}\{1+o_{\rm p}(1)\} 
		\end{align*}
		for some $C\in(0,1)$. Notice that $|\mathbb{E}_n\{g_{i,j_0}^2(\btheta_0)\} - \mathbb{E}\{g_{i,j_0}^2(\btheta_0)\}|= O_{\rm p}(n^{-1/2})$. By Condition \ref{A.ee}(b), we have $\mathbb{E}_n\{g_{i,j_0}^2(\btheta_0)\} \geq \mathbb{E}\{g_{i,j_0}^2(\btheta_0)\} - o_{\rm p}(1)\geq 2 K_3/3$ w.p.a.1 for $K_3$ specified in Condition \ref{A.ee}(b). Thus $0 \leq |\breve{\lambda}_{0,j_0}||\bar{g}_{j_0}(\btheta_0)| - 4^{-1} K_3 |\breve{\lambda}_{0,j_0}|^2$ w.p.a.1. Since $|\bar{g}_{j_0}(\btheta_0)| = O_{\rm p}(n^{-1/2})$, then $|\breve{\lambda}_{0,j_0}| = O_{\rm p}(n^{-1/2}) = o_{\rm p}\{(\log n)^{-1}n^{-1/\gamma}\}$. It then holds w.p.a.1 that 
		\begin{align} \label{eq:j_0}
			0 = \frac{1}{n} \sum_{i=1}^n \frac{g_{i,j_0}(\btheta_0)}{1+\breve{\lambda}_{0,j_0} g_{i,j_0}(\btheta_0)} - \breve{\eta}_{j_0} \,,
		\end{align}
		where $\breve{\eta}_{j_0} = \nu\rho'(|\breve\lambda_{j_0}|;\nu)\sgn(\breve{\lambda}_{j_0})$ if $\breve\lambda_{j_0}\neq0$ and $\breve{\eta}_{j_0} \in[-\nu\rho'(0^+), \nu\rho'(0^+)]$ if $\breve\lambda_{j_0}=0$. Due to $|\breve{\lambda}_{0,j_0} g_{i,j_0}(\btheta_0)| = o_{\rm p}(1)$, then 
		\begin{align*}
			\frac{1}{n} \sum_{i=1}^n \frac{g_{i,j_0}(\btheta_0)}{1+\breve{\lambda}_{0,j_0} g_{i,j_0}(\btheta_0)} = \bar{g}_{j_0}(\btheta_0) + R_{j_0}
		\end{align*}
		with
		\begin{align*}
			|R_{j_0}| = \bigg| \frac{1}{n} \sum_{i=1}^n \frac{\breve{\lambda}_{0,j_0} g^2_{i,j_0}(\btheta_0)}{1+\breve{\lambda}_{0,j_0} g_{i,j_0}(\btheta_0)} \bigg| \leq |\breve{\lambda}_{0,j_0}| \mathbb{E}_n\{g_{i,j_0}^2(\btheta_0)\}\{1+o_{\rm p}(1)\} \,.
		\end{align*}
		By Conditions \ref{A.ee}(a), we have $|R_{j_0}| = O_{\rm p}(|\breve{\lambda}_{0,j_0}|)
		= O_{\rm p}(n^{-1/2})$. Together with $|\bar{g}_{j_0}(\btheta_0)| = O_{\rm p}(n^{-1/2})$, \eqref{eq:j_0} leads to
		$|\breve{\eta}_{j_0}| = O_{\rm p}(n^{-1/2}) = o_{\rm p}(\nu)$. Then $\breve\lambda_{j_0}=0$ w.p.a.1, which implies $\breve{\blambda}_0 = \bzero$ w.p.a.1. In the sequel, we will show that $\breve{\blambda}_0$ is a local maximizer for $f_n(\blambda;\bthetazero)$ w.p.a.1.
		
		Firstly, define $\breve{\Lambda}_0^* = \{\blambda\in \R^r:|\blambda_{\mathcal{H}}|_2 \leq \varepsilon,\blambda_{\mathcal{H}^\c}=\bzero\}$ for some sufficiently small constant $\varepsilon>0$, where $j_0 \in \mathcal{H} \subset [r]$ with $1 < |\mathcal{H}|\leq \ell_n$. For $\breve{\blambda}_0$ defined before, we will prove $\breve{\blambda}_0 =\arg \max_{\blambda\in \breve{\Lambda}_0^*} f_n(\blambda;\bthetazero)$ w.p.a.1. Since $\breve{\blambda}_0 \in \breve{\Lambda}_0$ and $j_0 \in \mathcal{H}$, we know $\breve{\blambda}_0 \in \breve{\Lambda}_0^*$ for sufficiently large $n$. Restricted on $\blambda \in \breve{\Lambda}_0^*$, by the concavity of $f_n(\blambda;\bthetazero)$ w.r.t $\blambda$, it suffices to show that $\breve{\bfw} = \breve{\blambda}_{0,\mathcal{H}}=: (\breve{w}_{1},\ldots,\breve{w}_{|\mathcal{H}|})^\T \in \R^{|\mathcal{H}|}$ satisfies the equation
		\begin{align} \label{eq:1.order.con.1_case2}
			\bzero= \frac{1}{n} \sum_{i=1}^{n} \frac{\bfg_{i,\mathcal{H}}(\bthetazero)}{1+\breve{\bfw}^\T \bfg_{i,\mathcal{H}}(\bthetazero)} - \breve{\bseta}^*
		\end{align}
		w.p.a.1, where $\breve{\bseta}^*=(\breve\eta^*_1,\ldots,\breve\eta^*_{|\mathcal{H}|})^\T$ with $\breve{\eta}^*_j=\nu\rho'(|\breve{w}_j|;\nu)\sgn(\breve{w}_j)$ for $\breve{w}_j\neq0$ and $\breve{\eta}^*_j \in [-\nu\rho'(0^+),\nu\rho'(0^+)]$ for $\breve{w}_j=0$. Recall $j_0\in\mathcal{H}$. Without loss of generality, we assume $j_0$ is the first component in $\mathcal{H}$. By \eqref{eq:j_0}, we know  $0= n^{-1}\sum_{i=1}^{n} g_{i,j_0}(\bthetazero)/\{1+\breve{\bfw}^\T \bfg_{i,\mathcal{H}}(\bthetazero)\} - \breve{\eta}^*_{1}$ holds. Since $\breve{\blambda}_0 = \bzero$ w.p.a.1, then $\breve{\bfw}= \bzero$ w.p.a.1, which implies
		it holds w.p.a.1 that
		\begin{align*} 
			\frac{1}{n}\sum_{i=1}^n\frac{g_{i,j}(\bthetazero)}{1+\breve{\bfw}^\T \bfg_{i,\mathcal{H}}(\bthetazero)}=\bar g_j(\bthetazero) 
		\end{align*}
		for any $j \in \mathcal{H} \backslash \{j_0\}$. By the moderate deviation of self-normalized sums \citep{JingShaoWang2003}, 
		$
		|\bar{\bfg}(\bthetazero)|_{\infty}
		=O_{\rm p}(\alpha_n)$. Due to $\alpha_n = o(\nu)$, then
		\[
		\max_{j \in \mathcal{H} \backslash \{j_0\}}\bigg|\frac{1}{n}\sum_{i=1}^n\frac{g_{i,j}(\bthetazero)}{1+\breve{\bfw}^\T \bfg_{i,\mathcal{H}}(\bthetazero)}\bigg|\leq \nu\rho'(0^+)
		\]
		w.p.a.1, which implies \eqref{eq:1.order.con.1_case2} holds. Thus $\breve{\blambda}_0 =\arg \max_{\blambda\in \breve{\Lambda}_0^*} f_n(\blambda;\bthetazero)$ w.p.a.1.
		
		Secondly, define $\bar{\Lambda}_{0}=\{\blambda\in \R^r: |\blambda_{\mathcal{H}}-\breve\blambda_{0,\mathcal{H}}|_2 \leq O(\ell_n^{1/2} \alpha_n), |\blambda_{\mathcal{H}^\c}|_1\leq O(\ell_n \alpha_n) \}$. We will show $\breve{\blambda}_{0}=\arg \max_{\blambda\in \bar{\Lambda}_{0}} f_n(\blambda;\bthetazero)$ w.p.a.1. Recall $\max_{i\in[n],j\in[r]} |g_{i,j} (\bthetazero)| = O_{\rm p}(n^{1/\gamma})$ and $|\breve{\blambda}_{0}|_2=|\breve{\lambda}_{0,j_0}| = O_{\rm p}(n^{-1/2})$. Since $\ell_n\alpha_n=o(n^{-1/\gamma})$, we have
		\begin{align*}
			\sup_{i\in[n],\blambda \in \bar{\Lambda}_{0}}|\blambda^\T\bfg_{i}(\bthetazero)|
			\leq&\,\sup_{i\in[n],\blambda \in \bar{\Lambda}_{0}}|\blambda^\T_{\mathcal{H}}\bfg_{i,\mathcal{H}}(\bthetazero)|+\sup_{i\in[n],\blambda \in \bar{\Lambda}_{0}}|\blambda^\T_{\mathcal{H}^\c}\bfg_{i,\mathcal{H}^\c}(\bthetazero)| \\
			\leq&\,\sup_{i\in[n],\blambda \in \bar{\Lambda}_{0}} |\blambda _{\mathcal{H}}|_2|\bfg_{i,\mathcal{H}}(\bthetazero)|_2 + \sup_{i\in[n],\blambda \in \bar{\Lambda}_{0}} \max_{j \in [r]}|g_{i,j}(\bthetazero)| |\blambda^\T_{\mathcal{H}^\c}|_{1} =o_{\rm p}(1) \,.
		\end{align*}
		For any $\blambda \in \bar{\Lambda}_{0}$, denote by $\mathring\blambda=(\blambda^\T_{\mathcal{H}},\bzero^\T)^\T$ the projection of $\blambda=(\blambda^\T_{\mathcal{H}},\blambda^\T_{\mathcal{H}^\c})^\T$ onto $\breve{\Lambda}_0^*$. Write $\blambda=(\lambda_1,\ldots,\lambda_r)^\T$. By the Taylor expansion, it holds that
		\[
		f_n(\blambda;\bthetazero)-f_n(\mathring\blambda;\bthetazero)=\frac{1}{n}\sum_{i=1}^n\frac{\bfg_{i}(\bthetazero)^\T (\blambda-\mathring\blambda)}{1+\blambda_*^\T\bfg_{i}(\bthetazero)}-\sum_{j\in\mathcal{H}^\c}P_{\nu}(|\lambda_j|)\,,
		\]
		where $\blambda_*$ is on the jointing line between $\blambda$ and $\mathring\blambda$. Let $\check{\blambda}_{0}=\arg \max_{\blambda\in \bar\Lambda_{0}} f_n(\blambda;\bthetazero)$. Due to $\breve\blambda_{0} \in {\rm int} (\bar\Lambda_{0})$, then $f_n(\breve{\blambda}_0;\btheta_0)\leq f_n(\check{\blambda}_0;\btheta_0)$. For any $\blambda \in \bar\Lambda_{0}$, due to $\sum_{j\in\mathcal{H}^\c}P_{\nu}(|\lambda_j|) \geq \nu\rho'(0^+) |\blambda_{\mathcal{H}^\c}|_1$ and
		\begin{align*}
			\bigg|\frac{1}{n}\sum_{i=1}^n\frac{\bfg_{i}(\bthetazero)^\T (\blambda-\mathring\blambda)}{1+\blambda_*^\T\bfg_{i}(\bthetazero)}\bigg|
			=&\,\bigg|\blambda_{\mathcal{H}^\c}^\T\bar\bfg_{\mathcal{H}^\c}(\bthetazero)-\frac{1}{n}\sum_{i=1}^n \frac{\blambda_*^\T\bfg_{i}(\bthetazero)\bfg_{i,\mathcal{H}^\c}(\bthetazero)^\T\blambda_{\mathcal{H}^\c}}{1+\blambda_*^\T\bfg_{i}(\bthetazero)} \bigg|\\
			\leq&\, |\bar\bfg_{\mathcal{H}^\c}(\bthetazero)|_\infty|\blambda_{\mathcal{H}^\c}|_1+  \frac{1}{n}\sum_{i=1}^n\sum_{j=1}^r\sum_{k\in\mathcal{H}^\c}| \lambda_{*,j}g_{i,j}(\bthetazero)\lambda_kg_{i,k}(\bthetazero) |\{1+o_{\rm p}(1)\} \\
			\leq&\, O_{\rm p}(\alpha_n) \cdot  |\blambda_{\mathcal{H}^\c}|_1+|\blambda_{\mathcal{H}^\c}|_1|\blambda_*|_1 \{1+o_{\rm p}(1)\}\max_{j \in [r]}\mathbb{E}_n\{|g_{i,j}(\bthetazero)|^2\} \\
			\leq&\, O_{\rm p}(\ell_n\alpha_n) \cdot |\blambda_{\mathcal{H}^\c}|_1 \,,
		\end{align*}
		then  
		$
		f_n(\blambda;\bthetazero)-f_n(\mathring\blambda;\bthetazero)  \leq  \{-\nu\rho'(0^+)+ O_{\rm p}(\ell_n\alpha_n)\}|\blambda_{\mathcal{H}^\c}|_1$ for any $\blambda\in\bar{\Lambda}_0$, where the term $O_{\rm p}(\ell_n\alpha_n)$ holds uniformly over $\blambda\in\bar{\Lambda}_0$. Since $\ell_n \alpha_n=o(\nu)$, we have $\check{\blambda}_{0,\mathcal{H}^\c}=\bzero$ w.p.a.1, which implies $\check{\blambda}_0\in{\rm int}(\breve{\Lambda}_0^*)$ w.p.a.1. Recall $\breve{\blambda}_0 =\arg \max_{\blambda\in \breve{\Lambda}_0^*} f_n(\blambda;\bthetazero)$ w.p.a.1. Then $f_n(\breve{\blambda}_0;\btheta_0)\geq f_n(\check{\blambda}_0;\btheta_0)$ w.p.a.1. 
		Therefore, $f_n(\breve{\blambda}_0;\btheta_0) = f_n(\check{\blambda}_0;\btheta_0)$ w.p.a.1. 
		By the concavity of $f_n(\blambda;\bthetazero)$ w.r.t $\blambda$, we have $\check{\blambda}_0=\breve{\blambda}_0$ w.p.a.1, which indicates that $\bzero$ is a local maximizer for $f_n(\blambda;\btheta_0)$ w.p.a.1.  
		\hfill $\Box$

		\subsection{Proof of Lemma \ref{l.lam.thetan.hat}}\label{sec:pflemma3}
		
		Same as the proof of Lemma \ref{l.lam.theta0}, we only need to show that there exists a local maximizer satisfying the results stated in the lemma. Recall $\mathcal{M}_{\hbthetan}^*=\{j\in[r]: |\bar{g}_j(\hbthetan)| \geq  C_*\nu \rho'(0^{+}) \}$ for some $C_* \in (0,1)$, and $\mathbb{P}(\max _{\btheta \in \bTheta:\,|\btheta-\bthetazero|_2 \leq c_n}|\mathcal{M}_{\btheta}^*|\leq\ell_n)\rightarrow1$ for some $c_n\rightarrow 0$ satisfying $\nu c_n^{-1} \rightarrow 0$. For $\tilde{c} \in (C_*,1)$ given in Condition \ref{A.g_subgra.1}(a), write $\mathcal{M}_{\hbthetan}:=\mathcal{M}_{\hbthetan}(\tilde{c})=\{ j\in[r]:|\bar{g}_j(\hbthetan)| \geq \tilde{c}\nu\rho'(0^{+}) \}$. By Proposition \ref{pro.cons}, $|\hat{\btheta}_n-\btheta_0|_\infty=O_{\rm p}(\nu)$. Notice that $p$ is fixed. Then $|\hat{\btheta}_n-\btheta_0|_2=O_{\rm p}(\nu)$ which implies $\ell_n \geq |\mathcal{M}_{\hbthetan}^*| \geq |\mathcal{M}_{\hbthetan}|$ w.p.a.1. Restricted on $\mathcal{M}_{\hbthetan}$, we select $\delta_n$ satisfying $\delta_n=o(\ell_n^{-1/2} n^{-1/\gamma})$ and $\ell_n^{1/2} \alpha_n=o(\delta_n) $, which can be guaranteed by $\ell_n\alpha_n=o(n^{-1/\gamma})$. Let $\Lambda_n=\{\blambda\in \R^{r}:|\blambda_{\mathcal{M}_{\hbthetan}}|_2\leq \delta_n, \blambda_{\mathcal{M}^\c_{\hbthetan}}=\bzero \}$ and $\tilde{\blambda}_n=(\tilde{\lambda}_{n,1},\ldots,\tilde{\lambda}_{n,r})^\T=\arg \max _{\blambda \in \Lambda_n} f_n(\blambda; \hbthetan)$. %By Condition \ref{A.ee}(a), $\max_{i\in[n], j\in[r]} |g_{i,j} (\hbthetan)| = O_{\rm p}(n^{1/\gamma})$, which implies that $\max_{i\in[n]} |\bfg_{i,\mathcal{M}_{\hbthetan}} (\hbthetan)|_2 = O_{\rm p}(\ell_n^{1/2} n^{1/\gamma})$. Then $\max_{i\in[n]} |\bar \blambda^\T_n \bfg_{i}(\hbthetan)|=o_{\rm p}(1)$. Write $\bar \blambda_n$. 
		By the Taylor expansion, we have
		\begin{align} \label{eq:lamd_n}
			0=f_n(\bzero;\hbthetan) \leq f_n(\tilde{\blambda}_n;\hbthetan)
			=\tilde{\blambda}^\T_n \bar{\bfg}(\hbthetan)- \frac{1}{2n} \sum_{i=1}^{n} \frac{\tilde{\blambda}^\T_n \bfg_{i} (\hbthetan)^{\otimes2} \tilde{\blambda}_n}{\{1+C\tilde{\blambda}^\T_n \bfg_{i} (\hbthetan)\}^2} - \sum_{j=1}^{r} P_{\nu}(|\tilde{\lambda}_{n,j}|)
		\end{align}
		for some $C \in (0,1)$. By Proposition \ref{pro.cons}, Lemma \ref{l.V.hat} and Condition \ref{A.ee}(b), if $\log r=o(n^{1/3})$, $\ell_n\nu^2=o(1)$ and $\ell_n\alpha_n=o[\min\{\nu,n^{-1/\gamma}\}]$, we have $\lambda_{\min}\{\widehat{\bfV}_{\mathcal{M}_{\hbthetan}} (\hbthetan)\}$ is uniformly bounded away from zero w.p.a.1. Therefore, it holds w.p.a.1 that
		$$
		0 \leq \tilde{\blambda}_{n,\mathcal{M}_{\hbthetan}}^\T \big[\bar{\bfg}_{\mathcal{M}_{\hbthetan}}(\hbthetan)-\nu \rho'(0^+) \sgn\{\bar{\bfg}_{\mathcal{M}_{\hbthetan}}(\hbthetan)\}\big] - 4^{-1}K_3 |\tilde{\blambda}_{n,\mathcal{M}_{\hbthetan}}|_2^2$$ with $K_3$ specified in Condition \ref{A.ee}(b), 
		which implies $|\tilde\blambda_{n,\mathcal{M}_{\hbthetan}}|_2 \leq 4K_3^{-1} |\bar{\bfg}_{\mathcal{M}_{\hbthetan}}(\hbthetan)-\nu \rho'(0^+) \sgn\{\bar{\bfg}_{\mathcal{M}_{\hbthetan}}(\hbthetan)\}|_2$ w.p.a.1. %In the sequel, we will show
		%\begin{align} \label{eq:g_bar}
		%|\bar{\bfg}_{\mathcal{M}_{\hbthetan}}(\hbthetan)-\nu \rho'(0^+) \sgn\{\bar{\bfg}_{\mathcal{M}_{\hbthetan}}(\hbthetan)\}|_2=O_{\rm p}(\ell_n^{1/2} \alpha_n) \,,
		%\end{align}
		
		%Our first step is to specify the order of $|\bar{\bfg}_{\mathcal{M}_{\hbthetan}}(\hbthetan)-\nu \rho'(0^+) \sgn\{\bar{\bfg}_{\mathcal{M}_{\hbthetan}}(\hbthetan)\}|_2$. 
		Select $\blambda_n^* \in \R^r$ satisfying $\blambda^*_{n,\mathcal{M}^\c_{\hbthetan}}=\bzero$ and $$\blambda^*_{n,\mathcal{M}_{\hbthetan}}=\frac{\delta_n [\bar{\bfg}_{\mathcal{M}_{\hbthetan}}(\hbthetan)-\nu \rho'(0^+) \sgn\{\bar{\bfg}_{\mathcal{M}_{\hbthetan}}(\hbthetan)\}]}{|\bar{\bfg}_{\mathcal{M}_{\hbthetan}}(\hbthetan)-\nu \rho'(0^+) \sgn\{\bar{\bfg}_{\mathcal{M}_{\hbthetan}}(\hbthetan)\}|_2}\,.$$ Then $\blambda_n^* \in \Lambda_n$. As shown in the proof of Proposition \ref{pro.cons}, $\max_{\blambda \in \wLambadn(\bthetazero)}f_n(\blambda;\bthetazero) = O_{\rm p}(\ell_n \alpha_n^2)=o_{\rm p}(\delta_n^2)$, which implies $\max_{\blambda \in \wLambadn(\hbthetan)}f_n(\blambda;\hbthetan)=o_{\rm p}(\delta_n^2)$. Write $\blambda_n^*=(\lambda_{n,1}^*,\ldots,\lambda_{n,r}^*)^\T$. Notice that $\Lambda_n\subset\hat{\Lambda}_n(\hat{\btheta}_n)$ w.p.a.1. By the Taylor expansion, it holds w.p.a.1 that
		\begin{align*}
			o_{\rm p}(\delta_n^2)=\max_{\blambda \in \wLambadn(\hbthetan)} f_n(\blambda;\hbthetan)
			&\geq \frac{1}{n} \sum_{i=1}^{n} \log \{ 1+\blambda_{n,\mathcal{M}_{\hbthetan}}^{*,\T} \bfg_{i,\mathcal{M}_{\hbthetan}} (\hbthetan) \} -\sum_{j \in \mathcal{M}_{\hbthetan}} P_{\nu}(|\lambda_{n,j}^*|) \\
			&=\blambda_{n,\mathcal{M}_{\hbthetan}}^{*,\T} \bar{\bfg}_{\mathcal{M}_{\hbthetan}}(\hbthetan)- \frac{1}{2n} \sum_{i=1}^{n} \frac{\blambda_{n,\mathcal{M}_{\hbthetan}}^{*,\T} \bfg_{i,\mathcal{M}_{\hbthetan}}(\hbthetan)^{\otimes2} \blambda_{n,\mathcal{M}_{\hbthetan}}^*}{\{1+\bar C \blambda_{n,\mathcal{M}_{\hbthetan}}^{*,\T} \bfg_{i,\mathcal{M}_{\hbthetan}}(\hbthetan)\}^2} \\
			&~~~ - \sum_{j \in \mathcal{M}_{\hbthetan}} \nu \rho'(0^+) |\lambda_{n,j}^*| - \frac{1}{2} \sum_{j \in \mathcal{M}_{\hbthetan}} \nu\rho''(c_j |\lambda_{n,j}^*|;\nu)|\lambda_{n,j}^*|^2 \\
			&\geq \blambda_{n,\mathcal{M}_{\hbthetan}}^{*,\T} \big\{\bar{\bfg}_{\mathcal{M}_{\hbthetan}}(\hbthetan)-\nu \rho'(0^+) \sgn(\blambda_{n,\mathcal{M}_{\hbthetan}}^*)\big\} - C\delta_n^2 \{1+o_{\rm p}(1)\}
		\end{align*}
		for some $\bar C, c_j \in (0,1)$, where the last inequality follows from the condition that $P_{\nu}(\cdot)$ has bounded second-order derivative around $0$. For any $j \in \mathcal{M}_{\hbthetan}$, we have $\sgn(\lambda_{n,j}^*)=\sgn\{\bar g_j(\hbthetan)\}$ if $|\bar {g}_j(\hbthetan)| > \nu \rho'(0^{+})$, and $\bar g_j(\hbthetan)-\nu \rho'(0^+) \sgn\{\bar g_j(\hbthetan)\} = 0 = \lambda_{n,j}^*$ if $|\bar {g}_j(\hbthetan)| = \nu \rho'(0^{+})$. Thus, $$\lambda_{n,j}^* \big\{\bar g_j(\hbthetan) - \nu \rho'(0^+) \sgn(\lambda_{n,j}^*)\big\} = \lambda_{n,j}^* \big[\bar g_j(\hbthetan) - \nu \rho'(0^+) \sgn\{\bar g_j(\hbthetan)\}\big]$$ for any $j \in \mathcal{M}_{\hbthetan}$ with $|\bar {g}_j(\hbthetan)| \geq \nu \rho'(0^{+})$. By Condition \ref{A.g_subgra.1}(a), $\{j \in [r]: \tilde{c} \nu\rho'(0^{+}) \leq |\bar {g}_j(\hbthetan)| < \nu\rho'(0^{+})\} = \emptyset$ w.p.a.1. Recall $\mathcal{M}_{\hbthetan}=\{j \in [r]: |\bar {g}_j(\hbthetan)| \geq \tilde{c} \nu \rho'(0^{+})\}$. We then have w.p.a.1 that
		\begin{align*}
			o_{\rm p}(\delta_n^2)\geq&\, \blambda_{n,\mathcal{M}_{\hbthetan}}^{*,\T} \big\{\bar{\bfg}_{\mathcal{M}_{\hbthetan}}(\hbthetan)-\nu \rho'(0^+) \sgn\{\bar{\bfg}_{\mathcal{M}_{\hbthetan}}(\hbthetan)\big\} - C\delta_n^2 \{1+o_{\rm p}(1)\}	\\
			=&\,\delta_n |\bar{\bfg}_{\mathcal{M}_{\hbthetan}}(\hbthetan)-\nu \rho'(0^+) \sgn\{\bar{\bfg}_{\mathcal{M}_{\hbthetan}}(\hbthetan)\}|_2 - C \delta_n^2 \{1+o_{\rm p}(1)\}\,.
		\end{align*}
		Thus, %we can obtain
		$
		|\bar{\bfg}_{\mathcal{M}_{\hbthetan}}(\hbthetan)-\nu \rho'(0^+) \sgn\{\bar{\bfg}_{\mathcal{M}_{\hbthetan}}(\hbthetan)\}|_2 =O_{\rm p}(\delta_n)
		$.
		For any $\epsilon_n \rightarrow 0$, select $\blambda_n^{**}$ such that $\blambda_{n,\mathcal{M}_{\hbthetan}}^{**}=\epsilon_n [\bar{\bfg}_{\mathcal{M}_{\hbthetan}}(\hbthetan)-\nu \rho'(0^+) \sgn\{\bar{\bfg}_{\mathcal{M}_{\hbthetan}}(\hbthetan)\}]$ and $\blambda^{**}_{n,\mathcal{M}^\c_{\hbthetan}}=\bzero$. Then $|\blambda_n^{**}|_2=o_{\rm p}(\delta_n)$. Due to $f_n(\blambda_n^{**};\hbthetan) \leq \max_{\blambda \in \wLambadn(\hbthetan)}f_n(\blambda;\hbthetan) \leq \max_{\blambda \in \wLambadn(\bthetazero)}f_n(\blambda;\bthetazero)=O_{\rm p}(\ell_n \alpha_n^2)$, using the same arguments given above, we have 
		\begin{align*}
			&\epsilon_n |\bar{\bfg}_{\mathcal{M}_{\hbthetan}}(\hbthetan)-\nu \rho'(0^+) \sgn\{\bar{\bfg}_{\mathcal{M}_{\hbthetan}}(\hbthetan)\}|^2_2 \\
			&~~~~~~~~~~~~~~~- C\epsilon_n^2|\bar{\bfg}_{\mathcal{M}_{\hbthetan}}(\hbthetan)-\nu \rho'(0^+) \sgn\{\bar{\bfg}_{\mathcal{M}_{\hbthetan}}(\hbthetan)\}|^2_2\{1+o_{\rm p}(1)\}=O_{\rm p}(\ell_n \alpha_n^2)\,.
		\end{align*}
		Hence, $\epsilon_n |\bar{\bfg}_{\mathcal{M}_{\hbthetan}}(\hbthetan)-\nu \rho'(0^+) \sgn\{\bar{\bfg}_{\mathcal{M}_{\hbthetan}}(\hbthetan)\}|^2_2=O_{\rm p}(\ell_n \alpha_n^2)$. Since we can select arbitrary slow $\epsilon_n \rightarrow 0$, it holds that
		\begin{align} \label{eq:g_bar}
			|\bar{\bfg}_{\mathcal{M}_{\hbthetan}}(\hbthetan)-\nu \rho'(0^+) \sgn\{\bar{\bfg}_{\mathcal{M}_{\hbthetan}}(\hbthetan)\}|_2=O_{\rm p}(\ell_n^{1/2} \alpha_n) \,,
		\end{align}
		which implies 
		$
		|\tilde{\blambda}_n|_2 = |\tilde{\blambda}_{n,\mathcal{M}_{\hbthetan}}|_2=O_{\rm p}(\ell_n^{1/2} \alpha_n) = o_{\rm p}(\delta_n) 
		$. %Our second step is to show that $\sgn (\tilde{\lambda}_{n,j})=\sgn\{\bar{g}_j(\hbthetan)\}$ w.p.a.1 for any $j \in \mathcal{M}_{\hbthetan}$ with $\tilde{\lambda}_{n,j} \neq 0$. 
		Write $\tilde{\blambda}_{n,\mathcal{M}_{\hbthetan}}=(\tilde{\lambda}_{n,1},\ldots,\tilde{\lambda}_{n,|\mathcal{M}_{\hbthetan}|})^\T$. %{\color{blue}Recall $\tilde{\blambda}_n\in{\rm int}(\Lambda_n)$.??} 
		We have w.p.a.1 that
		$$\bzero= \frac{1}{n}\sum_{i=1}^{n} \frac{\bfg_{i,\mathcal{M}_{\hbthetan}}(\hbthetan)}{1+\tilde{\blambda}_{n,\mathcal{M}_{\hbthetan}}^\T \bfg_{i,\mathcal{M}_{\hbthetan}}(\hbthetan)} - \tilde{\bseta}\,,$$ 
		where $\tilde{\bseta}=(\tilde\eta_1,\ldots,\tilde\eta_{|\mathcal{M}_{\hbthetan}|})^\T$ with $\tilde\eta_j=\nu\rho'(|\tilde\lambda_{n,j}|;\nu)\sgn(\tilde{\lambda}_{n,j})$ for $\tilde\lambda_{n,j}\neq0$ and $\tilde\eta_j\in[-\nu\rho'(0^+),\nu\rho'(0^+)]$ for $\tilde\lambda_{n,j}=0$. Identical to \eqref{eq:rjlemma2}, we have $\tilde{\bseta}=\bar{\bfg}_{\mathcal{M}_{\hbthetan}}(\hbthetan)+\bfR$ for some $|\mathcal{M}_{\hbthetan}|$-dimensional vector $\bfR$. Applying the same arguments for deriving the rate of $R_j$ in \eqref{eq:rjbd}, it holds that  $|\bfR|_{\infty}=O_{\rm p}(\ell_n^{1/2} \alpha_n)$. Since $\ell_n\alpha_n=o(\nu) $, we then have $\sgn (\tilde{\lambda}_{n,j})=\sgn\{\bar{g}_j(\hbthetan)\}$ for any $j \in \mathcal{M}_{\hbthetan}$ with $\tilde{\lambda}_{n,j} \neq 0$ w.p.a.1. Using the arguments in Section \ref{sec:pflemma2} for showing $\tilde\blambda_{0}$ is a local maximizer for $f_n(\blambda;\bthetazero)$ w.p.a.1, we can prove $\tilde\blambda_n$ is a local maximizer for $f_n(\blambda;\hbthetan)$ w.p.a.1, which implies $\hat{\blambda}(\hbthetan)=\tilde{\blambda}_n$ w.p.a.1. We then have Lemma \ref{l.lam.thetan.hat}.    \hfill $\Box$
		
		\subsection{Proof of Lemma \ref{l.envlope}}\label{sec:pflemma4}
		
		Recall $\hat\blambda(\btheta)=\arg\max_{\blambda \in \wLambadn(\btheta)} f_n(\blambda;\btheta)$. Then $\hbthetan$ and $\hat{\blambda}(\hbthetan) =(\hat\lambda_1,\ldots,\hat\lambda_r)^\T$ satisfy %$\nabla_{\blambda} f_n\{\hat{\blambda}(\hbthetan); \hbthetan\}=\bzero$, that is,
		\begin{align} \label{eq:fn_lambda_hat}
			\bzero=\frac{1}{n} \sum_{i=1}^{n} \frac{\bfg_{i}(\hbthetan)}{1+\hat{\blambda}(\hbthetan)^\T \bfg_{i} ( \hbthetan)}-\hat{\bseta} \,,
		\end{align}
		where $\hat{\bseta}=(\hat{\eta}_{1},\ldots,\hat{\eta}_{r})^\T$ with $\hat{\eta}_j=\nu \rho' (|\hat{\lambda}_j|;\nu) \sgn(\hat{\lambda}_j)$ for $\hat{\lambda}_j\neq0$ and $\hat{\eta}_j \in [-\nu \rho'(0^{+}),\nu \rho'(0^{+})]$ for $\hat{\lambda}_j=0$. Recall $\mathcal{R}_n={\rm supp}\{\hat{\blambda}(\hbthetan)\}$. Restricted on $\mathcal{R}_n$, for any $\btheta \in \bTheta$ and $\bzeta=(\zeta_{1},\ldots,\zeta_{|\Rn|})^\T \in \R^{|\Rn|}$ with each $\zeta_j \neq 0$, define
		$$
		\bfm(\bzeta,\btheta)=\frac{1}{n} \sum _{i=1}^{n} \frac{\bfg_{i,\Rn}(\btheta)}{1+\bzeta^\T \bfg_{i,\Rn}(\btheta)}-\bfw\,,$$ 
		where $\bfw=(w_{1},\ldots,w_{|\Rn|})^\T$ with $w_j=\nu \rho' (|\zeta_j|;\nu) \sgn(\zeta_j)$. From \eqref{eq:fn_lambda_hat}, we know $\hat{\blambda}_\Rn(\hbthetan)$ and $\hbthetan$ satisfy $\bfm\{\hat{\blambda}_\Rn(\hbthetan), \hbthetan\}= \bzero$. By the implicit function theorem [Theorem 9.28 of \cite{Rudin1976}], for all $\btheta$ in a small neighborhood of $\hbthetan$, denoted by $\mathcal{U}(\hbthetan)$, there exists a $\bzeta(\btheta)$ such that $\bfm\{\bzeta(\btheta),\btheta\}=\bzero$, $\bzeta(\hbthetan)=\hat{\blambda}_\Rn(\hbthetan)$ and $\bzeta(\btheta)$ is continuously differentiable in $\btheta\in \mathcal{U}(\hbthetan)$. By Condition \ref{A.g_subgra.1}(b), the event $\E=\{\max_{j \in \mathcal{R}_n^\c} |\hat{\eta}_j| < \nu\rho'(0^+)\}$ holds w.p.a.1. Restricted on $\E$, let $\varsigma_n=\nu\rho'(0^+) - \max_{j \in \mathcal{R}_n^\c} |\hat{\eta}_j|$ and define $\bTheta_*=\{\btheta \in \mathcal{U}(\hbthetan):\, |\btheta-\hbthetan|_1 \leq o[\min\{\varsigma_n, \chi_n\}], |\bzeta(\btheta)-\bzeta(\hbthetan)|_1 \leq o [\min\{\varsigma_n, \ell_n^{1/2} \alpha_n\}]\}$ for some $\chi_n>0$. Since all the components of $\bzeta(\hbthetan)$ are nonzero and $\bzeta(\btheta)$ is continuously differentiable in $\hbthetan$, we can select sufficiently small $\chi_n$ such that all the components of $\bzeta(\btheta)$ are nonzero for any $\btheta \in \bTheta_*$. For any $\btheta \in \bTheta_*$, let $\tilde\blambda(\btheta)= \{\tilde\lambda_1(\btheta), \ldots, \tilde\lambda_r(\btheta)\}^\T \in \R^r$ satisfy $\tilde\blambda_\Rn(\btheta)=\bzeta(\btheta)$ and $\tilde\blambda_{\mathcal{R}_n^\c}(\btheta)=\bzero$. Since $\bfm\{\bzeta(\btheta),\btheta\}=\bzero$, $\tilde\blambda_\Rn(\btheta)=\bzeta(\btheta)$ and $\tilde\blambda_{\mathcal{R}_n^\c}(\btheta)=\bzero$ for any $\btheta \in \bTheta_*$, then
		$$
		0 = \frac{1}{n}\sum _{i=1}^{n} \frac{g_{i,j}(\btheta)}{1+\tilde\blambda(\btheta)^\T \bfg_{i}(\btheta)} - \nu \rho' \{|\tilde{\lambda}_j(\btheta)|;\nu\} \sgn\{\tilde{\lambda}_j(\btheta)\}$$ 
		for any $j \in \mathcal{R}_n$. For any $\btheta \in \bTheta_*$ and $j \in \mathcal{R}_n^\c$, by the Taylor expansion, we have
		\begin{align} \label{eq:gtheta}
			&\,\frac{1}{n} \sum _{i=1}^{n} \frac{g_{i,j}(\btheta)}{1+\tilde\blambda(\btheta)^\T \bfg_{i}(\btheta)} \notag \\	
			=&\, \frac{1}{n} \sum _{i=1}^{n} \frac{g_{i,j}(\hbthetan)}{1+\tilde\blambda(\btheta)^\T \bfg_{i}(\hbthetan)} +
			\bigg[ \frac{1}{n} \sum_{i=1}^{n} \frac{\{\nabla_{\btheta} g_{i,j}(\check{\btheta})\}^\T}{1+\tilde\blambda(\btheta)^\T \bfg_{i}(\check{\btheta})}- \frac{1}{n} \sum_{i=1}^{n} \frac{g_{i,j}(\check{\btheta}) \tilde\blambda(\btheta)^\T \nabla_{\btheta}\bfg_{i}(\check{\btheta})}{\{1+\tilde\blambda(\btheta)^\T \bfg_{i}(\check{\btheta})\}^2} \bigg] (\btheta-\hbthetan) \notag \\
			=&\, \frac{1}{n} \sum _{i=1}^{n} \frac{g_{i,j}(\hbthetan)}{1+\hat{\blambda}(\hbthetan)^\T \bfg_{i}(\hbthetan)} -\bigg[\frac{1}{n} \sum _{i=1}^{n} \frac{g_{i,j}(\hbthetan) \bfg_{i}(\hbthetan)^\T }{\{1+\check{\blambda}^\T \bfg_{i}(\hbthetan)\}^2} \bigg] \{\tilde\blambda(\btheta)-\hat{\blambda}(\hbthetan)\} \\
			&\,+\bigg[ \frac{1}{n} \sum_{i=1}^{n} \frac{\{\nabla_{\btheta} g_{i,j}(\check{\btheta})\}^\T}{1+\tilde\blambda(\btheta)^\T\bfg_{i}(\check{\btheta})} - \frac{1}{n} \sum_{i=1}^{n} \frac{g_{i,j}(\check{\btheta})\tilde\blambda(\btheta)^\T \nabla_{\btheta}\bfg_{i}(\check{\btheta})}{\{1+\tilde\blambda(\btheta)^\T \bfg_{i}(\check{\btheta})\}^2} \bigg] (\btheta-\hbthetan) \notag \,,
		\end{align}
		where $\check{\btheta}$ is lying on the jointing line between $\btheta$ and $\hbthetan$, and $\check{\blambda}$ is lying on the jointing line between $\tilde\blambda(\btheta)$ and $\hat{\blambda}(\hbthetan)$. By Lemma \ref{l.lam.thetan.hat}, $|\hat{\blambda}(\hbthetan)|_2=O_{\rm p}(\ell_n^{1/2} \alpha_n)$ and $|\mathcal{R}_n|\leq \ell_n$ w.p.a.1. Then $|\tilde{\blambda}(\btheta)|_2 = |\bzeta(\btheta)|_2 \leq |\bzeta(\hbthetan)|_2 + |\bzeta(\btheta)-\bzeta(\hbthetan)|_2 = O_{\rm p}(\ell_n^{1/2} \alpha_n)$, which implies $|\check{\blambda}|_2 = O_{\rm p}(\ell_n^{1/2} \alpha_n)$. Together with Condition \ref{A.ee}(a) and $\ell_n \alpha_n=o(n^{-1/\gamma})$, it yields that $\max_{i \in [n]} \{ |\tilde{\blambda}(\btheta)^\T \bfg_{i}(\check{\btheta})| + |\check{\blambda}^\T \bfg_{i}(\hat{\btheta}_n)|\} = o_{\rm p}(1)$. By Conditions \ref{A.ee}(a) and \ref{A.ee}(c), we have 
		\begin{align*}
			\max_{j \in \mathcal{R}_n^\c}\bigg| \frac{1}{n} \sum _{i=1}^{n} \frac{g_{i,j}(\hbthetan) \bfg_{i}(\hbthetan) }{\{1+\check{\blambda}^\T \bfg_{i}(\hbthetan)\}^2} \bigg |_\infty = O_{\rm p}(1) =\max_{j \in \mathcal{R}_n^\c}\bigg| \frac{1}{n} \sum_{i=1}^{n} \frac{\nabla_{\btheta} g_{i,j}(\check{\btheta})}{1+\tilde\blambda(\btheta)^\T\bfg_{i}(\check{\btheta})} \bigg|_\infty \,.
		\end{align*}
		It follows from the Cauchy-Schwarz inequality that
		\begin{align*}
			&\,\max_{j \in \mathcal{R}_n^\c}\bigg|\frac{1}{n} \sum_{i=1}^{n} \frac{g_{i,j}(\check{\btheta}) \tilde\blambda(\btheta)^\T \nabla_{\btheta}\bfg_{i}(\check{\btheta})}{\{1+\tilde{\blambda}(\btheta)^\T \bfg_{i}(\check{\btheta})\}^2} \bigg|_\infty  \\
			\leq&\, \{1+o_{\rm p}(1)\} \cdot\max_{j \in \mathcal{R}_n^\c,k\in[p]}\bigg\{\frac{1}{n} \sum_{i=1}^{n}  \sum_{l \in \Rn} |g_{i,j}(\check{\btheta})||\tilde\lambda_l(\btheta)| \bigg|\frac{\partial g_{i,l}(\check{\btheta})}{\partial \theta_{k}} \bigg| \bigg\} \\
			\leq&\, \{1+o_{\rm p}(1)\} \cdot\max_{j \in \mathcal{R}_n^\c,k\in[p]}\bigg[\mathbb{E}_n^{1/2}\{g_{i,j}^2(\check{\btheta})\} \sum_{l \in \Rn}|\tilde\lambda_l(\btheta)|  \mathbb{E}_n^{1/2}\bigg\{\bigg|\frac{\partial g_{i,l}(\check{\btheta})}{\partial \theta_{k}} \bigg|^2\bigg\} \bigg] \\	
			\leq &\, |\tilde\blambda(\btheta)|_1 \cdot O_{\rm p}(1) \leq \ell_n^{1/2}|\tilde\blambda(\btheta)|_2\cdot O_{{\rm p}}(1)= O_{\rm p}(\ell_n\alpha_n) = o_{\rm p}(1) \,.
		\end{align*}
		By \eqref{eq:gtheta}, for any $\btheta\in\bTheta_*$, we know
		\begin{align*}
			\frac{1}{n} \sum _{i=1}^{n} \frac{g_{i,j}(\btheta)}{1+\tilde\blambda(\btheta)^\T \bfg_{i}(\btheta)} 	
			%=&\, \frac{1}{n} \sum _{i=1}^{n} \frac{g_{i,j}(\hbthetan)}{1+\hat{\blambda}(\hbthetan)^\T \bfg_{i}(\hbthetan)} + O_{\rm p}(1) \cdot |\tilde\blambda(\btheta)-\hat{\blambda}(\hbthetan)|_1 + O_{\rm p}(1) \cdot |\btheta-\hbthetan|_1 \\
			=&\, \frac{1}{n} \sum _{i=1}^{n} \frac{g_{i,j}(\hbthetan)}{1+\hat{\blambda}(\hbthetan)^\T \bfg_{i}(\hbthetan)} + O_{\rm p}(1) \cdot |\bzeta(\btheta)-\bzeta(\hbthetan)|_1 + O_{\rm p}(1) \cdot |\btheta-\hbthetan|_1 \\
			=&\, \hat{\eta}_j + \varsigma_n \cdot o_{\rm p}(1)
		\end{align*}
		holds uniformly over $j \in \mathcal{R}_n^\c$. Due to $\P(\E) \rightarrow 1$, we have 
		$$
		\max_{j \in \mathcal{R}_n^\c}\bigg| \frac{1}{n}\sum _{i=1}^{n} \frac{g_{i,j}(\btheta)}{1+\tilde\blambda(\btheta)^\T \bfg_{i}(\btheta)} \bigg| \leq \nu\rho'(0^+)$$ w.p.a.1. 
		Therefore, $\tilde\blambda(\btheta)$ and $\btheta$ satisfy the score equation $\nabla_{\blambda} f_n\{\tilde\blambda(\btheta); \btheta\}=\bzero$ for any $\btheta \in \bTheta_*$ w.p.a.1. By the concavity of $f_n(\blambda;\btheta)$ w.r.t $\blambda$, we have $\tilde\blambda(\btheta)=\arg\max_{\blambda \in \wLambadn(\btheta)} f_n(\blambda;\btheta)=\hat{\blambda}(\btheta)$ for any $\btheta \in \bTheta_*$ w.p.a.1. 
		%By the concavity of $f_n(\blambda;\btheta)$ w.r.t $\blambda$, we have $\hat{\blambda}(\btheta)=\bar\blambda(\btheta)$ for any $\btheta \in \bTheta_*$ w.p.a.1. 
		Hence, $\hat{\blambda}(\btheta)$ is continuously differentiable at $\hbthetan$ and $ [\nabla_{\btheta} \hat{\blambda}(\hbthetan)]_{\mathcal{R}_n^\c,[p]}=\bzero$ w.p.a.1. %We complete the proof of this lemma.
		\hfill $\Box$
		
		\subsection{Proof of Lemma \ref{l.ee}}\label{sec:pflemma5}
		
		The proof is almost identical to that of Lemma 2 in \cite{Changetal2018}. From Lemma \ref{l.lam.thetan.hat}, we have $|\hat{\blambda}|_2=O_{\rm p}(\ell_n^{1/2} \alpha_n)$. Recall $p$ is fixed in our current setting. We only need to replace the convergence rate of $|\hat{\blambda}|_2$ in the proof of Lemma 2 in \cite{Changetal2018} by $O_{\rm p}(\ell_n^{1/2} \alpha_n)$ and also set $(s,\omega_n)$ there as $(p,1)$ and all the arguments still hold. \hfill $\Box$

		\subsection{Proof of Lemma \ref{l.Gamma.hat}}\label{sec:pflemma6}
		
		The proof is almost identical to that of Lemma 3 in \cite{Changetal2018}. Since $p$ is fixed, we only need to replace $\{\omega_n,\varpi_n,b_n^{1/(2\beta)},s\}$ in the proof of Lemma 3 in \cite{Changetal2018} by $(1,1,\nu,p)$ and all the arguments still hold. \hfill $\Box$

		\subsection{Proof of Lemma \ref{l.norm}}\label{sec:pflemma7}
		
		%The proof is almost identical to that of Lemma 4 in \cite{ChangTangWu2018}. Recall $p$ is fixed in our current setting. We only need to replace $\{\varrho_n,\omega_n,\varpi_n,b_n^{1/(2\beta)},s\}$ in the proof of Lemma 4 in \cite{ChangTangWu2018} by $(1,1,1,\nu,p)$. \hfill $\Box$

		Recall $\bGamma_{\mF}(\bthetazero)=\mE \{\nabla_{\btheta} \bfg_{i,\mF}(\bthetazero) \}$ and $\bfV_{\mF} (\bthetazero)= \mE \{ \bfg_{i,\mF} (\bthetazero)^{\otimes2} \}$. For any $\bft \in \R^p$ with $|\bft|_2=1$, let $Z_{i,\mF}=\bft^\T \bfH_{\mF}^{-1/2} \bGamma_{\mF}(\bthetazero)^\T \bfV^{-1}_{\mF} (\bthetazero) \bfg_{i,\mF} (\bthetazero)$ with $\bfH_{\mF}=\{\bGamma_{\mF}(\bthetazero)^\T \bfV^{-1/2}_{\mF} (\bthetazero)\}^{\otimes2}$. Write $G_{\mF}= \mE_n(Z_{i,\mF})$ and $\hat{G}_{\mF}= \bft^\T \widehat{\bfH}_{\mF}^{-1/2} \widehat{\bGamma}_{\mF}(\hbthetan)^\T \widehat{\bfV}^{-1}_{\mF} (\hbthetan) \bar{\bfg}_{\mF} (\bthetazero)$. It follows from the Berry-Esseen inequality that %we have
		$$
		\sup_{u \in \R} |\P(n^{1/2} G_{\mF} \leq u) -\Phi(u)| \leq Cn^{-1/2} \mE(|Z_{i,\mF}|^3)$$ for some universal constant $C>0$. By the Cauchy-Schwarz inequality, %we have
		\begin{align*}
			|Z_{i,\mF}|^2
			\leq&\,|\bfV^{-1/2}_{\mF}(\bthetazero) \bGamma_{\mF}(\bthetazero) \bfH_{\mF}^{-1/2} \bft|_2^2\cdot |\bfV^{-1/2}_{\mF}(\bthetazero) \bfg_{i,\mF}(\bthetazero)|_2^2\\
			\leq&\, \lambda_{\min} ^{-1}\{\bfV_{\mF}(\bthetazero)\} |\bfg_{i,\mF}(\bthetazero)|_2^2\leq K_3^{-1}|\bfg_{i,\mF}(\bthetazero)|_2^2
		\end{align*}
		for $K_3$ given in Condition \ref{A.ee}(b). 
		By the Jensen's inequality, Condition \ref{A.ee}(a) yields 
		$
		\mE\{|\bfg_{i,\mF}(\bthetazero)|_2^3\}\leq K_2^{3/\gamma}\ell_n^{3/2}$ for $K_2$ and $\gamma$ given in Condition \ref{A.ee}(a), which implies 
		$
		\mE(|Z_{i,\mF}|^3) \leq K_3^{-3/2}\mE\{|\bfg_{i,\mF}(\bthetazero)|_2^3\} \leq K_2^{3/\gamma}K_3^{-3/2}\ell_n^{3/2}
		$. If $\ell_n=o(n^{1/3})$, we have
		$$ 
		\sup_{\mathcal{F} \in \mathscr{F}} \sup_{u \in \R}|\P(n^{1/2} G_{\mF} \leq u) -\Phi(u)| \rightarrow 0$$ 
		as $n\rightarrow\infty$.  
		By Conditions \ref{A.ee}(b) and \ref{A.ee2} and Lemmas \ref{l.V.hat} and \ref{l.Gamma.hat}, it holds that  
		$\sup_{\mathcal{F} \in \mathscr{F}}|n^{1/2}(\hat{G}_{\mF}-G_{\mF})|= O_{\rm p}\{\ell_n \nu (\log r)^{1/2}\} + O_{\rm p}\{\ell_n^{3/2} \alpha_n(\log r)^{1/2}\}$. 
		For any constant $\delta>0$, due to  
		$
		\P (n^{1/2} \hat{G}_{\mF} \leq u ) - \Phi(u)
		\leq \P (n^{1/2} G_{\mF} \leq u+\delta) + \P\{|n^{1/2}(\hat{G}_{\mF}-G_{\mF})|
		\geq \delta \}  - \Phi(u)$	
		and 
		$
		\P(n^{1/2} \hat{G}_{\mF} \leq u) - \Phi(u)
		\geq \P (n^{1/2} G_{\mF} \leq u-\delta) - \P\{|n^{1/2}(\hat{G}_{\mF}-G_{\mF})|
		\geq \delta \}  - \Phi(u)$, it holds that
		\begin{align*}
			\sup_{\mathcal{F} \in \mathscr{F}} \sup_{u \in \R} |\P(n^{1/2}\hat{G}_{\mF} \leq u) - \Phi(u)|\leq&\, \sup_{\mathcal{F} \in \mathscr{F}} \sup_{u \in \R}|\P(n^{1/2} G_{\mF} \leq u ) - \Phi(u)|+ \sup_{\mathcal{F} \in \mathscr{F}} \P\{|n^{1/2}(\hat{G}_{\mF}-G_{\mF})|
			\geq \delta \} \\
			&+ \sup_{u \in \R} |\Phi(u+\delta)-\Phi(u-\delta)|\,.
		\end{align*}
		Notice that $\sup_{\mathcal{F} \in \mathscr{F}}|n^{1/2}(\hat{G}_{\mF}-G_{\mF})|=o_{\rm p}(1)$ and $\sup_{u \in \R} |\Phi(u+\delta)-\Phi(u-\delta)| \leq(2\pi^{-1})^{1/2}\delta$. Then it holds that $\limsup_{n\rightarrow\infty}\sup_{\mathcal{F} \in \mathscr{F}} \sup_{u \in \R} |\P(n^{1/2}\hat{G}_{\mF} \leq u) - \Phi(u)|\leq (2\pi^{-1})^{1/2}\delta$. Due to the arbitrary selection of $\delta>0$, we have $\sup_{\mathcal{F} \in \mathscr{F}} \sup_{u \in \R} |\P(n^{1/2}\hat{G}_{\mF} \leq u) - \Phi(u)|\rightarrow0$ as $n\rightarrow\infty$. \hfill $\Box$
		
		\subsection{Proof of Lemma \ref{l.lam.theta}} \label{sec:pflemma8}
		
		Recall $\mathcal{C}_1=\{\btheta\in\bTheta:|\btheta-\hat{\btheta}_n|_2\leq \alpha_n\}$. For any $\btheta \in \mathcal{C}_1$, same as the proof of Lemma \ref{l.lam.theta0}, we only need to show that there exists a local maximizer satisfying the results stated in the lemma. For $\tilde{c}$ specified in Condition \ref{A.g_subgra.1}(a), we select $c \in (\tilde{c},1)$. We select $\delta_n$ satisfying $\delta_n=o(\ell_n^{-1/2} n^{-1/\gamma})$ and $\ell_n^{1/2} \alpha_n=o(\delta_n) $, which can be guaranteed by $\ell_n\alpha_n=o(n^{-1/\gamma})$. For each $\btheta \in \mathcal{C}_1$, define $\Lambda_{\btheta}=\{\blambda\in \R^{r}:\,|\blambda_{\mathcal{M}_{\btheta}(c)}|_2\leq \delta_n, \blambda_{\mathcal{M}_{\btheta}(c)^\c}=\bzero \}$ and $\tilde \blambda_{\btheta}=\arg \max _{\blambda \in \Lambda_{\btheta}} f_n(\blambda; \btheta)$. %Recall $\alpha_n=n^{-1/2}(\log r)^{1/2}=o(\nu)$. 
		Similar to \eqref{eq:lamd_n} and the arguments below \eqref{eq:lamd_n},
		if $\log r =o(n^{1/3})$, $\ell_n \nu^2=o(1)$ and $\ell_n\alpha_n=o[\min\{\nu,n^{-1/\gamma}\}]$, we have
		$|\tilde \blambda_{\btheta,\mathcal{M}_{\btheta}(c)}|_2 \leq 4K_3^{-1} |\bar{\bfg}_{\mathcal{M}_{\btheta}(c)}(\btheta)-\nu \rho'(0^+) \sgn\{\bar{\bfg}_{\mathcal{M}_{\btheta}(c)}(\btheta)\}|_2$ for any $\btheta \in \mathcal{C}_1$ w.p.a.1, where $K_3$ is specified in Condition \ref{A.ee}(b).
		%Our first step is to derive the order of $|\bar{\bfg}_{\mathcal{M}_{\btheta}(c)}(\btheta)-\nu \rho'(0^+) \sgn\{\bar{\bfg}_{\mathcal{M}_{\btheta}(c)}(\btheta)\}|_2$. 
		Notice that
		\begin{align*}
			|\bar{\bfg}_{\mathcal{M}_{\btheta}(c)}(\btheta)-\nu \rho'(0^+) \sgn\{\bar{\bfg}_{\mathcal{M}_{\btheta}(c)}(\btheta)\}|_2
			\leq &\, \underbrace{|\bar{\bfg}_{\mathcal{M}_{\btheta}(c) \bigcap \mathcal{M}_{\hbthetan}(\tilde{c})}(\btheta)-\nu \rho'(0^+) \sgn\{\bar{\bfg}_{\mathcal{M}_{\btheta}(c) \bigcap \mathcal{M}_{\hbthetan}(\tilde{c})}(\btheta)\}|_2}_{T_{1,\btheta}} \\
			&+\underbrace{|\bar{\bfg}_{\mathcal{M}_{\btheta}(c) \bigcap \mathcal{M}_{\hbthetan}(\tilde{c})^\c}(\btheta)-\nu \rho'(0^+) \sgn\{\bar{\bfg}_{\mathcal{M}_{\btheta}(c) \bigcap \mathcal{M}_{\hbthetan}(\tilde{c})^\c}(\btheta)\}|_2}_{T_{2,\btheta}} \,.
		\end{align*}
		By the Taylor expansion and Condition \ref{A.ee}(c), we have $\sup_{\btheta \in \mathcal{C}_1} |\bar{\bfg}(\btheta) - \bar{\bfg}(\hbthetan)|_\infty = O_{\rm p}(\alpha_n)$, which implies $\sup_{\btheta \in \mathcal{C}_1}|\bar{\bfg}_{\mathcal{M}_{\hbthetan}(\tilde{c})}(\btheta) - \bar{\bfg}_{\mathcal{M}_{\hbthetan}(\tilde{c})}(\hbthetan)|_2 = O_{\rm p}(\ell_n^{1/2} \alpha_n)$. Due to $\alpha_n=o(\nu)$ and $|\bar{g}_j(\hbthetan)| \geq \tilde{c} \nu \rho'(0^+)$ for any $j \in \mathcal{M}_{\hbthetan}(\tilde{c})$, we then have $ \sgn\{\bar{\bfg}_{\mathcal{M}_{\hbthetan}(\tilde{c})}(\btheta)\}=\sgn\{\bar{\bfg}_{\mathcal{M}_{\hbthetan}(\tilde{c})}(\hbthetan)\} $ for any $\btheta \in \mathcal{C}_1$ w.p.a.1. By the triangle inequality and \eqref{eq:g_bar}, we have w.p.a.1 that
		\begin{align*}
			\sup_{\btheta\in\mathcal{C}_1}T_{1,\btheta} 
			\leq&\,\sup_{\btheta\in\mathcal{C}_1}|\bar{\bfg}_{\mathcal{M}_{\hbthetan}(\tilde{c})}(\hbthetan)-\nu \rho'(0^+) \sgn\{\bar{\bfg}_{\mathcal{M}_{\hbthetan}(\tilde{c})}(\hbthetan)\}|_2 \\
			&+  \sup_{\btheta\in\mathcal{C}_1}|\bar{\bfg}_{\mathcal{M}_{\hbthetan}(\tilde{c})}(\btheta) - \bar{\bfg}_{\mathcal{M}_{\hbthetan}(\tilde{c})}(\hbthetan)|_2 \\
			=&~O_{\rm p}(\ell_n^{1/2} \alpha_n)\,.
		\end{align*}
		For any $j \in \mathcal{M}_{\btheta}(c) \bigcap \mathcal{M}_{\hbthetan}(\tilde{c})^\c$, we have  $|\bar{g}_j(\btheta)| \geq c \nu \rho'(0^+)$ and $|\bar{g}_j(\hbthetan)| < \tilde{c} \nu \rho'(0^+)$. Due to $c \in (\tilde{c},1)$ and $\sup_{\btheta \in \mathcal{C}_1}|\bar{\bfg}(\btheta) - \bar{\bfg}(\hbthetan)|_\infty = o_{\rm p}(\nu)$, then $ \mathcal{M}_{\btheta}(c) \bigcap \mathcal{M}_{\hbthetan}(\tilde{c})^\c = \emptyset$ for any $\btheta \in \mathcal{C}_1$ w.p.a.1, which implies $T_{2,\btheta}=0$ for any $\btheta\in\mathcal{C}_1$ w.p.a.1. Hence, 
		\begin{align} \label{eq:g_bar_theta}
			\sup_{\btheta \in \mathcal{C}_1}|\bar{\bfg}_{\mathcal{M}_{\btheta}(c)}(\btheta)-\nu \rho'(0^+) \sgn\{\bar{\bfg}_{\mathcal{M}_{\btheta}(c)}(\btheta)\}|_2 = O_{\rm p}(\ell_n^{1/2} \alpha_n)  \,.
		\end{align}
		Then $\sup_{\btheta \in \mathcal{C}_1}|\tilde \blambda_{\btheta}|_2 = \sup_{\btheta \in \mathcal{C}_1}|\tilde \blambda_{\btheta,\mathcal{M}_{\btheta}(c)}|_2 = O_{\rm p}(\ell_n^{1/2} \alpha_n) =o_{\rm p}(\delta_n)$. Write $\tilde\blambda_{\btheta}= (\tilde\lambda_{\btheta,1},\ldots,\tilde\lambda_{\btheta,r})^\T$. Our next step is to show $\sgn (\tilde{\lambda}_{\btheta,j})=\sgn\{\bar{g}_j(\btheta)\}$ for any $\btheta \in \mathcal{C}_1$ and $j \in \mathcal{M}_{\btheta}(c)$ with $\tilde{\lambda}_{\btheta,j} \neq 0$ w.p.a.1. Its proof is almost identical to that in Section \ref{sec:pflemma3} for proving $\sgn (\tilde{\lambda}_{n,j})=\sgn\{\bar{g}_j(\hbthetan)\}$ for any $j \in \mathcal{M}_{\hbthetan}(\tilde{c})$ with $\tilde{\lambda}_{n,j} \neq 0$ w.p.a.1. We only need to replace $\{\tilde \blambda_n,\mathcal{M}_{\hat{\btheta}_n}(\tilde{c})\}$ there by $\{\tilde \blambda_{\btheta},\mathcal{M}_{\btheta}(c)\}$ and all the arguments still hold uniformly over $\btheta \in \mathcal{C}_1$. Using the same arguments stated in the proof of Lemma \ref{l.lam.theta0} for showing $\tilde\blambda_0$ is a local maximizer for $f_n(\blambda;\btheta_0)$ w.p.a.1, we can also prove $\tilde\blambda_{\btheta}$ is a local maximizer of $f_n(\blambda; \btheta)$ for any $\btheta \in \mathcal{C}_1$ w.p.a.1, which implies $\hat \blambda(\btheta) = \tilde{\blambda}_{\btheta}$ for any $\btheta \in \mathcal{C}_1$ w.p.a.1. We then have Lemma \ref{l.lam.theta}. \hfill $\Box$
		
		\subsection{Proof of Lemma \ref{l.diff.lam.hat}} \label{sec:pflemma9}
		Recall $\hat{\blambda}(\btheta)=\arg \max _{\blambda \in \wLambadn(\btheta)} f_n(\blambda;\btheta)$ and $\mathcal{C}_1=\{\btheta\in\bTheta:|\btheta-\hat{\btheta}_n|_2\leq \alpha_n\}$. Then $\btheta$ and  $\hat{\blambda}(\btheta)=\{\hat{\lambda}_{1}(\btheta),\ldots,\hat{\lambda}_{r}(\btheta)\}^\T$ satisfy
		\begin{align} \label{eq:fn_lambda}
			\bzero=\frac{1}{n} \sum_{i=1}^{n} \frac{\bfg_{i}(\btheta)}{1+\hat{\blambda}(\btheta)^\T \bfg_{i} (\btheta)}-\hat{\bseta}(\btheta) \,,
		\end{align}
		where $\hat{\bseta}(\btheta)=\{\hat{\eta}_{1}(\btheta),\ldots, \hat{\eta}_{r}(\btheta)\}^\T$ with $\hat{\eta}_j(\btheta)=\nu \rho' \{|\hat{\lambda}_j(\btheta)|;\nu\} \sgn\{\hat{\lambda}_j(\btheta)\}$ for $\hat{\lambda}_j(\btheta) \neq 0$ and $\hat{\eta}_j(\btheta) \in [-\nu\rho'(0^{+}),\nu \rho'(0^{+})]$ for $\hat{\lambda}_j(\btheta)=0$. Recall $\mathcal{R}(\btheta)={\rm supp}\{\hat{\blambda}(\btheta)\}$. For any $\btheta \in \mathcal{C}_1$,  restricted on $\mathcal{R}(\btheta)$, define
		$$	\bfm_{\btheta}(\bzeta,\bvartheta)=\frac{1}{n}\sum _{i=1}^{n}\frac{\bfg_{i,\Rtheta}(\bvartheta)}{1+\bzeta^\T \bfg_{i,\Rtheta}(\bvartheta)}-\bfw $$
		for any $\bvartheta \in \bTheta$ and $\bzeta=\{\zeta_{1},\ldots,\zeta_{|\Rtheta|}\}^\T \in \R^{|\Rtheta|}$ with each $\zeta_j \neq 0$, where $\bfw=\{w_{1},\ldots,w_{|\Rtheta|}\}^\T$ with $w_j=\nu \rho' (|\zeta_j|;\nu) \sgn(\zeta_j)$. From \eqref{eq:fn_lambda}, we know $\hat{\blambda}_\Rtheta (\btheta) $ and $\btheta$ satisfy $\bfm_{\btheta}\{\hat{\blambda}_\Rtheta(\btheta), \btheta\}= \bzero$. By the implicit function theorem [Theorem 9.28 of \cite{Rudin1976}], for all $\bvartheta$ in a small neighborhood of $\btheta$, denoted by $\mathcal{U}(\btheta)$, there exists a $\bzeta_{\btheta}(\bvartheta)$ such that $\bfm_{\btheta}\{\bzeta_{\btheta}(\bvartheta),\bvartheta\}=\bzero$, $\bzeta_{\btheta}(\btheta)=\hat{\blambda}_\Rtheta(\btheta)$ and $\bzeta_{\btheta}(\bvartheta)$ is continuously differentiable in $\bvartheta \in \mathcal{U}(\btheta)$. By Condition \ref{A.Pro2}(a),  the event $\E = \bigcap_{\btheta \in \mathcal{C}_1} \{\max_{j \in \Rtheta^\c} |\hat{\eta}_j(\btheta)| < \nu\rho'(0^+)\}$ holds w.p.a.1.
		Restricted on $\E$, let $\varsigma_n=\nu\rho'(0^+) - \sup_{\btheta \in \mathcal{C}_1}\max_{j \in \Rtheta^\c} |\hat{\eta}_j(\btheta)|$ and define $\bTheta_*(\btheta)=\{\bvartheta \in \mathcal{U}(\btheta): |\bvartheta-\btheta|_1 \leq o[\min\{\varsigma_n,  \chi_n(\btheta)\}], |\bzeta_{\btheta}(\bvartheta)-\bzeta_{\btheta}(\btheta)|_1 \leq o [\min\{\varsigma_n, \ell_n^{1/2} \alpha_n\}] \}$ for some $\chi_n(\btheta)>0$. Since all the components of $\bzeta_{\btheta}(\btheta)$ are nonzero and $\bzeta_{\btheta}(\bvartheta)$ is continuously differentiable in $\bvartheta$, we can select sufficiently small $\chi_n(\btheta)$ such that all the components of $\bzeta_{\btheta}(\bvartheta)$ are nonzero for any $\bvartheta \in \bTheta_*(\btheta)$. For any $\bvartheta \in \bTheta_*(\btheta)$, let $\tilde\blambda_{\btheta}(\bvartheta) \in \R^r$ satisfy $\tilde\blambda_{\btheta,\Rtheta}(\bvartheta)=\bzeta_{\btheta}(\bvartheta)$ and $\tilde\blambda_{\btheta,\Rtheta^\c}(\bvartheta)=\bzero$. By Lemma \ref{l.lam.theta}, $\sup_{\btheta \in \mathcal{C}_1}|\hat{\blambda}(\btheta)|_2 =O_{\rm p}(\ell_n^{1/2} \alpha_n)$ and $\sup_{\btheta \in \mathcal{C}_1}|\mathcal{R}(\btheta)| \leq \ell_n$ w.p.a.1, which imply 
		\begin{align*}
			\sup_{\btheta \in \mathcal{C}_1}\sup_{ \bvartheta \in \bTheta_*(\btheta) }|\tilde\blambda_{\btheta}(\bvartheta)|_2 \leq \sup_{\btheta \in \mathcal{C}_1}\sup_{ \bvartheta \in \bTheta_*(\btheta) }|\bzeta_{\btheta}(\bvartheta)-\bzeta_{\btheta}(\btheta)|_2+\sup_{\btheta \in \mathcal{C}_1}|\bzeta_{\btheta}(\btheta)|_2=O_{\rm p}(\ell_n^{1/2} \alpha_n)\,.
		\end{align*}
		Using the same arguments in the proof of Lemma \ref{l.envlope} for proving that $\tilde\blambda(\btheta)$ and $\btheta$ satisfy the score equation $\nabla_{\blambda} f_n\{\tilde\blambda(\btheta); \btheta\}=\bzero$ w.p.a.1 there, we can prove $\nabla_{\blambda} f_n\{\tilde\blambda_{\btheta}(\bvartheta); \bvartheta\}=\bzero$ for any $\btheta \in \mathcal{C}_1$ and $\bvartheta \in \bTheta_*(\btheta)$ w.p.a.1. By the concavity of $f_n(\blambda;\bvartheta)$ w.r.t $\blambda$, we have $\tilde\blambda_{\btheta}(\bvartheta)=\hat{\blambda}(\bvartheta)=\arg\max_{\blambda \in \wLambadn(\bvartheta)} f_n(\blambda;\bvartheta)$ for any $\btheta \in \mathcal{C}_1$ and $\bvartheta \in \bTheta_*(\btheta)$ w.p.a.1. Recall $\tilde\blambda_{\btheta}(\bvartheta)$ is continuously differentiable in $\bvartheta\in\mathcal{U}(\btheta)$ for any $\btheta \in \mathcal{C}_1$. Hence, $\hat{\blambda}(\btheta)$ is continuously differentiable at $\btheta$ and $[\nabla_{\btheta} \hat{\blambda}(\btheta)]_{\Rtheta^\c,[p]}=\bzero$ for any $\btheta \in \mathcal{C}_1$ w.p.a.1. Write $\hat{\blambda}_\Rtheta(\btheta) = \{\tilde{\lambda}_1(\btheta),\ldots,\tilde{\lambda}_{|\Rtheta|}(\btheta)\}^\T$. Since $\bzeta_{\btheta}(\btheta)=\hat{\blambda}_\Rtheta(\btheta)$, it holds that
		\begin{align*}
			[\nabla_{\btheta} \hat{\blambda}(\btheta)]_{\Rtheta,[p]}
			&=\nabla_{\bvartheta}\bzeta_{\btheta}(\bvartheta) \big |_{\bvartheta=\btheta}
			=-\bigg\{\frac{\partial\bfm_{\btheta}(\bzeta,\bvartheta)}{\partial \bzeta}\bigg\}^{-1}\frac{\partial\bfm_{\btheta}(\bzeta,\bvartheta)}{\partial \bvartheta} \bigg |_{\bvartheta=\btheta,\,\bzeta=\bzeta_{\btheta}(\btheta)}\\
			&=\bigg(\frac{1}{n} \sum _{i=1}^{n}\frac{\bfg_{i,\Rtheta}(\btheta)^{\otimes2}}{\{1+\hat{\blambda}_\Rtheta(\btheta)^\T\bfg_{i,\Rtheta}(\btheta)\}^2} +\nu \diag[\rho''\{|\tilde\lambda_{1}(\btheta)|;\nu \},\ldots,\rho''\{|\tilde\lambda_{|\Rtheta|}(\btheta)|;\nu \}]\bigg)^{-1}\\
			&\qquad \times \bigg\{\frac{1}{n} \sum_{i=1}^{n} \frac{ [\nabla_{\btheta} \bfg_{i}(\btheta)]_{\Rtheta,[p]}}{1+\hat{\blambda}_\Rtheta(\btheta)^\T\bfg_{i,\Rtheta}(\btheta)}-\frac{1}{n} \sum_{i=1}^{n} \frac{\bfg_{i,\Rtheta}(\btheta)\hat{\blambda}_\Rtheta(\btheta)^\T [\nabla_{\btheta} \bfg_{i}(\btheta)]_{\Rtheta,[p]}}{\{1+\hat{\blambda}_\Rtheta(\btheta)^\T\bfg_{i,\Rtheta}(\btheta)\}^2} \bigg\} \,.
		\end{align*}
		We complete the proof of Lemma \ref{l.diff.lam.hat}.
		\hfill $\Box$
		
		\subsection{Proof of Lemma \ref{l.supp}}\label{sec:pflemma10}
		
		Recall $\hat\blambda(\btheta)=\arg\max_{\blambda \in \hat{\Lambda}_n(\btheta)}f_n(\blambda;\btheta )$, $\mathcal{R}_n={\rm supp}\{\hat{\blambda}(\hbthetan)\}$ and $\mathcal{C}_1=\{\btheta\in\bTheta:|\btheta-\hat{\btheta}_n|_2\leq \alpha_n\}$. By Lemma \ref{l.lam.thetan.hat}, $|\mathcal{R}_n|\leq \ell_n$ w.p.a.1. Select $\delta_n$ satisfying $\delta_n=o(\ell_n^{-1/2} n^{-1/\gamma})$ and $\ell_n^{1/2} \alpha_n=o(\delta_n) $, which can be guaranteed by $\ell_n\alpha_n=o(n^{-1/\gamma})$. For any $\btheta \in \mathcal{C}_1$, let $\tilde \blambda(\btheta)=\arg \max _{\blambda \in \check\Lambda_n} f_n(\blambda; \btheta)$, where $\check\Lambda_n=\{\blambda\in \R^{r}:|\blambda_\Rn|_2\leq \delta_n, \blambda_{\mathcal{R}_n^\c}=\bzero \}$. Similar to \eqref{eq:lamd_n} and the arguments below \eqref{eq:lamd_n},
		if $\log r =o(n^{1/3})$, $\ell_n\alpha_n=o[\min\{\nu,n^{-1/\gamma}\}]$ and $\ell_n \nu^2=o(1)$, we have $|\tilde \blambda_{\mathcal{R}_n}(\btheta)|_2 \leq 4K_3^{-1} |\bar{\bfg}_{\mathcal{R}_n}(\btheta)-\nu \rho'(0^+) \sgn\{\bar{\bfg}_{\mathcal{R}_n}(\btheta)\}|_2$ for any $\btheta \in \mathcal{C}_1$ w.p.a.1, where $K_3$ is specified in Condition \ref{A.ee}(b). By Lemma \ref{l.lam.thetan.hat}, we have $\mathcal{R}_n \subset \mathcal{M}_{\hbthetan}(\tilde{c})$ w.p.a.1, where $\tilde{c}$ is specified in Condition {\rm \ref{A.g_subgra.1}(a)}. Using the arguments for deriving \eqref{eq:g_bar_theta}, we have $$
		\sup_{\btheta \in \mathcal{C}_1}|\bar{\bfg}_\Rn(\btheta)-\nu \rho'(0^+) \sgn\{\bar{\bfg}_\Rn(\btheta)\}|_2 = O_{\rm p}(\ell_n^{1/2} \alpha_n)\,,$$ 
		which implies $\sup_{\btheta \in \mathcal{C}_1}|\tilde \blambda_\Rn(\btheta)|_2 = O_{\rm p}(\ell_n^{1/2} \alpha_n)=o_{\rm p}(\delta_n)$. Write $\tilde\blambda_{\mathcal{R}_n}(\btheta)=\{\dot\lambda_1(\btheta), \ldots,  \dot\lambda_{|\mathcal{R}_n|}(\btheta)\}^\T$. By the first-order condition, for any $\btheta \in \mathcal{C}_1$, we have 
		$$
		\bzero = \frac{1}{n}\sum_{i=1}^{n} \frac{\bfg_{i,\mathcal{R}_n}(\btheta)}{1+\tilde \blambda_{\mathcal{R}_n}(\btheta)^\T \bfg_{i,\mathcal{R}_n}(\btheta)} - \tilde\bseta(\btheta)\,,$$ 
		where $\tilde\bseta(\btheta)=\{\tilde\eta_1(\btheta), \ldots,\tilde\eta_{|\mathcal{R}_n|}(\btheta)\}^\T$ with $\tilde\eta_j(\btheta)=\nu\rho'\{|\dot\lambda_j(\btheta)|;\nu\}\sgn\{\dot\lambda_j(\btheta)\}$ for $\dot{\lambda}_j(\btheta)\neq 0$ and  $\tilde\eta_j(\btheta) \in [-\nu\rho'(0^+), \nu\rho'(0^+)]$ for $\dot{\lambda}_j(\btheta)= 0$. Using the same arguments for addressing the remainder terms in \eqref{eq:gtheta}, for any $\btheta \in \mathcal{C}_1$ and $j\in\mathcal{R}_n^\c$, it holds that  
		$$
		\frac{1}{n}\sum _{i=1}^{n} \frac{g_{i,j}(\btheta)}{1+\tilde\blambda(\btheta)^\T \bfg_{i}(\btheta)}	
		=\frac{1}{n}\sum _{i=1}^{n} \frac{g_{i,j}(\hbthetan)}{1+\hat{\blambda}(\hbthetan)^\T \bfg_{i}(\hbthetan)} + o_{\rm p}(\nu)\,,$$ 
		where the term $o_{\rm p}(\nu)$ holds uniformly over $\btheta \in \mathcal{C}_1$ and $j \in \mathcal{R}_n^\c$.  
		Together with Condition \ref{A.Pro2}(a), we have that 
		$$
		\sup_{\btheta \in \mathcal{C}_1} \max_{j \in \mathcal{R}_n^\c}\bigg|\frac{1}{n}\sum _{i=1}^{n} \frac{g_{i,j}(\btheta)}{1+\tilde\blambda(\btheta)^\T \bfg_{i}(\btheta)}\bigg| \leq \nu\rho'(0^+)$$ w.p.a.1. 
		Therefore, $\tilde \blambda(\btheta)$ and $\btheta$ satisfy the score equation $\nabla_{\blambda} f_n\{\tilde\blambda(\btheta); \btheta\}=\bzero$ for any $\btheta \in \mathcal{C}_1$ w.p.a.1. By the concavity of $f_n(\blambda;\btheta)$ w.r.t $\blambda$, it holds that $\tilde\blambda(\btheta)=\hat{\blambda}(\btheta)=\arg\max_{\blambda \in \wLambadn(\btheta)} f_n(\blambda;\btheta)$ for any $\btheta \in \mathcal{C}_1$ w.p.a.1, which implies ${\rm supp}\{\hat{\blambda}(\btheta)\} \subset {\rm supp}\{\hat{\blambda}(\hbthetan)\}$ for any $\btheta \in \mathcal{C}_1$ w.p.a.1. Select $\btheta^* \in \mathcal{C}_1$ such that $|{\rm supp}\{\hat \blambda(\btheta^*)\}| \leq |{\rm supp}\{\hat \blambda(\btheta)\}|$ for any $\btheta \in \mathcal{C}_1$, and define $\mathcal{B}_2(\btheta^*,2\alpha_n) = \{\btheta \in\bTheta:\,|\btheta-\btheta^*|_2 \leq 2\alpha_n \}$. Using the same arguments above for proving ${\rm supp}\{\hat\blambda(\btheta)\} \subset {\rm supp}\{\hat \blambda(\hat\btheta_n)\}$ for any $\btheta \in \mathcal{C}_1$ w.p.a.1, we have ${\rm supp} \{\hat\blambda(\btheta)\} \subset {\rm supp} \{\hat\blambda(\btheta^*)\}$ for any $\btheta \in \mathcal{B}_2(\btheta^*,2\alpha_n)$ w.p.a.1. Since $\mathcal{C}_1 \subset \mathcal{B}_2(\btheta^*,2\alpha_n)$, then ${\rm supp} \{\hat\blambda(\btheta)\} \subset {\rm supp}\{\hat \blambda(\btheta^*)\}$ for any $\btheta \in \mathcal{C}_1$ w.p.a.1. Due to $|{\rm supp}\{\hat \blambda(\btheta^*)\}| \leq |{\rm supp}\{\hat \blambda(\btheta)\}|$ for any $\btheta \in \mathcal{C}_1$, we have ${\rm supp}\{\hat \blambda(\btheta)\}={\rm supp}\{\hat \blambda(\btheta^*)\}$ for any $\btheta \in \mathcal{C}_1$ w.p.a.1. %Hence, ${\rm supp} \{\hat\blambda(\hat\btheta_n)\} \subset{\rm supp}\{\hat\blambda(\btheta)\}$ for any $\btheta \in \mathcal{B}_{2}(\hat{\btheta}_n,\alpha_n)$ w.p.a.1. 
		We complete the proof of Lemma \ref{l.supp}.
		\hfill $\Box$
		
		\subsection{Proof of Lemma \ref{l.fourterms}} \label{sec.pflemma11}
		
		%Recall $\mathcal{C}_1=\{\btheta\in\bTheta:|\btheta-\hat{\btheta}_n|_2\leq \alpha_n\}$. 
		By Lemma \ref{l.lam.theta}, we have $\sup_{\btheta \in \mathcal{C}_1}|\mathcal{R}(\btheta)|\leq \ell_n$ w.p.a.1 and $\sup_{\btheta \in \mathcal{C}_1}|\hat{\blambda}(\btheta)|_2=O_{\rm p}(\ell_n^{1/2} \alpha_n)$. Under Condition \ref{A.ee}(a), if $\ell_n\alpha_n=o(n^{-1/\gamma})$, then $\sup_{\btheta \in \mathcal{C}_1}\max _{ i \in [n]}|\hat{\blambda}(\btheta)^\T\bfg_{i}(\btheta)|=o_{\rm p}(1)$. Write $\hat{\blambda}(\btheta)=\{\hat\lambda_1(\btheta),\ldots,\hat\lambda_r(\btheta)\}^\T$ and $\bft=(t_1,\ldots,t_p)^\T$. For $T_{\btheta,1}$, by the Cauchy-Schwarz inequality and Condition \ref{A.ee}(c), we have
		\begin{align*}
			|\bft^\T T_{\btheta,1} \bft|
			\leq&\, \frac{1}{n} \sum_{i=1}^{n} \bigg \{\sum_{k=1}^p \sum_{j \in \Rtheta}  t_k \frac{\partial g_{i,j}(\btheta)}{\partial \theta_k} \hat{\lambda}_j(\btheta) \bigg\}^2  \cdot \{1+o_{\rm p}(1)\} \\
			\leq &\, \{1+o_{\rm p}(1)\} \cdot \ell_np\cdot |\bft|_2^2 |\hat{\blambda}(\btheta)|_2^2 \cdot \max_{j \in \Rtheta} \max_{k \in [p]}  \frac{1}{n} \sum_{i=1}^{n} \bigg|\frac{\partial g_{i,j}(\btheta)}{\partial \theta_k} \bigg|^2 =|\bft|_2^2 \cdot O_{\rm p}(\ell_n^2 \alpha_n^2)
		\end{align*}
		holds uniformly over $\btheta \in \mathcal{C}_1$ and $\bft \in \mathbb{R}^p$.
		For $T_{\btheta,3}$, by Condition \ref{A.ee}(c), we have
		\begin{align*}
			|\bft^\T T_{\btheta,3} \bft|
			&= \bigg| \sum_{k_1,k_2=1}^p  t_{k_1}t_{k_2} \frac{1}{n} \sum_{i=1}^{n} \frac{1}{1+\hat{\blambda}_\Rtheta(\btheta)^\T\bfg_{i,\Rtheta} ( \btheta)} \bigg \{\sum_{j \in \Rtheta} \frac{\partial^2 g_{i,j}(\btheta)}{\partial \theta_{k_1} \partial \theta_{k_2}}\hat{\lambda}_j(\btheta) \bigg \} \bigg| \\
			&\leq \{1+o_{\rm p}(1)\} \cdot |\bft|_1^2 |\hat{\blambda}(\btheta)|_{1} \max_{j \in \Rtheta} \max_{k_1,k_2 \in [p]}\frac{1}{n} \sum_{i=1}^{n} \Big|\frac{\partial^2 g_{i,j}(\btheta)}{\partial \theta_{k_1} \partial \theta_{k_2}} \Big|= |\bft|_2^2 \cdot O_{\rm p}(\ell_n \alpha_n) 
		\end{align*}
		holds uniformly over $\btheta \in \mathcal{C}_1$ and $\bft \in \mathbb{R}^p$. Let $\hat\blambda_{\Rtheta}(\btheta)=\{\tilde{\lambda}_1(\btheta),\ldots,\tilde{\lambda}_{|\Rtheta|}(\btheta)\}^\T$. %By Lemma \ref{l.diff.lam.hat}, 
		%\begin{align*}
		%	[\nabla_{\btheta} \hat{\blambda}(\btheta)]_{\Rtheta,[p]}
		%	=&\,\bigg(\frac{1}{n} \sum _{i=1}^{n}\frac{\bfg_{i,\Rtheta}(\btheta)^{\otimes2}}{\{1+\hat{\blambda}_\Rtheta(\btheta)^\T\bfg_{i,\Rtheta}(\btheta)\}^2} +\nu \diag[\rho''\{|\tilde\lambda_{1}(\btheta)|;\nu \},\ldots,\rho''\{|\tilde\lambda_{|\Rtheta|}(\btheta)|;\nu \}]\bigg)^{-1}\\
		%	\qquad &\times \bigg\{\frac{1}{n} \sum_{i=1}^{n} \frac{[\nabla_{\btheta} \bfg_{i}(\btheta)]_{\Rtheta,[p]}}{1+\hat{\blambda}_\Rtheta(\btheta)^\T\bfg_{i,\Rtheta}(\btheta)}-\frac{1}{n} \sum_{i=1}^{n} \frac{\bfg_{i,\Rtheta}(\btheta)\hat{\blambda}_\Rtheta(\btheta)^\T [\nabla_{\btheta} \bfg_{i}(\btheta)]_{\Rtheta,[p]}}{\{1+\hat{\blambda}_\Rtheta(\btheta)^\T\bfg_{i,\Rtheta}(\btheta)\}^2} \bigg\} \,.
		%\end{align*}
		By the Cauchy-Schwarz inequality, Conditions \ref{A.ee}(a) and \ref{A.ee}(c), if $\log r=o(n^{1/3})$, $\ell_n\alpha_n=o[\min\{\nu,n^{-1/\gamma}\}]$ and $\ell_n\nu^2=o(1)$, then
		\begin{align}
			&\bigg| \bigg (\frac{1}{n} \sum_{i=1}^{n} \frac{\bfg_{i,\Rtheta}(\btheta)\hat{\blambda}_\Rtheta(\btheta)^\T [\nabla_{\btheta} \bfg_{i} (\btheta)]_{\Rtheta,[p]}}{\{1+\hat{\blambda}_\Rtheta(\btheta)^\T\bfg_{i,\Rtheta}(\btheta)\}^2} \bigg) \bft \bigg|^2_2 \notag \\
			&~~~~~~~~\leq\{1+o_{\rm p}(1)\}\cdot\sum_{j \in \Rtheta} \bigg \{\frac{1}{n} \sum_{i=1}^{n} |g_{i,j}(\btheta)| \sum_{l \in \Rtheta} \sum_{k=1}^p |\hat{\lambda}_l(\btheta)|  \bigg|\frac{\partial g_{i,l}(\btheta)}{\partial \theta_{k}}\bigg|  |t_k| \bigg\}^2   \notag \\
			&~~~~~~~~\leq \{1+o_{\rm p}(1)\}\cdot \sum_{j \in \Rtheta} \bigg\{\frac{1}{n} \sum_{i=1}^{n} |g_{i,j}(\btheta)|^2 \bigg\}
			\bigg[\frac{1}{n} \sum_{i=1}^{n} \bigg\{ \sum_{l \in \Rtheta} \sum_{k=1}^p |\hat{\lambda}_l(\btheta)| \bigg|\frac{\partial g_{i,l}(\btheta)}{\partial \theta_{k}}\bigg| |t_k| \bigg\}^2 \bigg] \notag \\
			&~~~~~~~~\leq \{1+o_{\rm p}(1)\}\cdot |\bft|_2^2 |\hat{\blambda}(\btheta)|_2^2 \cdot\ell_np \cdot \max_{j \in \Rtheta} \max_{k \in [p]}  \frac{1}{n} \sum_{i=1}^{n} \bigg|\frac{\partial g_{i,j}(\btheta)}{\partial \theta_k} \bigg|^2  \cdot \sum_{j \in \Rtheta} \bigg\{\frac{1}{n} \sum_{i=1}^{n} |g_{i,j}(\btheta)|^2 \bigg\}
			\notag \\
			&~~~~~~~~= |\bft|^2_2 \cdot O_{\rm p}(\ell_n^3 \alpha_n^2) \label{eq:a2:1}
		\end{align}
		holds uniformly over $\btheta \in \mathcal{C}_1$ and $\bft \in \mathbb{R}^p$. Recall $\alpha_n=o(\nu)$. By Proposition  \ref{pro.cons}, Lemma \ref{l.V.hat} and Condition \ref{A.ee}(b), if $\log r =o(n^{1/3})$, $\ell_n\alpha_n=o[\min\{\nu,n^{-1/\gamma}\}]$ and $\ell_n \nu^2=o(1)$, we know that  $\inf_{\btheta \in \mathcal{C}_1}\lambda_{\min}\{\widehat{\bfV}_\Rtheta(\btheta)\}$ and $\sup_{\btheta \in \mathcal{C}_1}\lambda_{\max}\{\widehat{\bfV}_\Rtheta(\btheta)\}$ are uniformly bounded away from zero and infinity w.p.a.1. 
		Using the same arguments in the proof of Lemma \ref{l.ee}, if $\log r=o(n^{1/3})$, $\ell_n\alpha_n=o[\min\{\nu,n^{-1/\gamma}\}]$ and $\ell_n\nu^2=o(1)$, it holds that
		\begin{align}
			&\sup_{\btheta \in \mathcal{C}_1} \bigg \|\frac{1}{n} \sum _{i=1}^{n}\frac{\bfg_{i,\Rtheta}(\btheta)^{\otimes2}}{\{1+\hat{\blambda}_\Rtheta(\btheta)^\T\bfg_{i,\Rtheta}(\btheta)\}^2} - \widehat{\bfV}_\Rtheta(\btheta)\bigg \|_2 =O_{\rm p}(\ell_nn^{1/\gamma}\alpha_n) \,, \notag\\ 
			&\sup_{\btheta \in \mathcal{C}_1}\bigg|\bigg \{\frac{1}{n} \sum_{i=1}^{n} \frac{[\nabla_{\btheta} \bfg_{i} (\btheta)]_{\Rtheta,[p]}}{1+\hat{\blambda}_\Rtheta(\btheta)^\T\bfg_{i,\Rtheta}(\btheta)} - \widehat{\bGamma}_\Rtheta(\btheta)\bigg\} \bft\bigg|_2 = |\bft|_2 \cdot O_{\rm p}(\ell_n \alpha_n) \,. \label{eq:a2:2}
		\end{align}
		By Condition \ref{A.ee2} and the same arguments in the proof of Lemma \ref{l.Gamma.hat}, if $\log r=o(n^{1/3})$, $\ell_n\alpha_n=o[\min\{\nu,n^{-1/\gamma}\}]$ and $\ell_n\nu^2=o(1)$, we have $\sup_{\btheta \in \mathcal{C}_1}\|\widehat{\bGamma}_\Rtheta(\btheta)\|_2=O_{\rm p}(1)$. Notice that  $\sup_{\btheta \in \mathcal{C}_1}\|\widehat{\bfV}^{-1}_\Rtheta(\btheta)\|_2=O_{\rm p}(1)$ and 
		$
		\sup_{\btheta \in \mathcal{C}_1}\|\nu \diag[\rho''\{|\tilde\lambda_{1}(\btheta)|;\nu \},\ldots,\rho''\{|\tilde\lambda_{|\Rtheta|}(\btheta)|;\nu \}]\|_2=O_{\rm p}(\nu)
		$. Thus, 
		\begin{align}
			&\,\sup_{\btheta \in \mathcal{C}_1}\bigg\| \bigg(\frac{1}{n} \sum _{i=1}^{n}\frac{\bfg_{i,\Rtheta}(\btheta)^{\otimes2}}{\{1+\hat{\blambda}_\Rtheta(\btheta)^\T\bfg_{i,\Rtheta}(\btheta)\}^2} +\nu\diag[\rho''\{|\tilde\lambda_{1}(\btheta)|;\nu \},\ldots,\rho''\{|\tilde\lambda_{|\Rtheta|}(\btheta)|;\nu \}]\bigg)^{-1}  \notag \\ &~~~~~~~~~~~~~~~~~~~~~~~-\widehat{\bfV}^{-1}_\Rtheta(\btheta) \bigg\|_2 = O_{\rm p}(\ell_nn^{1/\gamma}\alpha_n) + O_{\rm p}(\nu) \label{eq:a2:3} \,.
		\end{align}
		Combining \eqref{eq:a2:1}, \eqref{eq:a2:2} and \eqref{eq:a2:3}, by Lemma \ref{l.diff.lam.hat}, we know $\sup_{\btheta \in \mathcal{C}_1}|[\nabla_{\btheta} \hat{\blambda}(\btheta)]_{\Rtheta,[p]} \bft|_2 = |\bft|_2 \cdot O_{\rm p}(1)$ holds uniformly over $\bft \in \mathbb{R}^p$, which implies 
		\begin{align*}
			\sup_{\btheta \in \mathcal{C}_1}|\bft^\T T_{\btheta,2} \bft | 
			\leq&\, \sup_{\btheta \in \mathcal{C}_1}\bigg|\bigg(\frac{1}{n} \sum_{i=1}^{n} \frac{ \bfg_{i,\Rtheta}(\btheta) \hat{\blambda}_\Rtheta(\btheta)^\T [\nabla_{\btheta} \bfg_{i} (\btheta)]_{\Rtheta,[p]} }{\{1+\hat{\blambda}_\Rtheta(\btheta)^\T\bfg_{i,\Rtheta}(\btheta)\}^2}\bigg) \bft \bigg|_2 |[\nabla_{\btheta} \hat{\blambda}(\btheta)]_{\Rtheta,[p]} \bft|_2 \\
			=&\,|\bft|_2^2 \cdot O_{\rm p}(\ell_n^{3/2} \alpha_n) 
		\end{align*}
		holds uniformly over $\bft \in \mathbb{R}^p$. For $T_{\btheta,4}$, by \eqref{eq:a2:1}, \eqref{eq:a2:2} and \eqref{eq:a2:3}, Lemma \ref{l.diff.lam.hat} implies 
		\begin{align*}
			\bft^\T T_{\btheta,4} \bft
			= &\,\bft^\T \bigg\{ \frac{1}{n} \sum_{i=1}^{n} \frac{[\nabla_{\btheta} \bfg_{i} (\btheta)]_{\Rtheta,[p]}^\T}{1+\hat{\blambda}_\Rtheta(\btheta)^\T\bfg_{i,\Rtheta} (\btheta)}\bigg\}  [\nabla_{\btheta} \hat{\blambda}(\btheta)]_{\Rtheta,[p]}\bft  \\
			=&\,\bft^\T \{\widehat{\bGamma}_\Rtheta(\btheta)^\T \widehat{\bfV}^{-1/2}_\Rtheta(\btheta)\}^{\otimes2} \bft + |\bft|_2^2 \cdot \{O_{\rm p}(\ell_n^{3/2} \alpha_n) + O_{\rm p}(\ell_nn^{1/\gamma}\alpha_n)  + O_{\rm p}(\nu) \}
		\end{align*}
		holds uniformly over $\btheta \in \mathcal{C}_1$ and $\bft \in \mathbb{R}^p$. We then obtain the result by \eqref{eq:fourterms}. $\hfill\Box$
		
		\subsection{Proof of Lemma \ref{l.h.cons}} \label{sec:pflemma12}
		Denote by $\P_{\mathcal{X}_n}(\cdot)$ and $\mathbb{E}_{\mathcal{X}_n}(\cdot)$, respectively, the conditional probability and conditional expectation given $\mathcal{X}_n$. For any integer $k \geq 1$, recall $\hat{\bzeta}_{k+1} = N_{k}^{-1} \sum_{i=1}^{N_{k}} \omega^{k}_i \bfh(\btheta^{k}_i)$ only depends on the $N_k$ samples $\{\btheta^{k}_1,\ldots,\btheta^{k}_{N_k}\}$ generated from the proposal distribution with density $\varphi(\btheta\,;\hat{\bzeta}_{k})$, where $\omega^{k}_i = \pi^\dag(\btheta^{k}_i\,|\,\mathcal{X}_n) / \varphi(\btheta^{k}_i\,;\hat{\bzeta}_{k})$. 
		%if $\btheta^{k}_i \in \bTheta$ and $\omega^{k}_i = 0$ if $\btheta^{k}_i \notin \bTheta$. 
		Thus, the random sequence $\{\hat{\bzeta}_{k}\}_{k\geq1}$ forms a Markov chain. Recall that $\bTheta \subset \R^p$ is a compact set with fixed $p$. Since $\sup_{\btheta \in \bTheta} |\bfh(\btheta)|_\infty \leq K_9$ for some universal constant $K_9>0$, and $\varphi(\btheta\,; \bzeta)$ is positive and continuous on $(\btheta, \bzeta) \in \bTheta \times \R^s$, there exists a positive and continuous function $\varrho(\cdot)$ such that 
		$$
		\sup_{\btheta \in \bTheta} \frac{\pi^\dag(\btheta\,|\,\mathcal{X}_n)|\bfh(\btheta)|_\infty}{\varphi(\btheta\,; \bzeta)} \leq \varrho(\bzeta)
		$$
		for any $\bzeta \in \R^s$. %Define the $\sigma$-algebra $\mathcal{G}_k = \sigma(\btheta_1^1,\ldots, \btheta_{N_1}^1,\ldots, \btheta_1^{k-1},\ldots, \btheta_{N_{k-1}}^{k-1})$.  
		Since $\pi^\dag(\btheta\,|\,\mathcal{X}_n) = 0$ for any $\btheta \notin \bTheta$, then $$
		\sup_{\btheta \in \bTheta^\c} \frac{\pi^\dag(\btheta\,|\,\mathcal{X}_n)|\bfh(\btheta)|_\infty}{\varphi(\btheta\,; \bzeta)} = 0 
		$$
		for any $\bzeta \in \R^s$. 
		Notice that $\mE(\hat{\bzeta}_{k+1} \,|\, \hat{\bzeta}_{k}) = \bzeta^* = \mE_{\btheta \sim \pi^\dag}\{\bfh(\btheta)\}$. Write $\hat{\bzeta}_{k+1} = (\hat{\zeta}_{k+1, 1}, \ldots, \hat{\zeta}_{k+1, s})^\T$ and $\bzeta^* = (\zeta^*_1,\ldots, \zeta^*_s)^\T $. For any $\varepsilon>0$, by %the Markov property of $\{\hat{\bvartheta}_{k}\}_{k\geq1}$ and 
		the Hoeffding's inequality, we have 
		\begin{align} \label{eq:Hoe_inq}
			%\P(|\hat{\bvartheta}_{k+1} - \bvartheta^*|_2 > \varepsilon \,\big|\, \mathcal{G}_k) =
			\P\big(|\hat{\bzeta}_{k+1} - \bzeta^*|_\infty > \varepsilon \,\big|\, \hat{\bzeta}_k \big) 
			\leq s \max_{j \in [s]} 
			\P\big(|\hat{\zeta}_{k+1,j} - \zeta^*_j| > \varepsilon \,\big|\, \hat{\bzeta}_k \big)
			\leq 2s\exp\bigg\{-\frac{N_{k}\varepsilon^2}{2\varrho^2(\hat{\bzeta}_k)}\bigg\} \,. 
		\end{align}
		Let $C_\varepsilon = \sup_{\bzeta\in\R^s:\, |\bzeta - \bzeta^*|_\infty \leq \varepsilon} \varrho^2(\bzeta)$. By \eqref{eq:Hoe_inq}, we have  
		\begin{align} 
			\P_{\mathcal{X}_n}\big(|\hat{\bzeta}_{k+1} - \bzeta^*|_\infty > \varepsilon, |\hat{\bzeta}_{k} - \bzeta^*|_\infty \leq \varepsilon\big) 
			&= \mE_{\mathcal{X}_n}\big[I\big(|\hat{\bzeta}_{k} - \bzeta^*|_\infty \leq \varepsilon \big) \mE\big\{I\big(|\hat{\bzeta}_{k+1} - \bzeta^*|_\infty > \varepsilon\big) \,\big|\, \hat{\bzeta}_{k}\big\}\big] \notag \\ 
			&\leq 2s\exp\bigg(-\frac{N_{k}\varepsilon^2}{2C_\varepsilon} \bigg) \P_{\mathcal{X}_n}\big(|\hat{\bzeta}_{k} - \bzeta^*|_\infty \leq \varepsilon\big) \label{eq:P_cap} \,,
		\end{align}
		which implies 
		\begin{align*}
			\P\big(|\hat{\bzeta}_{k+1} - \bzeta^*|_\infty > \varepsilon \,\big|\, |\hat{\bzeta}_{k} - \bzeta^*|_\infty \leq \varepsilon\big) \leq 2s\exp\bigg(-\frac{N_{k}\varepsilon^2}{2C_\varepsilon} \bigg) \,.
		\end{align*}
		For any integer $k'\geq k$, by the Markov property of $\{\hat{\bzeta}_{k}\}_{k\geq1}$, it then holds that
		\begin{align*}
			\P_{\mathcal{X}_n}\bigg(\bigcap_{t=k}^{k'} \{|\hat{\bzeta}_{t+1} - \bzeta^*|_\infty \leq \varepsilon\} \bigg)
			&= \P_{\mathcal{X}_n}\big(|\hat{\bzeta}_{k+1} - \bzeta^*|_\infty \leq \varepsilon\big) \prod_{t=k+1}^{k'} \P\big(|\hat{\bzeta}_{t+1} - \bzeta^*|_\infty \leq \varepsilon \,\big|\, |\hat{\bzeta}_t - \bzeta^*|_\infty \leq \varepsilon\big) \\
			&\geq \P_{\mathcal{X}_n}\big(|\hat{\bzeta}_{k+1} - \bzeta^*|_\infty \leq \varepsilon\big) \prod_{t=k+1}^{k'} \bigg\{1 - 2s\exp\bigg(-\frac{N_{t}\varepsilon^2}{2 C_\varepsilon}\bigg) \bigg\} \,.
		\end{align*}
		Letting $k' \rightarrow \infty$, then
		$$
		\P_{\mathcal{X}_n}\bigg(\bigcap^\infty_{t= k} \{|\hat{\bzeta}_{t+1} - \bzeta^*|_\infty \leq \varepsilon\} \bigg) 
		\geq \P_{\mathcal{X}_n}\big(|\hat{\bzeta}_{k+1} - \bzeta^*|_\infty \leq \varepsilon\big) \prod^\infty_{t= k+1} \bigg\{1 - 2s\exp\bigg(-\frac{N_{t}\varepsilon^2}{2 C_\varepsilon}\bigg) \bigg\} \,.
		$$
		Since $s$ is fixed and  $\sum_{k=1}^{\infty}\exp(-CN_k) < \infty$ for any $C>0$, we have 
		\begin{align} \label{eq:P_cap_2}
			\liminf_{k \rightarrow \infty} \P_{\mathcal{X}_n}\bigg(\bigcap^\infty_{t= k} \{|\hat{\bzeta}_{t+1} - \bzeta^*|_\infty \leq \varepsilon\} \bigg) \geq \liminf_{k \rightarrow \infty} \P_{\mathcal{X}_n}\big(|\hat{\bzeta}_{k+1} - \bzeta^*|_\infty \leq \varepsilon\big) \,.
		\end{align} 
		For any $z>0$, let $\bar{C}_z = \sup_{\bzeta\in \R^s:\,|\bzeta|_\infty \leq z} \varrho^2(\bzeta)$. Using the same arguments for \eqref{eq:P_cap}, it holds that 
		$$
		\P_{\mathcal{X}_n}\big(|\hat{\bzeta}_{k+1} - \bzeta^*|_\infty > \varepsilon, |\hat{\bzeta}_{k}|_\infty \leq z\big) \leq 2s\exp\bigg(-\frac{N_{k}\varepsilon^2}{2C_z}\bigg) \P_{\mathcal{X}_n}\big(|\hat{\bzeta}_{k}|_\infty \leq z\big) \leq 2s\exp\bigg(-\frac{N_{k}\varepsilon^2}{2C_z}\bigg) \,.
		$$
		By the Markov's inequality and triangle inequality,   
		\begin{align}
			\P_{\mathcal{X}_n}\big(|\hat{\bzeta}_{k}|_\infty > z\big) &\leq  z^{-1}\mE_{\mathcal{X}_n}\big(|\hat{\bzeta}_{k}|_\infty\big)\leq  z^{-1}\mE_{\mathcal{X}_n}\bigg\{\frac{1}{N_{k-1}} \sum_{i=1}^{N_{k-1}}\omega^{k-1}_i |\bfh(\btheta^{k-1}_i)|_\infty \bigg\} \notag\\
			& = z^{-1} \int_{\R^p}  \frac{\pi^\dag(\btheta\,|\,\mathcal{X}_n)}{\varphi(\btheta\,;\hat{\bzeta}_{k-1})}|\bfh(\btheta)|_\infty \varphi(\btheta\,;\hat{\bzeta}_{k-1}) \,{\rm d}\btheta  = z^{-1} \mE_{\btheta \sim \pi^\dag }\{|\bfh(\btheta)|_\infty\} \label{eq:P_tail} \,, 
		\end{align} 
		%where $\mE_{\pi^\dag}(|\bfh(\btheta)|_\infty)$ is the expectation of $|\bfh(\cdot)|_\infty$ w.r.t $\pi^\dag(\cdot\,|\,\mathcal{X}_n)$. 
		which implies 
		$\P_{\mathcal{X}_n}(|\hat{\bzeta}_{k+1} - \bzeta^*|_\infty > \varepsilon) \leq 2s\exp\{-(2C_z)^{-1}N_{k}\varepsilon^2\} +  z^{-1}\mE_{\btheta \sim \pi^\dag}\{|\bfh(\btheta)|_\infty\}$. Due to $N_k \rightarrow \infty$ as $k \rightarrow \infty$, we know $\limsup_{k \rightarrow \infty} \P_{\mathcal{X}_n}(|\hat{\bzeta}_{k+1} - \bzeta^*|_\infty > \varepsilon) \leq z^{-1}\mE_{\btheta \sim \pi^\dag}\{|\bfh(\btheta)|_\infty\}$. Notice that $\sup_{\btheta \in \bTheta} |\bfh(\btheta)|_\infty \leq K_9$ for some universal constant $K_9>0$. Letting $z \rightarrow \infty$, it holds that $\limsup_{k \rightarrow \infty} \P_{\mathcal{X}_n}(|\hat{\bzeta}_{k+1} - \bzeta^*|_\infty > \varepsilon) = 0$, which implies $\liminf_{k \rightarrow \infty} \P_{\mathcal{X}_n}(|\hat{\bzeta}_{k+1} - \bzeta^*|_\infty \leq \varepsilon) = 1$. Together with \eqref{eq:P_cap_2}, we have 
		$$
		\liminf_{k \rightarrow \infty} \P_{\mathcal{X}_n}\bigg(\bigcap^\infty_{t = k} \{|\hat{\bzeta}_{t+1} - \bzeta^*|_\infty \leq \varepsilon\}\bigg) = 1 \,.
		$$
		Since $\varepsilon>0$ is arbitrary, we then obtain that, conditional on $\mathcal{X}_n$, $|\hat{\bzeta}_{k+1} - \bzeta^*|_\infty \rightarrow 0$ almost surely as $k \rightarrow \infty$. We complete the proof of Lemma \ref{l.h.cons}. 
		\hfill $\Box$

		\subsection{Proof of Lemma \ref{l.three_res}} \label{sec:pflemma14}
		Denote by $\P_{\mathcal{X}_n}(\cdot)$ and $\mathbb{E}_{\mathcal{X}_n}(\cdot)$, respectively, the conditional probability and conditional expectation given $\mathcal{X}_n$. Recall $\bzeta^* = \mE_{\btheta \sim \pi^\dag}\{\bfh(\btheta)\}$. Let $
		\widehat{\bfZ}_{k+1}= N_k^{-1} \sum_{i=1}^{N_k} \btheta^k_i \pi^\dag(\btheta^k_i\,|\,\mathcal{X}_n) / \varphi(\btheta^k_i\,;\bzeta^*)$ for any integer $k \geq 2$ and 
		$$
		\bfZ(\bzeta)=\mE_{\btheta \sim\varphi(\cdot\,;\,\bzeta)} \bigg\{\frac{\btheta \pi^\dag(\btheta\,|\,\mathcal{X}_n)}{\varphi(\btheta\,; \bzeta^*)} \bigg\} = \int_{\R^p} \frac{\btheta \pi^\dag(\btheta\,|\,\mathcal{X}_n)}{\varphi(\btheta\,; \bzeta^*)} \varphi(\btheta\,;\bzeta) \,{\rm d} \btheta
		$$ for any $\bzeta \in \R^s$, where $\{\btheta^1_1, \ldots,  \btheta^1_{N_1}, \ldots,\btheta^k_1,\ldots, \btheta^k_{N_k}\}$ are generated via Algorithm {\rm \ref{alg2}}. 
		
		Our first step is to show that conditional on $\mathcal{X}_n$, we have $|\widehat{\bfZ}_{k+1} - \bfZ(\hat{\bzeta}_{k})|_\infty \rightarrow 0$ almost surely as $k \rightarrow \infty$. Notice that $\bTheta \subset \R^p$ is a compact set with fixed $p$ and  $\pi^\dag(\btheta\,|\,\mathcal{X}_n) =0$ for any $\btheta \notin \bTheta$. Since $\varphi(\btheta\,; \bzeta)$ is positive and continuous on $(\btheta, \bzeta) \in \bTheta \times \R^s$, we know
		\begin{align} \label{eq:rio_bound}
			\sup_{\btheta \in \bTheta}  \frac{\pi^\dag(\btheta\,|\,\mathcal{X}_n)|\btheta|_\infty}{\varphi(\btheta\,; \bzeta^*)} \leq \tilde{C} ~~\textrm{and}~~ \sup_{\btheta \in \bTheta^\c}  \frac{\pi^\dag(\btheta\,|\,\mathcal{X}_n)|\btheta|_\infty}{\varphi(\btheta\,; \bzeta^*)} = 0 
		\end{align}
		for some universal constant $\tilde{C}>0$.  Recall that $\widehat{\bfZ}_{k+1}$ depends on the $N_k$ samples $\{\btheta^k_1,\ldots,\btheta^k_{N_k}\}$ generated from the proposal distribution with density $\varphi(\btheta\,;\hat{\bzeta}_{k})$. Then $\mE(\widehat{\bfZ}_{k+1} \,|\, \hat{\bzeta}_{k}) = \bfZ(\hat{\bzeta}_{k})$. For any $\varepsilon>0$, using the same arguments for \eqref{eq:Hoe_inq}, we have 
		\begin{align} \label{eq:P_cond_Z}
			\P\big\{|\widehat{\bfZ}_{k+1} - \bfZ(\hat{\bzeta}_{k})|_\infty > \varepsilon \,\big|\, \hat{\bzeta}_{k}\big\}
			%		\leq \P(|\widehat{\bfZ}_{k+1} - \bfZ(\hat{\bvartheta}_{k})|_2 > \varepsilon \,\big|\, \hat{\bvartheta}_{k} = \bvartheta) 
			\leq 2p\exp\bigg(-\frac{N_{k}\varepsilon^2}{2 \tilde{C}^2}\bigg) \,.
		\end{align}
		%It then holds that
		%\begin{align}
		%	\P\{|\widehat{\bfZ}_{k+1} - \bfZ(\hat{\bvartheta}_{k})|_\infty > \varepsilon, |\hat{\bvartheta}_{k} - \bvartheta^*|_\infty \leq \varepsilon\}
		%	&= \mE\big[I\big(|\hat{\bvartheta}_{k} - \bvartheta^*|_\infty \leq \varepsilon \big) \mE\big\{I\big(|\widehat{\bfZ}_{k+1} - \bfZ(\hat{\bvartheta}_{k})|_\infty > \varepsilon\big) \,\big|\, \hat{\bvartheta}_{k}\big\}\big] \notag \\ 
		%	&\leq 2p\exp\bigg(-\frac{N_{k}\varepsilon^2}{2\tilde{C}^2} \bigg) \P(|\hat{\bvartheta}_{k} - \bvartheta^*|_\infty \leq \varepsilon) \label{eq:P_cap_Z} \,,	
		%\end{align}
		%which implies 
		%\begin{align*}
		%\P\big\{|\widehat{\bfZ}_{k+1} - \bfZ(\hat{\bvartheta}_{k})|_\infty > \varepsilon \,\big|\, |\hat{\bvartheta}_k - \bvartheta^*|_\infty \leq \varepsilon \big\} \leq 2p\exp\bigg(-\frac{N_{k}\varepsilon^2}{2\tilde{C}^2} \bigg) \,.
		%\end{align*}
		Define the event $A_{k+1} = \{|\widehat{\bfZ}_{k+1} - \bfZ(\hat{\bzeta}_{k})|_\infty \leq \varepsilon, |\hat{\bzeta}_{k+1} - \bzeta^*|_\infty \leq \varepsilon\}$. Note that $A_{k} \in \sigma(\btheta_1^{k-1},\ldots, \btheta_{N_{k-1}}^{k-1}, \hat{\bzeta}_{k-1})$ and the conditional joint distribution of $(\widehat{\bfZ}_{k+1},\hat{\bzeta}_{k+1})$ given $\mathcal{X}_n$ is fully determined by $\hat{\bzeta}_{k}$. By \eqref{eq:Hoe_inq} and  \eqref{eq:P_cond_Z}, it holds that 
		\begin{align*}
			\P_{\mathcal{X}_n}\big(A^\c_{k+1} \cap A_{k}\big) 
			&= \mE_{\mathcal{X}_n}\big[\mE\big\{I(A^\c_{k+1}) I(A_{k}) \,\big|\, \btheta_1^{k-1},\ldots, \btheta_{N_{k-1}}^{k-1}, \hat{\bzeta}_{k-1},  \hat{\bzeta}_{k}\big\}\big] 
			= \mE_{\mathcal{X}_n}\big[I(A_{k}) \mE\big\{I(A^\c_{k+1}) \,\big|\, \hat{\bzeta}_{k}\big\}\big] \\
			&\leq \mE_{\mathcal{X}_n}\big\{I(A_{k}) \mE\big[I\big\{|\widehat{\bfZ}_{k+1} - \bfZ(\hat{\bzeta}_{k})|_\infty > \varepsilon\big\} \,\big|\, \hat{\bzeta}_{k} \big] \big\} + \mE_{\mathcal{X}_n}\big[I(A_{k}) \mE\big\{I(|\hat{\bzeta}_{k+1} - \bzeta^*|_\infty > \varepsilon) \,\big|\, \hat{\bzeta}_{k} \big\} \big]\\
			& = \mE_{\mathcal{X}_n}\big[I(A_{k}) \P\big\{|\widehat{\bfZ}_{k+1} - \bfZ(\hat{\bzeta}_{k})|_\infty > \varepsilon  \,\big|\, \hat{\bzeta}_{k} \big\} \big] + \mE_{\mathcal{X}_n}\big\{I(A_{k}) \P\big(|\hat{\bzeta}_{k+1} - \bzeta^*|_\infty > \varepsilon \,\big|\, \hat{\bzeta}_{k} \big) \big\} \\
			&\leq \bigg\{ 2p\exp\bigg(-\frac{N_{k}\varepsilon^2}{2\tilde{C}^2} \bigg) + 2s\exp\bigg(-\frac{N_{k}\varepsilon^2}{2C_\varepsilon} \bigg) \bigg\} \P_{\mathcal{X}_n}(A_{k}) \,,
		\end{align*}
		where $C_\varepsilon = \sup_{\bzeta\in\R^s:\, |\bzeta - \bzeta^*|_\infty \leq \varepsilon} \varrho^2(\bzeta)$ with the function $\varrho(\cdot)$ specified in the proof of Lemma \ref{l.h.cons}. Then
		$$
		\P_{\mathcal{X}_n}\big(A^\c_{k+1} \,\big|\, A_{k} \big) 
		\leq 2p\exp\bigg(-\frac{N_{k}\varepsilon^2}{2\tilde{C}^2} \bigg) + 2s\exp\bigg(-\frac{N_{k}\varepsilon^2}{2C_\varepsilon} \bigg) 
		\leq 2(p+s)\exp\bigg(-\frac{N_{k}\varepsilon^2}{\check{C}_\varepsilon} \bigg)
		$$ 
		for some $\check{C}_\varepsilon >0$ depending on $\varepsilon$. For any integer $k'\geq k$, by the Markov property of $\{(\widehat{\bfZ}_{k}, \hat{\bzeta}_{k})\}_{k\geq 2}$, it then holds that
		\begin{align*}
			\P_{\mathcal{X}_n}\bigg(\bigcap_{t=k}^{k'} A_{t+1} \bigg)
			= \P_{\mathcal{X}_n}(A_{k+1}) \prod_{t=k+1}^{k'} \P\big(A_{t+1} \,\big|\, A_{t}\big)
			\geq \P_{\mathcal{X}_n}(A_{k+1}) \prod_{t=k+1}^{k'} \bigg\{1 - 2(p+s)\exp \bigg(-\frac{N_{t}\varepsilon^2}{\check{C}_\varepsilon}\bigg) \bigg\} \,.
		\end{align*}
		Letting $k' \rightarrow \infty$, then
		$$
		\P_{\mathcal{X}_n}\bigg(\bigcap^\infty_{t=k} A_{t+1} \bigg) 
		\geq \P_{\mathcal{X}_n}(A_{k+1}) \prod^\infty_{t = k+1} \bigg\{1 - 2(p+s)\exp \bigg(-\frac{N_{t}\varepsilon^2}{\check{C}_\varepsilon}\bigg) \bigg\} \,.
		$$
		Since $p$ and $s$ are fixed constants and $\sum_{k=1}^{\infty}\exp(-CN_k) < \infty$ for any $C>0$, we have 
		\begin{align} \label{eq:P_cap_Z_2}
			\liminf_{k \rightarrow \infty} \P_{\mathcal{X}_n}\bigg(\bigcap^\infty_{t=k} A_{t+1} \bigg) &\geq \liminf_{k \rightarrow \infty} \P_{\mathcal{X}_n}(A_{k+1}) \notag \\
			&\geq 1 - \limsup_{k \rightarrow \infty} \P_{\mathcal{X}_n}\big\{|\widehat{\bfZ}_{k+1}-\bfZ(\hat{\bzeta}_{k})|_\infty > \varepsilon\big\} - \limsup_{k \rightarrow \infty}\P_{\mathcal{X}_n}\big(|\hat{\bzeta}_{k+1} - \bzeta^*|_\infty > \varepsilon\big) \notag \\
			&= 1 - \limsup_{k \rightarrow \infty} \P_{\mathcal{X}_n}\big\{|\widehat{\bfZ}_{k+1}-\bfZ(\hat{\bzeta}_{k})|_\infty > \varepsilon\big\} \,,
		\end{align}
		where the last step is due to the fact $\limsup_{k \rightarrow \infty} \P_{\mathcal{X}_n}(|\hat{\bzeta}_{k+1} - \bzeta^*|_\infty > \varepsilon) = 0$ as shown in the proof of Lemma \ref{l.h.cons}. For any $z>0$, by \eqref{eq:P_cond_Z}, it holds that 
		\begin{align*}
			\P_{\mathcal{X}_n}\big\{|\widehat{\bfZ}_{k+1} - \bfZ(\hat{\bzeta}_{k})|_\infty > \varepsilon, |\hat{\bzeta}_{k}|_\infty \leq z\big\}
			&= \mE_{\mathcal{X}_n}\big\{I\big(|\hat{\bzeta}_{k}|_\infty \leq z \big) \mE\big[I\big\{|\widehat{\bfZ}_{k+1} - \bfZ(\hat{\bzeta}_{k})|_\infty > \varepsilon\big\} \,\big|\, \hat{\bzeta}_{k}\big]\big\} \\
			&\leq 2p\exp\bigg(-\frac{N_{k}\varepsilon^2}{2\tilde{C}^2} \bigg) \P_{\mathcal{X}_n}\big(|\hat{\bzeta}_{k}|_\infty \leq z\big) 
			\leq 2p\exp\bigg(-\frac{N_{k}\varepsilon^2}{2\tilde{C}^2} \bigg) \,.
		\end{align*}
		Together with \eqref{eq:P_tail}, we have
		$$
		\P_{\mathcal{X}_n}\{|\widehat{\bfZ}_{k+1} - \bfZ(\hat{\bzeta}_{k})|_\infty > \varepsilon\} \leq 2p\exp\bigg(-\frac{N_{k}\varepsilon^2}{2\tilde{C}^2} \bigg) +  z^{-1}\mE_{\btheta \sim\pi^\dag}\{|\bfh(\btheta)|_\infty\} \,.
		$$
		Due to $N_k \rightarrow \infty$ as $k \rightarrow \infty$, we know $\limsup_{k \rightarrow \infty} \P_{\mathcal{X}_n}\{|\widehat{\bfZ}_{k+1} - \bfZ(\hat{\bzeta}_{k})|_\infty > \varepsilon\} \leq z^{-1}\mE_{\btheta \sim\pi^\dag}\{|\bfh(\btheta)|_\infty\}$. Notice that $\sup_{\btheta \in \bTheta} |\bfh(\btheta)|_\infty \leq K_9$ for some universal constant $K_9>0$. Letting $z \rightarrow \infty$, it holds that $\limsup_{k \rightarrow \infty} \P_{\mathcal{X}_n}\{|\widehat{\bfZ}_{k+1} - \bfZ(\hat{\bzeta}_{k})|_\infty > \varepsilon\} = 0$. Together with \eqref{eq:P_cap_Z_2}, we have 
		$$
		\liminf_{k \rightarrow \infty} \P_{\mathcal{X}_n}\bigg[\bigcap^\infty_{t=k} \big\{ |\widehat{\bfZ}_{t+1} - \bfZ(\hat{\bzeta}_{t})|_\infty \leq \varepsilon \big\}\bigg]
		\geq \liminf_{k \rightarrow \infty} \P_{\mathcal{X}_n}\bigg(\bigcap^\infty_{t=k} A_{t+1} \bigg) =1 \,.
		$$
		Since $\varepsilon>0$ is arbitrary, conditional on $\mathcal{X}_n$, we have $|\widehat{\bfZ}_{k+1} - \bfZ(\hat{\bzeta}_{k})|_\infty \rightarrow 0$ almost surely as $k \rightarrow \infty$.  
		
		Our second step is to show that conditional on $\mathcal{X}_n$, we have $|\bfZ(\hat{\bzeta}_{k}) - \bfZ(\bzeta^*)|_\infty \rightarrow 0$ almost surely as $k \rightarrow \infty$. By \eqref{eq:rio_bound}, we have $|\bfZ(\bzeta)|_\infty < \infty$ for any $\bzeta \in \R^s$, and
		$$
		|\bfZ(\hat{\bzeta}_{k}) - \bfZ(\bzeta^*)|_\infty 
		\leq \int_{\R^p} \frac{\pi^\dag(\btheta\,|\,\mathcal{X}_n)|\btheta|_\infty}{\varphi(\btheta\,; \bzeta^*)} |\varphi(\btheta\,; \hat{\bzeta}_{k}) - \varphi(\btheta\,; \bzeta^*)| \,{\rm d}\btheta 
		\leq \tilde{C}\int_{\bTheta}|\varphi(\btheta\,; \hat{\bzeta}_{k}) - \varphi(\btheta\,; \bzeta^*)| \,{\rm d}\btheta \,.
		$$
		For some sufficiently large $M>0$, since conditional on $\mathcal{X}_n$, we have $|\hat{\bzeta}_{k} - \bzeta^*|_\infty \rightarrow 0$ almost surely as $k \rightarrow \infty$, then for any $\epsilon >0$ there exists a sufficiently large integer $k_\epsilon$ such that 
		$
		\P_{\mathcal{X}_n}(\mathcal{A}) \leq \epsilon
		$ with $\mathcal{A} = \bigcup_{t=k_\epsilon}^\infty \{|\hat{\bzeta}_{t} - \bzeta^*|_\infty > M\}$. 
		Define a compact set $\mathcal{B}=\{\bzeta \in \R^s: |\bzeta - \bzeta^*|_\infty \leq M \} $. Recall $\bTheta \subset \R^p$ is a compact set with fixed $p$. Due to the continuity of $\varphi(\btheta\,; \bzeta)$, we know $\varphi(\btheta\,; \bzeta)$ is uniformly continuous on $(\btheta\,; \bzeta) \in \bTheta \times \mathcal{B}$. For any $\varepsilon>0$, there exists $\delta(\varepsilon)>0$ such that $|\varphi(\btheta_1\,; \bzeta_1) - \varphi(\btheta_2\,; \bzeta_2)| < \tilde{C}^{-1} \varepsilon / \mathbb{L}(\bTheta)$ for any $(\btheta_1, \bzeta_1), (\btheta_2, \bzeta_2)  \in \bTheta \times \mathcal{B}$ satisfying $|\btheta_1 - \btheta_2|_\infty \leq \delta(\varepsilon)$ and $|\bzeta_1 - \bzeta_2|_\infty \leq \delta(\varepsilon)$, where $\mathbb{L}(\cdot)$ is the Lebesgue measure on $\R^p$. Since 
		\begin{align*}
			\big\{|\bfZ(\hat{\bzeta}_{t}) - \bfZ(\bzeta^*)|_\infty > \varepsilon , \mathcal{A}^\c\big\} 
			&\subset \bigg\{\int_{\bTheta}|\varphi(\btheta\,; \hat{\bzeta}_{t}) - \varphi(\btheta\,; \bzeta^*)| \,{\rm d}\btheta > \frac{\varepsilon }{\tilde{C}}, \mathcal{A}^\c \bigg\}  \\
			&\subset \big\{ |\hat{\bzeta}_{t} - \bzeta^*|_\infty >  \delta(\varepsilon), \mathcal{A}^\c\big\} \subset \big\{ |\hat{\bzeta}_{t} - \bzeta^*|_\infty >  \delta(\varepsilon)\big\} \,,
		\end{align*}
		we then have 
		\begin{align*}
			\limsup_{k \to \infty}\P_{\mathcal{X}_n}\bigg[\bigcup_{t=k}^\infty \big\{|\bfZ(\hat{\bzeta}_{t}) - \bfZ(\bzeta^*)|_\infty > \varepsilon\big\}  \bigg] 
			\leq&~ \P_{\mathcal{X}_n}(\mathcal{A}) + \limsup_{k \to \infty}\ \P_{\mathcal{X}_n}\bigg[\bigcup_{t=k}^\infty \big\{|\hat{\bzeta}_{t} - \bzeta^*|_\infty >  \delta(\varepsilon)  \big\} \bigg] \\
			=&~ \mathbb{P}_{\mathcal{X}_n}(\mathcal{A}) \leq \epsilon \,,
		\end{align*}
		where the second step is due to the fact that conditional on $\mathcal{X}_n$ we have $|\hat{\bzeta}_{k} - \bzeta^*|_\infty \rightarrow 0$ almost surely as $k \rightarrow \infty$. Letting $\epsilon \to 0$, we know that conditional on $\mathcal{X}_n$, we have $|\bfZ(\hat{\bzeta}_{k}) - \bfZ(\bzeta^*)|_\infty \rightarrow 0$ almost surely as $k \rightarrow \infty$.
		
		Our third step is to show that conditional on $\mathcal{X}_n$, we have $|\widehat{\mE}^*_{\pi^\dag,K}(\btheta) - \mE_{\btheta \sim \pi^\dag}(\btheta)|_\infty \rightarrow 0$ almost surely as $K \rightarrow \infty$. By the triangle inequality,  $|\widehat{\bfZ}_{k+1} - \bfZ(\bzeta^*)|_\infty \leq |\widehat{\bfZ}_{k+1} - \bfZ(\hat{\bzeta}_{k})|_\infty + |\bfZ(\hat{\bzeta}_{k}) - \bfZ(\bzeta^*)|_\infty $. Based on the results shown in Steps 1 and 2 above, it holds that conditional on $\mathcal{X}_n$, we have $|\widehat{\bfZ}_{k+1} - \bfZ(\bzeta^*)|_\infty \rightarrow 0$ almost surely as $k \rightarrow \infty$. Notice that $\bfZ(\bzeta^*)=\mE_{\btheta \sim\pi^\dag}(\btheta)$ and
		$$
		\widehat{\mE}^*_{\pi^\dag,K}(\btheta) 
		= \frac{1}{S_K} \sum_{k=1}^{K} \sum_{i=1}^{N_k}  \frac{\pi^\dag(\btheta^k_i\,|\,\mathcal{X}_n)}{\varphi(\btheta^k_i\,;\bzeta^*) } \btheta^k_i 
		= \frac{1}{S_K} \sum_{k=1}^{K} N_k \widehat{\bfZ}_{k+1}
		$$ 
		with $S_K=N_1+\cdots+N_K$. Notice that conditional on $\mathcal{X}_n$, $|\widehat{\bfZ}_{k+1} - \bfZ(\bzeta^*)|_\infty \rightarrow 0$ almost surely as $k \rightarrow \infty$. Given a constant $\varepsilon>0$,  for any $\epsilon>0$ there exists a sufficiently large integer $\tilde{k}_\epsilon$ such that $\P_{\mathcal{X}_n}(\mathcal{C}) \leq \epsilon$ with $\mathcal{C} = \bigcup_{k=\tilde{k}_\epsilon}^\infty\{|\widehat{\bfZ}_{k+1} - \bfZ(\bzeta^*)|_\infty > \varepsilon/2 \}$. Due to $S_K=N_1+\cdots+N_K$ with $N_K \rightarrow \infty$ as $K \rightarrow \infty$ and
		\begin{align*}
			&\big\{|\widehat{\mE}^*_{\pi^\dag,t}(\btheta) - \mE_{\btheta \sim \pi^\dag}(\btheta)|_\infty > \varepsilon, \mathcal{C}^\c\big\} \\
			&~~~~~~~\subset \bigg\{\frac{1}{S_t}\sum_{k=1}^{\tilde{k}_\epsilon}N_k |\widehat{\bfZ}_{k+1} - \bfZ(\bzeta^*)|_\infty >  \frac{\varepsilon}{2}, \mathcal{C}^\c \bigg\}  
			\bigcup \bigg\{\frac{1}{S_t}\sum_{k=\tilde{k}_\epsilon+1}^{t}N_k |\widehat{\bfZ}_{k+1} - \bfZ(\bzeta^*)|_\infty >  \frac{\varepsilon}{2}, \mathcal{C}^\c \bigg\} \\
			&~~~~~~~\subset\bigg\{\frac{1}{S_t}\sum_{k=1}^{\tilde{k}_\epsilon}N_k |\widehat{\bfZ}_{k+1} - \bfZ(\bzeta^*)|_\infty >  \frac{\varepsilon}{2} \bigg\}
		\end{align*}
		for any integer $t > \tilde{k}_\epsilon$, we then have
		\begin{align*}
			&\limsup_{K \to \infty}\P_{\mathcal{X}_n} \bigg[\bigcup_{t=K}^\infty \big\{|\widehat{\mE}^*_{\pi^\dag,t}(\btheta) - \mE_{\btheta \sim \pi^\dag}(\btheta)|_\infty > \varepsilon\big\} \bigg] \\
			&~~~~~\leq \P_{\mathcal{X}_n}(\mathcal{C}) + \limsup_{K \to \infty}\P_{\mathcal{X}_n} \bigg[\bigcup_{t=K}^\infty \bigg\{\frac{1}{S_t}\sum_{k=1}^{\tilde{k}_\epsilon}N_k |\widehat{\bfZ}_{k+1} - \bfZ(\bzeta^*)|_\infty >  \frac{\varepsilon}{2} \bigg\}\bigg] \,.
		\end{align*}
		Notice that 
		\[
		\bigcup_{t=K}^\infty \bigg\{\frac{1}{S_t}\sum_{k=1}^{\tilde{k}_\epsilon}N_k |\widehat{\bfZ}_{k+1} - \bfZ(\bzeta^*)|_\infty >  \frac{\varepsilon}{2} \bigg\}=\bigg\{\frac{1}{S_K}\sum_{k=1}^{\tilde{k}_\epsilon}N_k |\widehat{\bfZ}_{k+1} - \bfZ(\bzeta^*)|_\infty >  \frac{\varepsilon}{2} \bigg\}\]
		and  $S_K^{-1} \sum_{k=1}^{\tilde{k}_\epsilon} N_k |\widehat{\bfZ}_{k+1} - \bfZ(\bzeta^*)|_\infty = o_{\rm p}(1)$ as $K \rightarrow \infty$ for given $(\epsilon,\varepsilon)$. Then  
		\[
		\limsup_{K \to \infty}\P_{\mathcal{X}_n} \bigg[\bigcup_{t=K}^\infty \bigg\{\frac{1}{S_t}\sum_{k=1}^{\tilde{k}_\epsilon}N_k |\widehat{\bfZ}_{k+1} - \bfZ(\bzeta^*)|_\infty >  \frac{\varepsilon}{2} \bigg\}\bigg]=0\,,
		\]
		which implies 
		\begin{align*}
			\limsup_{K \to \infty}\P_{\mathcal{X}_n} \bigg[\bigcup_{t=K}^\infty \big\{|\widehat{\mE}^*_{\pi^\dag,t}(\btheta) - \mE_{\btheta \sim \pi^\dag}(\btheta)|_\infty > \varepsilon\big\} \bigg] \leq \epsilon \,.
		\end{align*}
		Letting $\epsilon \to 0$, we know that conditional on $\mathcal{X}_n$, $|\widehat{\mE}^*_{\pi^\dag,K}(\btheta) - \mE_{\btheta \sim \pi^\dag}(\btheta)|_\infty \rightarrow 0$ almost surely as $K \rightarrow \infty$.
		We complete the proof of Lemma \ref{l.three_res}. 
		\hfill $\Box$

	\end{document}